%% file: CERN-2017-001-M_web.tex
\documentclass[twoside]{cernyrep}
\usepackage{texnames}
\usepackage[T1]{fontenc}
\usepackage{color}
\usepackage{placeins}
\usepackage{tikz}		
\usepackage{tgheros}
\usepackage{setspace}				
\usepackage{enumerate}
\usepackage[labelfont={bf},size={small}]{caption}
\usepackage{subcaption}		
\usepackage{siunitx}
\usepackage{wasysym}
\usepackage[bookmarks, colorlinks=true, pdftex, linkcolor=blue, citecolor=blue, urlcolor=blue]{hyperref}
\usepackage{rotating}
\usepackage{tikz}
\usepackage{adjustbox}
\usepackage{blindtext}

\usepackage[version=4]{mhchem}
\usepackage{xcolor}
\usepackage{colortbl}
\usepackage{booktabs}
\usepackage{tabu}
\usepackage{lscape}
\usepackage{siunitx}
\usepackage{url}
\usepackage{lineno}

\usepackage{subcaption}
\usepackage{graphicx}

\usepackage{float}
\usepackage{subfloat}
\usepackage[toc,page]{appendix}
\usepackage{chapterbib}
\usepackage{filecontents}

\usepackage{color}

\usetikzlibrary{decorations}
\usetikzlibrary{decorations.pathmorphing}
\usetikzlibrary{decorations.pathreplacing}
\usetikzlibrary{arrows,backgrounds,automata}
\usetikzlibrary{positioning}
\usepackage{rotating}
\usetikzlibrary{shapes.arrows}
\usepgflibrary{plotmarks}
\usetikzlibrary{trees}
\usetikzlibrary{patterns}
\usetikzlibrary{snakes}
\everymath{\displaystyle}
\usepackage{varwidth}

\usepackage{hyperref}
\hypersetup{colorlinks = false}

\begin{document}
\newcommand{\hlr}[1]{\textcolor{red}{#1}}
\newcommand{\BLKP}{
  \ifthenelse{\isodd{\value{page}}}{\relax}{\mbox{}\thispagestyle{empty}\newpage}}
\newcommand*{\UMVperm} {\ensuremath{\USP\text{MV/m}}}
\newcommand*{\UKilp} {\ensuremath{\USP\text{Kilpatrick}}}
\newcommand*{\Udeg} {\ensuremath{{}^{\degree}}}
\newcommand{\D}{\displaystyle}
\newcommand{\meqn}[3]{\noindent\parbox{9cm}{\begin{align*} #1 \end{align*}}\hspace{1cm}\parbox{5cm}{\raggedright\bf #2}\hfill\parbox[b]{8mm}{\begin{equation} \label{#3} \end{equation}}}

\definecolor{vcol}{rgb}{0.9,0.3,0.3}

\include{title}
\pagestyle{plain}
\pagenumbering{roman}
\setcounter{page}{1}

\clubpenalty = 10000
\widowpenalty = 10000
\displaywidowpenalty = 10000

\include{FrontBackMaterial/00_title}



\include{FrontBackMaterial/00_frontmatter}

\BLKP
\cleardoublepage

\pagenumbering{roman}
\setcounter{page}{3}
\cleardoublepage
\pagestyle{fancy}
\pagenumbering{arabic}
\setcounter{page}{1}
\renewcommand{\floatpagefraction}{0.9}
\renewcommand*\thesection{\thechapter.\arabic{section}}

\include{Chapters/ExecutiveSummary}
\include{Chapters/Introduction}

\include{Chapters/BeamRequirements}

\include{Chapters/IonSource}

\include{Chapters/FrontendAndLinac}

\include{Chapters/InjectionTransferLine}

\include{Chapters/BeamDynamics}

\include{Chapters/Extraction}

\include{Chapters/Beamlines}

\include{Chapters/EA}

\include{Chapters/Vacuum}

\include{Chapters/Infrastructure}

\include{Chapters/MO}

\include{Chapters/Controls}
\include{Chapters/RP}

\include{Chapters/Safety}
\include{Chapters/Planning}

\include{Chapters/Cost}
\include{Chapters/Risk}

\include{Chapters/PointsFlagged}

\begin{appendices}
\renewcommand{\appendixname}{Appendix}
\fancyhead[RE]{\textsc{\appendixname\ \thechapter}}
\include{Appendices/appendix_Introduction}

\include{Appendices/appendix_AlternativeUseCases}

\include{Appendices/appendix_EA}

\end{appendices}



\include{FrontBackMaterial/10_backmatter}
\end{document}

%% file: FrontBackMaterial/00_title.tex
\thispagestyle{empty}
\setlength{\unitlength}{1mm}
\begin{picture}(0.001,0.001)
\put(-16,8){CERN Yellow Reports: Monographs}
\put(-16,2){Volume 1/2017}
\put(110,8){CERN-2017-001-M}
\put(30,-80){\LARGE\bfseries
              Feasibility Study for BioLEIR}
\put(48,-95){\Large 
}

\put(-5,-140){\Large Editors:}
\put(12,-140){\Large S. Ghithan}
\put(12,-147){\Large G. Roy}
\put(12,-154){\Large S. Schuh}
\put(60,-250){\includegraphics{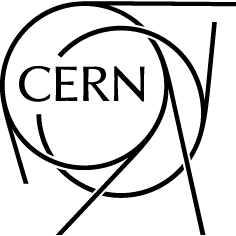}}
\end{picture}
\newpage

\thispagestyle{empty}
\mbox{}
\vfill

\begin{flushleft}
CERN Yellow Reports: Monographs\\
Published by CERN, CH-1211 Geneva 23, Switzerland\\[3mm]

\begin{tabular}{@{}l@{~}l}
  ISBN & 978--92--9083--440--3 (paperback) \\
  ISBN & 978--92--9083--441--0 (PDF) \\
  ISSN & 2519-8068 (Print)\\ 
  ISSN & 2519-8076 (Online)\\ 
  DOI & \url{https://doi.org/10.23731/CYRM-2017-001}\\
\end{tabular}\\[3mm]
Accepted for publication by the CERN Report Editorial Board (CREB) on 10 March 2017\\
Available online at \url{http://publishing.cern.ch/} and \url{http://cds.cern.ch/}\\[3mm]

Copyright \copyright{} CERN, 2017\\[1mm]
\raisebox{-1mm}{\includegraphics[height=12pt]{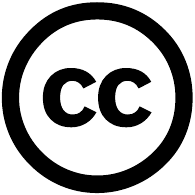}}
 Creative Commons Attribution 4.0\\[1mm]
Knowledge transfer is an integral part of CERN's mission.\\[1mm]
CERN publishes this report Open Access under the Creative Commons Attribution 4.0 license\\
(\url{http://creativecommons.org/licenses/by/4.0/}) in order to permit its wide dissemination and use.\\
The submission of a contribution to a CERN Yellow Report series shall be deemed to constitute the contributor's agreement to this copyright and license statement. Contributors are requested to obtain any clearances that may be necessary for this purpose.\\[5mm]

This volume is indexed in: CERN Document Server (CDS), INSPIRE.\\[5mm]

This volume should be cited as:\\[1mm]

Feasibility Study for BioLEIR, edited by S. Ghithan, G. Roy and S. Schuh, CERN Yellow Reports: Monographs Vol. 1/2017, CERN-2017-001-M (CERN, Geneva, 2017). https//doi.org/10.23731/CYRM-2017-001\end{flushleft}

%% file: FrontBackMaterial/00_frontmatter.tex
\begin{flushleft}
\mbox{}\\[1mm]
\bfseries\LARGE Abstract\\[1cm]
\end{flushleft}

\begin{quotation}
The biomedical community asked CERN to investigate the possibility to transform the Low Energy Ion Ring (LEIR) accelerator into a multidisciplinary, biomedical research facility (BioLEIR) that could provide ample, high-quality beams of a range of light ions suitable for clinically oriented fundamental research on cell cultures and for radiation instrumentation development. BioLEIR would be operated when LEIR is not providing heavy ions for the CERN physics programme. The study group was mandated to write a Feasibility Study Report, using high-level engineering estimates based on previous experience, with the aim to:
\begin{itemize}
	\item collect the requirements for such a facility from the biomedical community in close collaboration with the International Strategy Committee for CERN Medical Applications; 
    \item determine a coherent set of beam parameters, based on the requirements;
    \item explore whether the beam requirements can be met throughout the facility, from the source to the biomedical end-stations;
    \item perform a feasibility
    study of the facility, taking into consideration the overall CERN schedules and programmes;
	\item favour simplicity and robustness of the facility design, while minimizing the cost of maintenance and operation;
    \item establish a high-level costing of material and personnel needed for project implementation;
    \item describe the preferred installation scenario;
    \item perform a high-level risk analysis for the project;
    \item identify the areas of potential difficulty, and the required R\&D should the study go ahead and become a project.
\end{itemize}
The study found no technical show-stopper for the BioLEIR facility.

BioLEIR could be operational in 2021, after CERN Long Shutdown 2 (LS2), provided a decision is taken without delay and the required, experienced personnel and funds are made available by mid-2017 at the latest.
A minimum of 3.5 years are necessary to design, procure, test, install and commission all elements required for first beam to BioLEIR. Should we miss the LS2 window of opportunity, and on the basis of the current CERN schedule, the BioLEIR facility could only become operational after LS3 in 2026.
\end{quotation}
\vfill
\vskip 1.em
\begin{center}
Geneva, Switzerland \\
March 2017\\
\end{center}


\newpage\BLKP
\begin{flushleft}
\mbox{}\\[1mm]
\bfseries\LARGE Contributors\\[1cm]
\end{flushleft}
\label{sec:contributors}

\begin{flushleft}
D.~Abler$^{1,2}$, T.~Amin$^{1,3}$, J.~Axensalva\footnotemark[1], S.~Baird\footnotemark[1], M.~Battistin\footnotemark[1],
N.~Bellegarde\footnotemark[1], J.~Borburgh\footnotemark[1], A.~Broche\footnotemark[1], C.~Carli\footnotemark[1], E.~Carlier\footnotemark[1], O.~Choisnet\footnotemark[1],
J.-M.~Cravero\footnotemark[1], M.~Di~Castro\footnotemark[1], J.~Devine\footnotemark[1], M.~Dosanjh\footnotemark[1], A.~Dworak\footnotemark[1], R.~Froeschl\footnotemark[1], J.M.~Garland\footnotemark[1], A.~Garonna$^{1,4}$, S.~Ghithan\footnotemark[1], S.~Gilardoni\footnotemark[1], B.~Goddard\footnotemark[1], A.~Gutierrez\footnotemark[1], A~Huschauer\footnotemark[1], E.~Jensen\footnotemark[1], S.~Jensen\footnotemark[1],
J.~Jowett\footnotemark[1],
B.~Jones$^{1,5}$, V.~Kain\footnotemark[1], R.~Kersevan\footnotemark[1], D.~K\"uchler\footnotemark[1], J.-B.~Lallement\footnotemark[1], M.~Lazzaroni\footnotemark[1], A.~Lombardi\footnotemark[1],  R.~Lopez\footnotemark[1],
D.~Manglunki\footnotemark[1], D.~Mcfarlane\footnotemark[1], A.~Milanese\footnotemark[1],  Y.~Muttoni\footnotemark[1],
D.~Nicosia\footnotemark[1], P.~Ninin\footnotemark[1], M.~Nonis\footnotemark[1], J.~Osborne\footnotemark[1], S.~Pasinelli\footnotemark[1], 
R.~Rata\footnotemark[1], G.~Riddone\footnotemark[1], G.~Roy\footnotemark[1], I.~Ruehl\footnotemark[1], S.~Schuh\footnotemark[1], R.~Scrivens\footnotemark[1], M.~Silari\footnotemark[1], R.~Steerenberg\footnotemark[1],
M.~Tavlet\footnotemark[1], V.~Toivanen$^{1,6}$, G.~Tranquille\footnotemark[1], A.~Tursun\footnotemark[1], F.~Valentini\footnotemark[1], M.~Wilhelmsson\footnotemark[1]       
\end{flushleft}

\begin{flushleft}
\mbox{}\\[1mm]
\bfseries\LARGE Editorial Board Members
\end{flushleft}
\label{Sec:EditorialBoard}

\begin{flushleft}
C.~Carli\footnotemark[1], M.~Dosanjh\footnotemark[1], 
S.~Ghithan\footnotemark[1], B.~Goddard\footnotemark[1],\\ D.~Manglunki\footnotemark[1], G.~Roy\footnotemark[1], 
S.~Schuh\footnotemark[1], R.~Steerenberg\footnotemark[1] 
\end{flushleft}

\begin{flushleft}
\mbox{}\\[10mm]
This document is building on earlier work done by  Daniel Abler, Christian Carli, Manjit Dosanjh, 
Adriano Garonna, Steve Myers and Ken Peach, following initial ideas by Ugo Amaldi and Manjit Dosanjh.
\end{flushleft}

\begin{flushleft}
\mbox{}\\[2mm]
This work has been encouraged by the International Strategy Committee for CERN Medical Applications.
\end{flushleft}

\begin{flushleft}
\footnotetext[1]{~CERN - European Organization for Nuclear Research, Geneva, Switzerland}
\footnotetext[2]{~University of Bern, Bern, Switzerland}
\footnotetext[3]{~University of Huddersfield, Huddersfield, UK}
\footnotetext[4]{~TERA Foundation, Novara, Italy}
\footnotetext[5]{~University of Oxford, Oxford, UK}
\footnotetext[6]{~University of Caen Normandy, Caen, France}
\end{flushleft}
\newpage\BLKP

\begingroup\baselineskip.99\baselineskip
\tableofcontents
\endgroup
\newpage
\addcontentsline{toc}{chapter}{List of Tables}
\markboth{\textsc{Tables}}{\textsc{Tables}}
\listoftables
\listoffigures
\newpage
\begin{flushleft}
\mbox{}\\[1mm]
\bfseries\LARGE Acronyms and abbreviations\\[1cm]
\end{flushleft}
\addcontentsline{toc}{chapter}{Acronyms and abbreviations}
\markboth{Acronyms and abbreviations}{Acronyms and abbreviations}

\begin{longtable}{@{}ll}
1D & One-Dimensional\\
2D & Two-Dimensional\\
3D & Three-Dimensional\\
4D & Four-Dimensional\\
AC & Alternating Current\\
AD & Antiproton Decelerator at CERN\\
ADCs & Analog-to-Digital Converters\\
AISHa & Advanced Ion Source for Hadron therapy\\
AP & Access Point\\
ATEX & ATmosph\'eres EXplosibles\\
AWAKE & proton driven plasma wakefield acceleration experiment at CERN\\
B & magnetic field\\
BCTs & Beam Current Transformers\\
BE & Beams department, CERN\\
BE-CO & Controls group, CERN\\
BE-ICS & Industrial Controls and Safety group, CERN\\
BE-OP & Operations group, CERN\\
BED & Biological Effective Dose\\
$\beta_x$ & horizontal beta function, one of the Twiss parameters\\
Biolab & Biological laboratory\\
BNC & Bayonet Neill-Concelman connector \\
BPM & Beam Position Monitor\\
BSM & Bunch Shape Monitor\\
BTV & Beam Television screen\\
C1 & Counting room for H2 and V irradiation rooms\\
C2 & Counting room for H1 irradiation room\\
CC & Command Console\\
CCD & Charge-Coupled Device\\
CERN & European Organization for Nuclear Research, Geneva\\
CHARM & CERN High energy AcceleRator Mixed field\\
CMASC & CERN Medical Applications Steering Committee\\
CNAO & National Centre of Oncological hadron therapy, Italy\\
CT & Computer Tomography\\
DC & Direct Current \\
 DCCT & Direct Current to Current Transformer\\
DDD & Direct Diode Detection principle\\
DESY & Deutsches Elektronen-Synchrotron, Germany\\
DNA & DeoxyriboNucleic Acid\\
DSO & Departmental Safety Officer, CERN\\ 
DTL & Drift Tube Linac \\
$D_x$ & Horizontal Dispersion function\\
E & Electric field\\
E/A & Energy over Mass ratio \\
EBIS & Electron Beam Ion Source\\
EBIS-SC & SuperConducting high performance Electron Beam Ion Source\\
ECR & Electron Cyclotron Resonance\\
ECRIS & Electron Cyclotron Resonance Ion Source\\
EIS & Elements Important for Safety\\
ELENA & Extra Low ENergy Antiproton ring, CERN\\
EN & Engineering department, CERN\\
EN-ACE & Alignment Coordination and Engineering group, CERN\\
EN-CV & Cooling and Ventilation group, CERN\\
EN-EA & Experimental Areas group, CERN\\
EN-EL & Electrical Engineering group, CERN\\
EN-HE & Handling Engineering group, CERN\\
EN-MME & Mechanical \& Materials Engineering group, CERN\\
EN-STI & Sources, Targets and Interactions group, CERN\\
EN-STI-ECE & Equipment Controls and Electronics group, CERN\\
ENLIGHT & European Network for LIGht ion Hadron Therapy\\
ES & Electrostatic Septum\\
ESRF & European Synchrotron Radiation Facility, Grenoble, France\\
ESS & Energy Selection System\\
EYETS & Extended Year End Technical Stop\\
FGC & Function Generator Controller\\
Frankfurt MEDEBIS & Prototype of an injector for a synchrotron dedicated for cancer therapy\\
FISC & SCintillating FIbre monitor\\
FLUKA & FLUktuierende KAskade, Monte Carlo simulation package \\
FODO & Magnet structure consisting of Focusing lens, Drift, Defocusing Lens and Drift\\
FTE & Full-Time Equivalent\\
FWHM & Full Width at Half Maximum\\
GBAR & Gravitational Behaviour of Anti-hydrogen at Rest experiment at CERN\\
GCR & Galactic Cosmic Ray\\
GEANT4 & GEometry ANd Tracking, Monte Carlo simulation package\\
GEM & Gas Electron Multiplier\\
GHMC & Gunma university Heavy ion Medical Centre, Japan\\
GSI & GSI Helmholtzzentrum f\"{u}r Schwerionenforschung GmbH, Darmstadt, Germany\\
GS/s &  Giga samples per second\\
H & Horizontal \\
H1 & First Horizontal beamline\\
H2 & Second Horizontal beamline\\
HE & High-Energy\\
HEP & High Energy Physics\\
HIE & High Intensity and Energy\\
HIT & Heidelberg Ion Therapy center, Heidelberg, Germany \\
HMI & Human Machine Interfaces\\
HSE & Occupational health \& Safety and Environmental protection unit, CERN\\
HSE-RP & Radiation Protection group, CERN\\
HSE-SEE & Safety Engineering and Environmental protection group, CERN\\
HV & High Voltage\\
HW & HardWare\\
HZB & Helmholtz-Zentrum, Berlin\\
ICRE & International Conference on Translational Research in radio-oncology\\ 
ICRU & International Commission for Radiation Units and measurements\\
IH & Inter-digital H-structure\\
InCA & Injector Controls Architecture\\
INFN-LNS &  Istituto Nazionale di Fisica Nucleare - Laboratori Nazionali del Sud, Italy\\
IR & Irradiation Room\\
ISC & International Strategy Committee for CMASC\\
ISO & International Organization for Standardization \\
ISOLDE & Isotope Separator On Line DEvice, experiment at CERN \\
ISR &  Intersecting Storage Rings, CERN\\
IT & Information Technology\\
JINR & Joint Institute for Nuclear Research, Dubna, Russia\\
JYFL & Jyv\"askyl\"an Yliopiston Fysiikan Laitos, Jyv\"askyl\"a, Finland\\
kVp & Peak kiloVoltage\\
LEAR & Low Energy Antiproton Ring at CERN\\
LEBT & Low-Energy Beam Transport \\
LEIR & Low Energy Ion Ring at CERN\\
LEP & Large Electron--Positron collider at CERN \\
LET & Linear Energy Transfer\\
LHC & Large Hadron Collider at CERN \\
LHCf & Large Hadron Collider forward experiment at CERN\\
LHeC & Large Hadron electron Collider at CERN\\
Linac & Linear accelerator \\
LINAC3 & Linear accelerator 3 at CERN\\
LINAC4 & Linear accelerator 4 at CERN\\
LINAC5 & Linear accelerator 5 at CERN\\
LIU & LHC Injector Upgrade programme\\
LMAIS & Liquid Metal Alloy Ion Source\\
LQ & Linear Quadratic model\\
LS2 & Long Shutdown 2, CERN\\
LS3 & Long Shutdown 3, CERN\\
LS4 & Long Shutdown 4, CERN\\
LV & Low Voltage\\
MAD-X & Methodical Accelerator Design computer program, version X,\\
MC & Monte Carlo simulation\\
MCP & Micro-Channel Plate\\
MEBT & Medium-Energy Beam Transport \\
MedAustron & MedAustron particle accelerator, Austria\\
MeV/u & Mega Electron-Volts per Nucleon (kinetic Energy per Nucleon)\\
MIVOC & Metal Ions from VOlatile Compounds \\
MLC & Multi Leaf Collimator\\
MR & Magnetic Resonance\\
MSE & Magnet Septum Extraction\\
MST & Magnetic Septum Thin\\
MTP & Medium-Term Plan\\
MWPC & Multi-Wire Proportional Chamber\\
NA & North Area\\
NB & Nota Bene, Take Notice\\
NEG & Non-Evaporable Getter\\
NPL & National Physical Laboratory, UK\\
nTOF & neutron Time-Of-Flight experiment at CERN \\
OASIS & Open Analogue Signal Information System\\
OC1 & Orbit Corrector 1\\
OC2 & Orbit Corrector 2\\
PAD & Personnel Access Device\\
PC & Power Converter\\
PFWs & Pole-Face Windings\\
PHE & Physics for Health in Europe\\
PIMMS & Proton-Ion Medical Machine Study\\
PMI & air filled plastic ionization chamber\\
PMQ & Permanent Magnet Quadrupole\\
PPM & Pulse-to-Pulse Modulation\\
PPS & Personnel Protection System\\
PS & Proton Synchrotron at CERN \\
PSB & Proton Synchrotron Booster at CERN \\
PSI & Paul Scherrer Institute, Switzerland\\
PTW & Physikalisch-Technische Werkst\"atten\\
PTW MP3 & IBA blue phantom\\
PY & Person-Years\\
QA & Quality Assurance\\
Q/A & Charge over Mass ratio\\
Q-D & Quadrupole-Driven\\
RADED & RADiation Effects Facility, Finland\\
RBE & Relative Biological Effectiveness\\
R\&D & Research \& Development \\
REMUS & Radiation and Environment Monitoring Unified Supervision\\ 
RF & Radio Frequency \\
RF-ID & Radio-Frequency IDentification\\
RF-KO & RF Knock-Out\\
RFQ & Radio-Frequency Quadrupole \\
RGA & Residual-Gas Analyser\\
RH & Relative Humidity\\
rms & root mean square \\
RP & Radiation Protection\\
RS232 & standard for serial communication transmission of data\\
S & normalized sextupole Strength\\
SEE & Single Event Effect\\
SEM-grids & Secondary Emission beam profile Monitors\\
SEU & Single Event Upset\\
SF & Surviving Factor\\
SIL & Safety Integrity Level\\
SMB & Site Management and Buildings department, CERN\\
SOBP & Spread Out Bragg Peak\\
SR-C & CERN Regulation concerning Chemical hazards\\
SSD & Source to Surface Distance\\
SVT &  Science, Vacuum Techniques\\
TDR & Technical Design Report\\
TE & Technology department, CERN\\
TE-VSC & Vacuum, Surfaces and Coatings group, CERN\\
TEPCs & Tissue Equivalent Proportional Counters\\
T.m & Tesla.meter\\
ToF & Time-of-Flight\\
TPC & Time Projection Chamber\\
UCL & Universit\'e Catholique Louvain-la-Neuve, Belgium\\
UTP &  Unshielded Twisted Pair\\
UV & UltraViolet\\
V & Vertical beamline\\
VME & Versa Module Europa (computer bus standard) \\
$X_0$ & radiation length\\
XBOX & X-band high RF power test stand, CERN\\
YETS & Year End Technical Stop\\

\end{longtable}
\newpage

%% file: Chapters/ExecutiveSummary.tex
\chapter{Executive Summary}
\label{Chap:Exec}

The biomedical community has asked CERN
to investigate the possibility to transform the Low Energy Ion Ring (LEIR) accelerator into a multidisciplinary, biomedical research facility (BioLEIR) that could provide ample, high-quality beams of a range of light ions suitable for clinically oriented, fundamental research on cell cultures and for radiation instrumentation development. \\

The study group was mandated by the CERN management to undertake a preliminary study and write a Feasibility Study Report with minimal resources, using high-level engineering estimates based on previous experience, with the aim to:
\begin{itemize}
	\item collect the requirements for such a facility from the biomedical community in close collaboration with the International Strategy Committee for CERN Medical Applications; 
    \item determine a coherent set of beam parameters, based on the requirements;
    \item explore whether the beam requirements can be met throughout the facility, from the source to the biomedical end-stations;
    \item perform a feasiblity
    study of the facility, taking into consideration the overall CERN schedules and programmes;
	\item favour simplicity and robustness of the facility design, while minimizing the cost of maintenance and operation;
    \item establish a high-level costing of material and personnel needed for project implementation;
    \item describe the preferred installation scenario;
    \item perform a high-level risk analysis for the project;
    \item identify the areas of potential difficulty, and the required R\&D should the study go ahead and become a project.
\end{itemize}

The LEIR machine is needed as pre-injector for the LHC and North Area ion physics programmes. The heavy ion programmes are very important at CERN, yet production beams are not continuously delivered to the LHC or North Area throughout the year and the LEIR machine has some unexploited capacities.
LEIR could accelerate light ions up to the few hundred MeV/u energies required for clinical use. Therefore, LEIR would be a suitable machine to provide beams for biomedical research.

The present LEIR machine uses fast beam extraction to the next accelerator of the ion chain eventually leading to the LHC. To provide beam for a biomedical research facility, a new slow extraction must be installed. A short transfer line will bring the ions to two horizontal beamlines with energies in the range of 50 to 440\,MeV/u for most light ions of interest. A vertical beamline with energies up to 70\,MeV/u will complete the facility.

BioLEIR will have a dedicated light ion frontend with its own light ion source, RFQ and linac (LINAC5). LINAC3 which is producing heavy ions for the North Area and the LHC, will hence not be affected by LINAC5 and BioLEIR operations. Once LINAC5 is commissioned, the baseline operation will allow sending beam to BioLEIR during the whole year in parallel with LHC physics, with switching time scales of less than 10 minutes. LINAC5 will be housed in the space that is occupied by LINAC2 (currently producing protons for the LHC) once it is decommissioned and removed, thereby saving cost on infrastructure and building.

The delivery of the facility is foreseen in 3 stages, integrated with the overall CERN accelerator and consolidation schedules, and gradually increasing the functionality, as well as spreading the investment over 5-7 years.

Stage 1 will consist of a new slow extraction and 3 experimental beamlines which will be supplied with Argon and Oxygen, and possibly Carbon, ions from LINAC3. In this stage, 4 full months of beam time per year will be available for BioLEIR experiments. The minimum lead time for design, installation and commissioning of a slow extraction and the experimental beamlines is 3.5 years. A control irradiator shall be available within the BioLEIR experimental area.

During Stage 1, characterization of the new light ion source, together with the design, preparation and construction of LINAC5 and the connecting transfer line to LEIR will take place in preparation of Stage 2.

Stage 2 of BioLEIR will have the new light ion source and LINAC5 operational, and will deliver the full range of light ions to BioLEIR at maximum energies of 250\,MeV/u. The intensities available at Stage 2 will be limited to 10$^8$\,ions/s and 10$^{9}$\,protons/s. Stage 2 allows complete study of all Relative Biological Effectiveness (RBE) in-cell effects and is considered the baseline for BioLEIR. It can be ready at the earliest about 1 year after LINAC2 has been decommissioned.

Stage 3 will have upgraded LEIR power converters to allow the desired clinical light ion energies of 440\,MeV/u. Due to the higher energies available in Stage 3, the shielding thickness of the horizontal irradiation rooms will need to be dimensioned accordingly. This clinical energy allows the full penetration depth of particles in humanoid phantoms. Alternatively, the upgraded shielding would allow an increase of beam intensities for biophysics and (micro-) dosimetry studies at the cost of reducing beam energies.

Finally, looking beyond Stage 3, interleaved operation of BioLEIR with LHC and North Area ion physics remains an option for the future as a way to increase beam time for biomedical experiments even further, should that be needed. However, the additional cost for interleaved operations and the entailed operational complexity is estimated to be substantial and a careful cost-benefit analysis shall be made before deciding for such an upgrade.

The cost of the full BioLEIR facility (Stage 1, 2 and 3) is estimated at 28.7\,MCHF and 119 person-years with the following break-down:

Stage 1 is estimated to cost about 15\,MCHF and 61\,person-years and includes a new slow extraction from LEIR, as well as all three new beamlines and the full corresponding experimental area. In this stage, the lightest ions accessible with LINAC3 are available to BioLEIR during about 4 months per year.

Stage 2 is projected to cost around 13\,MCHF and 55\,person-years and will have a new light ion injector chain that can produce the full range of light ions requested by the biomedical community.

During Stage 1 and 2 the ion energies will be limited to 250\,MeV/u.

Stage 3 will allow ion energies up to maximum values of 440\,MeV/u and cost an additional 1\,MCHF and 3 person-years.

The cost for radiation shielding, magnets, power converters, as well as electrical power and cooling requirements scale with the maximum ion energies requested. In  this document we assume maximum ion energies of 440\,MeV/u, which is consistent with the realisation of the facility up to and including Stage 3. If the decision can be made upfront to limit the maximum ion energies to 250\,MeV/u, and to forego Stage 3 entirely, additional savings beyond the cost of Stage 3 itself, could be realised for Stage 1 and Stage 2 by reducing the scope for radiation shielding, magnets, power converters, electrical power and cooling systems.
The Project cost at reduced scope could be estimated at a next project stage, but the reduction is expected to be at the level of 20\% of the total cost.\\

It is instructive to compare the cost of the BioLEIR facility with the cost of an equivalent biomedical facility if it were built from green-field elsewhere. A survey of clinical facilities in Europe that are capable of providing different ions and protons found a construction cost at a conservative level of 140\,MCHF (excluding personnel cost).
In order to account for the higher cost of a clinical facility, we scale this cost down by about 30\% to a level of 100\,MCHF needed to establish a centre dedicated to R\&D with a scientific outcome comparable to that which can be reached with BioLEIR.\\

Assuming a project start without delay, with funds and experienced resources committed by mid-2017 at the latest, Stage 1 of BioLEIR could come on-line in 2021, after LS2. Stage 2 of BioLEIR can be operational 2 years after LS2, in 2023. Finally, Stage 3 of the facility can be made available any time when LEIR is not running for a short period of time (e.g. during a Year End Technical Stop).

Should the start date be postponed to later than mid-2017, the project will miss the window for installation of the slow extraction in LEIR during LS2. The next possible window would be LS3 (2025) with first beams to BioLEIR in 2026 - at which time the facility could come on-line with Stage 1, Stage 2 and Stage 3 directly.\\

The study found no technical show-stopper for the BioLEIR facility. This document provides a first estimate of required personnel and material cost. Once a decision regarding implementation of BioLEIR is taken, a refinement of the design and estimations shall be undertaken.

%% file: Chapters/Introduction.tex
\chapter{Introduction to the BioLEIR project proposal}
\label{Chap:Intro}
\section{BioLEIR: a biomedical research facility hosted at CERN}
\label{Sec:Intro_Motivation}
Cancer is a critical societal issue. Worldwide, in 2012, 14.1 million people were diagnosed with, and 8.2 million had died of cancer. The annual global incidence is expected to rise to as many as 25 million cases in 2035. Worldwide, over 32 million people live with cancer, mostly untreated~\cite{Ferlay:2013}. Radiation therapy is an essential component of effective cancer control. In Europe, about half of the number of patients diagnosed with cancer are treated with radiation therapy and every other cured patient undergoes radiation therapy as part of their treatment. The main aim of radiation therapy is to deliver a maximally effective dose of radiation to a designated tumour site while sparing the surrounding healthy tissues as much as possible.

Charged hadrons (protons, Carbon ions and other ions) have a unique depth-dose distribution (Bragg peak) and unique radiobiological properties (higher radiobiological effectiveness, RBE), which make them an actively pursued development in modern cancer radiotherapy. 
Advanced radiotherapy using proton, Carbon ion or other ion beams (hadrontherapy) to deliver a dose to the tumour, has gained huge momentum over the last two decades. Many new centres have been built, and many more are under construction. Currently, just  over 50 centres are in operation, in construction or in planning stage and numbers are expected to double by 2020. Many of these hadrontherapy centres are driven by financial considerations and are heavily concentrating on treating large patient numbers.

Compared with photon beam radiotherapy, hadrontherapy allows more selective deposition of radiation dose in various cancers while reducing dose to surrounding healthy tissues. Heavier ions (e.g. Carbon ions), in addition, due to the higher density of ionisation events along the particle track, exhibit a higher relative biological effectiveness than X-rays or protons, making them prime candidates for the treatment of also radio-resistant tumours.

Still, these advantages are felt to be preliminary, based on scattered physical and biological studies, spread over many years, and performed in heterogeneous centres under different conditions, resulting in significant systematic uncertainties. There is also criticism on the lack of large-scale clinical or preclinical studies data to support assumptions on the effectiveness of hadrontherapy.

Consequently, to fully utilize the benefits of particle therapy, a concerted research effort is called for to provide the biological and physical data sets and innovative diagnostics and treatment tools to help derive the optimal use of this advanced therapy.

A dedicated centre for physics and fundamental, biomedical research, offering extended blocks of beam time, with a variety of ions and energies provided, is urgently needed, as the existing hadrontherapy centres do not have sufficient beam time available for the basic research efforts needed.  

This need for an open-access biomedical facility dedicated to medically-oriented research for optimising hadrontherapy was first raised at the European Network for  Light Ion Hadron Therapy (ENLIGHT) meeting in Oropa, Italy in 2005~\cite{Oropa:2005} and later in the 2010 Physics for Health workshop\cite{ICTR:2012}, where members of the biomedical and physics community jointly asked CERN to take the lead on this initiative. In 2012, a brainstorming meeting suggested the possibility of modifying the existing Low Energy Ion Ring (LEIR) accelerator to establish BioLEIR~\cite{meeting:2012}.\\\\\\

\subsection{The biomedical motivation}
\label{SubSec:Intro_Motivation_Biomedical}
Particle irradiation and the impact in vitro have been reported in a number of publications, resulting in a large range of Relative Biological Effectiveness (RBE) data from a number of different cell lines and endpoints. These studies have been very useful in confirming to a large extent the hypotheses of the effect of particle irradiation. Nonetheless, the heterogeneity between the different studies makes it hard to combine the obtained data and to draw definitive final conclusions.

The aim of BioLEIR is to provide a unique facility with a range of ion beams and energies suitable for multidisciplinary medical physics oriented research. The facility would be capable of providing a range of different ion species such (e.g. He, Li, Be, B, C, N, O) for the purposes of: 

\begin{itemize}
\item In situ and remote detector development of nuclear activation products around the Bragg peak positions, with real time imaging and tracking capabilities to assess simulated tumours with organ motion compensation.
\item Measurements and predictions of the ballistic characteristics of the various ionic beams in humanoid phantoms, with special attention to 3D ionisation density and dose distributions and to advanced simulation techniques (GEANT4, GEANT-DNA, FLUKA). 
\item Systematic radiobiology experiments, simulation and modelling studies to RBE, its maximum efficiency points, non-linear modifications with dose and ionisation clustering compared to standard hospital megavoltage photon exposures in a suitable panel of biologically well-characterised cell lines (see appendix~\ref{App:Intro_A} for more details).
\item Massive data storage and analysis of the experiments performed within and across each of the above categories.
\end{itemize}

The above themes need to be systematically explored under standardised dosimetry and laboratory conditions, for a range of cancer and normal cells, over a more extensive experimental range than that in the past, in order to obtain greater uniformity and statistical accuracy, essential for radiotherapy applications.
A concerted effort by different research groups working together at a dedicated facility, systematically assembling a complete data set of the particle energy spectrum and the spectrum of absorbed energy for a range of detectors and in different biological systems (i.e. cell cultures) would provide the much needed systematic and basic biomedical understanding of effects relevant for safe and efficient application of advanced radiotherapy.\\

A dedicated centre will not only provide the necessary beam time in large time blocks, but also foster closer collaborations between research teams from different countries to rapidly move the field of hadrontherapy forward. Creating a scenario similar to the large collaborations found in high-energy physics, where many teams from different institutions and many countries work side by side on specific pieces of a puzzle, pursuing one common goal. Such a facility will serve to optimise hadrontherapy and help current and future hadrontherapy centres worldwide.
Building upon existing structures, CERN can provide the community at a reduced cost with the much-needed opportunity to perform fundamental research to support the application of particle beams to the health sector. CERN has a strong tradition in hosting international collaborations, a scientific and technical support infrastructure, and an excellence in science achieved over the many years of operation, making it the ideal place where such a project could be initiated and propagated. CERN is a centre of competences and expertise of people and working side by side with researchers from many other areas of physical sciences, computer science, and mathematics is expected to spark new ideas that can lead to new approaches to the existing problems.


\section{An overview of the BioLEIR project}
\label{Sec:Intro_ProjectOverview}

The establishment of a multidisciplinary research facility to investigate biomedical applications of high energy ion beams at CERN would require conversion of the LEIR synchrotron.  Although LEIR has to be maintained and operated for the Heavy Ion physics programmes, production beams are not continuously delivered to the LHC or North Area throughout the year and the LEIR machine has some unexploited capacities. The energies of the ions that can be accelerated in LEIR are similar to those required for clinical use (a few hundred MeV/u) which makes LEIR suited to provide beams needed for better understanding and optimization of hadrontherapy.

The present LEIR uses only fast beam extraction to the next accelerator of the ion chain eventually leading to the LHC. To provide beam for a biomedical research facility, a new slow extraction needs to be installed. A short transfer line brings the ions to two horizontal beamlines with energies in the range of 50 to 440\,MeV/u for most light ions of interest. A vertical beamline with energies up to 70\,MeV/u completes the facility.
BioLEIR is expected to have a dedicated light ion frontend with its light ion source, RFQ and linac (LINAC5),  optimized for ions of interest for biomedical experiments (protons, light ions such as Lithium, Boron and Carbon, and perhaps heavier ions like Oxygen and Neon).

Other solutions have also been studied, among which was the option to enhance the existing LINAC3 with a light ion source and respective frontend elements. This option was ruled out early on, as a dedicated, new light ion frontend represents less risk to the present-day Heavy Ion programmes than a modification of the present-day LINAC3. In addition, a new light ion frontend allows the option for biomedical experiments to take place 
in time-sharing mode with the LHC and North Area Heavy Ion programmes.

Assuming a project kick-off in 2017 with full resources allocated, the delivery of the facility is foreseen in several stages, as follows:\\\\
\textbf{Preparation phase (2017-2020):}
\vspace{-1mm}
	\begin{itemize}
		\item Procurement and characterization of the light ion source as input for the final LINAC5 design.
        \item LEIR beam dynamics studies (Oxygen during Xenon run in 2017).
        \item Integration studies.
       \item Preparation of the experimental facility (emptying of hall 150, infrastructure preparations).
        \item Design, procurement, testing and commissioning of all machine elements needed for Stage~1.
        \item Start work on all elements needed for Stage 2 (LINAC5, injection transfer line)
	\end{itemize}
\textbf{Stage 1 (2021-2022): running with ions from LINAC3 up to 250\,MeV/u:}
\vspace{-1mm}
    \begin{itemize}
	    \item The new slow extraction \& 3 new beamlines are available.
        \item LINAC3 remains the injector for LEIR, producing  lightest accessible ions - Argon and Oxygen. A single ion source is connected to LINAC3, and BioLEIR will operate outside the periods of Heavy Ion programmes in the LHC and the North Area.
        \item The full Biolab is available.
        \item Continue the work on all machine elements needed for Stage~2.
		\item LINAC2 dismantling starts in 2021
        \item Installation and commissioning of LINAC5 starts in 2022. 
        \item The light ion source continues preparatory beam development for the different light ion species.
    \end{itemize}
\textbf{Stage 2 (2023- ): running with ions from LINAC5 up to 250\,MeV/u:}
\vspace{-1mm}
    \begin{itemize}
		\item The new, full LINAC5 injection is available.
		\item Switching between LINAC3 and LINAC5 operation is possible within minutes, and BioLEIR beam is available from Easter to early November.
     \end{itemize}
\textbf{Stage 3 (2024- ): running with ions from LINAC5 up to 440\,MeV/u:}
\vspace{-1mm}
    \begin{itemize}
		\item The LEIR power converters have been upgraded to allow higher ion energies.
     \end{itemize}

At Stage 1, a control irradiator in the form of a Cs-137 source will be available within the BioLEIR experimental area. At a later stage, a 6\,MeV control irradiation facility could be provided, as close as possible to the experimental area in the South Hall.\\

At a later stage, the user community might request to upgrade the injector chain to allow interleaved operation of BioLEIR with LHC and North Area ion physics, in order to increase beam time for biomedical experiments even further. However, the cost to upgrade injector chain elements to pulse-to-pulse modulation represents a significant cost and the entailed operational complexity is estimated to be substantial. A careful cost-benefit analysis shall be made before deciding for such an upgrade if deemed necessary.

\section{Potential interest for light ions to the CERN research programme and the existing user community}
\label{Sec:Intro_NonBioResearch}
In addition to the necessity of light ions for biomedical research, light ions could also be of interest to the CERN research programme and the existing CERN user community at large. A first, non-comprehensive feedback and interest has been gauged and found to be mainly around the accelerator ion programmes, as well as around detector development. A brief overview is given here, more details are found in appendix~\ref{App:Intro_AlternativeUseCases}.

\subsection{Accelerator ion programmes}
\label{SubSec:Intro_NonBioResearch_Accelerators}
Although nothing is presently foreseen in the future LHC programme, there has been some interest in the LHC community in, for example, Argon-Argon, proton-Argon and proton-Nitrogen collisions. The availability of deuterons is of potential interest for electron-deuteron collisions at the LHeC. Different types of light ions might also be of interest to physics programmes currently underway in the North Area, for example at NA61. There is interest to study atmospheric cosmic rays for which ions from Iron down to Nitrogen could be of interest. Similarly, the community studying the effect of radiation on electronics used in space exploration could be interested in Iron ions. Should BioLEIR become a reality, the availability of light ions could spark research ideas from the CERN physics community and such interest could be investigated in more detail at that time.

\subsection{Additional potential use cases}
\label{SubSec:Intro_NonBioResearch_Detectors}
A non-exhaustive list of potentially interesting research at the BioLEIR facility beyond biomedical research includes the following areas:
\begin{itemize}
\item  Detector development and testing
	\begin{itemize}
	\item further development of reliability and accuracy of dosimetry techniques
    \item characterization of particle physics detector technologies in clinical ion beams
    \item testing of detectors suitable for beam monitoring and dosimetry
	\end{itemize}
\item Real-time tumour tracking and dose delivery in hadrontherapy
\item Nuclear physics studies and benchmarking of simulation codes
\item Radiation damage to electronics and microelectronics 
\item Shielding and induced radioactivity in accelerator and beamline components
\end{itemize}

\section{Strategic importance of the BioLEIR facility}
\label{Sec:Intro_Strategy}
According to the ISC, an international panel of experts in the field, the BioLEIR facility at CERN would address a number of key strategic issues that the biomedical community is facing today. The ISC reckons that the BioLEIR facility is:
\begin{itemize}
	\item Urgently required because of increasing use of ion beams in medicine for ablative and oncology therapeutic purposes;
    \item Essential in order to inform policy makers, e.g. International Commission for Radiation Units and Measurements (ICRU) and other similar national and international bodies that recommend treatment policies;
    \item More likely to produce answers to difficult research questions, faster and at a lower cost, by sharing international expertise and also by using as much existing infrastructure as possible.
\end{itemize}

The ISC recognizes that the envisaged research at BioLEIR shall be an extension of work done elsewhere, in a complementary manner. Examples include:
\begin{itemize}
	\item Cooperation with existing university hospital based facilities, holding animal research licences, that can perform \textsl{in vivo} experiments;
    \item Pooling of data from different countries;
    \item Use of data emerging on molecular structure and mechanisms from other sophisticated facilities, such as the Diamond Synchrotron (UK), ESRF (France) and DESY (Germany);
    \item Use of the best available analytical genetic and molecular techniques in association with collaborating universities worldwide.
\end{itemize}

BioLEIR will both enhance and complement the few existing or planned "part time" beam lines for this kind of multidisciplinary research, by providing far longer running periods and more research facilities than are available in existing clinical centres, where clinical priority demands a large proportion of the available beam time. A pan-European collaborative network of treatment and research centres is envisaged for experimental design and beam allocation in an effective, open, complementary and concerted way. 

The long-term goal, to which BioLEIR would contribute, is the optimisation of particle beams applied in cancer and other medical conditions, as well as in diagnostic medicine, all leading to societal healthcare benefits. This would include improved tissue dose allocation as well as provide data that would impact on the design of a future optimal medical particle accelerator.

The beam energy reach, the capability of providing beam time for most of the year, the lack of clinical imperatives, make LEIR an ideal accelerator for conversion into BioLEIR. Given the existing infrastructure needed to host a multidisciplinary research community, the highly specialised personnel, and the fact that LEIR is already in operation, the shared cost of establishing a research facility dedicated to medical applications will be significantly lower at CERN than other stand-alone options elsewhere. Such a research facility will effectively enhance global efforts on building capacity, raise awareness and provide knowledge sharing in the field of advanced radiation therapy.

The CERN ethos would provide a highly stimulating environment for medical research applications. Analogous to how research programmes have been successfully decided by CERN in the field of particle physics, biomedical experiments at BioLEIR will be presented by external institutes or international collaborations and selected on a merit basis by an independent expert committee in order to ensure rigorous assessments and to benefit from the culture of scientific openness, collaboration across national borders and attracting experts from a variety of fields. BioLEIR will become a hub for interdisciplinary exchange, offering Research and Development opportunities and complementing work done at other facilities, further enhancing the societal impact derived from particle physics research. BioLEIR will also provide unique educational and research training opportunities, in doctoral and post-doctoral positions, for young researchers who will engage with collaborators at CERN and will later join other teams in Europe and across the world.

At the end of the 1990's, the Proton Ion Medical Machine Study (PIMMS) -- brought forward by CERN, TERA and MedAustron, using the knowledge of the time -- has led to the realization of two hadron-therapy centres in Europe: CNAO in Pavia (IT) and MedAustron in Wiener Neustadt (AT). These successes are due to the enormous value for the medical community of the know-how put by CERN at the disposal of the partners through a few key experts. Today, BioLEIR is  the opportunity to open new possibilities of research to the biomedical community and have an even larger reach and long-term impact.

%% file: Chapters/BeamRequirements.tex
\chapter{Beam Specification Parameters}
\label{Chap:BeamParameters}

\section{Background and approach}
\label{Sec:BeamParameters_Background}
The goal of this chapter is to establish the full set of operating parameters for a non-clinical biomedical research facility at CERN. It shall be noted that work on cells (2D and in future possibly 3D) and inert materials (such as detectors and humanoid tissue equivalent phantoms) are envisaged, while work on animals or patients are excluded. The operating parameters are defined by the irradiation requirements set out at the biological and dosimetry end-stations, which aim to use standard and novel techniques in laboratory conditions that replicate clinical particle therapy. Those requirements are then transposed back through the entire accelerator chain all the way to the source with the aim of ensuring that the full accelerator chain design is based on a common set of agreed upon parameters.

The requirements at the irradiation points are defined by the biomedical user community for the variety of research topics the community is interested in. They have been initially produced by a working group established under the umbrella of the International Strategy Committee (ISC) of the CERN Medical Applications Strategy Committee (CMASC). These requirements have been further fine-tuned in exchanges with the key stakeholders of the user community at national laboratories involved in non-clinical biomedical research (CNAO, HIT, MedAustron, PSI), as well as in discussions, for example at the ICTR-PHE 2016 and Divonne Brainstorming Meetings for Medical Applications, among others~\cite{PHE:2010, Abler:2012, AblerSurvey:2012, ICTR:2012, meeting:2012, Abler:2013, AblerPhD:2013, Dosanjh:2013, Brainstorming:2014, ICTR-PHE:2014, Brainstorming:2016, ICTR:2016}.

The parameters defining an ideal non-clinical, biomedical research facility are listed below in their entirety together with some reasons for their motivation. Section~\ref{Sec:BeamParameters_WhatCanCERNProvide} discusses to what extent the BioLEIR research facility can satisfy the ideal requirements, distinguishing "must-have" from "nice-to-have", as well as feasible requirements from non-feasible requests. The chapter closes with a summary of the resulting BioLEIR beam parameters.

\section{Requirements for an ideal, non-clinical, biomedical research facility}
\label{Sec:BeamParameters_IdealParameters}

The ISC Task Force on BioLEIR requirements~\footnote{The ISC task forces to establish biological and physics requirements for BioLEIR were composed of Philippe Lambin (chair), Kevin Prise, Jorg Pwelke and Wolfgang Dorr for the biology part, and of Alberto Delguerra, Alejandro Mazal, Marco Schippers and Michael Waligorski (chair) for the physics part.} produced a summary of the requisites for an ideal biomedical non-clinical research laboratory. The presentation~\cite{ISCBioLEIRTaskForce} builds the foundation for this chapter.

\subsection{Irradiation field parameters}
\label{SubSec:BeamParameters_IdealParameters_IrrField}
The ideal biomedical irradiation laboratory has at least one horizontal and one vertical beamline for biomedical research and possibly one more horizontal beamline for research more related to (micro-)dosimetry development and fragmentation studies. The vertical beamline allows us to treat cells in the growth medium where the cells can be kept in aqueous phase medium, with or without addition of radiation modifiers, and will be upon the culture medium base at the lowest point of the sample holder. By irradiating through the thickness of the sample holder, cells receive the full intended dose as electron equilibrium exists at the cellular level. This also avoids irradiation of cells through air or a liquid of variable thickness due to evaporation and surface tension effects. 

The following requirements are considered ideal for horizontal \& vertical beamlines for a non-clinical biomedical research facility:

\subsubsection{Ion species}
\label{SubSubSec:BeamParameters_IdealParameters_IonSpecies}
\begin{itemize}
  \item There is no need for a complete range of ions of all possible $Z$ numbers, because the range of energies that cover the tissue depths required for human radiotherapy, and their event size, are given by the lighter ions.
  \item The lighter ions suitable for radiobiology and radiotherapy are H, He, Li, Be, B, C, N and O.
  \item Heavier ions useful for research in biophysics (such as fragmentation), dosimetry, medical detector development, radiation protection and some aspects of fundamental radiobiology such as cosmic simulations, include Ne, Ar, Fe, Ar, Kr, Xe, Pb and U.
  \item Ion species shall be changeable within an hour or two for most experiments but mixed ion irradiation may be suggested as a future development, where changes within a few minutes might be required.
  \item In the future, it may be of interest to provide a radioactive source for work  on radionuclides. This is a low priority requirement.
\end{itemize}

\subsubsection{Ion energies}
\label{SubSubSec:BeamParameters_IdealParameters_IonEnergies}
The complete range of energies associated with clinical beams are required, namely up to 250\,MeV/u for protons and 440\,MeV/u for Carbon ions. Although the radiobiological properties must be tested at all energies (and with mixtures of high and low energies as occurs due to spreading out of Bragg peak positions), the most interesting increment in bioeffectiveness occurs at energies below 70\,MeV/u but measurements at higher energies are necessary as a reference to prevent systematic errors due to facility specific beam characteristics.\\

The requirements ion energy depend on the research being pursued and can be subdivided as:
\begin{itemize}
  \item Lower energies for radiobiology: $<$\,70\,MeV/u, where biological effects increase sharply.
  \item Medium energies for studies of medical applications in radiotherapy beam ballistics, dose distributions and dosimetry. Measurements at medium energies are necessary as a reference and to prevent systematic errors in RBE determination at low energies: 50 -- 500\,MeV/u.
  \item High energies for fundamental biophysics and dosimetry: $>$\,1\,GeV/u.
\end{itemize}
The energies shall be modifiable:
\begin{itemize}
  \item in a fast (10-15\,ms), uncomplicated way (spill by spill, if possible - fill by fill, if not);
  \item directly in synchrotron mode;
  \item with optional range shifters that would allow better comparison with output from cyclotron based therapy.
\end{itemize}

\subsubsection{Beam size}
\label{SubSubSec:BeamParameters_IdealParameters_BeamSize}
Beam dimensions of the following order would be useful:
\begin{itemize}
  \item Broad beams: a cross section of minimum 50$\times$50\,mm$^2$ at a distance of at least 30\,cm from the collimating systems is suggested. Further divergence with distance produces larger cross-sectional areas which in clinical situations increase to 200$\times$200\,mm$^2$ at distances of 1\,m or more. In the context of conventional radiotherapy (electrons \& photons) the geometric field size is generally defined with reference to a predetermined distance (Source to Surface Distance SSD), typically at the level of 100\,cm.
  \item Pencil beams: 5 -- 10\,mm FWHM, with good knowledge of divergence and focus point characteristics, as well as scanning capability covering the broad beam area.
\end{itemize}

These two beam size options, together with a particle fluence that can be varied over several orders of magnitude, could be useful to investigate the effect of (locally) high dose rates. In addition, physical data for Monte Carlo (MC) simulation and therapy planning shall be gathered using water and tissue equivalent phantoms, at all relevant energies and beam sizes, for mono-energetic beams as well as beams with a controlled spectrum of energies typical for clinical synchrotrons (momentum spreads of the order of $\pm$\,0.15\%). 

\subsubsection{Beam uniformity} \label{SubSubSec:BeamParameters_IdealParameters_Uniformity}
Beam uniformity satisfying the following constraints is desired:
\begin{itemize}
  \item The typical requirement of delivering a dose with 1\% accuracy requires a beam uniformity (flatness and symmetry) of 1-2\% over the entire beam dimension.
  \item Online monitoring of beam uniformity at the level of 0.5\%.
  \item It is desirable to have the capability of switching off the beam if beam uniformity deteriorates beyond the 2\% level during an experiment.
\end{itemize}

\subsubsection{Beam intensity and stability} \label{SubSubSec:BeamParameters_IdealParameters_Intensity}
The beam intensity shall be continuously variable from around 10$^9$\,ions/s downwards. The higher intensities would be for radiotherapy \& detector development, the lower intensities may be required for some radiobiology and radiation protection experiments. However, during a typical radiotherapy spill, the intensity should remain at 10$^9$\,ions/s which then shall be tailored in a continuous way for physics or other specific experiments. A beam stability of 2-3 (Max/Avg) shall be maintained during the spill~\cite{PrivCommPeters}.

For certain studies in the dosimetry domain, intensities above 10$^9$~ions/s could be useful. Fragmentation studies would see reduced data taking times with intensities at the level of 10$^{10}$\,ions/s, whereas radiation damage studies could make good use of intensities of 10$^{11}$\,ions/s.

\subsubsection{Dose delivery} \label{SubSubSec:BeamParameters_IdealParameters_DoseDelivery}
It is desirable to administer the dose in two ways:
\begin{itemize}
   \item Time driven: applying the beam during a pre-calculated duration. This method relies on a constant beam intensity.
   \item Dose driven: applying the beam and irradiating until a pre-set dose is achieved. This method is independent of dose rate, yet dose rate fluctuations should remain below the threshold of influencing dose monitor response and timing.
\end{itemize}

\subsubsection{Beam purity}
\label{SubSubSec:BeamParameters_IdealParameters_BeamPurity}
It is desirable to have a primary beam as pure as possible, with a minimal beam purity of 10$^{-4}$~\cite{PrivCommPeters}. Material at the end-user station that could lead to neutron and low energy gamma rays through beam-induced nuclear activation, shall be kept minimal. If needed, the proportion of neutron and $\gamma$-ray contamination at the irradiation spot shall be characterized with respect to energy and technique, for example with the use of scattering foils.

\subsubsection{Beam-delivery for horizontal and vertical beams} \label{SubSubSec:BeamParameters_IdealParameters_Delivery}
\begin{itemize}
  \item Spread Out Bragg Peak (SOBP): it can be realized by use of similar techniques as in clinical applications, namely -
  \begin{itemize}
    \item passive: wobbler or wedge filters.
    \item active: a commercial scanning head would be useful as long as it remains compatible with other electronics.
  \end{itemize}
  \item Building flat SOBPs: 
  \begin{itemize}
    \item Ripple filter (3 or 5\,mm).
    \item Ridge filters for most common ions in passive mode (passive SOBP).
    \item Multi Leaf Collimator (MLC) for the study and development of conformal dose distributions using MLCs. MLCs as part of an experiment shall be on the horizontal beamline only.
  \end{itemize}
  \item Pencil beam:
  \begin{itemize}
    \item should be available on request.
    \item require good knowledge of FWHM, divergence and focus point.
    \item should have multi-spot capabilities rather than use of a single pencil beam.
  \end{itemize}
  \item Beam gating:
  \begin{itemize} 
    \item beam gating as a technique for isolation of motion of target volumes (tumors) situated in anatomical locations which are susceptible to movement i.e lungs, with respect to the beam. This would be interesting for testing of methods and would be applicable to the horizontal beamline only.
   \end{itemize}
\end{itemize}

\subsubsection{Requirements for dosimetry and fluence monitoring} \label{SubSubSec:BeamParameters_IdealParameters_Dosimetry}
Dose measurements must be possible at low and high dose rates, as well as for low and high Low Energy Transfer (LET) particles. 
\begin{itemize}
  \item Basic dosimetry: ionization chambers and other techniques, solid state, calorimetry and chemical systems, as required in a medical clinic. The standards required by Swiss/French national reference laboratories for absorbed doses should be applied \textit {(IAEA TRS-398, Code of practice for the determination of absorbed doses in external beam therapy, section 4.2.1 gives recommendations for ion chambers).}
  \item Dose measurement precision at the isocenter should be within 2\%, in reference to a nationally defined dose either in France or Switzerland.
  \item Measurements of LET is also required (Roose chambers).
  \item A position sensitive monitor for beam position and shape measurements.
  \item Study of microdosimetric properties of the clinical beams. 
  \item Testing of new microdosimeters, and even sub-microdosimeters.
\end{itemize}

\subsubsection{Biomedical infrastructure} \label{SubSubSec:BeamParameters_IdealParameters_BiomedicalInfrastructure}
An ideal biomedical research facility shall house a state-of-the-art biomedical infrastructure. In particular, a dedicated, on-site tissue culture laboratory with the following basic requirements is desired:
\begin{itemize}
  \item Biologically restricted areas.
  \item At least four incubators.
  \item Three flowbenches for handling of sterile samples.
  \item One fumehood.
  \item Refrigerators for storage of growth media.
  \item Freezer (-20$^{\circ}$C) for cell cultures.
  \item Freezer (-80$^{\circ}$C) for longer storage of cells and biological samples.
  \item Two centrifuges for separation techniques.
  \item Cell counters to confirm experimental cell numbers per mm$^3$.
  \item Autoclaver.
  \item Special glassware dishwasher to obtain clean surfaces.
  \item Two standard microscopes to confirm cell morphology and exclude contamination.
  \item Gas supply (including O$_2$ and CO$_2$ for optimal cell growth in the incubators).
  \item Water sterilizer.
  \item At least 10 meters of bench space.
  \item Standard biolab consumables: gloves, pipettes etc.
  \item One UV microscope for advanced molecular studies.
  \item Biohazard disposal area - could be further disposed at a nearby hospital.
\end{itemize}

\subsubsection{Other equipment and aspects} \label{SubSubSec:BeamParameters_IdealParameters_Other}
Non-beam related, essential equipment for a biomedical research facility which tests bioeffectiveness of high linear energy transfer (LET) radiations include:
\begin{itemize}
  \item Access to a reference radiation source with a variable dose rate 0.5 -- 5\,Gy/min:
  \begin{itemize}
    \item preferably an X-ray reference unit capable of delivering constant dose rates, comparable to those found in current clinical settings:
    \begin{itemize}
      \item a basic linac operating at 4 -- 6\,MeV, that delivers low LET radiation, as used in hospitals.
      \item a 250\,kVp (peak kilovoltage) X-ray unit with good filtration of low energy photons (which have higher LET values).
    \end{itemize}
    \item less ideal would be a Co-60 reference unit. 
Ideally operating at 2\,Gy/min, it requires an incremental source - sample distance reduction with time to compensate for radiation decay.
  \end{itemize}
  \item At each isocenter it shall be necessary to have good handling facilities:
  \begin{itemize}
    \item remotely controlled and logged 3D robotic movement of sample holders with sub-mm precision (0.01\,mm). 
    \item possibly a small conveyor belt or robot to allow multiple cells to be irradiated in rapid succession.
    \item water phantom positioning (PTW MP3 or similar).
    \item orthogonal laser guides for exact re-positioning and other optical x-y-z 3D registration system.
    \item video surveillance during irradiation with capability of high-resolution photos of the sample.
    \item data transfer systems.
    \item power supplies.
  \end{itemize}
  \item Radiation safety equipment to detect contamination and activation.
  \item Access to the experiment shall be possible within several seconds of switching off the beam.
\end{itemize}

\subsubsection{Mechanical infrastructure: experimental setup} \label{SubSubSec:BeamParameters_IdealParameters_ExpSetup}
The experimental setup shall be designed to be modular and flexible, such that it can be easily adapted to incorporate future research topics and technology development.

\subsubsection{Services \& support} \label{SubSubSec:BeamParameters_IdealParameters_Services}
A number of general services shall be available in close vicinity of the three irradiation spots:
\begin{itemize}
  \item Access to machine shop facilities for:
  \begin{itemize}
    \item building custom sample holders.
    \item production of custom phantoms.
  \end{itemize}
  These should be agreed upon by a steering committee and obtained separately by order.
  \item Counting hut with ample cabling possibilities for experimental measurement station (BNC, RS232, Ethernet etc.).
  \item Ten offices for visiting groups with space for 2 -- 4 researchers in each, in close proximity to the end-user station facility.
  \item A (part-time) Site Manager for BioLEIR, ideally a medical physicist seconded to BioLEIR from the biomedical collaboration trained in dosimetry, who:
  \begin{itemize}
     \item is responsible for the dose calibration.
     \item instructs on use of equipment.
     \item collaborates with users.
     \item liaise with accelerator staff and workshop.
     \item designs and maintains dosimetry and control equipment, power supplies, communication networks, data transfer etc.
     \item performs his own research on site.
   \end{itemize}
\end{itemize}

\section{Discussion of the feasibility of ideal parameters at the Bio\-LEIR facility}
\label{Sec:BeamParameters_WhatCanCERNProvide}
In this section, the feasibility of all biomedical irradiation facility parameters are discussed.

Table~\ref{Tab:BeamParameters_WhatCanCERNProvide_IonIntensities} shows the desired ion intensities on target for different ion species and their corresponding charge intensities within the LEIR accelerator, as well as the currents required at the ion source output. For this document, only the ion intensity on target was used, rather than a required dose. In the final facility, a prescribed dose rather than an ion intensity over a specific period of time, is applied to a line of cells.

A simple calculation was done starting from the desired ion intensities on target and calculating back the necessary ion beam currents out of the source, taking into account the transmission percentages at different stages of the machine, as indicated in table~\ref{Tab:BeamParameters_WhatCanCERNProvide_Transmissions}.

\begin{table}[!htb]
\caption{Ion intensities on target, with their corresponding ion and charge intensities at the different machine stages, all the way back to the ion source. The assumed transmission efficiencies are listed in table~\ref{Tab:BeamParameters_WhatCanCERNProvide_Transmissions}.}
\label{Tab:BeamParameters_WhatCanCERNProvide_IonIntensities}
\centering
\resizebox{\textwidth}{!}{%
\begin{tabular}{l r c r r c r r r r}
\hline
\rule{0pt}{3ex}\textbf{Ion} 	& \textbf{Mass}	& \textbf{Charge}	& \textbf{Energy}	& \textbf{Energy per}	& \textbf{Charge state}	& \textbf{Current out}		& \textbf{Charges/Cycle}	& \textbf{Charges/Cycle}	& \textbf{Current out} 	\\
\textbf{species}	&	\textbf{}	&	\textbf{}		& \textbf{[MeV/u]}		& \textbf{nucleon}	& \textbf{at source}	& \textbf{of source} & \textbf{at source}	& \textbf{bef linac}		& \textbf{of linac}\\
\textbf{}	&	\textbf{}	&	\textbf{}		& \textbf{}		& \textbf{[J]}	& \textbf{}	& \textbf{[e$\mu$A]} & \textbf{}	& \textbf{}		& \textbf{[e$\mu$A]}\\
\hline
\rule{0pt}{3ex}Pb	& 208	& 54	& 72.25	& 1.16$\times 10^{-11}$	& 29	& 7.49	& 5.05$\times 10^{+11}$	& 6.58$\times 10^{+11}$	& 8.79\\
Ar	& 40 &	11 &	124.36 &	1.99$\times 10^{-11}$ &	11 &	13.95 &	1.92$\times 10^{+11}$ &	1.34$\times 10^{+11}$ &	8.79 \\
\hline
\rule{0pt}{3ex}O &	16 &	8 &	250-440 &	4.01$\times 10^{-11}$ &	4 &	6.98 &	6.97$\times 10^{+10}$ &	9.75$\times 10^{+10}$ &	8.79 \\
C &	12 &	6 &	250-440 &	4.01$\times 10^{-11}$ &	3 &	0.01 &	5.23$\times 10^{+07}$ &	7.32$\times 10^{+07}$ &	0.0088 \\
C &	12 &	6 &	250-440 &	4.01$\times 10^{-11}$ &	3 &	0.70 &	5.23$\times 10^{+09}$ &	7.32$\times 10^{+09}$ &	0.88 \\
C &	12 &	6 &	250-440 &	4.01$\times 10^{-11}$ &	3 &	69.77 &	5.23$\times 10^{+11}$ &	7.32$\times 10^{+11}$ &	87.91\\
He & 4 &	2 &	250-440 &	4.01$\times 10^{-11}$ &	1 &	6.98 &	1.74$\times 10^{+10}$ &	2.44$\times 10^{+10}$ &	8.79 \\
H$^{3+}$ &3 &	1 &	250-440 &	4.01$\times 10^{-11}$ &	1 &	0.47 &	1.74$\times 10^{+09}$ &	1.22$\times 10^{+09}$ &	0.88 \\
H$^{3+}$ &3 &	1 &	250-440 &	4.01$\times 10^{-11}$ &	1 &	46.51 &	1.74$\times 10^{+11}$ &	1.22$\times 10^{+11}$ &	87.91\\
H$^{3+}$ &3 &	1 &	250-440 &	4.01$\times 10^{-11}$ &	1 &	465.14 &1.74$\times 10^{+12}$ &	1.22$\times 10^{+12}$ &	879.12\\
\hline
\hline
\end{tabular}
}
\vskip 1.5em
\resizebox{\textwidth}{!}{%
\begin{tabular}{l r r r r r r r r}
\hline
\rule{0pt}{3ex}\textbf{Ion} 	& \textbf{Charges/Cycle}	& \textbf{Charges/Cycle}	& \textbf{Charges/Cycle}	& \textbf{Charges/Cycle}	& \textbf{Ions/Cycle}	& \textbf{Ions/Cycle} 	& \textbf{Ions/Cycle}	& \textbf{Nominal Ions/s}\\
\textbf{species}	&	\textbf{before}	& \textbf{before}	& \textbf{stacking/}	& \textbf{in LEIR}	&  \textbf{in LEIR}	&  \textbf{after}	&  \textbf{on Target} &  \textbf{on Target}\\
\textbf{}	&	\textbf{Transferline}	& \textbf{injection}	& \textbf{ramping}	& \textbf{}	&  \textbf{}	&  \textbf{Extraction}	&  \textbf{} &  \textbf{}\\
\hline
\rule{0pt}{3ex}Pb & 5.93$\times 10^{+11}$ & 5.33$\times 10^{+11}$	& 2.67$\times 10^{+11}$	& 2.40$\times 10^{+11}$	& 4.44$\times 10^{+09}$	& 2.67$\times 10^{+09}$	& 2.4$\times 10^{+09}$	& $10^{+09}$ \\
Ar & 1.21$\times 10^{+11}$ &	1.09$\times 10^{+11}$ &	5.43$\times 10^{+10}$ &	4.89$\times 10^{+10}$ &	4.44$\times 10^{+09}$ &	2.67$\times 10^{+09}$ &	2.4$\times 10^{+09}$ &	$10^{+09}$ \\
\hline
\rule{0pt}{3ex}O & 8.78$\times 10^{+10}$ &	7.90$\times 10^{+10}$ &	3.95$\times 10^{+10}$ &	3.56$\times 10^{+10}$ &	4.44$\times 10^{+09}$ &	2.67$\times 10^{+09}$ &	2.4$\times 10^{+09}$ &	$10^{+09}$ \\
C & 6.58$\times 10^{+07}$ &	5.93$\times 10^{+07}$ &	2.96$\times 10^{+07}$ &	2.67$\times 10^{+07}$ &	4.44$\times 10^{+06}$ &	2.67$\times 10^{+06}$ &	2.4$\times 10^{+06}$ &	$10^{+06}$ \\
C & 6.58$\times 10^{+09}$ &	5.93$\times 10^{+09}$ &	2.96$\times 10^{+09}$ &	2.67$\times 10^{+09}$ &	4.44$\times 10^{+08}$ &	2.67$\times 10^{+08}$ &	2.4$\times 10^{+08}$ &	$10^{+08}$ \\
C & 6.58$\times 10^{+11}$ &	5.93$\times 10^{+11}$ &	2.96$\times 10^{+11}$ &	2.67$\times 10^{+11}$ &	4.44$\times 10^{+10}$ &	2.67$\times 10^{+10}$ &	2.4$\times 10^{+10}$ &	$10^{+10}$ \\
He & 2.19$\times 10^{+10}$ &	1.98$\times 10^{+10}$ &	9.88$\times 10^{+09}$ &	8.89$\times 10^{+09}$ &	4.44$\times 10^{+09}$ &	2.67$\times 10^{+09}$ &	2.4$\times 10^{+09}$ & $10^{+09}$ \\
H$^{3+}$& 1.10$\times 10^{+09}$ &	9.88$\times 10^{+08}$ &	4.94$\times 10^{+08}$ &	4.44$\times 10^{+08}$ &	4.44$\times 10^{+08}$ &	2.67$\times 10^{+08}$ &	2.4$\times 10^{+08}$ &	$10^{+08}$ \\
H$^{3+}$& 1.10$\times 10^{+11}$ &	9.88$\times 10^{+10}$ &	4.94$\times 10^{+10}$ &	4.44$\times 10^{+10}$ &	4.44$\times 10^{+10}$ &	2.67$\times 10^{+10}$ &	2.4$\times 10^{+10}$ &	$10^{+10}$ \\
H$^{3+}$& 1.10$\times 10^{+12}$ &	9.88$\times 10^{+11}$ &	4.94$\times 10^{+11}$ &	4.44$\times 10^{+11}$ &	4.44$\times 10^{+11}$ &	2.67$\times 10^{+11}$ &	2.4$\times 10^{+11}$ &	$10^{+11}$ \\
\hline
\hline
\end{tabular}
}
\end{table}

\begin{table}[!htb]
\centering
\caption{Assumed transmission efficiencies in the different BioLEIR machine elements.}
\label{Tab:BeamParameters_WhatCanCERNProvide_Transmissions}
\begin{tabular}{l r}
\hline
\rule{0pt}{3ex}
\textbf{System} & \textbf{Transmission efficiency [\%]} \\
\hline
\rule{0pt}{3ex}Ion source to linac & 70 \\
linac & 90 \\
LEIR injection transfer line & 90 \\
LEIR injection & 50 \\
LEIR stacking/ramping & 90 \\
Extraction & 60 \\
Beamlines & 90 \\
\hline
\end{tabular}
\end{table}

We conclude from the ion intensity (table~\ref{Tab:BeamParameters_WhatCanCERNProvide_IonIntensities}) that BioLEIR can produce the desired ion intensities on target. As discussed in more detail in chapter~\ref{Chap:Source}, the baseline light ion source Supernanogan ECRIS by Pantechnik is expected to be able to deliver all required ion currents. Extensive operative experience with the LEIR accelerator with Lead and Argon ions showed that charges/cycle of the order of 2.4$\times$10$^{11}$ are, in principle, possible. There may be smaller operational differences with different ion species, yet it is expected that the required high-priority ion intensities on target, as specified in table~\ref{Tab:BeamParameters_IdealParameters_FullListOfRequirements}, can be produced. Carbon ions at intensity levels of 10$^{10}$, as well as protons at intensities of 10$^{11}$ may be at or just above the limit of current LEIR accelerator stability. Such high intensity beams are considered a nice-to-have request in order to reduce sample or detector irradiation time. Even though these high ion and proton fluxes are, in principle, possible to be produced at BioLEIR, it must be noted that should they be required for a baseline solution, proper, much more expensive and massive shielding thicknesses around the irradiation rooms will be necessary (see chapter~\ref{Chap:RP} for details on shielding requirements). 

For the purpose of costing a baseline biomedical facility, the ion intensities have been restricted to maximally 10$^{10}$ for protons, and 10$^8$ for ions. More details on the trade-off between maximum energies and maximum particle intensities and their corresponding necessary shielding requirements are found in chapter~\ref{Chap:RP}.

The biomedical requirements are split into high priority and low to medium priority ones, according to their definition in table \ref{Tab:BeamParameters_IdealParameters_FullListOfRequirements}. They are listed in their respective tables \ref{Tab:BeamParameters_Operations_MachineRequirementsLowImportance} and~\ref{Tab:BeamParameters_Operations_MachineRequirementsHighImportance} and, where useful, links to the relevant chapters on more detailed technical discussions are given in the comment area for each requirement.


\subsection{Ion species}
\label{SubSec:BeamParameters_WhatCanCERNProvide_IonSpecies}
One lower priority requirement for ion species 
is access to heavier ions for more advanced biophysics (such as fragmentation studies), dosimetry, detector development, radiation protection and  radiobiology. In particular, the following heavy ions are suggested: Ne, Ar, Fe, Ar, Kr, Xe, Pb and U.

\begin{landscape}
\begin{longtable}{m{2.2cm} | m{2.8cm} | m{5.5cm} | m{4.2cm} | m{1.7cm} | m{4.5cm} }
\caption{Requirements for an ideal biomedical research facility.}
\label{Tab:BeamParameters_IdealParameters_FullListOfRequirements}\\
\hline
\bf{Requ.} & \bf{Title} & \bf{Description} & \bf{Details} & \bf{Priority} & \bf{Comment} \\
		\endfirsthead
		
		\multicolumn{6}{c}%
		{{\bfseries \tablename\ \thetable{} -- continued from previous page}} \\
        \hline
		\bf{Requ.} & \bf{Title} & \bf{Description} & \bf{Details} & \bf{Priority} & \bf{Comment}\\
		\endhead
		
		\multicolumn{6}{r}{{Continued on next page}}
		\endfoot
		
		\hline
		\endlastfoot
		
        \hline \hline
        \rule{0pt}{3ex}R0 Beamlines & 1 horizontal;  1 vertical & & & HIGH & \\
        \hline \hline
        \rule{0pt}{3ex}R1 Ion species & & & & & \\
        R1.1 & light ions & H, He, Li, Be, B, C, N, O & suitable for radiobiology \& radiotherapy &  HIGH &  \\ 
        \hline
        R1.2 & heavy ions & Ne, Ar, Fe, Kr, Xe, Pb, U & biophysics, dosimetry, medical detector development, radiation protection and fundamental radiobiology &  LOW & \\ 
        \hline
        R1.3 &  ion species change & changeable within 1-2 hours &  & HIGH &  potential development: mixed ion irradiation with species change within a few minutes\\ 
        \hline
        R1.4 &  radioactive ion beam &  a potential future development for work on radionuclides &  & LOW &  \\ 

        \hline \hline
        \rule{0pt}{3ex}R2 Ion energies & & & & &\\
        R2.1 & low &  $<$\,70\,MeV/u & for vertical beam setup &  HIGH & \\ 
        \hline
        R2.2 & medium & 50-440\,MeV/u & 440\,MeV/u: maximal clinical therapeutic particle energies &  HIGH &   \\ 
        \hline
        R2.3 & high & $>$\,1\,GeV/u & for fundamental biophysics and dosimetry & LOW & \\ 
        \hline \hline
        \rule{0pt}{3ex}R3 Energy change & & & & &\\
        R3.1 & fast and simple & 10-15\,ms & spill-by-spill & HIGH & \\ 
        \hline
        R3.2 & in synchrotron mode &  &  & HIGH & \\ 
        \hline
        R3.3 & optional range shifters & & & HIGH & allow better comparison with output from cyclotron based therapy\\ 
        \hline \hline
        \rule{0pt}{3ex}R4 Beam uniformity & 1-2\,\% & intensity uniformity & flatness and symmetry & HIGH & Feedback interlock\\
        \hline \hline
        \rule{0pt}{3ex}R5.1 Beam intensity & & & & &\\
        R5.1.1  & 10$^9$\,ions/s & continuously variable downwards & clinical & HIGH &\\
        R5.1.2 & 10$^{10}$\,ions/s & & fragmentation studies & MEDIUM &\\
        R5.1.3 & 10$^{11}$\,ions/s & & radiation damage studies & LOW &\\
        \hline
        \rule{0pt}{3ex}R5.2 Dose delivery & & & & &\\
        R5.2.1 & time-driven & & & HIGH &\\
        R5.2.2 & dose-driven & & & LOW &\\
        \hline \hline
        \rule{0pt}{3ex}R6 Beam purity & 10$^{-4}$ & & & HIGH &\\
        \hline \hline
        \rule{0pt}{3ex}R7 Beam delivery & & & & &\\
        R7.1 Broad beams & & $>$50$\times$50\,mm$^2$ at 30\,cm SSD &  & HIGH &\\ 
        R7.1.1 & SOBP & passive: wobbler/wedge filters & & HIGH &\\
        R7.1.2 & SOBP & active: commercial scanning head & & HIGH &\\
        \hline
        \rule{0pt}{3ex}R7.2 Flat SOBP & & & & & \\
        R7.2.1 & ripple filter: 3 or 5\,mm & & & HIGH &\\
        R7.2.2 & & ridge filter & & HIGH &\\
        R7.2.3 & & multi leaf collimator & & HIGH &\\
        \hline
        \rule{0pt}{3ex}R7.3 Pencil beams & & 5-10\,mm FWHM & deflection capability to cover area of R7.1 & HIGH &\\            
        R7.3.1 & pencil beam & good FWHM knowledge, divergence and focal point & & HIGH &\\
        R7.3.2 & pencil beam & multi-spot capabilities & & MEDIUM &\\
        \hline
        \rule{0pt}{3ex}R7.4 & Beam gating & & & LOW & \\
        \hline \hline
        \rule{0pt}{3ex}R8 Dosimetry and fluence monitoring & & & & &\\
        R8.1 & basic dosimetry & precision at isocenter: 2\% & reference to a nationally defined dose & HIGH &\\
        R8.2 & LET measurement & Roose chambers & & HIGH &\\
        R8.3 & microdosimetry & & & HIGH &\\
        \hline \hline
        \rule{0pt}{3ex}R9 Non-beam related equipment & & & & &\\
        R9.1 Biomedical infrastructure & & & & HIGH &\\
        R9.2. Reference radiation sources & variable dose rate: 0.5-5\,Gy/min & & & &\\
        R9.2.1.1 & X-ray reference unit & basic linac at 5\,MeV & & HIGH &\\
        R9.2.1.2 & & 250\,kVp X-ray unit & & MEDIUM &\\
        R9.2.2 & Co-60 reference unit & & & MEDIUM &\\
        \hline
        \rule{0pt}{3ex}R9.3. Equipment at each isocenter & & & & &\\
        R9.3.1 & 3D robotic movement & sub-mm precision & & HIGH &\\
        R9.3.2 & water phantom & PTW MP3 or similar & & HIGH &\\
        R9.3.3 & optical 3D positioning system & orthogonal laser guides or equivalent & & HIGH &\\
        R9.3.4 & video surveillance & & & HIGH &\\
        R9.3.5 & data transfer systems & & & HIGH &\\
        R9.3.6 & power supplies & & & HIGH &\\
        \hline 
\end{longtable}
\end{landscape}

\begin{landscape}
\begin{longtable}{m{2 cm} m{7.5 cm} m{10.5 cm}}
\caption{Facility requirements with low or medium priority.}
\label{Tab:BeamParameters_Operations_MachineRequirementsLowImportance}\\
\hline
        \bf{Requ.} & \bf{Description} & \bf{Comment}\\
        \hline \hline
        R1.2 & ion species: Ne, Ar, Fe, Kr, Xe, Pb, U & Acceleration of heavy ions at LEIR is possible, yet only to energies lower than clinical energies due to magnet limitations, see chapter \ref{Chap:LEIR}. Access to heavy ions at clinical energies could be possible elsewhere at CERN; they are extracted to the North Area. \\ 
        R1.3.1 & Potential development: mixed ion irradiation with species change within a few minutes. & Ad-hoc non obvious implementation, outside the scope of the current study. \\
        R1.4 &  Potential development: radioactive ion beam. & For work on radionuclides.\\ 
        \hline \hline
       R2.3 & ion energy: $>$\,1\,GeV/u & Those ion energies are not accessible at LEIR due to magnet limitations (see chapter~\ref{Chap:LEIR}), yet of course exist elsewhere at CERN. \\ 
        \hline \hline
        R5.1.2 & 10$^{10}$\,ions/s & fragmentation studies\\
        R5.1.3 & 10$^{11}$\,ions/s & radiation damage \\
        R5.2.2 & dose-driven dose delivery & \\
       \hline \hline
        R7.3.2 & pencil beam: multi-spot capabilities & \\
        R7.4 & beam gating &\\
        \hline \hline
        R9.1 & biomedical infrastructure as specified &\\
        R9.2.1.2 & X-ray reference unit: 250\,kVp Xray unit &\\
        R9.2.2 & reference unit: Co-60 reference unit &\\
\hline \hline
\end{longtable}
\end{landscape}

\begin{landscape}
\begin{longtable}{m{2cm} m{7.5cm} m{10.5cm}}
\caption{Facility requirements with high priority.}
\label{Tab:BeamParameters_Operations_MachineRequirementsHighImportance}\\
\hline
        \bf{Requ.} & \bf{Description} & \bf{Comment}\\
        \hline \hline
        R1.1 & ion Species: H, He, Li, Be, B, C, N, O  & Particular interest in CERN providing exotic beams \\ 
        \hline
        R1.3 &  ion species changeable within 1-2 hours & Input from source setup \& RF configuration setup\\
        \hline \hline
        R2.1 & ion energy: $<$\,70\,MeV/u for vertical beam & Study of shielding and integration \\
        \hline
        R2.2 & ion energy: 50-440\,MeV/u & 440\,MeV/u: maximal clinical therapeutic particle energies \\
        \hline \hline
        R3.1 & energy change: 10-15\,ms, spill-by-spill & \\ 
        \hline
        R3.2 & energy change in synchrotron mode & \\ 
        \hline
        R3.3 & energy range shifters &  \\ 
        \hline \hline
        R4 & beam intensity, flatness and symmetry uniformity: 1-2\% & \\
        \hline \hline
        R5.1.1 & beam intensity: 10$^9$\,ions/s continuously variable downwards & \\
        \hline
        R5.2.1 & time-driven dose delivery & \\
        \hline \hline
        R6 & beam purity: 10$^{-4}$ & \\
        \hline \hline
        R7.1 & broad beams $>$\,50$\times$50\,mm$^2$ at 30\,cm SSD & \\ 
        R7.1.1 & wobbler/wedge filters & \\
        R7.1.2 & commercial scanning head & \\
        R7.2.1 & ripple filter: 3 or 5\,mm & \\
        R7.2.2 & ridge filter& \\
        R7.2.3 & multi leaf collimator &\\
        \hline
        R7.3 & pencil beams: 5-10\,mm FWHM \& deflection capability to cover $>$\,50$\times$50\,mm$^2$ at 30\,cm SSD & \\
        R7.3.1 & pencil beams: good FWHM knowledge, divergence and focal point & \\
        \hline \hline
        R8.1 & dose measurement at isocenter: 2\% &\\
        R8.2 & LET measurement w/ Roose chambers &\\
        R8.3 & microdosimetry & \\
        \hline \hline
        R9.1 & biomedical infrastructure as specified & \\
        R9.2.1.1 & X-ray reference unit: basic linac at 5\,MeV & \\
        R9.3 & equipment at each isocenter as specified & \\
\hline \hline
\end{longtable}
\end{landscape}

It shall be noted that this requirement cannot be easily realized at the BioLEIR facility located at the LEIR accelerator, in a transparent way to the existing research programme, because:
\begin{itemize}
  \item the new light ion linac (LINAC5) will be optimized for light ions up to Oxygen (see chapter~\ref{Chap:Linac}).
  \item the existing heavy ion linac (LINAC3) could accelerate those ions to the required energies, yet an extraction at the clinical energies at the LEIR ring is not possible with the current setup, due to magnet limitations as detailed in chapter~\ref{Chap:LEIR}.
  \item heavy ion extraction is into the LHC or to the CERN North Area.
\end{itemize}

\subsection{Ion energies}
\label{SubSec:BeamParameters_WhatCanCERNProvide_IonEnergies}

\subsubsection{Access to high energy ions ($>$1\,GeV/u)}
\label{SubSubSec:BeamParameters_WhatCanCERNProvide_IonEnergies_HighEnergy}

Another lower-priority requirement is the access to high ion energies (section~\ref{SubSubSec:BeamParameters_IdealParameters_IonEnergies}), greater than 1\,GeV/u. Measurements at higher energies would be useful as a reference to prevent systematic errors due to site specific beam characteristics.

This requirement cannot be realized at the BioLEIR facility as it is currently contemplated at the LEIR accelerator because:
\begin{itemize}
  \item a planned power upgrade for the LEIR accelerator magnets will allow LEIR to accelerate light ions (up to Oxygen) up to a limit of 440\,MeV/u (see section~\ref{Sec:Power_upgrade_LEIR}).
  \item after the power upgrade, the magnets installed at the LEIR accelerator will be operating at their limit. A desire to accelerate to higher energies at LEIR will necessitate more powerful magnets.

  \item ions are accelerated to higher energies and extracted elsewhere at CERN, for example in the CERN North Area. However, the fixed target research lines in the North Area have to date not been considered for biomedical research and considerations for this are beyond the scope of the reflections for the BioLEIR project.
\end{itemize}

\subsubsection{Energy switching}
\label{SubSubSec:BeamParameters_WhatCanCERNProvide_EnergySwitching}
Another required parameter concerns fast, uncomplicated energy switching in order for Bragg peaks to cover clinically relevant target dimensions. It is considered fast to switch the beam energy within 10 - 15\,ms. The energy change shall be achieved with two modalities: directly at the synchrotron level and passively with a range shifter. Magnet hysteresis must be taken into account when exploiting the synchrotron to regulate the irradiation beam energy. The simplest way to achieve this requirement, would be to accept a beam energy switching time of a few seconds as it would allow to set the LEIR accelerator to a new beam energy for a new cycle (cycling time of a few seconds). This method is considered quite sufficient for cell irradiations. However, should faster energy switching be desired, it can be achieved passively with a range shifter which would be installed just at the start of the new beamlines, before the LEIR shielding wall.

\subsubsection{Dose delivery}
\label{SubSubSec:BeamParameters_WhatCanCERNProvide_DoseDelivery}
Administering the dose in a time-driven way is easier to implement and therefore most likely the method to start with. Further on, it can be aimed to implement the lower priority dose-driven approach.

\subsubsection{Beam purity and beam delivery}
\label{SubSubSec:BeamParameters_WhatCanCERNProvide_BeamPurity}
In order to maintain the required beam purity at the level of 10$^{-4}$, interaction with the beam shall be avoided as much as possible. For those instances where specific beam delivery methods require interaction with the beam, those interactions take place upstream right after the slow extraction, before the LEIR shielding wall, such as not to influence the beam quality negatively. The question of whether the required beam purity can be achieved remains open and needs study.
%
%
\subsection{Operational feasibility aspects on access to heavy ions and high-energy ions}
\label{SubSec:BeamParameters_Operations}
Scenarios could be imagined where all requirements voiced by the biomedical user community are satisfied. However, a number of operational aspects shall be taken into account when discussing meeting the request to have access to heavy ions, as well as access to higher energy ions at the BioLEIR facility:
\begin{itemize}
  \item Creation of BioLEIR facility shall remain transparent to the existing CERN physics programme, notably the heavy ion physics programme for which LEIR is used as a pre-accelerator.
  \item Increasing the accessible magnetic field at LEIR to accelerate ions to higher energies (i.e. 440\,MeV/u) than currently accessible entails considerable cost. The potential impact on the 
  CERN heavy ion physics programme must be taken into consideration.

  \item Potential access to heavy ions as well as higher energy ions at the Proton Synchrotron (PS) or North Area could be investigated, yet is considered outside the scope of the BioLEIR study.
\end{itemize}

\section{Summary of the BioLEIR beam parameters}
\label{Sec:BeamParameters_ParameterSummary}
Table~\ref{Tab:BeamParameters_Operations_MachineRequirementsHighImportance} summarizes the high priority requirements and shows that the planned potential transformation of the LEIR accelerator at CERN could well satisfy all high priority requirements for a non-clinical biomedical research facility.
No detriment to the pre-existing CERN research programmes could be identified.

In short summary, the LEIR accelerator at the BioLEIR facility shall provide a beam for a range of light ions from protons up to at least Oxygen for the purposes of biomedical research. An extension up to and beyond Iron would additionally allow experiments relevant to space research as well as in fundamental dosimetry and radiobiology - however, such an extension is not part of the currently planned ion range for BioLEIR. The physical beam parameters of the BioLEIR facility will be as similar as possible to the achievable beam parameters as implemented in already existing facilities based on the Proton-Ion Medical Machine Study (PIMMS)~\cite{PIMMS}: HIT~\cite{HIT}, CNAO~\cite{CNAO} and MedAustron~\cite{MedAustron}. This shall facilitate collaboration between all facilities providing beam for non-clinical biomedical research.\\

The BioLEIR biomedical end-station shall have 3 experimental beamlines: 
\begin{itemize}
   \item Two horizontal beamlines with an energy range between 50\,MeV/u and the full clinical energies of up to $\approx$440\,MeV/u:
   \begin{itemize}
      \item one horizontal beamline, H1, shall be used preferentially for biomedical experiments;
      \item the other horizontal beamline, H2, could be used for hardware developments: dosimetry and detector development.
   \end{itemize}
\item One vertical beamline, V, with energies up to 70\,MeV/u.
\end{itemize}


The lower priority biomedical requests will a priori not be satisfied at the BioLEIR facility. Table~\ref{Tab:BeamParameters_BioLEIR_BeamParameters} summarizes the beam parameters with which the BioLEIR facility is planned to operate.

\begin{table}[!htb]
\centering
\caption{Summary of the operational beam parameters for the BioLEIR facility.}
\label{Tab:BeamParameters_BioLEIR_BeamParameters}
\begin{tabular}{l r c c}
\hline
\rule{0pt}{3ex}\textbf{Ion species} & & proton & Helium to Argon \\
\textbf{Energies}  & H: & 50-250 & 50-440 \\
\textnormal{[MeV/u]} &  V: & $<$70 & $<$70\\
\textbf{Intensities} & & & \\
\textnormal{[ions/s on target]} 
  & H: & 10$^{9}$-10$^{10}$ & 10$^{8}$-10$^{9}$\\
  & V: & 10$^{8}$& 10$^{8}$\\
\hline
\end{tabular}
\end{table}

%% file: Chapters/IonSource.tex
\chapter{Light Ion Source}
\label{Chap:Source}

\begin{figure}[ht]
\centering
\includegraphics[width=0.4\textwidth]{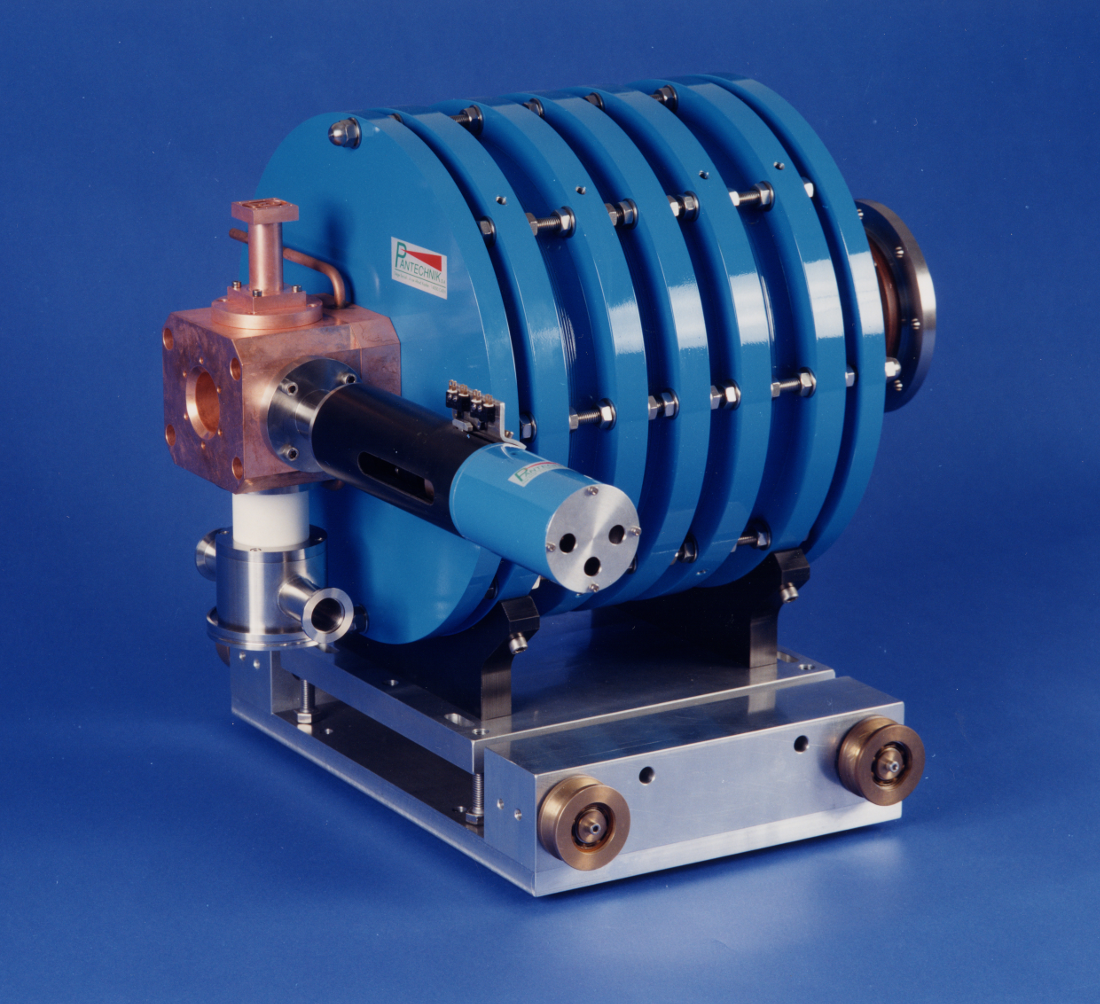}
\includegraphics[width=0.4\textwidth]{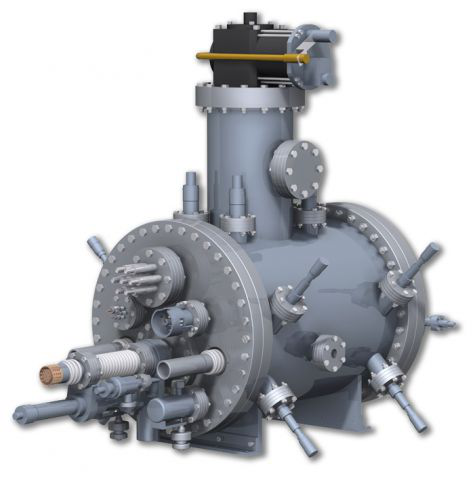}
\caption{The Supernanogan ECRIS by Pantechnik S.A. (left) \cite{pantechnik} 
and the Dresden EBIS-SC by DREEBIT GmbH (right) \cite{dreebit}.}
\label{Fig:Source_Supernanogan_ebis-sc}
\end{figure}

Many different ion source types exist for the production of ions from a wide
variety of chemical elements. For multiply charged ions, a handful of suitable 
ion sources exist. In particular, the Electron Cyclotron Resonance Ion Source
(ECRIS) and the Electron Beam Ion Source (EBIS) are widely adopted in 
accelerator facilities around the world. Currently all of the operational 
ion beam therapy facilities and the new facilities under construction 
that provide ions heavier than protons use ECR ion sources \cite{kitagawa10}.
EBIS ion sources are currently not used in clinical ion beam therapy in any of the 
existing facilities but this option has been the topic of many discussions 
and several published studies spanning the last 35~years
\cite{hamm79,becker92,kester96,becker98,zschornack07,zschornack09,
zschornack10,zschornack11,schmidt12}. As such, we focus on these 
two ion sources and compare their features and performance with the 
requirements set by the radiobiological applications. The commercially 
available compact all-permanent magnet Supernanogan ECRIS by Pantechnik S.A 
(figures~\ref{Fig:Source_Supernanogan_ebis-sc}) \cite{pantechnik} has been chosen 
to represent the ECR ion sources, as it is widely used in existing and 
under-construction medical ion beam facilities. In addition, it shares 
many similar features with other ECR ion sources that are used for the 
same purpose (e.g. the all-permanent magnet KeiGM at GHMC, Japan
\cite{kitagawa10}). The Dresden EBIS-SC, a compact liquid cryogen free 
EBIS by DREEBIT GmbH (figure~\ref{Fig:Source_Supernanogan_ebis-sc}) \cite{dreebit}, 
has been developed and marketed for medical ion beam therapy, and represents 
the commercially available compact option for EBIS. 
The main technical specifications of these two ion sources are summarized 
in table~\ref{Tab:Source_Ion_Sources}.

\begin{table}[ht]
\centering
\caption{Main technical specifications of the Supernanogan ECRIS and 
the Dresden EBIS-SC. The information has been collected from references 
\cite{zschornack09,zschornack10,pantechnik,dreebit,ciavola08_cnao,sun12,nakagawa13}. For the Dresden EBIS-SC the values in parenthesis show the 
maximum reported values if they differ from the nominal values reported 
for the commercially available device.}
\label{Tab:Source_Ion_Sources}
{\tabulinesep=1.2mm
\begin{tabu}{l l l }
\toprule
\textbf{Feature} & \textbf{Supernanogan ECRIS} & \textbf{Dresden EBIS-SC} \\
\midrule
RF frequency 		& $14.5$\,GHz & - \\
RF power 			& $600$\,W & - \\
Magnet type			& RT permanent magnets & Superconducting \\
		            &  & LHe/LN free \\
Magnetic field 		& $B_{\text{inj,ext}} = 1.2$\,T 	& $B_{\text{max}} = 6$\,T \\
					& $B_{\text{min}} = 0.37$\,T & \\
					& $B_{\text{rad}} = 1.1$\,T 	& \\
Ion extraction 		& Max. $30$\,kV & N/A (a few tens of kV) \\
Electron beam current 	& - & $0.5$\,A ($1$~A) \\
Electron beam energy 	& - & $20$\,keV ($30$\,keV) \\
Electron density in trap 	& - & $>1000$\,A/cm$^2$ \\
Trap length 				& - & $0.2$\,m 	\\
Trap capacity 				& - & $2.0 \times 10^{10}$\,e \\
Pulse length 		& external chopper 	& $1 - 100$\,$\mu$s \\
             		& (DC beam) 	&  \\
Repetition rate 	& external chopper 	& up to hundreds of Hz \\
                	& (DC beam) 	&  \\
\bottomrule
\end{tabu}
}
\end{table}

\section{Requested beam elements}
\label{Sec:Source_RequestedBeamElements}

The proposed biomedical facility requests light ions ranging from protons to Oxygen to be
injected into the LEIR synchrotron. Out of these elements, ion beams of protons, Helium, 
Carbon, Nitrogen and
Oxygen are routinely produced in accelerator facilities using ECRIS and EBIS ion sources. The aforementioned
elements are gaseous at room temperature (or have gaseous compounds, like \ce{CO} and \ce{CO_2} for
Carbon), and as such are easy to produce with these types of ion sources. Lithium beams have been
produced with ECR ion sources e.g. at JINR, Dubna, and GSI, Darmstadt, from metallic \ce{Li} and
\ce{LiF} using resistively heated ovens \cite{bogomolov99,tinschert08}. Boron beams have been 
produced with ECR ion sources e.g. at JINR, Dubna, and JYFL, Jyv\"askyl\"a, with the MIVOC method
\cite{koivisto_mivoc} using m-carborane compound (\ce{C2H12B10}) \cite{bogomolov99,koivisto_ecr2}.
Experiments of \ce{Li} and \ce{B} beam production with a Supernanogan ECR ion source are currently
being performed at HZB, Berlin \cite{josh14}. Beryllium ion beams are rarely produced due to their
highly toxic properties, which are detrimental to the use and operational reliability of the ion
source. The beams accessible with the MIVOC method can also be produced with EBIS using the same
approach. However, metal beams can be produced with ECRIS sources using ovens, whereas EBIS sources usually require an external ion source injector of singly charged ions \cite{nakamura00,pikin06,sokolov10,thorn12,delahaye13}. With an ECRIS source the change from one element to another requires hours to days of retuning of the ion source. With an EBIS source, depending on the
injection scheme, it is possible to change elements on a pulse-to-pulse basis \cite{nakagawa13}.   

\section{Beam current, quality and stability}
\label{Sec:Source_BeamCurrent}

Table~\ref{Tab:Source_Supernanogan_Currents}~\cite{schlitt08} shows the measured ion beam currents of \ce{H_2^+}, \ce{H_3^+}, \ce{^3He+}, \ce{^{12}C^{4+}} and \ce{^{16}O^{6+}} that the ion sources at the HIT facility in Heidelberg are capable to deliver. These values correlate well with the beam currents reported for other ion beam therapy facilities \cite{kitagawa10}, indicating beam current levels from the ion source that are adequate to fulfill current treatment requirements. The actual beam currents are somewhat lower and depend on numerous details. For example, at HIT the routine operation is performed with $140$\,$\mu$A of \ce{^{12}C^{4+}} with Supernanogan ECR ion sources~\cite{schlitt08}. Some typical beam currents reported for the ion source are presented in table~\ref{Tab:Source_Supernanogan_Currents}. The ECR ion sources are usually operated in DC mode and the ion beam is pulsed after the ion source with an external beam chopper to the desired pulse pattern. HIT uses pulses of up to $300$\,$\mu$s length at up to $5$\,Hz repetition rate~\cite{schlitt08}. As a consequence, the desired properties of the pulse pattern do not influence the ion beam production, and the ion source is not a limiting factor if pattern variation is required. The pulsed beam can also be produced by pulsing the ion source RF input power. With this approach it is possible to take advantage of the afterglow effect, a high current burst of medium and high charge state ions lasting up to a few milliseconds at the end of the microwave pulse. During the burst the currents of the highest ion charge states can exceed the stable operation by a factor of $5$ or more (see e.g. \cite{tarvainen10}). However, the production of multiply charged ions require time scales from milliseconds to tens of milliseconds \cite{ropponen11}, limiting the repetition rate. For sub-millisecond pulses, additional pulse trimming is required. Table~\ref{Tab:Source_Ecris_Ebis_Comparison} presents the ions per $300$\,$\mu$s pulse of selected light ions produced with the Supernanogan. 

\begin{table}[!t]
\centering
\caption{Ion beam currents of various ion species reported for the Supernanogan ECR ion source by (a) Pantechnik S.A. \cite{pantechnik} and (b) HIT ion beam facility \cite{schlitt08}. Values from the preliminary LINAC3 light ion frontend report by D.~K\"uchler \cite{kuchler_linac3_li_front_end} are also included (c). The reported beam current values are in $\mu$A and the ion source is operated in DC mode.}
\label{Tab:Source_Supernanogan_Currents}
{\tabulinesep=1.2mm
\begin{tabu}{l l l l l l}
\toprule
\textbf{Ion} & \multicolumn{5}{c}{\textbf{Charge state}} \\
 & 1+ & 2+ & 4+ & 6+ & 8+ \\
\midrule
\ce{H} & 2000$^{\text{\,a}}$ & & & & \\
\ce{H_2} & \textgreater2000$^{\text{\,b}}$ & & & & \\
\ce{H_3} & 710$^{\text{\,b}}$ & & & & \\
\ce{^3He} & 840$^{\text{\,b}}$ & & & & \\
\ce{^4He} & 2000$^{\text{\,a,c}}$ & 1000$^{\text{\,a,c}}$ & & & \\
\ce{^{12}C} & 500$^{\text{\,c}}$ & 350$^{\text{\,c}}$ & 200$^{\text{\,a,b,c}}$ & 2.5$^{\text{\,a}}$ & \\
\ce{^{14}N} & 1000$^{\text{\,c}}$ & 200$^{\text{\,c}}$ & 100$^{\text{\,c}}$ & 10$^{\text{\,c}}$ & \\ 
\ce{^{16}O} & 1000$^{\text{\,c}}$ & 400$^{\text{\,c}}$ & 300$^{\text{\,c}}$ & 170$^{\text{\,b}}$ & \\ 
 & & & & (200$^{\text{\,c}}$) & \\ 
\bottomrule
\end{tabu}}
\end{table}

Compared with the ECRIS, the EBIS ion capacity is more limited and they usually provide lower extracted beam currents, with the exception of very highly or fully ionized medium to heavy elements. The EBIS ion source is normally operated in pulsed mode and the ion source performance in terms of beam current is defined by the ion source trap capacity, which describes the theoretical upper limit for the ion charge that can be produced for one pulse. The extracted ion charge usually amounts to $50-80$\% of the trap capacity, spread over the different extracted ion species \cite{sun12}. The maximum repetition rate is dictated by the time required to produce and fill the trap with the desired ion species (ionisation or breeding time), which increases with the desired ion charge state. The typical charge breeding times for EBIS start from a few milliseconds for the very lightest elements and exhibit an average linear increase of $10$\,ms per charge state with the heavier elements \cite{delahaye13}. However, the breeding time can be decreased with increasing electron beam current density and adjusting the electron gun voltage (current density and cross section dependencies on electron energy). This feature can be used to reduce the required time and increase the repetition rate.

The performance of EBIS sources has steadily improved over the last decades. For example, MEDEBIS, a prototype ion source for synchrotron injection and ion beam therapy developed in Frankfurt in the 1990's, can deliver $1\times10^9$ particles/s in $5$\,$\mu$s pulses at $5$\,Hz repetition rate. This results in a modest $5\times10^3$ ions per pulse, which was at the time considered to be too low for practical application in ion beam therapy \cite{kester96}. Currently, DREEBIT GmbH offers the Dresden EBIS-SC, which is marketed as a modern EBIS designed for ion beam therapy promising adequate beam currents for \ce{^1H+}, \ce{H_2^+}, \ce{^{12}C^{6+}}, \ce{^{16}O^{8+}} and other ions \cite{zschornack07,zschornack09,zschornack10,dreebit}. The Dresden EBIS-SC trap capacity is $C = 2\times10^{10}$\,\cite{zschornack10}. The measured performance is presented in table~\ref{Tab:Source_Ecris_Ebis_Comparison}. The pulse length is reported to be variable between $1-100$\,$\mu$s in the commercial device, even though experimental data is presented for pulse lengths up to $1$\,ms \cite{zschornack07,dreebit}. The reported ionisation times agree rather well with the aforementioned scaling. For example, the ionisation times for \ce{^{12}C^{4+}} and \ce{^{14}N^{7+}} are $10$ and $100$\,ms, respectively. For light elements up to Neon these characteristics translate to repetition rates ranging from a few tens of Hertz up to a few hundreds of Hertz, depending on the desired charge state. 

\begin{table}[!t]
\centering
\caption{Comparison of the Supernanogan ECRIS and the Dresden EBIS-SC in terms of produced ions per pulse. The data are extracted from references \cite{pantechnik,dreebit,schlitt08,kuchler_linac3_li_front_end,wenander_ebis_ecris}. The pulse lengths are $300$\,$\mu$s for Supernanogan ECRIS and 1 -- 1000\,$\mu$s for Dresden EBIS-SC.}
\label{Tab:Source_Ecris_Ebis_Comparison}
{\tabulinesep=1.2mm
\begin{tabu}{l l l l}
\toprule
\textbf{Ion}	& \textbf{Supernanogan ECRIS}	& \textbf{Dresden EBIS-SC} 	\\
\midrule
\ce{H^+} 			& $3.7 \times 10^{12}$			& $3.5 \times 10^{9}$	\\
\ce{H_2^+} 			& $>3.7 \times 10^{12}$			& $4.5 \times 10^{8}$	\\
\ce{H_3^+} 			& $1.3 \times 10^{12}$			& 						\\
\ce{^3He+}			& $1.6 \times 10^{12}$			& 						\\
\ce{^4He+}			& $3.7 \times 10^{12}$			& 						\\
\ce{^4He^{2+}}		& $9.4 \times 10^{11}$			& 						\\
\ce{^{12}C^{1+}}	& $9.4 \times 10^{11}$			&						\\
\ce{^{12}C^{2+}}	& $3.3 \times 10^{11}$			&						\\
\ce{^{12}C^{4+}}	& $9.3 \times 10^{10}$			& $6.6 \times 10^{8}$	\\
\ce{^{12}C^{5+}}	& 								& 						\\
\ce{^{12}C^{6+}}	& $7.8 \times 10^{8}$			& $2.4 \times 10^{8}$	\\
\ce{^{14}N^{1+}}	& $1.9 \times 10^{12}$			& 						\\
\ce{^{14}N^{2+}}	& $3.7 \times 10^{11}$			& 						\\
\ce{^{14}N^{4+}}	& $4.7 \times 10^{10}$			& 						\\
\ce{^{14}N^{5+}}	&								& $2.0 \times 10^{8}$	\\
\ce{^{14}N^{6+}}	& $3.1 \times 10^{9}$			& $7.0 \times 10^{7}$	\\
\ce{^{14}N^{7+}}	&								& $1.0 \times 10^{7}$	\\
\ce{^{16}O^{1+}}	& $1.9 \times 10^{12}$			& 						\\
\ce{^{16}O^{2+}}	& $3.7 \times 10^{11}$			& 						\\
\ce{^{16}O^{4+}}	& $1.4 \times 10^{11}$			& 						\\
\ce{^{16}O^{6+}}	& $5.3 \times 10^{10}$			& $2.1 \times 10^{8}$	\\
\ce{^{16}O^{7+}}	&								& $1.3 \times 10^{8}$	\\
\ce{^{16}O^{8+}}	& 								& $1.0 \times 10^{8}$	\\
\bottomrule
\end{tabu}}
\end{table}

Comparison of the achievable ions per pulse in commercially available compact ion sources (table~\ref{Tab:Source_Ecris_Ebis_Comparison}) shows that the Supernanogan ECRIS produces substantially higher initial beam intensities than the Dresden EBIS-SC. This trend is especially clear with all charge states at least up to Carbon, for which the high charge state data is available. With the lightest elements and medium charge states of the heavier ions the difference in ion current output is up to $2-3$ orders of magnitude. For \ce{^{12}C^{6+}} production, the advantage of the Supernanogan has decreased to roughly a factor of $3$. Ion beam therapy facilities normally use sub-millisecond pulses~\cite{schlitt08,iwata07}, and with EBIS repetition rates of a few Hz to a few hundred Hz for light ions, produced ion currents are not high enough to fill the desired pulse length with multiple EBIS pulses to achieve the total number of desired particles per pulse.

Beams produced with EBIS sources have substantially better beam quality than those produced with ECRIS - both, in terms of transverse beam emittance and spatial particle distribution, which leads to improved beam transmission. The ECR ion sources are known to exhibit strong triangular and hollow beam shape distortions, which are absent in the beams produced with EBIS sources~\cite{zschornack09,spadtke08}. For the HIT Supernanogan ECRIS, transverse non-normalized 4-rms emittances of up to $300$~mm.mrad have been reported, with $150-200$~mm.mrad being the normal emittance region for $^{12}$C$^{4+}$ beams with beam currents around $150$\,$\mu$A at $24$\,kV extraction voltage~\cite{schlitt08}. For the MedAustron Supernanogan (with the same extraction voltage) non-normalized 4-rms emittances of $150$~mm.mrad (horizontal) and $197$~mm.mrad (vertical) have been measured for a $470$\,$\mu$A \ce{H_{3}^+} beam~\cite{penescu13}. Somewhat higher emittance values of $292$~mm.mrad (horizontal) and $193$~mm.mrad (vertical) have been reported for the \ce{^{12}C^{4+}} beams produced with the $10$\,GHz KeiGM all-permanent magnet ECRIS with $30$\,kV extraction voltage~\cite{iwata07}. This ion source is used with an injector design developed for ion beam therapy at NIRS, Japan. 

In comparison, the measured transverse 4-rms emittance of the Dresden EBIS-A, a room temperature EBIS with somewhat lower performance than that of the Dresden EBIS-SC, is $<32$~mm.mrad (with extraction voltages around $20-30$\,kV)~\cite{zschornack09}. 

Ions that are not initially fully ionized (as is typically the case with injectors based on ECRISs) are typically stripped before synchrotron injection. Usually this is done at energies where the stripping efficiency is very high (several MeV/u), assuring low beam losses. However, stripping does increase the beam emittance and energy spread, leading to further beam quality degradation. Based on measurements performed at HIT, stripping of Carbon ions from charge state $4+$ to fully stripped $6+$ results in transverse emittance increases of 25\% and 9\%, in the horizontal and the vertical planes respectively~\cite{schlitt08}. Using EBIS that produces fully stripped ions, no further stripping would be required in the injector design, improving the beam quality and simplifying the injector design. The LEIR accelerator does not require injection of fully stripped ions, yet the use of higher charge states within LEIR has some advantages. For proton beams, the molecular hydrogen beam can be stripped at the end of the injector to provide proton beam insertion into LEIR, as is done in the currently operated medical facilities. Alternatively, the molecular beam can be accelerated and stripped into protons at the BioLEIR extraction. As the protons have substantially lower magnetic rigidity compared to the heavier elements, the molecular beam acceleration can be advantageous in LEIR as it allows higher magnetic fields in the magnets. 

To evaluate the expected differences in the accelerated beam currents, some estimations of the transmission efficiencies with the two types of ion sources are required. The original HIT injector transmission efficiency with Supernanogan ECRIS has been reported to be $>90$\% for the Low Energy Beam Transport (LEBT), $\sim30$\% for the Radio-Frequency Quadrupole (RFQ) and $>90$\% for the Inter-digital H (IH) linac, resulting in a typical total transmissions of 20-30\% due to an ion optics mismatch and manufacturing problems with the injector RFQ~\cite{schlitt08}. With similar RFQs a transmission of $70$\% has been reached, increasing the total achievable injector transmission to about $60$\%~\cite{emhofer10}. Consequently, the gain in beam intensity from an EBIS source at the end of the injector due to improved beam quality over ECRIS would not be expected to be more than a factor of $2-5$. 

Beam stability issues can be divided into two categories: (1) the short-term stability of the ion source and (2) the long-term stability and reproducibility. The short-term stability (current fluctuations within the beam pulses) is not critical for operation, even in the case of high intensity current peaks originating from the ion source, as any beam pulse fine structure from the frontend is lost during the acceleration inside the synchrotron. This is valid for all injector designs~\cite{kitagawa10}. The long-term stability and reproducibility is more important in order to ensure consistent irradiation and constant dose rates during operation. To assure this aspect, the beam current variations should be less than $10$\%, in which regard both the ECRIS and the EBIS have been reported to perform suitably well~\cite{kitagawa10,becker98,zschornack07,zschornack09}.

Table~\ref{Tab:Source_Ions_1_Per_4} presents a set of ion species from protons to Oxygen for a nominal charge over mass ratio of $Q/A=1/4$. Keeping in mind the operational limitations of the RFQ, the design value of $Q/A=1/4$ is treated as a lower limit. The maximum extraction voltage reported for the Supernanogan ECRIS extraction system by Pantechnik S.A. is $30$\,kV. Taking these constraints into account together with the wish for an operational margin, it is favorable to operate the source extraction at 5.0\,keV/u.

Ion sources can suffer from the production of ions from the residual gas, including previously created beams from earlier runs. With many ion requests overlapping on $Q/A=1/4$, the requirement for high purity at the target station will have to be evaluated (in most cases with measurements) for each of these types and the production schemes may be adjusted to increase purity at the expense of intensity.

\begin{table}[!t]
\centering
\caption{Suitable ions with charge over mass ratios of $Q/A \geq 1/4$ with isotopic abundances, charge over mass differences from the design value ($\Delta(Q/A)$) and ion source extraction voltages required to match the ion beam energies of 5.0\,keV/u and 7.5\,eV/u.}
\label{Tab:Source_Ions_1_Per_4}
{\tabulinesep=1.2mm
\begin{tabu}{l S[table-format=3.4] S[table-format=1.3] S[table-format=2.2] S[table-format=2.2] S[table-format=2.2]}
\toprule
\multicolumn{4}{l}{\textbf{Design Q/A = 1/4}} & \multicolumn{2}{c}{\textbf{V$_{\text{ext}}$~[kV] req. for E/A}} \\
\midrule
Ion & \multicolumn{1}{c}{Iso. abundance [\%]} &  Q/A & \multicolumn{1}{c}{$\Delta$(Q/A) [\%]} & \multicolumn{1}{c}{5.0~keV/u} & \multicolumn{1}{c}{7.5~keV/u} \\
\midrule
\ce{H^{1+}_{3}} (mol)	& 99.9885 	& 0.333 	& 33.33 	& 15.00	& 22.50 \\
\ce{^{4}_{2}He^{1+}} 	& 99.9999 	& 0.250 	& 0.00 		& 20.00	& 30.00 \\
\ce{^{7}_{3}Li^{2+}} 	& 92.41 	& 0.286 	& 14.29 	& 17.50	& 26.25 \\
\ce{^{9}_{5}Be^{3+}} 	& 100.0 	& 0.333 	& 33.33 	& 15.00	& 22.50 \\
\ce{^{11}_{5}B^{3+}} 	& 80.1 		& 0.273 	& 9.09 		& 18.33 & 27.50 \\
\ce{^{12}_{6}C^{3+}} 	& 98.93 	& 0.250 	& 0.00 		& 20.00	& 30.00 \\
\ce{^{14}_{7}N^{4+}} 	& 99.636 	& 0.286 	& 14.29 	& 17.50	& 26.25 \\
\ce{^{16}_{8}O^{4+}} 	& 99.757 	& 0.250 	& 0.00 		& 20.00	& 30.00 \\
\ce{^{19}_{9}F^{5+}} 	& 100.0 	& 0.263 	& 5.26 		& 19.00	& 28.50 \\
\ce{^{20}_{10}Ne^{5+}} 	& 90.48 	& 0.250 	& 0.00 		& 20.00	& 30.00 \\
\bottomrule
\end{tabu}}
\end{table}

\section{The AISHa source}
\label{Sec:Source_AISHA}

It has been suggested that future hadrontherapy facilities may require better performance in terms of beam current and beam quality than the commercially available all-permanent magnet ECR ion sources can currently provide~\cite{celona12}. To address this, a relatively compact $18$\,GHz ECR ion source design called AISHa (Advanced Ion Source for Hadrontherapy) has been developed at INFN-LNS. It has been designed to be a versatile, multipurpose ion source providing higher performance and wider range of ion species than the ion sources in existing clinical facilities, while featuring easy maintenance and operation to make it suitable for clinical use.

\section{Project staging and source operation}
\label{Sec:Source_Operation}

It would be beneficial that a source on a test stand would be available as soon as possible after the project start date, but latest during the preparation of Stage 1 of BioLEIR (see section~\ref {Sec:Intro_ProjectOverview}). It is essential that the source be fully characterized such its characteristics used as input requirements for the design of the subsequent frontend machine elements.

Next to the ion source(s) used for standard light ion production during Stage 2 of BioLEIR operations, it is useful to have an identical source at a test stand or a frontend branch on which new ion beams can be developed in parallel with normal operations. This is particularly interesting for developments of exotic beams (as needed e.g. for Beryllium), to improve existing ion beam performance, or to tune a source for the subsequent ion species planned for the BioLEIR experiments. Operation with two operational sources in parallel could be part of a later stage of BioLEIR.

The intensity and stability of the ion beam may depend on the element requested.
It is known that e.g. Carbon influences the long-term source
performance negatively. The periods of beam availability can be limited, especially for beams based on solid material, and down-times for source refills have to be taken into account in facility availability estimates.

\section{Expected source performance}
\label{Sec:Source_Performance}

In order to achieve the required ion intensities on target for the various ion species as specified in the facility requirements, and taking into account the different transmission efficiencies from the ion source to the target area (see table~\ref{Tab:BeamParameters_WhatCanCERNProvide_IonIntensities}), the Supernanogan source seems capable to deliver all required ion currents. The Supernanogan is therefore considered the baseline, proven solution for the light ion source for BioLEIR. Other solutions, most notably the AISHa source, could be of interest for BioLEIR, particularly considering a potential collaborative R\&D.

At this stage of the study, no in-house measured source output data is available for the BioLEIR baseline choice and emittance numbers are available from publications in mostly non-normalized ways. As specified in section~\ref{Sec:Linac_RFQ}, the frontend expects a 0.13\,mm.mrad rms normalized emittance which corresponds to 200\,mm.mrad unnormalized total emittance at the chosen source extraction voltage.

\section{Power and cooling requirements}
\label{Sec:Source_Infrastructure}

The Supernanogan source requires less than 50\,kW of electrical power for all subsystems. The cooling system has to provide demineralised water with a cooling power not more than the requested electrical power. For the AISHa source one can expect a higher power consumption (e.g. the GTS-LHC ion source consumes up to 200\,kW).

The cabling of the new ion source is estimated at 50\,kCHF. The electrical power requirement for the light ion source is less than the power currently supplied for LINAC2. Therefore, it is assumed that no additional investment beyond renewed cabling is required.

\section{Resource estimate}
\label{Sec:Source_Cost}

The cost of a Supernanogan ion source and all related subsystems, additional components required for metal ion beam production, ion beam analysis and extensive beam diagnostics, is estimated at about 1\,MCHF. This estimate includes a 4D pepperpot emittance scanner and its software.

Commercial compact electron beam ion sources are provided by DREEBIT GmbH. The product line-up includes a liquid cryogen free superconducting Dresden EBIS-SC ion source, which is marketed as being developed and aimed for medical particle therapy. A Dresden EBIS-SC and its subsystems (discussed in section~\ref{Sec:Source_BeamCurrent}), a MIVOC system and an external Liquid Metal Alloy Ion Source (LMAIS) for metal ion beam production are estimated at about 1.1\,MCHF.

For the AISHa source a very rough cost estimate of 1.5$\pm$0.5\,MCHF is assumed. The large uncertainty is due to the fact that not as much experience exists for AISHa as for commercially available sources.

The ancillary cost for generation of the higher risk ion species types is estimated at the 50\,kCHF level and includes mostly a glove box and items for chemical handling.  The amount of H$_2$ needed is much lower, more comparable to what is used for the ELENA source. Therefore, the gas safety system could be much simpler than the one implemented for LINAC4.

Table~\ref{Tab:Source_Costing_Material} summarizes the source options and their estimated cost, whereas table~\ref{Tab:Source_Costing_Material_Full} tallies the full estimated cost for the light ion source. It shall be noted that the Supernanogan source can be used as delivered in a standalone test-stand, as it has sufficient beam analysis and diagnostics included in its subsystems. The EBIS-SC full system does not include sufficient beam analysis and diagnostics and therefore, the BioLEIR beam instrumentation intended for the frontend would need to be available already for a source test-stand operation with an EBIS-SC.

The expected lead time for commercially available ion sources is 10 -- 18 months, indicating that a full ion source test bench setup can be made available within 12 -- 20 months from order date.

\begin{table}[!ht]
\centering
\caption{Cost estimates for different light ion source options for the BioLEIR frontend. The estimates are based on price quotations received in Euros in 2014, a conversion factor of 1.1 to CHF has been applied.}
\label{Tab:Source_Costing_Material}
\begin{tabular}{l l l }
\hline
\textbf{System} & \textbf{Units} &  \textbf{Cost estimate}\\ 
\textbf{} & \textbf{} & \textbf{[kCHF]} \\
\hline
\rule{0pt}{3ex}Supernanogan  &  \\
 & Supernanogan + subsystems  &   \\
 & incl. beam analysis \& diagnostics &  1000  \\
DREEBIT EBIS  &  &   \\
 & Dresden EBIS-SC + subsystems &   \\
 & MIVOC \& LMAIS & 1100  \\
 & (NB: no beam analysis \& diagnostics)   \\
AISHa &  & 1500$\pm$500  \\
\hline
\textbf{Baseline Supernanogan} & &\textbf{1000}\\
\hline
\end{tabular}
\end{table}

\begin{table}[!ht]
\centering
\caption{Material cost estimate for the light ion source. The uncertainty of this estimate is at the level of 10\%.}
\label{Tab:Source_Costing_Material_Full}
\begin{tabular}{l r}
\hline
\rule{0pt}{3ex}\textbf{System} & \textbf{Cost estimate} \\
\textbf{} & \textbf{[kCHF]}\\
\hline
\rule{0pt}{3ex}Supernanogan (Baseline)  &  1000\\
Cabling & 50\\
Ancillary equipment (chemical handling) & 50\\
\hline
\rule{0pt}{3ex}\textbf{Total Source} &\textbf{1100}\\
\hline
\end{tabular}
\end{table}

For installation of the source it is estimated that 4 FTE are needed during 6 months. The commissioning of the source is expected to take about 3 months with 2 persons. It is considered useful to start characterization of the light ion source as soon as the project goes ahead, as the source output characterization provides essential input for design optimization of LINAC5. Any further time available before LINAC5 comes on-line is useful for beam development for the different ion species that BioLEIR requires. It is estimated that 1 FTE will work on beam development for 4 years before LINAC5 is operational in 2023. Table~\ref{Tab:Source_Costing_FTE} summarizes these staffing estimates.

\begin{table}[ht]
\centering
\caption{Preliminary estimates of the integrated workforce needed for preparation, installation, commissioning and tuning of the new light ion source for BioLEIR, given in person-years.}
\label{Tab:Source_Costing_FTE}
\begin{tabular}{l r r}
\hline
\rule{0pt}{3ex}\textbf{System} & \textbf{Staff [PY]} & \textbf{Timeperiod}\\
\hline
\rule{0pt}{3ex}Test stand preparation & 0.5 & 2017\\
Installation & 2 & 2018\\
Commissioning & 0.5 & 2018\\
Beam Development & 4 & 2019-2022\\
\hline
\rule{0pt}{3ex}\textbf{Total Source [Personyears]} & \textbf{7}\\
\hline
\end{tabular}
\end{table}

One full-time ion source specialist is needed for operations and beam development in the first years of running of the facility. Once experience has been gained with operating the different ion species, it is expected that the source operations can be performed with 0.5~FTE. The material cost for operations is expected at the level of 150\,kCHF/year which covers development work for the different ion beams, safety considerations and one student. About 10\% of the source cost shall be allocated per year to build up a pool of spares and their replacement. This corresponds to about 100\,kCHF/year of maintenance cost.

%% file: Chapters/FrontendAndLinac.tex
\chapter{Light Ion Linac Design Using a Quasi-Alvarez DTL Accelerating Structure}
\label{Chap:Linac}

The purpose of LINAC5 is to accelerate the proposed light ion species for BioLEIR from Stage 2 of the facility onward. As described in chapter~\ref{Chap:Source}, all desired ions for biomedical research may be produced with a charge over mass ratio between $Q/A$\,=\,1/4 and $Q/A$\,=\,1/3 (relative to protons). The final energy of ions in LINAC5 is 4.2\,MeV/u which matches the injection energy of the LEIR accelerator.

LINAC5 is envisaged to be installed in the tunnel currently occupied by LINAC2 and as such is expected to use all the existing transfer lines to LEIR at minimal extra cost (see chapter~\ref{Chap:Transferline}). On removal of LINAC2, the space becomes available along with its existing power, cooling and access infrastructure. This has major cost saving advantages in the areas of civil engineering, transfer line component procurement, as well as power and cooling facilities.

\begin{figure}[!htb]
\centering
\includegraphics[width=0.8\columnwidth]{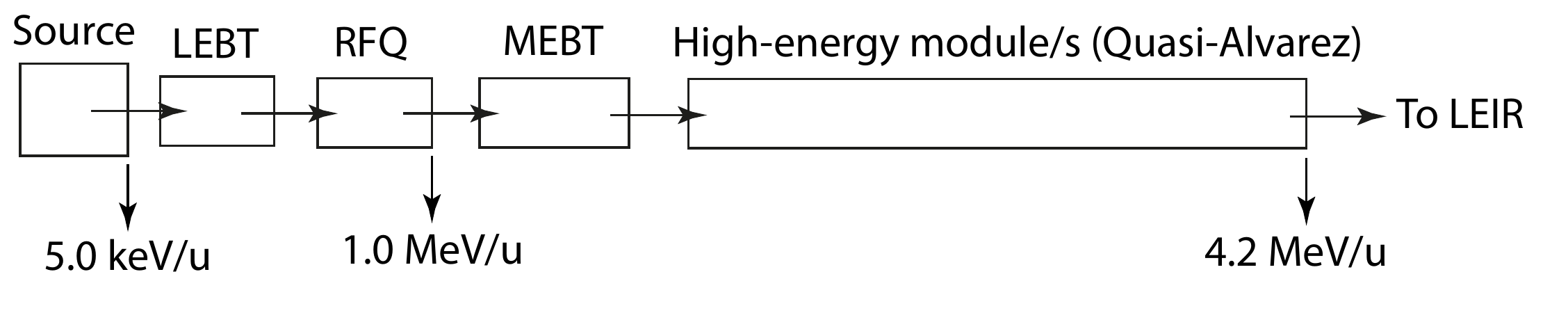}
\caption[]{Basic schematic layout of LINAC5 (not to scale) showing the source, the LEBT (Low Energy Beam Transport), the RFQ (Radio-Frequency Quadrupole), the MEBT (Medium Energy Beam Transport) and the Quasi-Alvarez DTL. The respective nominal output energies of the components are shown.}
\label{Fig:basic_linac_schem}
\end{figure}

A preliminary basic schematic layout of LINAC5 is shown in figure~\ref{Fig:basic_linac_schem}. The source and Low Energy Beam Transport (LEBT) section are followed by a Radio-Frequency Quadrupole (RFQ), a Medium Energy Beam Transport (MEBT) section and the high-energy accelerating structures. A Quasi-Alvarez DTL accelerating structure was designed in order to create a shorter linac than the more commonly used Drift-Tube Linac (DTL), but with more focusing control and better transmission than an Inter-digital H-structure (IH). Such a structure also has the potential for future energy upgrades for possible other uses due to its relatively superior ability for longitudinal beam control over an IH design. In this chapter, we present the preliminary design of the linac after the source including beam dynamics studies of the Quasi-Alvarez DTL. 

\vspace{-3mm}
\section{Low energy beam transport}
\label{Sec:Linac_LEBT}

The ions from the source pass into the LEBT section with a kinetic energy of 5.0\,keV/u and a maximum nominal beam current of 2\,mA. The primary purpose of an ion LEBT is to match the beam from the source to the following accelerating structures, separate the different charge states and to truncate the longitudinal dimensions of the beam. Typical designs include one or two solenoids for transverse focusing at low energies, a spectrometer for charge selection, an electro-static chopper for selecting the desired longitudinal part of the beam, steering and focusing magnets for transverse control and diagnostics for measuring beam current, emittance and position in the vacuum vessel. Charge selection can also be carried out by specific RFQ design, where tuning can be performed to reject all but the desired charge-state ions.
 
It is envisaged to use a LEBT design similar to those used at facilities such as MedAustron~\cite{medAus1}, HIT~\cite{HIT} and CNAO~\cite{CNAO} where such designs have proven reliability and good performance. An alternative, simple solution might be the passage of the beam from the source to the RFQ in a straight line. In this case, the spectrometer magnet is placed at a location such that it may be used for tuning the source with a separate diagnostics line and is switched during nominal operation. In this manner, the RFQ would perform the charge selection in nominal operation. This solution is realisable due to the relatively low nominal beam currents. Further beam dynamics studies of the proposed source and LEBT solutions are needed to choose between alternatives and verify a complete design.

\section{Radio-frequency quadrupole}
\label{Sec:Linac_RFQ}
 
A preliminary design for the RFQ was completed for the acceleration of ions with $Q/A$\,=\,1/4 from an energy of approximately 5\,keV/u to 1.0\,MeV/u, including preliminary RF design and machinability.

Two frequencies of 100\,MHz and 200\,MHz have been considered and, in a final global optimisation, the frequency of 200\,MHz has been selected. For ions with $Q/A$\,=\,1/4, the final energy is a trade-off between RFQ efficiency and dimension of the first cells of the structures which follow the RFQ, with a good compromise found at 1.0\,MeV/u.

\begin{figure}[!htb]
\centering
\includegraphics[width=0.8\columnwidth]{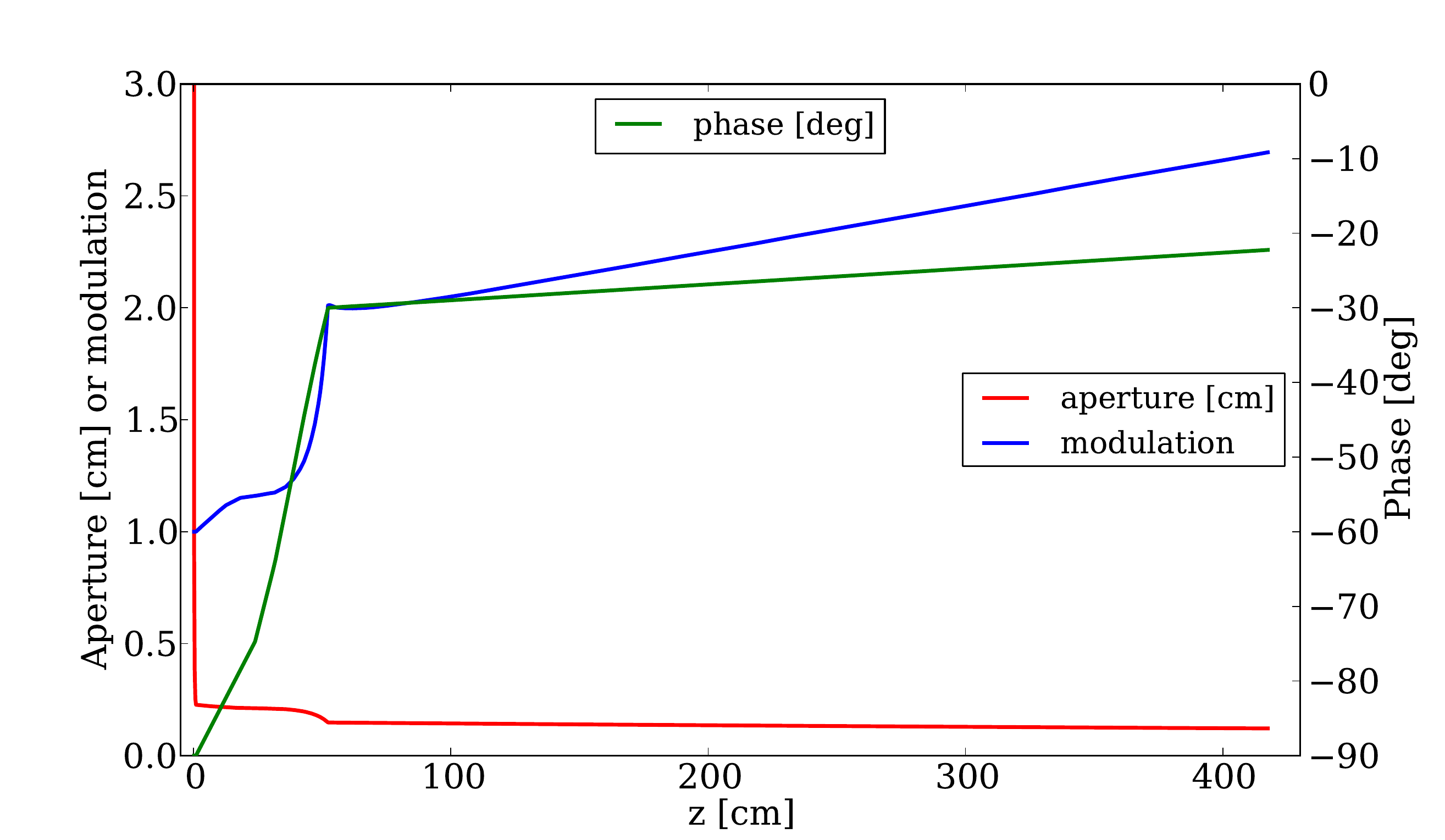}
\caption[]{The aperture (red), modulation (blue) and phase (green) as a function of the distance along the RFQ in the preliminary study.}
\label{Fig:RFQdat}
\end{figure}

The RFQ has been designed with an acceptance at zero current, that is 50\% higher than the expected emittance from the source (0.13\,mm.mrad rms normalized which corresponds to 200\,mm.mrad unnormalized total at the chosen source extraction voltage). The main parameters of the RFQ are summarized in table~\ref{Tab:Linac_RFQ_Parameters} and a plot of the parameters of aperture, modulation and phase are shown in figure~\ref{Fig:RFQdat}.

\begin{table}[!hbt]
\caption{Main parameters of the RFQ.}
\label{Tab:Linac_RFQ_Parameters}
	\centering
	\begin{tabular}{l c}
		\hline
		 \rule{0pt}{3ex}\bf{Parameter} & \bf{Value} \\
		 \hline
		  \rule{0pt}{3ex}Length [m] & 4 \\
		  RF frequency [MHz] & 202.56 \\
		  Maximum mass/charge ($A/Q$) & 4 \\
		  Input/output energy [MeV/u] & 0.005 to 1.0 \\
		  Vane voltage [kV] & 49 to 58 \\
		  Average aperture ($r_0$) [mm] & 2.3 \\
		  Transverse radius ($\rho$) [mm] & 1.8 \\
		  Max. modulation & 2.7 \\
		  Max. electric field on vane-tip [MV/m] & 27 \\
		  Output transverse emit. rms. norm. [$\pi$.mm.mrad] & 0.10 \\
		\hline
	\end{tabular}	
\end{table}

\section{Medium energy beam transport}
\label{Sec:Linac_MEBT}

A preliminary MEBT design was carried out to match the beam from the RFQ to the entrance of the Quasi-Alvarez structure and provide space for suitable diagnostics. In order to control the longitudinal phase spread from the exit of the RFQ, two bunching cavities are proposed. Simulations of the different ion species from the RFQ to the accelerating structure showed that the flexibility offered by two bunching cavities is advantageous. Four quadrupole magnets are also used to allow flexible matching before the accelerating structure. The main parameters of the MEBT line are shown in table~\ref{tab:mebtparams}.

\begin{table}[hbt]
\caption{Main parameters of the MEBT line connecting the RFQ to the Quasi-Alvarez accelerating module.}	
	\centering
	\begin{tabular}{l c}
		\hline
		 \rule{0pt}{3ex}\bf{Parameter} & \bf{Value} \\
		 \hline
		  \rule{0pt}{3ex}MEBT length [m] & 1.6 \\
		  Bore radius [cm] & 2 \\
		  Frequency [MHz] & 202.56 \\
		  Energy [MeV/u] & 1.0 \\
		  Buncher E0TL [MV.m] & 0.1 - 0.15 \\
		  Max. quad. grad. [Tm$^{-1}$] ($Q/A$ = 1/4) & 33 \\
		  Quad. length [cm] & 10 \\
		\hline
	\end{tabular}	
\label{tab:mebtparams}	
\end{table}

\section{Quasi-Alvarez DTL}
\label{Sec:Linac_QuasiAlvarez}

\begin{figure}[htb]
\centering
\includegraphics[width=0.5\columnwidth]{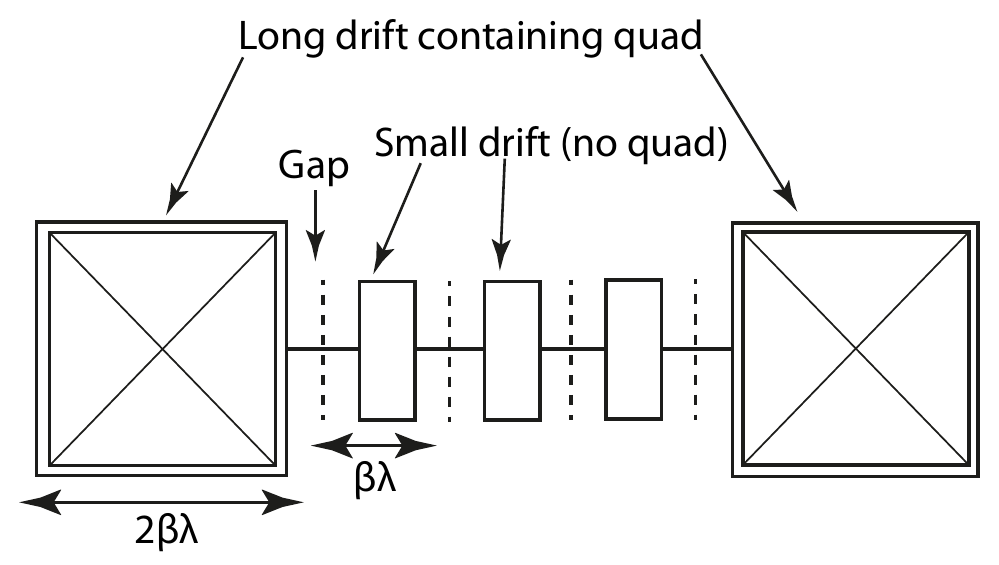}
\caption[]{Schematic representation of the Quasi-Alvarez DTL cell geometry. A given number of smaller drift tubes are periodically separated by one longer drift tube containing a quadrupole focussing magnet.}
\label{fig:Qa_schem1}
\end{figure}

In order to have an accelerating structure which is shorter than a DTL but has more focusing control and capability that an IH structure, we opted to design a Quasi-Alvarez DTL structure~\cite{QA-paper1,QA-paper2}. A Quasi-Alvarez configuration bares a close resemblance to that of a typical DTL where a series of drift-tubes with nominally zero RF field are separated longitudinally within a module by gaps containing an oscillating RF field, at a spacing of $\beta\lambda$. However, in a Quasi-Alvarez configuration, quadrupole magnets are not placed inside all drift tubes as in a DTL. Instead, a quadrupole is placed inside the $n^{th}$ drift-tube, where the periodicity $n$ is a free parameter of the design. A schematic representation of a Quasi-Alvarez structure is shown in figure~\ref{fig:Qa_schem1}. 

Using this concept it is possible for the designer to create a cell structure with a number of shorter drift-tubes at a longitudinal spacing of~$\beta\lambda$, separated by a longer drift-tube of length~$2\beta\lambda$ containing a quadrupole focussing magnet. Due to the fact that the shorter drift-tubes do not contain quadrupoles, the frequency of the Quasi-Alvarez structure may be increased beyond that what is usually suitable for a DTL where drift-tubes (and therefore~$\beta\lambda$) must be long enough to fit a quadrupole magnet inside. This allows the Quasi-Alvarez structure to be shorter than the conventional DTL for a given energy gain per unit length, provided the quadrupole magnets can be adequately designed given the space and beam focussing constraints.

In the current design, one accelerating module is proposed, where all the ion species are accelerated to the same energy per unit mass of 4.2  MeV/u. Due to the differing charge over mass ratio $Q/A$ (with respect to protons) for different ion species, the energy gain in the cavities of the structure is adjusted by the appropriate factor to preserve synchronicity, while the resulting total momentum/rigidity at the end of the linac depends on the Q/A.

\begin{figure}[!htb]
\centering
\includegraphics[width=0.7\columnwidth]{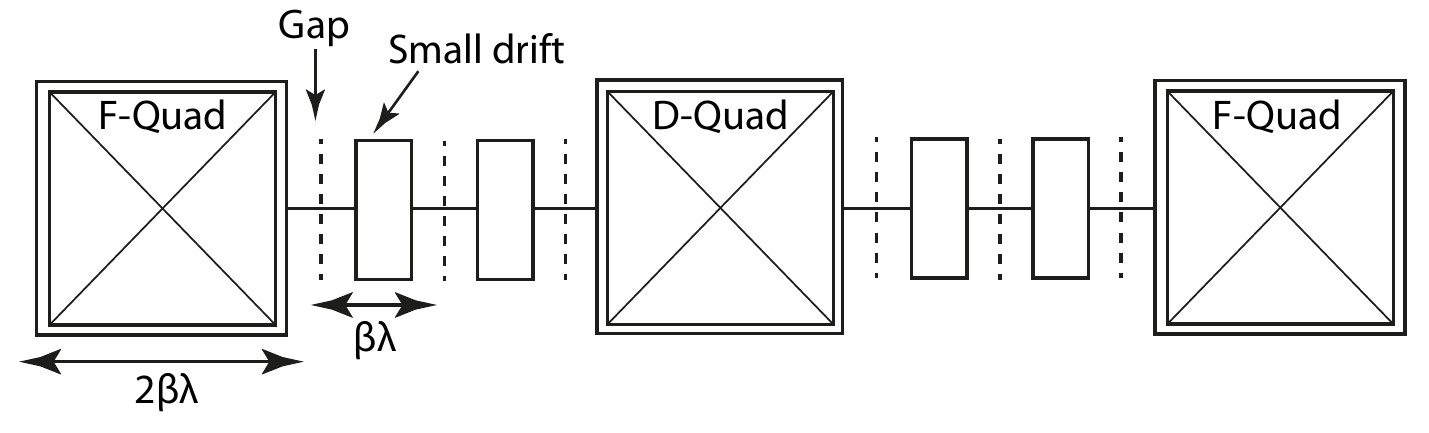}
\caption[]{Schematic representation of one period of the selected Quasi-Alvarez cell design.}
\label{fig:Qa_schem2}
\end{figure}

To balance the focusing strength required for the desired ion species with the length of the linac, the geometry was chosen such that large drift-tubes of length~$2\beta\lambda$ are separated by two smaller drift-tubes and three respective gaps, as shown in figure~\ref{fig:Qa_schem2}. Within the larger drift-tubes, the quadrupoles are placed in a FODO arrangement and we define one cell as being between neighbouring similar quadrupoles as shown in figure~\ref{fig:Qa_schem2}. 

The quadrupole dimensions and strengths were optimised to deliver good matched focusing with a realistic magnetic field strength for normal conducting magnets. The matching was performed as an optimisation over the ion species range $Q/A$ = 1/3 to 1/4, where each magnet has a constant field in order to allow the use of Permanent Magnet Quadrupoles (PMQs). It has been shown that PMQs can be utilised in light-ion accelerators with small drift-tube sizes, within low-beta accelerating structures and with similar field strengths to those proposed in this document~\cite{pmqs_l4, pmqs_lil}.

To make the small drift-tubes in the Quasi-Alvarez design as short as possible, we chose to double the frequency with respect to the RFQ, from 202.56\,MHz to 405.12\,MHz. This reduces~$\beta\lambda$ by a factor of two, and hence decreases the overall length. In order to reach the output energy with the shortest structure possible, the RF cavities were optimised in SuperFish~\cite{superfish}. Table~\ref{Tab:Linac_Cellparams2} shows the main parameters of the design.

It should be clearly noted that this scenario, containing only one accelerating module, is suitable for all proposed ion species where the energy per unit charge in~MeV/u is constant, yet the rigidity depends on $Q/A$. This design-induced difference in final momentum/rigidity requires that all transfer line magnets be adjusted for the different rigidities appropriately and hence an interleaved, pulse-to-pule modulation (PPM) scheme cannot be used.

\begin{table}[!hbt]
\caption{Main cell parameters of the Quasi-Alvarez DTL accelerating module.}	
\label{Tab:Linac_Cellparams2}	
	\centering
	\begin{tabular}{l c}
		\hline
		 \rule{0pt}{3ex}\bf{Parameter} & \bf{Value} \\
		 \hline
		  \rule{0pt}{3ex}Total length & 5.2 [m]\\
		  Bore radius & 0.6 [cm]\\
		  Tank diameter & 47.1 [cm] \\
	 	  Drit-tube outer radius & 4.5 [cm] \\
		  No. of cells/periods & 12 \\
		  Frequency & 405.12 [MHz] \\
		  Input/output $E_k$ & 1.0 - 4.2 [MeV/u] \\
		  g/l (start - end) & 0.18 - 0.25 \\
		 Kilpatrick (start - end) & 1.5 - 1.2 \\
		 E0  & 4.5 [MV/m] \\
		 E0TL/cell (start - end) & 0.09 - 0.24 [MV.m]\\
		 Cavity phase (nominal) & 20 [deg.]\\
		 Shunt impedance & 84 [MOhm/m]\\
		 Tank power & 1300 [kW] \\
		 Max. integ. quad. grad. & 6.4 [T]\\ 
		 Min. integ. quad. grad. & 4.7 [T]\\ 
		 Quad. length & 5 [cm] \\
		 (Acceptance)/(final emittance) & 1.5 \\
		\hline
	\end{tabular}	
\end{table}

\subsection{Longitudinal phase space painting into LEIR}
\label{SubSec:Linac_QuasiAlvarez_LongitudinalPhaseSpacePainting}
To allow longitudinal phase space painting of the beam into the LEIR machine, a bunch-to-bunch momentum variation of up to $\pm$1\% over the whole linac pulse is required. This is envisaged by installing a ramping cavity at the end of LINAC5, after the DTL structures, as in LINAC3~\cite{LHCDesignReport}. This ramping cavity would be similar in design and cost to a bunching cavity at approximately 150\,kCHF.

\subsection{Beam dynamics in MEBT and Quasi-Alvarez DTL}
\label{SubSec:Linac_QuasiAlvarez_BeamDynamics}

The beam dynamics were modeled from the exit of the RFQ to the end of the Quasi-Alvarez DTL module using the multi-particle tracking codes TraceWin~\cite{Tracewin} and Travel~\cite{Travel}. A beam of 1$\times10^{5}$ particles was created using a distribution derived from tracking particles through the RFQ as described in section~\ref{Sec:Linac_RFQ}; space charge was also modeled. The ion types at the charge over mass ratio extremities $Q/A$ = 1/3 and $Q/A$ = 1/4 were tracked independently. Transmission was 100\% for the preliminary, zero-error study cases. Given magnet alignment and other tolerances observed in the LINAC4 system that also uses PMQs~\cite{pmqs_l4}, we believe this preliminary design is capable of operating with a transmission greater than 90\% through the Quasi-Alvarez structures in the real machine.

\begin{figure}[!htb]
\centering
\includegraphics[width=0.8\columnwidth]{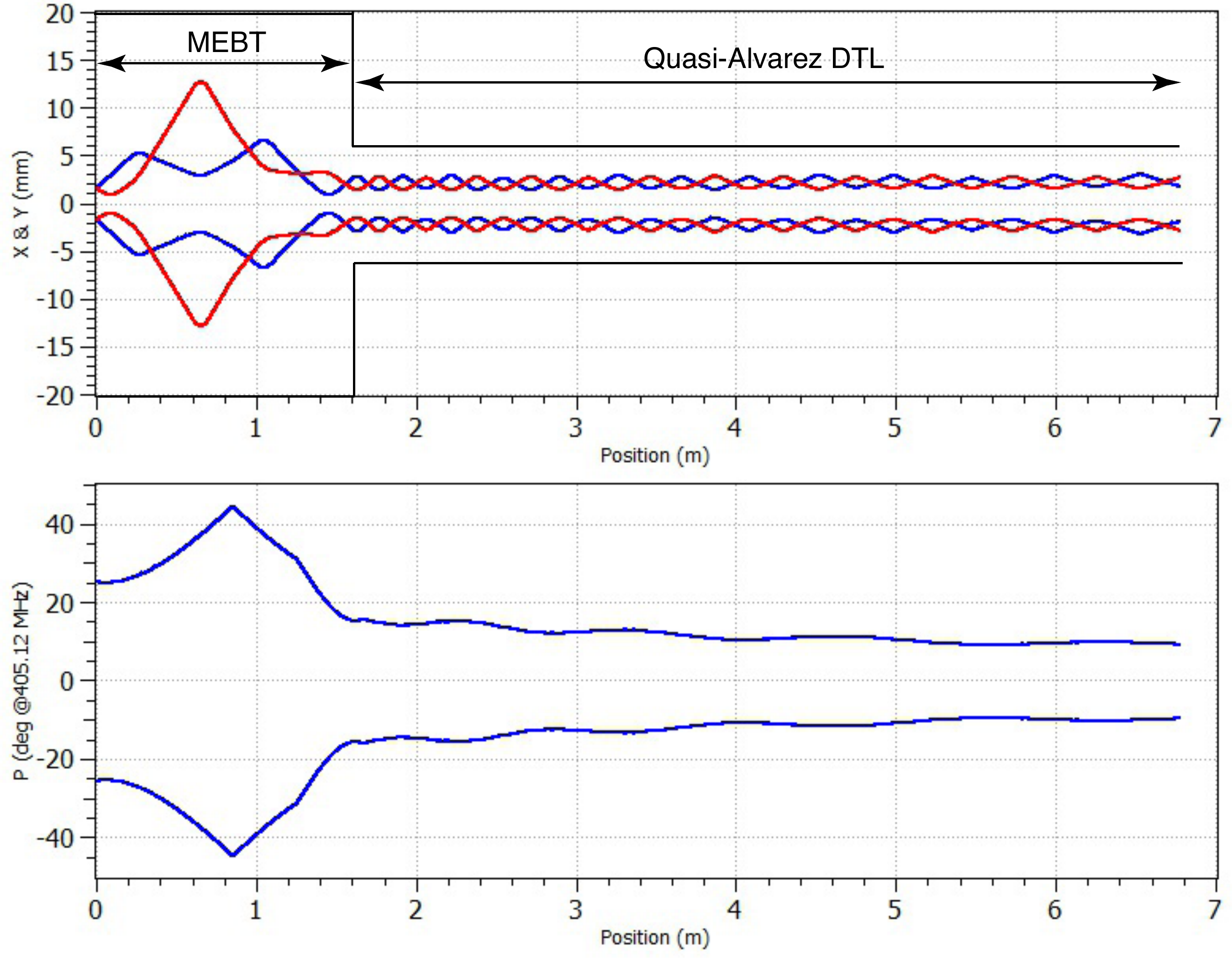}
\caption[]{The upper figure shows horizontal (blue) and vertical (red) $\sqrt{5}$rms beam envelopes and the beam-pipe aperture limit (black) from the end of the RFQ to the end of the Quasi-Alvarez DTL for ions with $Q/A$ = 1/4. The lower figure shows emittance evolution in the horizontal (blue) and vertical (red) planes.}
\label{fig:envXY1}
\end{figure}

\begin{figure}[!htb]
\centering
\includegraphics[width=0.8\columnwidth]{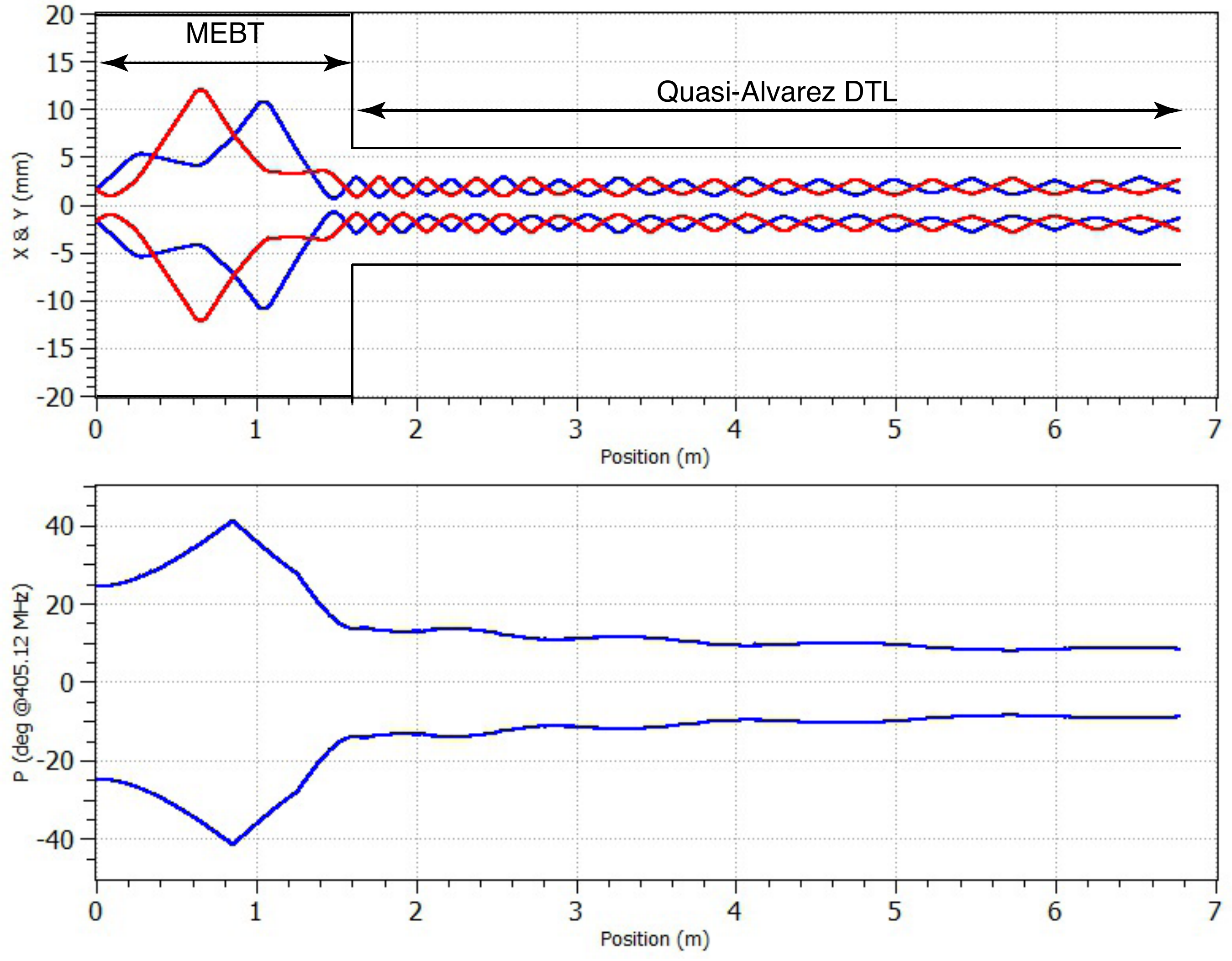}
\caption[]{The upper figure shows horizontal (blue) and vertical (red) $\sqrt{5}$rms beam envelopes and the beam-pipe aperture limit (black) from the end of the RFQ to the end of the Quasi-Alvarez DTL for ions with $Q/A$ = 1/3. The lower figure shows emittance evolution in the horizontal (blue) and vertical (red) planes.}
\label{fig:envXY2}
\end{figure}

\begin{figure}[!htb]
    \centering
    \begin{subfigure}{.4\linewidth}
       \hspace*{-8mm}\includegraphics[width=1.2\linewidth]{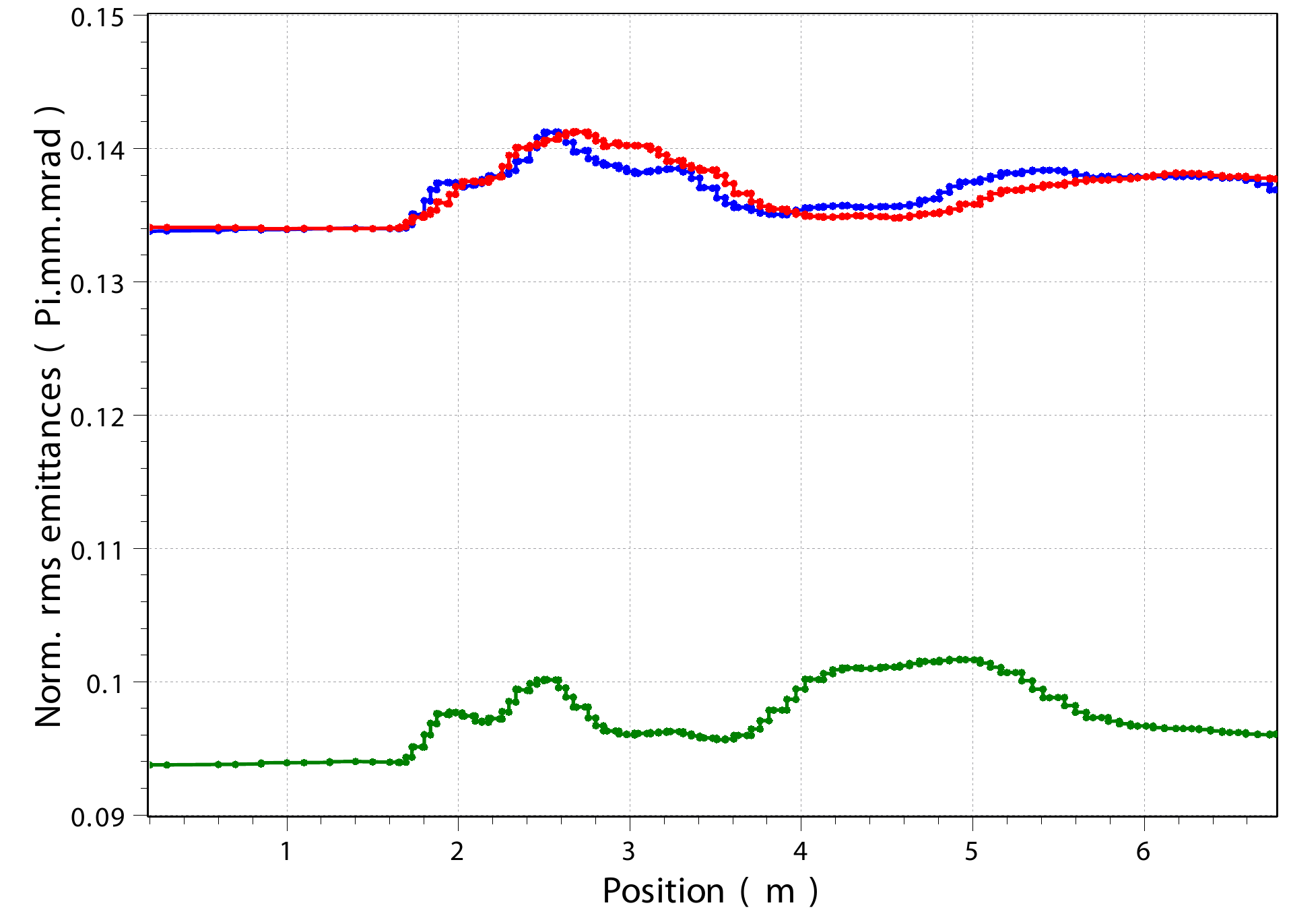}{}
        \caption{}
    \end{subfigure}
    \hskip4em
    \begin{subfigure}{.4\linewidth}
    \hspace*{-8mm}\includegraphics[width=1.2\linewidth]{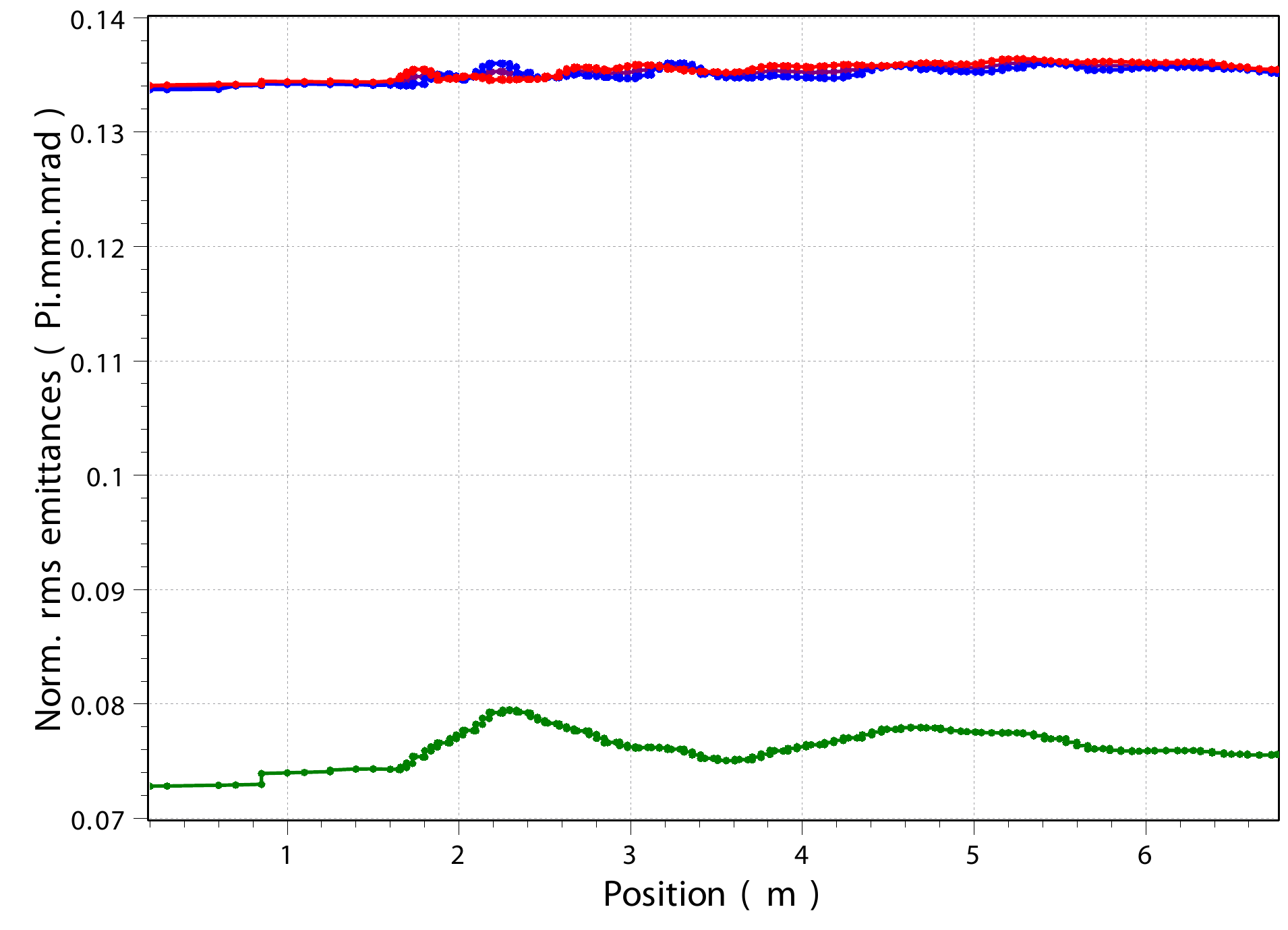}{}
    \caption{}
    \end{subfigure}
\caption[]{The horizontal (blue), vertical (red) and longitudinal (green) emittance evolution from the end of the RFQ to the end of the Quasi-Alvarez DTL for ions with $Q/A$ = 1/4 and 1/3 are shown respectively in (a) and (b).}
\label{fig:emit}
\end{figure}

The matched $\sqrt{5}$\,rms transverse beam envelopes and phase envelope for ions with $Q/A$ = 1/4 and 1/3 are shown respectively in figures~\ref{fig:envXY1} and~\ref{fig:envXY2}. The emittance evolution for ions with $Q/A$ = 1/4 and 1/3 are shown in figures~\ref{fig:emit} (a) and (b) respectively.

\subsection{Transmission estimates}
\label{SubSec:Linac_QuasiAlvarez_Transmission}

The transmission from the LEBT to the end of the RFQ is estimated to be around 70$\%$. Multi-particle simulations were used to estimate the transmission though the MEBT and the high-energy accelerating sections of LINAC5. Without using an error study model for the simulations, the transmission was above 99$\%$. However, we propose a conservative estimate of 90$\%$ transmission through the high-energy section once alignment, magnetic and RF field errors are considered. This gives a total proposed transmission from the source to the transfer lines after LINAC5 of around 60$\%$.

\subsection{Diagnostics, instrumentation and vacuum}
\label{SubSec:Linac_QuasiAlvarez_Diagnostics}

\begin{figure}[!htb]
\centering
\includegraphics[width=0.9\columnwidth]{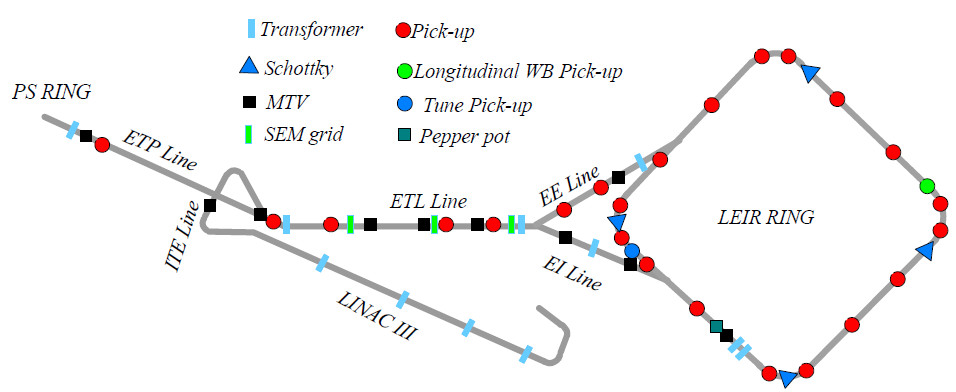}
\caption{Existing LEIR beam instrumentation.}
\label{Fig:Linac_BI}
\end{figure}

Most of the existing LEIR beam instrumentation can be reused for the BioLEIR facility (figure~\ref{Fig:Linac_BI}), except for new machine elements, like the new experimental beamlines as well for the new LINAC5.

It is envisaged that the following diagnostics are used to accurately measure, characterise and commission the beam:

\begin{itemize}
  \item Beam Current Transformers - BCTs (3): one in the LEBT, one in the MEBT and a third one at the end of the high-energy acceleration section.
  \item Beam Position Monitors - BPMs (6): two in the MEBT, three in the transfer lines between the end of LINAC5 and the existing transfer line magnet BHZ20 and one in between accelerating modules one and two (if this scenario is realised). These are used for steering and Time of Flight (ToF) measurements.
  \item Secondary Emission Monitors - SEM-grids (5-6): one in the LEBT, two in the MEBT and two or three in the transfer line after LINAC5. These are used for emittance reconstruction.
  \item Emittance meter (1): a full slit and grid emittance meter in the LEBT. One could also consider using the grid without the slit.
  \item Bunch Shape Monitor - BSM (1): one in the transfer line after LINAC5 to measure the longitudinal structure.
  \item Faraday cup (1-2): one in the LEBT and potentially one after LINAC5 in the transfer line. These are used for current measurement.
\end{itemize}

These diagnostics are very similar to the instrumentation that has been deployed for the LINAC4 project. Based on the real cost of these instruments, a total material cost of 1\,MCHF is estimated and detailed in table~\ref{Tab:Linac_BI}.

\begin{table}[ht]
\centering
\caption{Cost breakdown for LINAC5 beam monitors.}
\label{Tab:Linac_BI}
\begin{tabular}{l c r r}
\hline
\rule{0pt}{3ex}\textbf{Instrument} & \textbf{Number} & \textbf{Unit cost} & \textbf{Total cost} \\
\textbf{} & \textbf{} & \textbf{[kCHF]} & \textbf{[kCHF]} \\
\hline
\rule{0pt}{3ex}BCT  & 3 & 50 & 150 \\
BPM & 6 & 42 & 250 \\
SEM-grid & 6 & 42 & 250 \\
Emittance meter & 1 & 100 & 100\\
BSM & 1 & 200 & 200 \\
Faraday cup & 1 & 50 & 50\\
\hline
\rule{0pt}{3ex}\textbf{Total} & & &\textbf{1000}\\
\hline
\end{tabular}
\end{table}

\section{Power consumption and cooling requirements}
\label{Sec:Linac_PowerCV}

The proposed RF power consumption of the Quasi-Alvarez accelerating module can be seen in table~\ref{Tab:Linac_Cellparams2} as roughly 1.3\,MW. The quadrupole focusing magnets within the accelerating modules are proposed as PMQs, which need no electrical power or additional cooling beyond that of the RF cavities.

We estimate 1.5\,MW of RF power required for the high-energy accelerating part of the linac, 1\,MW for the RFQ and 0.5\,MW for the bunchers. For a repetition-rate of 2.4\,s and a bunch length of 200\,$\mu$s, the RF pulse is expected to be 1\,ms (conservative estimate). A duty cycle of 1\,ms/2.4\,s = 4.2$\times 10^{-4}$ and 4\,MW yields roughly 2\,kW of dissipated power to be cooled. The $\delta$T for the cavities depends on the mechanical design which is not available at this stage of the study and needs to be considered in the future.

For other magnets, the bunchers and the source, plus cooling of electronic racks, we believe that another 50\,kW of cooling is required and arrive at a conservative estimate of 100\,kW needing to be cooled at LINAC5. LINAC2 has currently 250\,kW of cooling power at its disposal which we expect to have access to.\\

\section{Resource estimate}
\label{Sec:Linac_Cost}

The cost estimate for LINAC5 excluding the source is given in table~\ref{Tab:Linac_Costing_Material}. It is based on recent experience with the LINAC4 construction. The preliminary integrated workforce estimates for design, construction and installation of LINAC5 is given in table~\ref{Tab:Linac_Costing_FTE}.

\begin{table}[ht]
\centering
\caption{Preliminary cost estimate for design, construction and installation of a light ion LINAC5. Total costs include prototyping and overheads where relevant, e.g. for tooling or design.}
\label{Tab:Linac_Costing_Material}
\begin{tabular}{l r r r}
\hline
\rule{0pt}{3ex}\textbf{System} & \textbf{Units} & \textbf{Unit cost} & \textbf{Total} \\
\textbf{} & \textbf{} & \textbf{[kCHF]} & \textbf{[kCHF]} \\
\hline
\rule{0pt}{3ex}RFQ  & 1 & 1000 & 1000 \\
Quasi-Alvarez HE section & 1 & 1000 & 1000 \\
LEBT \& MEBT girders/ancillary  & 1 & 200 & 200 \\
Bunchers & 3 & 150 & 450 \\
Magnets & multiple & - & 600 \\
Power converters & multiple & - & 940 \\
RF systems & multiple & - & 1700 \\
Diagnostic equipment & multiple& - & 1000\\
Layout/integration (person-months) & 6 & 10 & 60 \\
Alignment and survey (person-months) & 1  & 10 & 10 \\
Transport and handling (person-months) & 2  & 10 & 20 \\
\hline
\rule{0pt}{3ex}\textbf{Total} & & &\textbf{6980}\\
\hline
\end{tabular}
\end{table}

\begin{table}[ht]
\centering
\caption{Preliminary integrated personnel estimate for design, construction and installation of the new light ion linac for BioLEIR, given in person-years.}
\label{Tab:Linac_Costing_FTE}
\begin{tabular}{l r r}
\hline
\rule{0pt}{3ex}\textbf{System} & \textbf{Staff [PY]} & \textbf{Fellow [PY]} \\
\hline
\rule{0pt}{3ex}Beam dynamics & 1 & 2 \\
Magnets & 1 & 1\\
RF design & 4 & 3 \\
Power converters & 0.7 & 1.3 \\
Diagnostics & 1 & 2 \\
Power/control cabling & 2 & -\\
Installation coordination  & 1 & 1 \\
Beam preparation and commissioning & 1 & 1 \\
\hline
\rule{0pt}{3ex}\textbf{Linac Workforce [Person-years]} & \textbf{11.7} & \textbf{11.3}\\
\hline
\end{tabular}
\end{table}

It is estimated that 1~FTE are needed to maintain and operate LINAC5 with a yearly maintenance budget of about 370\,kCHF.

\section{Conclusions}
\label{Sec:Linac_Conclusions}
A linac solution for BioLEIR was developed using a Quasi-Alvarez DTL structure for the high-energy part of the linac. Matching was performed to comfortably meet the magnet and aperture constraints, the emittance growth is small and the transmission is acceptable and comparable to existing facilities. A power consumption of 1.3\,MW is similar to previously realised designs and should be obtainable with minimal added facilities. The total length of the current design from source to the end of the DTL section is estimated to be around 12 -- 13\,m and the final energy is 4.2\,MeV/u. More detailed studies should be carried out including error analysis, LEBT design and RFQ optimisation.

Major cost and staff saving elements are planned for LINAC5 due to the use of the existing LINAC2 tunnel (after its replacement by LINAC4). This includes no new buildings or structures, utilisation of much of the existing power and cooling facilities, access control and existing transfer lines to LEIR. The choice of PMQs within the Quasi-Alvarez high-energy accelerating modules further reduces the power consumption of the linac by not requiring electrical power or additional cooling.

%% file: Chapters/InjectionTransferLine.tex
\chapter{Injection Transfer Line}
\label{Chap:Transferline}

\begin{figure}[ht]
\centering\includegraphics[width=1\linewidth]{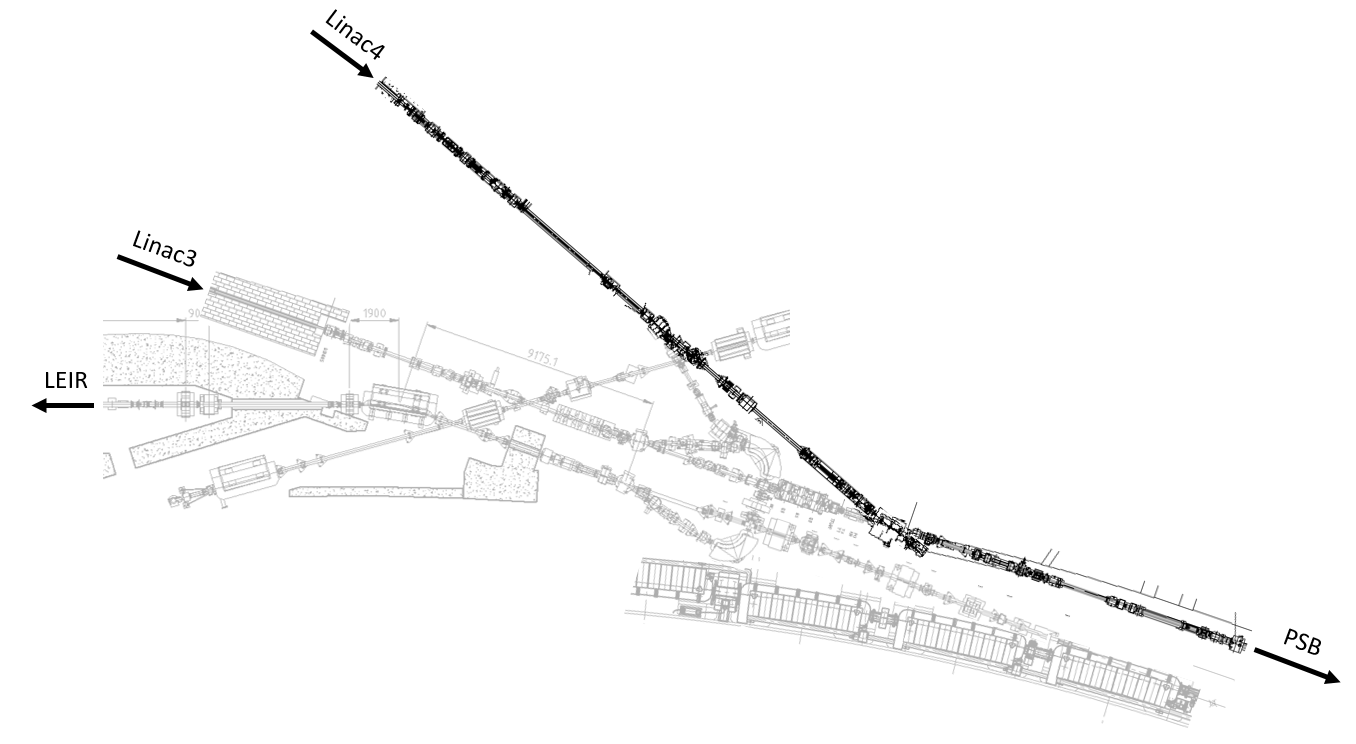}
\caption{Overview of the existing transfer of ions from LINAC3 to LEIR and of protons from LINAC2 to PSB (bold).}
\label{Fig:Transferline_01}
\end{figure}

LINAC2 injects protons into the Proton Synchrotron Booster (PSB) via the LT/LTB/BI transfer line, which passes close to the Proton Synchrotron (PS) ion injection region in sector 42. The injection of ions into LEIR is made via a 180$^\degree$ ITE turnaround loop into the bi-directional ETL transfer line, which is also used to inject ions from LEIR into the PS. An overview of the transfer of ions from LINAC3 to LEIR and of protons from LINAC2 to PSB are shown in figure ~\ref{Fig:Transferline_01}.

\section{Requirements and constraints}
\label{Sec:Transferline_Requirements}

Transfer of the ions from LINAC5 to LEIR is required for BioLEIR, using as far as possible the existing infrastructure. The main parameters of the beams to be transported from LINAC5 to LEIR are given in table~\ref{Tab:Transferline_Parameters}. 

\begin{table}[ht]
\centering
\caption{Main beam parameters for the LEIR injection beamline.}
\label{Tab:Transferline_Parameters}
\begin{tabular}{l l l}
\hline
\rule{0pt}{3ex}\textbf{Beam parameter} & \textbf{Unit} & \textbf{Value}\\
\hline
\rule{0pt}{3ex}Injection energy ($^{12}$C $^{6+}$) &	MeV/u &	4.2 \\
Injection rigidity &	T.m& 	0.59 \\
Injected beam emittance	& $\pi$.mm.mrad (normalised) &	0.9 \\ 
Total intensity	& ions	& $1\times10^{11}$ \\
Injection duration & ms & 0.2 \\
\hline
\end{tabular}
\end{table}

\section{Injection line from LINAC5 to LEIR}
\label{Sec:Transferline_Injection}

\begin{figure}[ht]
\centering\includegraphics[trim={5cm 0 0 0},clip, width=1.2\linewidth]{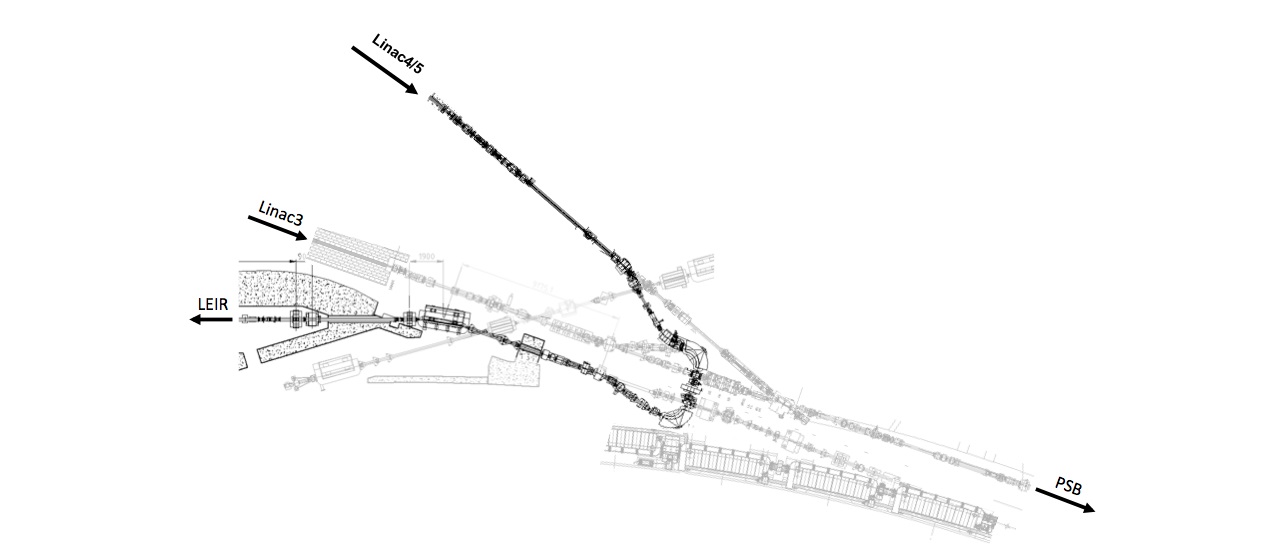}
\caption{Transfer of ions from LINAC5 to LEIR for BioLEIR, using the connection E0 between the LT line and the LTE line.}
\label{Fig:Transferline_02}
\end{figure}

The possibility to inject protons into the Low Energy Antiproton Ring (LEAR) from LINAC2 existed in the past, using the short section of transfer line (E0) linking LT and the ITE line at the level of the 180$^\degree$ turnaround loop. The elements of this short, unused E0 section line have been decommissioned and moved from the PS switchyard area into storage during the EYETS 2016/2017~\cite{Kuchler:2016}. The overall configuration of the line which would be used for injection of ions from LINAC5 into LEIR for BioLEIR is shown in figure~\ref{Fig:Transferline_02}.

A section of the LT line, between the LINAC4 connection and the switch to the LTE loop (i.e. between BHZ20 and BHZ25 inclusive), would need to be operable in Pulse-to-Pulse Modulation (PPM) mode, with fast changes of settings between different PSB users in order to be able to run BioLEIR while sending interleaved proton beams to the PSB from LINAC4. In particular, as LINAC4 operates with H$^{-}$, the polarity of the dipoles and most likely the quadrupoles as well, needs to flip between BioLEIR and regular LINAC4 cycles. This would impose an upgrade of any non-laminated magnets as well as of the power converters and controls, and possibly of the beam instrumentation controls to be compatible with the fast change of destination. 

\begin{figure}[ht]
\centering\includegraphics[width=1\linewidth]{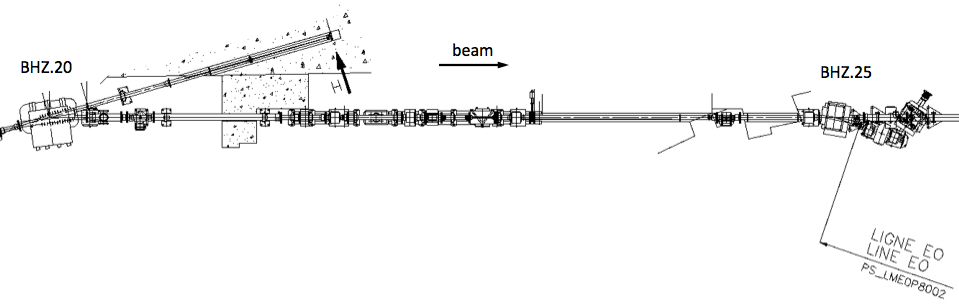}
\caption{Transfer of ions from LINAC5 to LEIR for BioLEIR, using the connection E0 between the LT line and the LTE line.}
\label{Fig:Transferline_03}
\end{figure}

The affected section of the LT transfer line is shown in figure~\ref{Fig:Transferline_03}. Two geometric horizontal dipoles are concerned, LT.BHZ20 and LT.BHZ25, both of laminated design BHZ. This magnet type (and the replacement to be installed for LINAC4) does not have enough horizontal aperture to accommodate beam from the two linacs, and needs replacing with a new magnet. There are also four quadrupoles, which are laminated of type MQNAI and can similarly be operated in pulsed mode with polarity flip. All elements would at least need new controls to allow rapid change and bipolar regulation.

When the LINAC4 is connected to the Booster, the LT.BHZ20 bending magnet is powered by a converter that can deliver bipolar current pulses with a flat-top of a few hundred milliseconds maximum duration. Consequently, to allow operation with extended flat-top duration or even in DC mode, a new power converter is required. A new power converter is also needed for LT.BHZ25. For these magnets, two SIRIUS\_4P-converters are foreseen to be installed. These converters are 4-quadrant power converters that are composed of 4 units connected in parallel. SIRIUS\_4P-converters are rated 80\,V and 800\,A in DC operation mode. In addition, the cabling of LT.BHZ25 has to be foreseen.

In the framework of the accelerators consolidation programme, the power converters of the four quadrupoles (LT.QFN50, QDN55, QFN60 and QDN65) are envisaged to be upgraded during LS2 with pulsed converters of MAXIDISCAP type. The MAXIDISCAP power converters are unipolar in current but can be upgraded to bipolar with the implementation of an additional polarity inverter that could be installed for BioLEIR. Moreover, if operation with 10\,Hz repetition rate is required, these converters have to be modified to MAXIDISCAP\_10Hz version. This 10\,Hz version of the MAXIIDSCAP power converter is currently in use in the LINAC3 machine.

Common use of this section of line may prove to be operationally very limiting, as it couples the operation of LINAC4 for protons to LINAC5 producing ions for BioLEIR. Further study is needed to decide whether the above proposal is really feasible. If not, then a short section of new transfer line is needed between BHZ20 and BHZ25, to link the LT line to the E0 line. This would add extra cost to the project, and possibly complications in terms of transport and handling, to access both beamlines.

The reuse of the existing short additional section of line E0 to connect the LT line to the LTE loop is not expected to be particularly problematic, although the beam trajectory passes far from the magnetic axis of the first E0 dipole BHN02. This requires a field-mapping of the dipole and development of a specific magnetic model. 

\begin{figure}[ht]
\centering\includegraphics[trim={0 0.1cm 0 0},clip, width=0.8\linewidth]{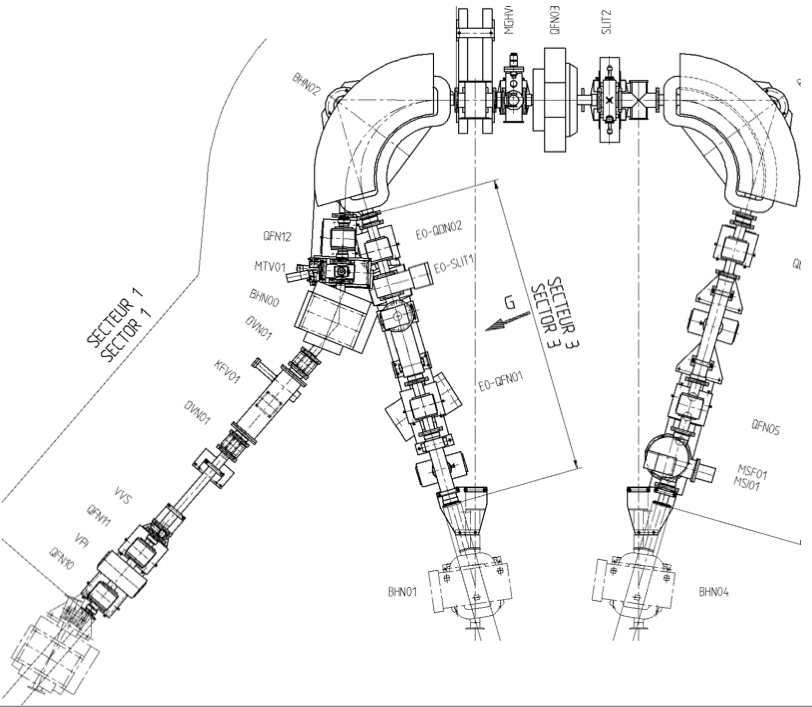}
\caption{E0LINE2 junction between LT line (not shown) and E0 loop. The off-axis trajectory through BHN02 is clearly seen, which presents issues of beam dynamics as the field in this region is far from a pure dipole.}
\label{Fig:Transferline_04}
\end{figure}

The layout of E0 and the junction with the LTE line is shown in figure~\ref{Fig:Transferline_04}. The power converters and cabling for the magnets in this section of E0 line need to be supplied. These comprise three quadrupoles of MQNAI type, and one geometric dipole BHN00 of MDX type. This magnet is solid and may be problematic if LINAC3 and LINAC5 need to use the E0 loop alternately.

As for the LT line, a SIRIUS\_2P (2 units in parallel) power converter operated in cycled mode is used to supply the dipole magnet, and MAXIDISCAP\_10Hz type converters are used to power the quadrupole magnets in fast pulsed operation mode with 10\,Hz maximum repetition rate. The power converter requirements for the injection lines are summarized in table~\ref{Tab:BeamLines_Converters_req}.

The final layout of the line can be optimised with the addition of correctors and quadrupoles, if needed, as the kicker magnet KFV01 (the LINAC3-LEAR head-tail clipper) is most likely not required. 

Note that this section of line (from BHZ25 to QFN12 included) has been decommissioned and moved into storage~\cite{Kuchler:2016}. The condition of all removed elements, supports and vacuum chambers need to be thoroughly examined before any potential re-installation for BioLEIR.

\section{Ion stripping and filtering}
\label{Sec:Transferline_4}

\begin{figure}[ht]
\centering\includegraphics[trim={0 0.1cm 0 0},clip, width=0.8\linewidth]{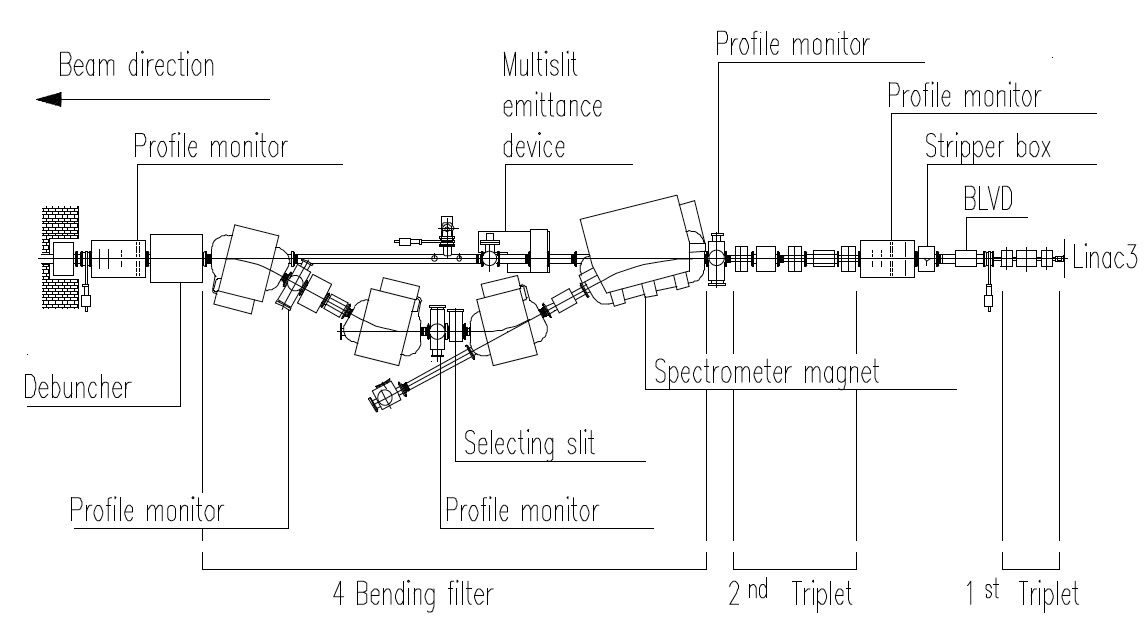}
\caption{Current 12-m-long filter line insertion downstream of LINAC3 for beam stripping and purification. A similar insertion downstream of LINAC5 is needed.}
\label{Fig:Transferline_05}
\end{figure}

A stripping foil is needed in the injection line to partially strip the 4.2~MeV/u ions. This can be based on the same stripper design as the one for LINAC3~\cite{Catalan-Lasheras:318968} or very similar, i.e. 75~$\mu$g/cm$^3$ of Carbon. For beam purity and characterisation, a downstream dogleg is probably required with a spectrometer, slits and instrumentation to remove any unwanted charge states, similar to the insertion used at the exit of LINAC3 (figure~\ref{Fig:Transferline_05}). It is assumed that this 12-m-long insertion can be accommodated in the line between the exit of LINAC5 and the junction with the LT line at BHZ20. It is assumed that four new 30$^\degree$ dipoles (one of BHZ1 type and three of BHZ2 type), together with new power converters, are required for this insertion. These are laminated and compatible with fast setting changes. The characteristics of the LINAC5 dogleg dipole are given in table~\ref{Tab:BeamLines_Dogleg_Dipoles}. 

If the design of the dipole magnets is done using laminated yokes and allow the operation in cycled mode, the power converters for the dipoles are of SIRIUS\_2P type.  
The requirements for the power converters for the whole injection lines are summarized in table~\ref{Tab:BeamLines_Converters_req}.
All power converters are controlled using standard CERN digital controller (FGC with Ethernet+) and interfaces (magnet interlocks, beam interlocks etc.).

\begin{table}[ht]
\caption{Filter line dipole parameters.}
\label{Tab:BeamLines_Dogleg_Dipoles}
\centering
\begin{tabular}{l l r r}
\hline
\rule{0pt}{3ex}\textbf{LINAC5 dogleg dipoles} & \textbf{Unit} & \textbf{Spectrometer} & \textbf{Short dipole} \\
\hline
\rule{0pt}{3ex}Magnet type & & BHZ2 & BHZ1 \\
Dipole bend angle & deg & 30 & 30  \\
Gap dipole field & T & 0.5 & 0.25 \\
Magnetic length & m & 0.6 & 1.2 \\
Integrated field & Tm & 0.3 & 0.3\\
Number installed (spare) &  & 3 (0) & 1 (0)\\
Magnets in series &  & 1 & 1  \\
Min. ramp time & s & 0.2 & 0.2 \\
Flat-top duration & s & 1.5 & 1.5 \\
Repetition rate & s & 2.5 & 2.5 \\
Nom. current & A & 400 & 400 \\
\hline
\end{tabular}
\end{table}

\begin{table}[!ht]
\caption{Injection line power converter parameters. The upgrade of the power converters for the LT quadrupoles is not mentioned here but taken into account for the cost estimate.}
\label{Tab:BeamLines_Converters_req}
\centering
\begin{tabular}{l l r r r r}
\hline
\rule{0pt}{3ex}\textbf{Parameter} & \textbf{Unit} & \textbf{Bend\_800A} & \textbf{Bend\_400A} & \textbf{Quad.}\\
\hline
\rule{0pt}{3ex}Converters (spare) &  & 2 (1) & 4 (0) & 3 (1)\\
Operation mode &   & cycled & cycled & fast pulsed \\
Nom. current & A & $\pm$800 & $\pm$400 & 200 \\
Max. voltage & V & 450 & 450 & 1000\\
Current precision & ppm & $\pm$200 & $\pm$100 & $\pm$1000 \\
AC input & & 3P-400Vac-125A & 3P-400Vac-63A & 1P-230Vac-10A \\
Dimensions & mm & 2000$\times$3200$\times$900 & 2000$\times$1800$\times$900 &10U$\times$19"$\times$650 \\
Cooling &  & Water & Water & Forced air \\
\hline
\end{tabular}
\end{table}

\section{Beam instrumentation}
\label{Sec:InjectionLines_BI}
The injection lines ITE, ETL (which is also used for ejection to the PS) and EI are equipped with current transformers for intensity measurement and scintillating screens for beam position and size measurement.

A total of 10 screens and their associated cameras (MTVs) are installed in the LEIR transfer lines. Seven are used to steer the injected beam, from which four can also be used at ejection. Depending on the radiation levels, CCD or Vidicon cameras are used. The MTV electronics hardware consists of a single VME 64x card capable of controlling the different types of positioning mechanisms for the screens, the adjustment of the illumination intensity, the different types of cameras (i.e. CCD or Vidicon tube) and the positioning of optical filters in front of the camera. Apart from the analogue video signal, the card provides also a digitised image.

The intensity of the injected beam is measured using fast current transformers. At the time of the LEIR project, the transformers in the EI and EE line were refurbished with new magnetic shielding and a water cooling system that protects the magnetic cores from overheating during bake-out. The ETL line is used for injection as well as ejection and the transformers can measure long current bunches as well as short, high-intensity bunches. These transformers are therefore equipped with two different electronics chains. All transformers are equipped with fast sampling ADCs; 10\,MHz for measurement of injection pulses or 200\,MHz for the ejected bunches. A calibration pulse is injected shortly before the passage of the beam and the beam currents are calculated after baseline restoration through digital integration.

The injection line is being equipped with nine new position pickups as part of the LHC Injector Upgrade (LIU) programme during EYETS/YETS 2016-18. This allows a much faster setting up of the beam transfer from the linacs to LEIR. No additional cost is expected for transfer line beam instrumentation at this point.

\section{Resource estimate}
\label{Sec:InjectionLines_Costing}

The estimated CERN personnel needs (Staff and Fellows) are given in table~\ref{Tab:InjLines_Costing_FTE}. The estimate includes the remaining conceptual design phase, engineering design, prototyping, construction, installation and commissioning. It assumes that the only new magnets needed are the four dipoles for the LINAC5 dogleg, and that the rest exists. This assumption of course requires validation by a full optics design study, and depends on whether the section of the LT line can be reused.

\begin{table}[ht]
\caption{Preliminary CERN personnel estimates for design, construction, installation and commissioning of extraction beamline elements. An extra 3.0 FTE are added if the common part of the LT line cannot be reused.} 
\label{Tab:InjLines_Costing_FTE}
\centering
\begin{tabular}{l r r}
\hline
\rule{0pt}{3ex}\textbf{System} & \textbf{Staff [PY]} & \textbf{Fellow [PY]} \\
\hline
\rule{0pt}{3ex}Beam dynamics & 0.5 & 1.0 \\
Magnets & 0.5 & 0.5\\
Power converters & 0.6 & 1.2 \\
Diagnostics & 1.0 & 1.0 \\
Power/control cabling & 0.1 & -\\
Water cooling & 0.1 & - \\
Installation coordination  & 0.3 & - \\
Beam preparation and commissioning & 0.5 & 1.0 \\
\hline
\rule{0pt}{3ex}\textbf{Total} & \textbf{3.6} &\textbf{4.7}\\
\hline
\end{tabular}
\end{table}

The number required  and assumed costs for the different equipment subsystems described above are given in table~\ref{Tab:InjLines_Costing_Material}. Note that in the absence of a detailed conceptual study, the numbers given are preliminary: the actual costs may differ from these estimates by as much as a factor of two. The costs include all non-CERN staffing such as Field Support Units.
No significant extra exploitation cost or staffing is required for the injection line, as most elements are already in operation. 

\begin{table}[ht]
\caption{Preliminary cost estimate for design, construction and installation of injection beamline elements including LINAC5 filter line and replacement of BHZ20 and BHZ25 dipole magnets. Total costs include prototyping and overheads where relevant, e.g. for tooling or design. Fellow costs are included in table~\ref{Tab:InjLines_Costing_FTE}. An extra 800 -- 1000\,kCHF needs to be added if the common part of the LT line cannot be reused.}
\label{Tab:InjLines_Costing_Material}
\centering
\begin{tabular}{l r r r}
\hline
\rule{0pt}{3ex}\textbf{System} & \textbf{Units} & \textbf{Unit cost} & \textbf{Total} \\
\textbf{} & \textbf{} & \textbf{[kCHF]} & \textbf{[kCHF]} \\
\hline
\rule{0pt}{3ex}H dipole magnets  & 6 & 75 & 450 \\
New dipole converters (4P+2P)  & 3+4 & 160+80 & 800 \\
Quadrupole converters (upgrade+new) & 4+3 & 10+15 & 85 \\
Filter line instrumentation & 1 & 100 & 100 \\
H dipole vacuum chambers & 6 & 10 & 60 \\
Drift chambers (m) & 20 & 5 & 100 \\
Dipole jacks & 12 & 5 & 60 \\
Magnet power cabling & 9 & 5 & 45 \\
Beam instrumentation cabling & 1 & 20 & 20\\
Magnet/converter water cooling & 4 & 5 & 20 \\
Layout/integration (man-months) & 2 & 10 & 20 \\
Alignment and survey (man-months) & 1  & 10 & 10 \\
Transport and handling (man-months) & 2  & 10 & 20 \\
\hline
\rule{0pt}{3ex}\textbf{Total} & & &\textbf{1790}\\
\hline
\end{tabular}
\end{table}

\section{Potential risks and remaining conceptual issues}
\label{Sec:InjLines_Risks}

The estimates are based only on a preliminary analysis without any actual design of the beamline. The main technical challenge is the conversion of part of the LT line to a PPM line, able to change polarity rapidly, which may not be feasible and which might need a 20\,m section of new transfer line to be designed and built.  This new line would require two dipoles, around four quadrupoles, at least one corrector per plane, some instrumentation for beam position and then vacuum systems, supports, cabling and other infrastructure. This would add at least another 800 -- 1000\,kCHF to the project cost, and about 3 person-years, if all elements need to be newly built.

Another item requiring attention is the recommissioning of the E0 line connecting to the ETH loop, and the field characterisation and proper beam transport through the fringe field of the BHN02 dipole. This is not yet demonstrated in terms of optics performance.

In addition, detailed beam dynamics studies are still required for the overall trajectory correction/aperture and the beam instrumentation systems. These are not expected to pose serious difficulties, but the assumptions need to be validated and basic specifications and engineering concepts need to be established. In particular, the reuse of the present LINAC3 stripping and filtering insertion needs to be validated.

%% file: Chapters/BeamDynamics.tex
\chapter{The LEIR Accelerator}
\label{Chap:LEIR}

\section{LEIR description}
\label{Sec:BeamDynamics_LeirDescription}

\begin{figure}[ht]
\centering\includegraphics[width=0.8\linewidth]{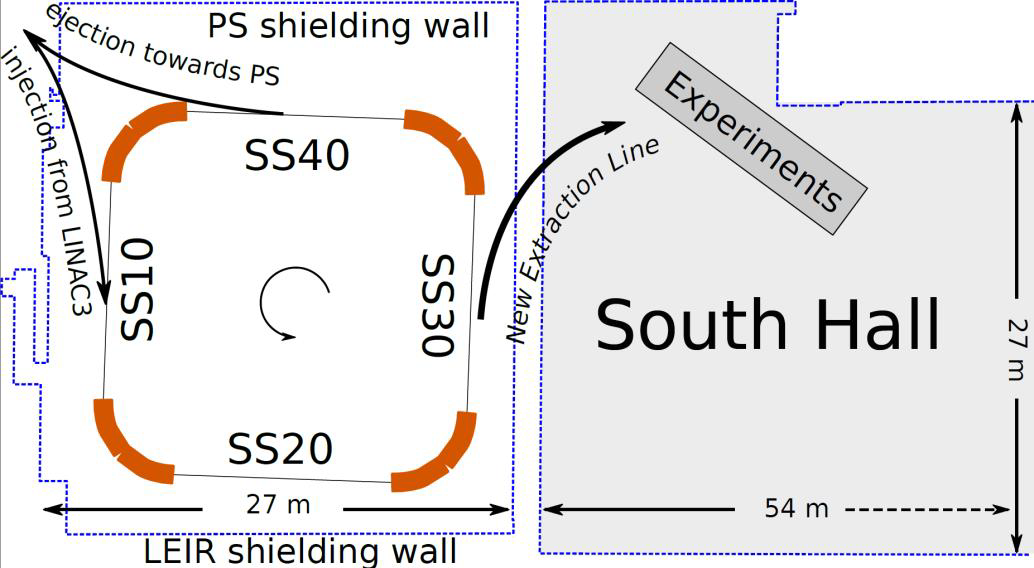}
\caption{Schematic view of LEIR and the adjacent areas with possible locations for experimental end-stations~\cite{Garonna:2014}.}
\label{Fig:BeamDynamics_Layout_LEIRLayout}
\end{figure}

The LEIR synchrotron has a circumference of about 80\,m and is composed of four 13\,m-long straight sections (SS) numbered 10 to 40 and four 90$^\degree$ bending magnets, as shown in figure~\ref{Fig:BeamDynamics_Layout_LEIRLayout}.

One of the straight sections (SS10) is used for injection, a second (SS20) houses the electron cooler, a third (SS30) has kicker magnets for fast extraction and the fourth (SS40) houses the extraction septum towards the Proton Synchrotron and the radiofrequency cavities. Focusing is provided by quadrupole doublets in the injection section (SS10) and the opposite section (SS30), by triplets in the electron cooling section (SS20) and in the section for current heavy ion extraction (SS40). Two families of four sextupole magnets are installed in SS10 and SS30 (with large dispersion of the nominal lattice for heavy ion operation) for chromaticity correction. In addition, two weaker sextupoles are present in SS40. The main bending magnets are composed of 4 blocks with the outer blocks equipped with pole face and backleg windings. These produce sextupolar and dipolar magnetic fields for orbit correction in both transverse planes~\cite{Garonna:2014, Benedikt:2004}.

LEIR is located in the CERN South Hall, next to the Proton Synchrotron ring. Only a small part of the South Hall is occupied by LEIR and most of the surface is currently being used as storage area. Clearing this space could provide approximately 1500\,m$^2$ for the installation of experimental beamlines and related infrastructure~\cite{Garonna:2014}, as shown in figure~\ref{Fig:BeamDynamics_Layout_LEIRLayout}.

\section{LEIR beam dynamics}
\label{Sec:BeamDynamics_LEIRBeamDynamics}

\begin{figure}[ht]
\centering\includegraphics[trim={0 0.4cm 0 1.3cm},clip, width=0.75\linewidth]{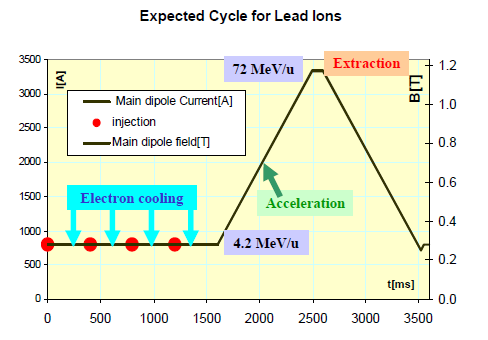}
\caption{LEIR cycle for Lead ions~\cite{Benedikt:2004}.}
\label{Fig:BeamDynamics_Layout_LEIRCycle}
\end{figure}

The LEIR cycle presently used for filling the LHC with Lead ion beams has a length of 3.6 s (figure~\ref{Fig:BeamDynamics_Layout_LEIRCycle}). In this scheme, seven long, low-intensity pulses of Lead ions at 4.2 MeV/u from the linear accelerator LINAC3. These pulses are spaced by 200\,ms, are accumulated in LEIR, by alternating a special stacking mechanism in transverse and longitudinal phase space and electron cooling. LEIR features a multi-turn injection with simultaneous stacking in momentum and in both transverse phase spaces. The nominal machine optics with the working point (Q$_x$, Q$_y$) = (1.82, 2.72) was tuned to optimize the injection efficiency (50-70\%).

\begin{figure}[ht]
\centering\includegraphics[width=0.8\linewidth]{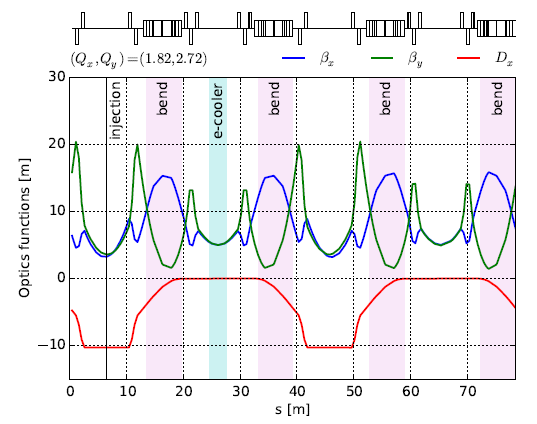}
\caption{Optics functions along the LEIR circumference. The solenoid of the electron cooler slightly perturbs the lattice symmetry~\cite{Bartosik:2016}.}
\label{Fig:BeamDynamics_Layout_LEIROpticsFunctions}
\end{figure}

Figure~\ref{Fig:BeamDynamics_Layout_LEIROpticsFunctions} shows the optics functions with their quasi-twofold periodicity around the 78\,m machine circumference of LEIR. The phase space volume of the injected and accumulated beam is reduced by electron cooling. Subsequently, the coasting beam is captured into two bunches using a double harmonic RF system in bunch lengthening mode, accelerated to a kinetic energy of 72.2\,MeV/u and extracted via fast extraction towards the Proton Synchrotron for further acceleration. LEIR has accumulated, cooled and stacked ion beams of Oxygen (O$^{4+}$), Lead (Pb$^{54+}$) and Argon (Ar$^{11+}$)~\cite{Bartosik:2016, Manglunki:2013}.

\section{Coupling due to the electron cooler and its compensation}
\label{Sec:BeamDynamics_EC}

The electron cooler strongly reduces the phase space volume of the injected beam, slightly decelerates the beam and adds it to the stack sitting at a slightly lower energy. Once seven LINAC3 pulses are accumulated and cooled, electron cooling is stopped. Electron cooling rate measurements showed that betatron functions around 5\,m and a finite dispersion were found to enhance cooling rates. Thus, the lattices presented here have been tuned to $\beta_x$ = $\beta_y$ = 5\,m, but in order to simplify operation with energy ramping during injection, zero dispersion D = 0 m has been chosen~\cite{Benedikt:2004}, as shown in figure~\ref{Fig:BeamDynamics_Layout_LEIROpticsFunctions}.

\begin{figure}[ht]
\centering\includegraphics[width=1.0\linewidth]{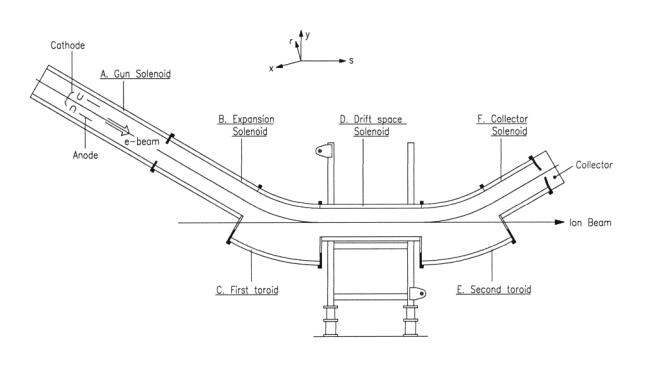}
\caption{General layout of the electron cooler~\cite{EC}.}
\label{Fig:BeamDynamics_EC_ElectronCooler}
\end{figure}

The principle of the electron cooler device is shown in figure~\ref{Fig:BeamDynamics_EC_ElectronCooler}. A gun provides a dense quasi monoenergetic electron beam. The beam is expanded by a dedicated solenoid and is bent by a toroid in order to merge with the circulating ion beam which is then cooled in the drift space. At the end of the drift space the electron beam is bent away from the ion beam and finally collected. The overall system is embedded in a longitudinal magnetic field aimed to counteract the electron beam space charge forces and to magnetize the electrons. The magnetic field in the drift space must be uniform in order to ensure a good cooling efficiency. The maximum magnetic field in the toroids and the drift space is 0.1 T~\cite{EC}.

A solenoidal field is necessary in order to focus the electron beam of the cooler. However, this affects the ion beam as well as introducing coupling. The electron cooler has been modelled as a solenoid with a length of 4.30\,m and with azimuthal symmetry. The coupling is compensated by two compensation solenoids with a length of 0.40 m on either side of the cooler installed at a distance of 3.15 m from the centre of the straight section. If skew quadrupolar fields are present in the region around the toroids, the residual coupling can be compensated efficiently by also using the skew-quadrupoles QSK21 and QSK22~\cite{Benedikt:2004}. 
Figure~~\ref{Fig:BeamDynamics_Layout_LEIROpticsFunctions} shows the optics functions, which were calculated using MADX~\cite{MADX} along the LEIR circumference when the electron cooler and its related compensation elements were off. The same optics functions were obtained when the electron cooler and its related compensation elements were powered and matching with elements all over the ring assured. This means that the coupling introduced by the electron cooler can be cancelled with the compensation elements (two compensation solenoids and two skew-quadrupoles).

BioLEIR will require a wide range of ion types to be stored and accelerated, which will require studies of the cooling time and loss rates for each species, and may require some optimisation of the ion charge state to be selected for injection from the linac.
Low intensities of ions that can be accumulated with a single injection from either LINAC3 or LINAC5, will probably not be limited by the electron cooling, but for higher intensities requiring more than one injection, the injection spacing and a charge-state selection will need to be studied and optimised.

\section{Beam instrumentation in the LEIR ring}
The LEIR ring is equipped with a ''semi-fast'' transformer measuring the intensity during multi-turn injection into the machine and a DC current transformer providing intensity measurement all along the accelerating cycle. The semi-fast transformer can measure every injection pulse (200\,$\mu$s) with a droop of less than 1\%, and has a time constant short enough to measure the next injection 200\,ms later, without any current offset.

The emittance and matching of the beam injected into LEIR is measured by means of a ''pepper pot'' device. This device consists of a molybdenum mask, in which a square matrix of 0.17\,mm holes with 1\,mm spacing is drilled. The surviving particles are observed on a scintillating screen located some 30\,cm further downstream and digitized using a CCD camera. By measuring the width and the centre of the distribution of each beamlet, it is possible to reconstruct the transverse phase space distributions of the injected beam.\\

Around the ring there are 32 ceramic-based electrostatic pickups which measure the closed orbit of the circulating beam and can also monitor the trajectory at injection for the correction of the coherent oscillations when the beam is injected. Eight horizontal and vertical pickups are placed in the bending magnets, eight combined pickups are installed in the straight sections and another two in the electron cooler. The electronics system consists of head amplifiers with three different gains followed by an analogue normaliser. The 32 averaged position signals are read out through a multiplexing ADC.

The transverse profiles are obtained using the ionisation profile monitors. Both monitors consist of electrodes separated by alumina plates and a two-stage micro-channel plate (MCP) for the amplification of the ionisation signal. This amplified signal is collected on a strip readout with a resolution of 1\,mm.
A scraper system for the destructive measurement of the beam profiles also exists but has never been used. It should be decided whether or not this system is required as the control system of the motors needs to be refurbished.

The Schottky pickups are non-perturbative measurement tools which provide a large set of information for either coasting or bunched beams. With a spectrum analyser one can deduce: the tune, the emittance, the momentum distribution and the revolution frequency. There are two Schottky pickups, consisting of a succession of short strip-line pickups, per transverse plane. For each plane one pickup of them is connected as travelling wave pickups suitable only for low energy particles at injection (relativistic $\beta$\,=\,0.095). The two other pickups are used in a parallel configuration, which work at all particle velocities.  The horizontal pickups are  summed to yield information on the longitudinal beam distribution.

The LEIR tune measurement system is based on the Direct Diode Detection (DDD) principle. In this scheme, electrode signals from a position pickup feed diode peak detectors, which down-mix the beam spectrum to the base-band. As most of the revolution frequency content is converted to DC, this large background is suppressed by series capacitors already before the first active stage of the analogue frontend, resulting in high system sensitivity and robustness against saturation due to large beam signals. The system, despite its simplicity, is capable of measuring tunes of bunched and debunched ion beams with small beam kicks without changing the system gain.

With the information at hand at this point, it is expected that the beam instrumentation of LEIR does not need any further modification or enhancement.

\section{Power converter upgrade for the LEIR accelerator magnets}
\label{Sec:Power_upgrade_LEIR}

The current power supply and water cooling of the main bending magnets limits LEIR operation to beam rigidities of 4.8\,T.m. However, their design limit is 6.7\,Tm. For fully stripped Carbon ions, this corresponds to maximal beam energies of 246\,MeV/u and 444\,MeV/u, respectively. 

A desired ion energy in LEIR of 440\,MeV/u requires the upgrade of the power converter for the bending magnet (ER.BHN) and the replacement of one power converter used for powering the quadrupole magnets (ER.QDN2040).

The estimated costs of power converter upgrade for the magnets ER.BHN and ER.QDN2040 are 600\,KCHF and 80\,kCHF, respectively. At this moment, no budget is foreseen in the Medium-Term Plan (MTP) for the consolidation of the LEIR power converters, consequently, the 680\,kCHF would have to be born by the BioLEIR project (table~\ref{Tab:CostEstimate_LEIR_upgrade}).

The future rms output current of the power converter for the bending magnet ER.BHN is envisaged to be $\sim$2200\,A. The envisaged thyristor based  topology is the same as the one used in the existing power converter. The losses in the water cooling system and the power drawn from the main power can be scaled from the existing power converter for the bending magnet ER.BHN. 

Any proof-of-principle studies for which higher energy protons are needed, could also be done with a proton energy limited to 250\,MeV/u and therefore would allow to reduce shielding dimensions in the irradiation rooms and hence reduce cost. It was therefore decided to limit the proton energy at BioLEIR to clinical energies of 250\,MeV/u and to design the irradiation room shielding for this maximum proton energy. A similar decision could be taken for the different ion species that are envisaged to be produced at BioLEIR - the ion beam energy could be limited to the energy available at Stage 2 of the facility (246\,MeV/u). In this case, the LEIR power-converter upgrade would not be needed which has a significant impact on the capital cost of the BioLEIR facility. The magnetic length of the main dipole in the extraction beamlines could be reduced by 25\%, which has an impact on the total cost for construction and installation of the extraction beamline elements. In addition, the total cost for shielding could be reduced and the cost for the LEIR power-converter upgrade could be saved ($\sim$1\,MCHF).

\begin{table}[!h]
\centering
\caption{Preliminary cost estimate for the power converter upgrade for the LEIR accelerator magnets, given in kCHF and integrated person-years.}
\label{Tab:CostEstimate_LEIR_upgrade}
\begin{tabular}{l l l c r r}
\hline
\textbf{} & \textbf{Cost} & \textbf{Deliverables} & \textbf{[PY]} & \textbf{Staff} & \textbf{Fellow/Technician} \\
\textbf{} & \textbf{[kCHF]} & \textbf{} & \textbf{} & \textbf{} & \textbf{} \\
\hline
\rule{0pt}{3ex}\textbf{Total LEIR PC Upgrade} & 680 & ER.BHN + SIRIUS 2P & 2 & 2/3 & 4/3\\
\hline
\end{tabular}
\end{table}

\section{List of outstanding topics to be investigated at LEIR}
\label{Sec:List_of_topics_LEIR}

Several aspects of LEIR operations shall be further investigated in a next stage of the BioLEIR project. They are listed here below:
\begin{enumerate}
\item Efficiency and stability of the injection system from LINAC3/5 to LEIR.
\vspace*{-1mm}
\item Intensity and stability for different ion species in LEIR.
\vspace*{-1mm}
\item Efficiency and stability of the slow extraction system (either quadrupole-driven method or   RF-knockout technique) towards the experimental area.
\vspace*{-1mm}
\item Efficiency and stability of the electron cooler for different ion species.
\vspace*{-1mm}
\item Efficiency and stability of the ejection system towards the PS with BioLEIR machine elements present.
\vspace*{-1mm}
\item Study of impact of BioLEIR devices on LHC beams.
\end{enumerate}

\section{Resource estimate}
\label{Sec:LEIR_Cost}
It is expected that 1.2~FTE (1~Fellow and 0.2~Staff) are needed to support the BioLEIR project with LEIR beam dynamics studies for the length of the project, covering the topics listed in section~\ref{Sec:List_of_topics_LEIR} and any other questions that may surface during the preparation and exploitation of BioLEIR.

It is assumed that the project preparation phase has a duration of 4 years, which corresponds to 4.8 integrated person-years for the project preparation; these resources will be counted in the capital cost for the project. After this, the 1.2~FTE are counted as yearly support.\\

The LEIR power converter upgrade is expected to cost 680\,kHCF and 2 person-years.

%% file: Chapters/Extraction.tex
\chapter{Beam Extraction from LEIR to BioLEIR}
\label{Chap:Extraction}

Slow extraction from LEIR to the experimental facility is required to provide the flexibility to accommodate the demanded variety of extracted beam characteristics, as detailed in table~\ref{Tab:Extraction_Intro_Species}. New electrostatic and magnetic septa are needed in SS30 to accomplish the extraction, two new bumper magnets are needed for the closed orbit distortion at the septa, and existing machine sextupoles can excite the third-integer resonance. Some changes to the machine vacuum chambers are needed to increase aperture in specific locations to accept the extracted beam.

\begin{table}[ht]
\centering
\caption{Ion species and main parameters for slow extraction. Note that the highest spill rates are not highest priority, as they are requested to shorten the irradiation time for (micro-)dosimetry detector development.}
\label{Tab:Extraction_Intro_Species}
\begin{tabular}{l l c}
\hline
\rule{0pt}{3ex}Ion species considered & & H, He, Li, B, C, O, N, Ne  \\
Reference species for study & & $^{12}$C$^{6+}$\\
Kinetic energy & MeV/u & 20-300 (H); 50-440 (C)\\
Spill rate & ions/s & $10^{5}-10^{11}$\\
Spill length & s & $1 - 10$\\
Normalised emittance & $\pi$.mm.mrad & $1.05$ \\
Physical emittance & $\pi$.mm.mrad & $5 - 25$ \\
Momentum spread &  & $\pm2-9\times10^{-4}$ \\
\hline
\end{tabular}
\end{table}

\section{Extraction mechanisms}
\label{Sec:Extraction_Mechanisms}

\begin{figure}[ht]
\centering\includegraphics[width=0.6\linewidth]{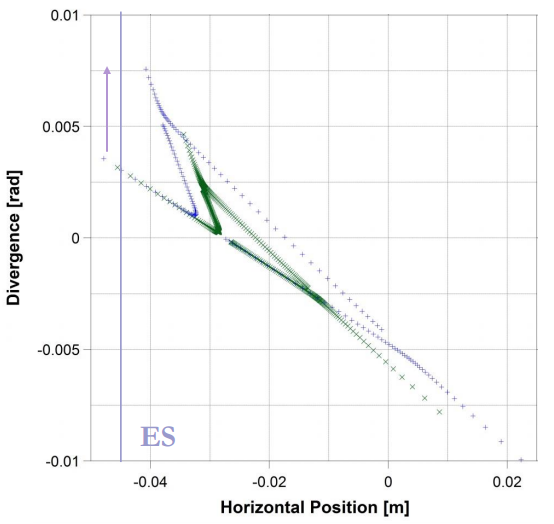}
\caption{Extraction separatrices for different momenta, showing overlap obtained through the Hardt condition and jump across the ES.}
\label{Fig:Extraction_Mechanisms_01}
\end{figure}

The horizontal phase space is distorted by excitation of correctly chosen sextupole lenses to a characteristic triangular shape (figure~\ref{Fig:Extraction_Mechanisms_01}) when the machine tune is close to a third-integer. The existing LEIR pole-face winding sextupoles can be used to generate this excitation. Particles outside the central triangular stable region grow rapidly in amplitude, returning each three turns to the same separatrix branch. They are extracted along the separatrix which crosses the thin Electrostatic Septum (ES), which in turn kicks them into the gap of the stronger magnetic septa and extracts them from the LEIR aperture into the experimental beamline. The spiral step is defined as the maximum three turn amplitude growth of a particle at the ES position. A slow closed orbit bump is needed to bring the beam close to the septa to limit the horizontal aperture excursions in the rest of the ring.

The momentum spread present in the beam produces a spread in angle of the extracted particles at the entrance to the electrostatic septum, resulting in increased extraction losses. With non-zero dispersion at the septum this effect can be compensated by arranging for the so-called Hardt condition~\cite{Hardt:1025914} to be met, by changing the machine chromaticity, which overlaps separatrices of different momenta at the ES. This is seen for LEIR in figure~\ref{Fig:Extraction_Mechanisms_01}.

The flux of extracted particles is linked to the rate at which particles become unstable. This process is carefully adjusted so that only a small fraction of the beam is extracted at each turn to give the required spill structure. The spill can be regulated by feed-back, feed-forward or a combination of the two \cite{JPARC:2009}.

The simplest method to bring the beam gradually to resonance is to modify the tune by varying the strength of one or more quadrupoles in the lattice (Quad Driven, Q-D). This method suffers, however, from the sensitivity to tune ripples, and from a variation of the characteristics of the extracted beam over the duration of the spill.

A more sophisticated and versatile alternative is the transverse RF-Knock-Out (RF-KO) method. Band-limited noise signals (up to 100~kHz width) are used to excite the beam in horizontal phase space. The kicks needed for 1$ - $10\,s spill lengths are moderate (5\,$\mu$rad maximum) and could be supplied by existing machine elements in combination with appropriate noise signal generators. In this method, the transverse and longitudinal characteristics of the beam basically remain unchanged through the spill.

\section{Machine lattice}
\label{Sec:Extraction_Lattice}

\begin{figure}[ht]
\centering\includegraphics[width=0.8\linewidth]{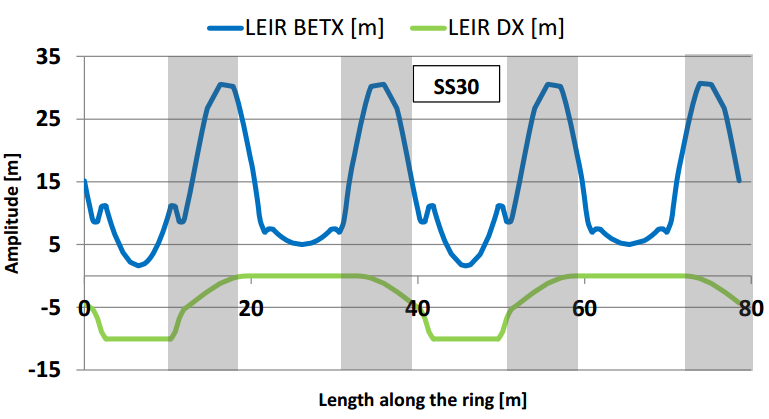}
\caption{Horizontal beta function ($\beta_x$) and dispersion ($D_x$) along the ring, the grey shaded areas correspond to the four 90$^\degree$ bends.}
\label{Fig:Extraction_Lattice_03}
\end{figure}

The most suitable optical configuration is a working point similar to the present LEIR working point for LHC operation~\cite{Carli:972662}, at 5/3 horizontal resonant tune and around 2.735 vertical tune. The horizontal ring optic functions are shown in figure~\ref{Fig:Extraction_Lattice_03}. Assuming a reasonable virtual normalized sextupole strength of S $= 1/2 \beta_{x}^{3/2}l_{s}k'_{s}=30\,$m$^{-1/2}$, the required horizontal chromaticity Q' to meet the Hardt condition is 9, with the vertical chromaticity not zero, but close to the natural value at -4.5 to keep the required sextupole strengths within the limits possible with the hardware. 

LEIR contains six sextupole families, with the characteristics shown in table~\ref{Tab:Extraction_Lattice_Sextupoles}. The integrated strength is defined as $\int{\frac{\delta^{2}B_{y}}{\delta^{2}x^{2}}} dl$.

\begin{table}[ht]
\centering
\caption{Sextupole families installed in LEIR.}
\label{Tab:Extraction_Lattice_Sextupoles}
\begin{tabular}{l c r r r r r r}
\hline
\rule{0pt}{3ex}\textbf{Family} & \textbf{Unit} & \textbf{SF1030} &\textbf{SD1030} &\textbf{S41} &\textbf{S42} &\textbf{SW01} &\textbf{SW02} \\
\hline
\rule{0pt}{3ex}$\beta_{x}$ & m & 10 & 10 & 7& 8& 19 & 24 \\
$\beta_{y}$ & m & 8 & 15 & 10& 11& 12 & 3 \\
D$_{x}$ & m &10 & 8 & 0& 0& 4 & 0 \\
Int. strength & T/m & 5.2&5.2&2.2&2.2&5.4&5.4\\
\hline
\end{tabular}
\end{table}

The dispersion at the chromaticity correction sextupoles SF1030 and SD1030 in sections 10 and 30 is large at 10\,m (each of these eight sextupoles is in fact independent with a dedicated power converter). At present, the Pole-Face Windings (PFWs) in the non-dispersive straights are grouped into two families: SW02 (XFW12, 21, 32 and 41) and SW01 (XFW11, 22, 31 and 42). Each family is connected in series to one power converter. 

In order to decouple the resonance excitation from the chromaticity correction, the sextupoles in SS10 and SS30 with large dispersion are only used for chromaticity correction. For excitation of the extraction resonance, it is preferred not to use elements inside the closed orbit distortion in SS30 to avoid large dispersion. The remaining independent free parameters are used to adjust the effective phase advance XFLS41, XFLS42, XFW41 (or XFW21) and XFW12 (or XFW32). Simulations showed that the optimum is to excite the extraction resonance with XFW41 at an integrated strength of $30-80\,$m$^{-2}$, generating spiral steps up to 11\,mm with a suitable orientation of the separatrices. This XFW41 circuit for the resonance excitation requires at least a recabling and a separate power converter.

\section{Extraction elements}
\label{Sec:Extraction_Elements}

\subsection{Closed orbit bumpers}
\label{SubSec:Extraction_Elements_Bumpers}

An orbit bump in SS30 is needed. The maximum bump needed to extract a beam with horizontal (geometric) emittance of 20\,$\pi$.mm.mrad is 40\,mm. For a normalized emittance of 0.9\,$\pi$.mm.mrad the configuration  corresponds to extracting a beam of 35\,MeV/u, assuming a 2.5\,$\sigma$ beam envelope.  The existing orbit correctors DEHV22 (in SS20), DWHV21 and DWHV22 (in bend 20) and DWHV31 and DWHV32 (in bend 30) are used in conjunction with two new orbit correctors (OC1 and OC2) placed in SS30. These need to provide about $70\times10^{-3}$\,T.m of integrated field, which is possible with a magnetic field of 
0.2\,T and 350\,mm magnetic length.

\subsection{Extraction septa}
\label{SubSec:Extraction_Elements_Septa}

The electrostatic septum (ES) previously used in LEAR (SEH11) is still available and could be reinstalled. It can create a minimum orbit separation of 10.2\,mm between the circulating beam and the extracted particles at the entrance of the first magnetic septum (MST). The presence of the fast extraction kicker in this straight section constrains the field length of the ES to only 733\,mm, requiring a relatively high field of 8.1\,MV/m. This is a technical challenge given the constraints on the LEIR vacuum imposed by Lead ion operation which imposes the use of a Titanium cathode, since anodised aluminium cathodes are excluded. It is necessary to consolidate the existing SEH11 spare cathode and high voltage deflectors, as well as to change the motorisation system to be bake-able and to be compatible with the existing remote displacement systems used for the septa in the PS complex. The motorisation system  uses a total of four independent motors to adjust the position and angle of both, cathode as well as septum (anode support).

For the thin magnetic extraction septum (MST) the spare of the LEIR injection septum (SMH11) could be installed. This septum is a DC septum outside vacuum. It could provide a kick of 56\,mrad and its entrance can be placed at -51\,mm. The position of the device is fixed and determined by the vacuum chamber around which the septum magnet is envisaged to be installed. Two spare coils are presently available, so that one spare coil for the injection septum as well as one for the MST would be available. This strategy does not significantly increase the risk of downtime from a septum problem for LEIR operation, as the coils are the elements which fail and need periodic replacement. A special extraction vacuum chamber (Y-chamber, see more details in chapter~\ref{Chap:Vacuum}) has to be designed and built. This vacuum chamber is part of the magnetic circuit of the MST septum, in the sense that the magnetic screen ($\sim$2\,mm of mu-metal) is integrated into the orbiting beam vacuum chamber and also the vacuum chamber incorporates the septum coil fixation. The extracted beam tube is very thin-walled and close to the septum ($\leqslant$1.5\,mm) to assure that the apparent septum thickness must not exceed the required 10\,mm; and 3\,mm thin walls for top and bottom are used to stay within the maximum vertical beam aperture. 

The second magnetic septum (MSE) provides a stronger kick of 104\,mrad. A dedicated, new 1-m-long septum outside vacuum is envisaged to be designed and built, together with its support and spare coils. The apparent septum thickness (including vacuum chambers) is 22\,mm and the integrated magnetic field is 0.7\,T.m. Its position is fixed by the extraction channel of the previously mentioned Y-chamber. The magnetic shielding is not part of the vacuum chamber, but clamped in a conventional manner to the septum magnet by the commonly used coil fixation clamps. A spare coil needs to be manufactured.

The electrostatic extraction septum blade can be slowly displaced for precise alignment in the horizontal plane, but should not reduce the machine acceptance for regular ion beam operation for LHC. This imposes the quasi fixed physical innermost positions of about -45\,mm for the ES. It is not feasible to rapidly displace the septum blade for different beam types - for this reason all the extraction scenarios have been validated with this fixed septum position.

The lengths, strengths and electrical parameters of the individual septa are given in~table~\ref{Tab:Extraction_septa}.

\begin{table}[ht]
\centering
\caption{Septum parameters.}
\label{Tab:Extraction_septa}
\begin{tabular}{l l r r r}
\hline
\rule{0pt}{3ex}\textbf{Parameter} & \textbf{Unit} & \textbf{ES} & \textbf{MST} & \textbf{MSE}\\
\hline
\rule{0pt}{3ex}Bend angle & mrad & 4 & 56  & 104 \\
B field & T & & 0.43 & 0.77 \\
Integrated B field & T.m & & 0.38 & 0.7 \\
E field & MV/m & 8.1 & &  \\
Field length L & mm & 733 & 880 & 910\\
Physical length & mm & 860 & 900 & 1050\\
Septum thickness & mm & 0.1 & 10 & 22\\
Nom. operating current & kA & & 1.9 & 1.9  \\
Nom. operating voltage & kV & 161 &  &  \\
Horizontal full aperture & mm & 20 & 156 & 93 \\
Vertical full aperture & mm & 70 & 46 & 40 \\
Inductance & mH & & 0.3 & 0.7 \\
Resistance & m$\Omega$ & & 12 & 17 \\
Peak power (I$^{2}$R) & kW & & 44 & 61 \\
Water cooling & l/min & & 42 & 60 \\
\hline
\end{tabular}
\end{table}

\subsection{Extraction instrumentation}
\label{SubSec:Extraction_Elements_BI}

Transverse diagnostics are needed at least at the entrance of the ES and MST, to be able to verify the profile and absolute position of the extracted beam separatrix, and to locate the extreme edge of the circulating beam. Given the low particle fluxes, these could be based on scintillating screens or Multi-Wire Proportional Chamber (MWPC) technology. The monitors may need to be incorporated in the septum tanks, for space reasons. In addition, sensitive beam loss monitors, for example using diamond technology, are also required at these elements, for loss and transmission optimisation. 

\subsection{Powering cycle and power converters}
\label{SubSec:Extraction_Elements_PC}

The septum magnets can be pulsed for energy economy, with a ramp-up/down time of the order of 1\rule{0pt}{3ex}s and a flat-top length of 1-10\,s. Fast changes of strength during an extraction are not considered. The minimum repetition period for the longest spill is 13.2\,s. The power supply for the electrostatic septum is part of the ES system.

New power converters are needed for the two orbit bumpers and the sextupole PFW circuit \cite{Chanel:209773} for the resonance excitation.
The overall number of power converters, and technical preliminary requirements are given in table~\ref{Tab:BeamLines_Converters_extraction}. 

\begin{table}[!ht]
\centering
\caption{Extraction power converter parameters. The output current precision is defined according to the maximum current. The dimensions are given for a single converter, except for the sextupole converter (*) where the dimensions are given for 1+1 units.}
\label{Tab:BeamLines_Converters_extraction}
\begin{tabular}{l l r r r r}
\hline
\rule{0pt}{3ex}\textbf{Parameter} & \textbf{Unit} & \textbf{MST+MSE} & \textbf{OC1/2} & \textbf{Sextupole}\\
\hline
\rule{0pt}{3ex}Converters (spare) &  & 2 (1) & 2 (1) & 1 (1)\\
Min. ramp time & s & 1 & 1 & 1 \\
Max. flat-top time & s & 10 & 10 & 10 \\
Max. period & s & 13.2 & 13.2 & 13.2 \\
Nom. current & A & 2000 & $\pm$50 & 200 \\
Max. voltage & V & 50 & 30 & 15\\
Current precision & ppm & $\pm$200 & $\pm$100 & $\pm$200 \\
AC input & & 3P-400Vac-32A & 3P-400Vac-4A & 3P-400Vac-6A \\
Dimensions & mm & 2000$\times$600$\times$900 & 4U$\times$1''$\times$650 &2000$\times$600$\times$900* \\
Cooling &  & Forced air & Forced air & Forced air \\
Peak power (I$^{2}$R) & kW & 100 & 1.5 & 3 \\
\hline
\end{tabular}
\end{table}

The MST and MSE septum magnets are powered independently by identical power converters based on the use of commercial power units connected in parallel. For each converter, six power units are connected in parallel. They will be controlled as a single current source. The power converter is a one-quadrant current source that can be cycled in order to reduce power consumption. The output current measurement is performed using high precision DCCT's. Each converter is fitted in one rack. 
The orbit correctors (OC1 and OC2) are powered by two CANCUN50 power converters. These converters are four-quadrants converters that can be operated in cycled operation mode. The output current is bipolar with accurate regulation at zero current.

In order to allow beam extraction using the extraction resonance, recabling of the PFW is needed and an additional converter is required to power the XFW41 sextupole independently. The associated power converter and its spare are commercial units installed in a single rack.  

All the power converters are controlled using digital control implemented with CERN standard controllers (FGC with Ethernet+) and interfaces (magnet interlock, beam interlock etc.). As all the requirements can be fulfilled with CERN standard power converters, no special designs are required and the cost and personnel needs are kept at a minimum.  

\subsection{Supports and alignment}
\label{SubSec:Extraction_Elements_Supports}

Dedicated septa magnet support structures have to be designed and manufactured, that allow the magnets to be withdrawn from the ring without losing their alignment. This feature is needed in case of a machine bake-out or for coil maintenance. The structures can be based on the design made for the ELENA injection septum.
The new correctors can be installed on support girders, with one XYZ translation stage per element.


\subsection{Cooling requirements}
\label{SubSec:Extraction_Elements_Cooling}
For both septa magnets, a dedicated cooling water manifold needs to be installed, that distributes the cooling water to the magnet coils and electrical connections. The flow meters and electro-valves are also installed on this manifold. The installed water cooling power for the septa needs to be at least 110\,kW.

The power converters are air cooled. Dissipation in ambient air is estimated to be around 30\,kW (of which 2$\times$14\,kW is for the MSE and MST).

\subsection{Layout and aperture}
\label{SubSec:Extraction_Elements_Layout}

\begin{figure}[ht]
\centering\includegraphics[width=1\linewidth]{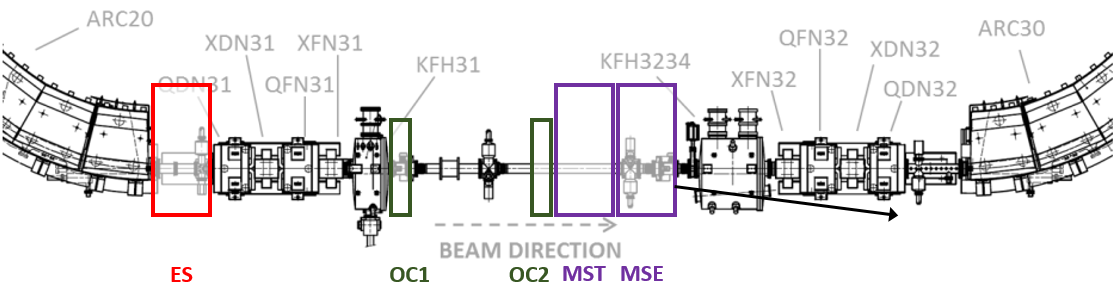}
\caption{Location of new extraction elements in SS30.}
\label{Fig:Extraction_Elements_Layout_04}
\end{figure}

The integration of the extraction septa into SS30 looks feasible with the specified strength and physical element lengths as shown in figure~\ref{Fig:Extraction_Elements_Layout_04}.

\begin{figure}[ht]
\centering\includegraphics[width=0.5\linewidth]{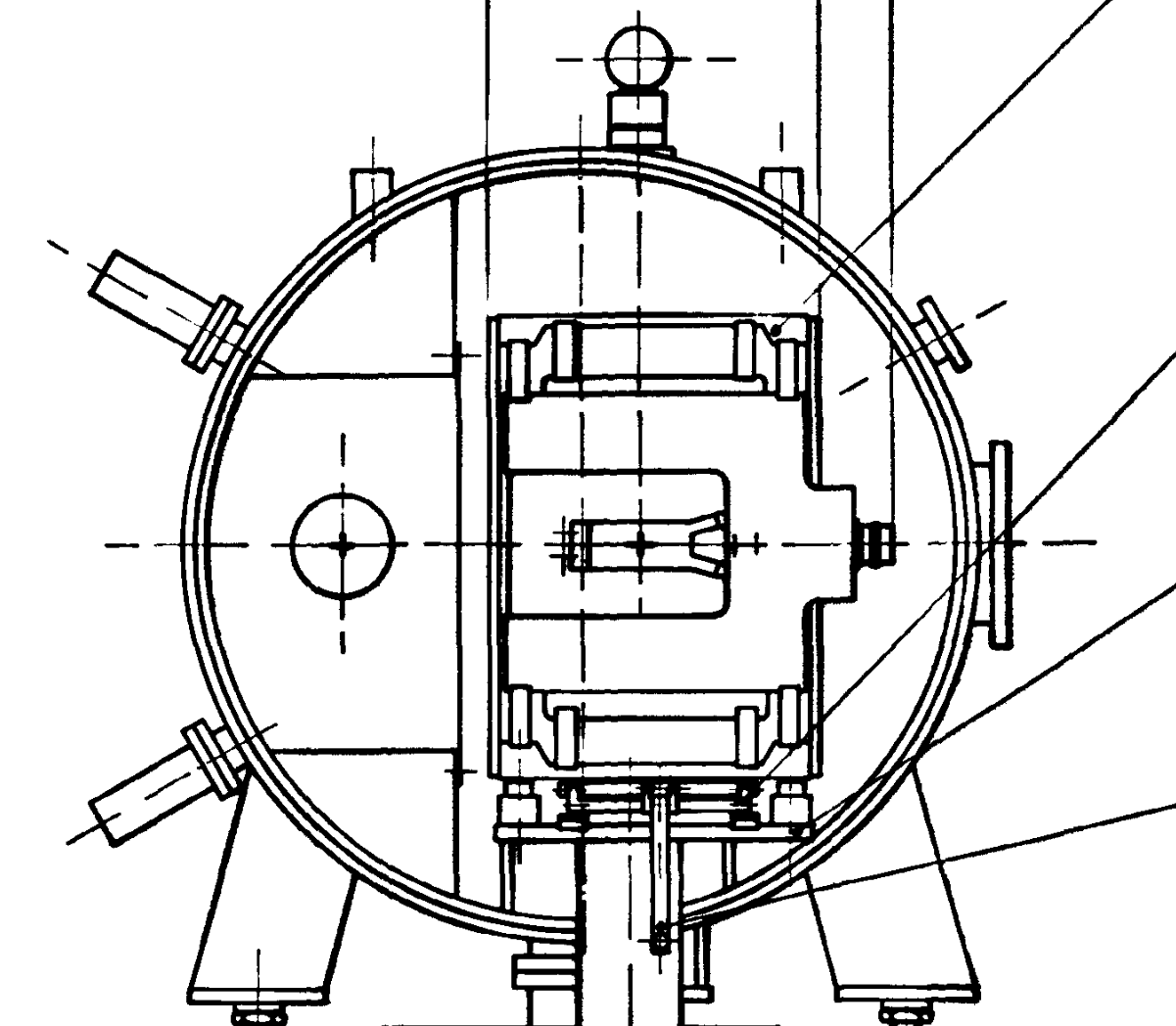}
\caption{Cross-section of KFH3234 in the SS30 straight section. The extracted beam needs to pass through the modified beampipe visible on the left of the picture.}
\label{Fig:Extraction_Elements_Layout_05}
\end{figure}

The beampipe from the ES through the quadrupole-sextupole doublet (aperture: $\pm$72.5\,m) and KFH31 (aperture: $\pm$73.5\,mm) up to the MST, has to be horizontally sufficiently large  to house both the circulating and the extracted beam. The doublet beampipe could be enlarged to $\pm$84\,mm in the horizontal axis. The extracted beam also has to traverse the KFH3234 kicker tank (similar to beams extracted from LEAR in the past), which requires a modification of the beam passage on the tank, see~figure~\ref{Fig:Extraction_Elements_Layout_05}.

The extracted beam has to travel through KFH3234 with a 150\,mrad angle to pass the yoke of QFN32. An extraction beampipe with a radius of around 30\,mm can be accommodated.

\begin{figure}[ht]
\centering\includegraphics[width=0.9\linewidth]{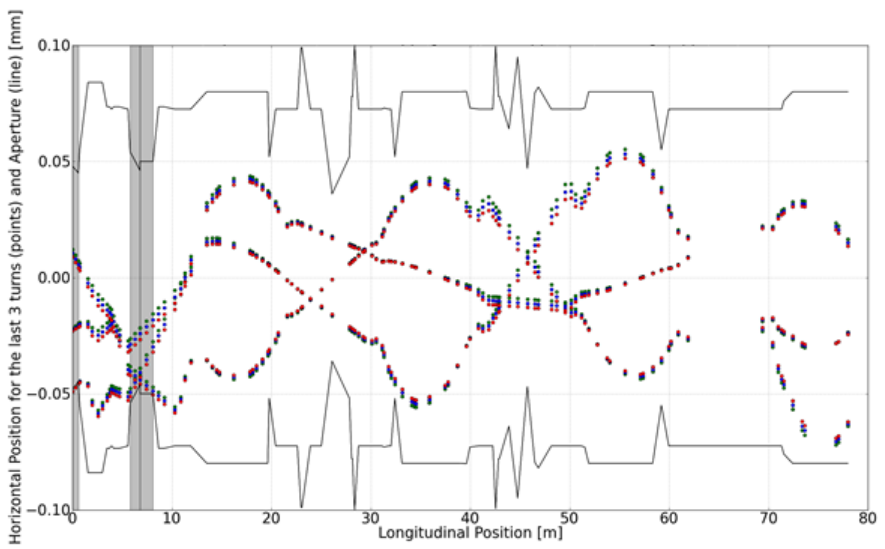}
\caption{Horizontal position for the last 3 turns of the outermost circulating beam (dots) and horizontal aperture (line) along the ring; 0~m corresponds to the entrance of the ES; the shaded areas represent the septa.}
\label{Fig:Extraction_Elements_Layout_06}
\end{figure}

The LEIR ring aperture for the extraction separatrices was checked with the nominal extraction element settings by tracking the last three turns of the extreme particles (figure~\ref{Fig:Extraction_Elements_Layout_06}).

\section{Comparison of extraction driving mechanisms}
\label{Sec:Extraction_Comparison}

\subsection{Quadrupole driven extraction}
\label{SubSec:Extraction_Comparison_QuadDriven}

For a quadrupole-driven (Q-D) slow extraction the tune is dynamically moved towards the third integer resonance and high emittance particles are extracted before low emittance ones, and, for a given emittance, different momenta are extracted at different times. It is foreseen to use the five quadrupole families of LEIR to vary the tune. Depending on the beam quality needed, a feed-forward system can be installed to mitigate the ripple effects of the power supplies and improve the spill quality~\cite{Kain:2207355}.

Two extreme beam conditions have been considered: high and low energy. Only the closed orbit distortions, the strength of the ES kick and the strength of the resonance excitation change; the physical septum positions, ring optics and MS strengths remain identical. For the high energy configuration, the last two correctors in bend 30 (DWHV31 and DWHV32) are powered to their limits. The assumed beam parameters and resulting extraction settings are shown in table~\ref{Tab:Extraction_Comparison_QuadDriven_Q-D}. 

\begin{table}[ht]
\centering
\caption{Assumed beam parameters and ES settings for Q-D extraction.}
\label{Tab:Extraction_Comparison_QuadDriven_Q-D}
\begin{tabular}{l c c c}
\hline
\rule{0pt}{3ex}\textbf{Parameter} & \textbf{Unit} & \textbf{Low energy} & \textbf{High energy} \\
\hline
\rule{0pt}{3ex}Energy & MeV/u & 20 & 440  \\
Emittance (physical) & $\pi$.mm.mrad & $0.2 - 25$ & $0.2 - 5$  \\
Momentum spread &  & $\pm9\times10^{-4}$ & $\pm2\times10^{-4}$  \\
\hline
\rule{0pt}{3ex}ES field & MV/m & 0.8 & 7.8  \\
ES kick & mrad & 7.3 & 3.9  \\
ES entrance position & mm & \multicolumn{2}{c}{-50.0}  \\
ES exit position & mm & \multicolumn{2}{c}{-46.4}  \\
\hline
\end{tabular}
\end{table}

\begin{figure}[ht]
\centering\includegraphics[width=0.8\linewidth]{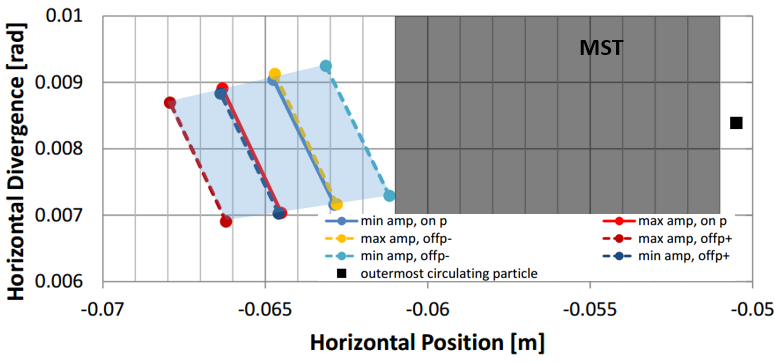}
\caption{Phase space plots of the extracted beam segments for the various configurations of
emittance and relative momentum at high energy at the MST entrance, for Q-D
extraction.}
\label{Fig:Extraction_Comparison_QuadDriven_08}
\end{figure}

The analysis at the extreme momentum offsets and particle amplitudes give a set of extreme separatrices which define the initial conditions of the transfer line, see~\cite{Garonna:1703468} for full details.  The transport of this 'bar of charge' through the MST and MSE has been checked, for both low and high energy. The off-momentum particles have a large offset in the MST septum due to the large dispersion, shown for the high energy case in figure~\ref{Fig:Extraction_Comparison_QuadDriven_08}. The horizontal Twiss parameters at the electrostatic extraction septum were computed from these results and give the values shown in table~\ref{Tab:Extraction_Comparison_QuadDriven_Twiss} for initial conditions for the transfer line matching.

\begin{table}[ht]
\centering
\caption{Initial Twiss parameters derived at the ES entrance.}
\label{Tab:Extraction_Comparison_QuadDriven_Twiss}
\begin{tabular}{l r r r r r}
\hline
\rule{0pt}{3ex}  & \textbf{$\epsilon_{rms}$~[$\pi$.mm.mrad]} & \textbf{$\beta$~[m]} & \textbf{$\alpha$} & \textbf{D~[m]} & \textbf{D'} \\
\hline
\rule{0pt}{3ex}Vertical & 0.6-4.2 & 15 & -2.8 & 0 & 0  \\
Horizontal & 2.0 & 15 & 0 & -4 & -1  \\
\hline
\end{tabular}
\end{table}

\subsection{RF knock-out extraction}
\label{SubSec:Extraction_Comparison_RFKO}

For RF knock-out (RF-KO) extraction, the emittance of the beam is increased such that particles gradually exceed the stability limit. Transverse kicks of a few $\mu$rad are needed, which could be provided by the existing damper in combination with appropriate signal generators. Assuming an RMS white noise kick of 2\,$\mu$rad, a beta function at the kicker of 5\,m and a revolution frequency of 2.8\,MHz, the RMS emittance growth would be around 30\,$\mu$m/s. For band-limited noise over a single sideband, a spectral deflection power density of 2.9\,nrad$^{2}$/Hz is needed.

To extract all momenta at the same time and with similar emittances, a small horizontal chromaticity was chosen. Again, the two extreme beam conditions were studied (high energy and low energy). The assumed beam conditions and resulting extraction settings are shown in table~\ref{Tab:Extraction_Comparison_RFKO_RF}. Again, the limiting factor is the aperture in the MST septum for the off-momentum particles.

\begin{table}[ht]
\centering
\caption{Assumed beam parameters and ES settings for RF-KO extraction.}
\label{Tab:Extraction_Comparison_RFKO_RF}
\begin{tabular}{l c c c}
\hline
\rule{0pt}{3ex}\textbf{Parameter} & \textbf{Unit} & \textbf{Low energy} & \textbf{High energy} \\
\hline
\rule{0pt}{3ex}Energy & MeV/u & 20 & 440  \\
Emittance (physical) & $\pi$.mm.mrad & $26$ & $6$  \\
Momentum spread &  & $\pm9\times10^{-4}$ & $\pm2\times10^{-4}$  \\
\hline
\rule{0pt}{3ex}ES field & MV/m & 0.7 & 6.9  \\
ES kick & mrad & 6.4 & 3.4  \\
ES entrance position & mm & \multicolumn{2}{c}{-49.0}  \\
ES exit position & mm & \multicolumn{2}{c}{-45.8}  \\
\hline
\end{tabular}
\end{table}

\section{Resource estimate}
\label{Sec:Extraction_Costing}

The estimated CERN personnel needs (Staff and Fellows) are given in table~\ref{Tab:Extraction_Costing_Manpower}. The estimates include the remaining conceptual design phase, engineering design, prototyping, construction and installation.

\begin{table}[ht]
\centering
\caption{Preliminary personnel estimate for design, construction, installation and commissioning of slow extraction elements.}
\label{Tab:Extraction_Costing_Manpower}
\begin{tabular}{l r r}
\hline
\rule{0pt}{3ex}\textbf{System} & \textbf{Staff [MY]} & \textbf{Fellow [MY]} \\
\hline
\rule{0pt}{3ex}Beam dynamics & 0.2 & 0.3 \\
Corrector magnets & 0.2 & 0.2\\
Septa & 1.3 & 1.0\\
Power converters & 0.3 & 0.6 \\
Beam diagnostics & 1.0 & 1.0 \\
Power/control cabling & 0.2 & -\\
Water cooling & 0.1 & - \\
Alignment and survey & 0.1 & - \\
Installation coordination  & 0.2 & - \\
Beam preparation and commissioning & 0.5 & 0.5 \\
Controls configuration and SW & 0.5 & 0.5 \\
\hline
\rule{0pt}{3ex}\textbf{Total} & \textbf{4.6} &\textbf{4.1}\\
\hline
\end{tabular}
\end{table}

The required numbers and assumed costs for the different equipment subsystems described in the sections above are summarized in table~\ref{Tab:Extraction_Costing_Material}. It shall be noted that some of the numbers given are extremely preliminary and the actual costs are likely to vary by as much as 50\% from the numbers given. Although the extraction concept is very solid and the feasibility clear, there are still large uncertainties on some technical systems like instrumentation. Total cost includes prototyping, whereas installation and one-off tooling costs are not included in the unit cost. The material costs include all non-CERN staff resources such as Field Support Units. Note that Fellows are not included in material cost, but are counted under CERN personnel.

No detailed annual material or personnel estimates for exploitation, nor for maintenance and operation, are given. This is typically of the order of $5-10$\% of the capital cost of an installation, and can be roughly estimated at about 70\,kCHF and 0.3\,FTE per year, spread across the technical groups concerned.

\begin{table}[ht]
\centering
\caption{Preliminary cost estimate for design, construction, installation and commissioning of extraction system elements. Total costs include prototyping and overheads where relevant, e.g. for tooling or design.}
\label{Tab:Extraction_Costing_Material}
\begin{tabular}{l r r r}
\hline
\rule{0pt}{3ex}\textbf{System} & \textbf{Units} & \textbf{Unit cost} & \textbf{Total} \\
\textbf{} & \textbf{} & \textbf{[kCHF]} & \textbf{[kCHF]} \\
\hline
\rule{0pt}{3ex}Electrostatic septum (incl. HV cabling) & 1 & 120 & 120 \\
MST magnetic septum (refurbish SMH11)  & 1 & 50 & 50 \\
New MSE magnetic septum \& support & 1 & 150 & 150 \\
Septum control electronics \& HV generator& 1 & 170 & 170 \\
KFH3234 modifications & 1 & 30 & 30 \\
New corrector magnets  & 3 & 15 & 60 \\
Corrector converters  & 3 & 10 & 30 \\
Septum converters  & 3 & 60 & 180 \\
Sextupole converters  & 2 & 15 & 30 \\
Extraction profile monitors & 3 & 35 & 140 \\
Beam loss monitors & 3 & 5 & 20 \\
Damper signal generator modifications & 1 & 30 & 30 \\
Magnet/septum supports \& alignment & 5 & 10 & 50 \\
Magnet/septum power cabling & 4 & 15 & 60 \\
Beam instrumentation cabling & 4 & 5 & 20\\
Water cooling (incl. septum manifold) & 4 & 10 & 60 \\
Layout/integration (man-months) & 2  & 10 & 20 \\
Alignment and survey (man-months) & 1  & 10 & 10 \\
\hline
\rule{0pt}{3ex}\textbf{Total} & & &\textbf{1230}\\
\hline
\end{tabular}
\end{table}

\section{Potential risks and remaining conceptual issues}
\label{Sec:Extraction_Risks}

For both extraction types with lowest energy beam, the spiral step with a positive off-momentum particle at maximum amplitude is at $0.6-0.7$\,mm very small. This results in increased beam loss at the ES. In addition, these off-momentum particles have a very large offset in the MST septum due to the large dispersion, and determine the minimum strength of the ES needed. A larger deflection from the ES would certainly useful. Such larger deflection would be available for beam energies significantly below 440\,MeV/u. This understanding needs to be verified with a full aperture model of the extraction region.

The optimum alignment of the ES blade changes with beam energy. A compromise must be defined such that beam losses on the septum remain acceptable for all energies expected at BioLEIR operations. This effect is more prominent for Q-D extraction due to the larger phase space area occupied by the extracted beam at the ES entrance. 

The optical effects (and resulting beta-beating) from the strong orbit correctors introduced in SS30 should be further investigated, and also the assumptions of adiabaticity when ramping the machine tune for Q-D extraction. The expected harmonic content of the spill can also be estimated from the ripple measurements of the LEIR power converters (in particular from the focusing quadrupoles and extraction sextupoles). This may be a key factor in deciding between RF-KO and Q-D extraction, although for the proposed use-cases it seems that precise dose determination is much more important than control of spill fluctuations in the few $10-100$\,Hz range.

The effects of amplitude detuning need to be incorporated into the model, as these can have significant effects on the form of the extraction separatrices. Information on the multipolar content of the main LEIR magnets is required for this, which may be available through machine studies.

The extraction scheme is based on the fact that the operational electrostatic field in the septum of 8.1\,MV/m can be achieved with a titanium cathode. This is still to be confirmed with prototyping and tests.  

Overall, the proposed extraction scheme is robust, and the parameters of the extraction equipment are reasonable. LEAR has already operated with slow extraction in the past, and no major issues are expected for BioLEIR.

%% file: Chapters/Beamlines.tex
\chapter{Beamlines to the BioLEIR Experimental Area}
\label{Chap:BeamLines}

Two horizontal and one vertical experimental beamlines are foreseen, see chapter~\ref{Chap:BeamParameters}. The vertical beamline is designed for maximum 75\,MeV/u, while the horizontal beamlines should deliver up to 440\,MeV/u (all energies are kinetic, and quoted for $^{12}$C$^{6+}$). The beamline minimum full aperture is approximately 80\,mm at all magnetic elements, to accommodate the expected beam envelopes. 

\begin{figure}[ht]
\centering\includegraphics[width=1.0\linewidth]{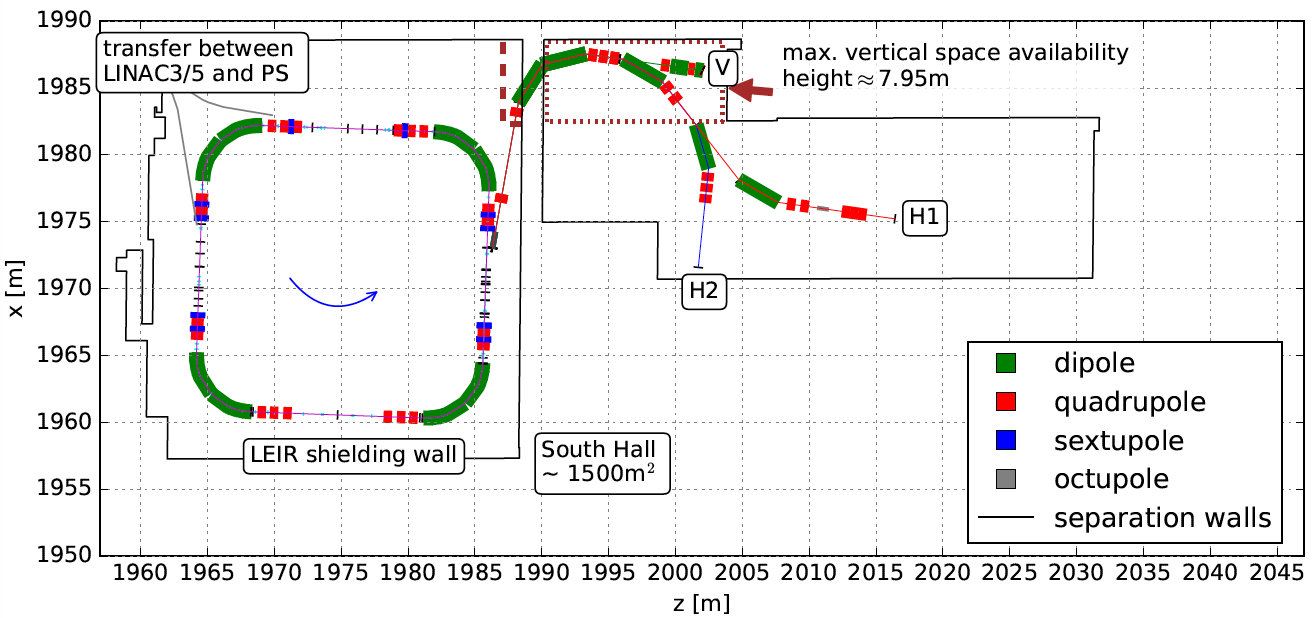}
\caption{Layout of LEIR and experimental beamlines. A possible layout for two horizontal beamlines is shown, with a vertical beamline branching from the common horizontal line before the final horizontal dipole. The vertical target plane is located about 6.5\,m above the LEIR floor level.}
\label{Fig:BeamLines_1.1}
\end{figure}

The layout of the beamlines is shown schematically in figure~\ref{Fig:BeamLines_1.1}. The common horizontal beamline has a 132$^\degree$ bending angle, accomplished with three identical dipoles, with a quadrupole triplet between the second and third dipoles, and an additional separated dipole for each beamline. The fourth dipole in the H1 beamline is inverted, as shown in figure~\ref{Fig:BeamLines_1.1}. The vertical beamline has a 90$^\degree$ bend, using a split 45$^\degree$ dipole pair with an interposed quadrupole. For the vertical beamline, a single quadrupole is used to make the bend achromatic, while for the horizontal beamlines a quadrupole triplet is used between the second and third dipoles in the common part of the beamline to cancel the horizontal dispersion coming from the LEIR ring. The third and fourth dipoles in the H2 beamline is equipped with 'Y' chambers to enable switching between the different experimental stations. Overall there are about 65\,m of beamline to construct.

\section{Requirements and constraints}
\label{Sec:BeamLines_Requ}

Optically, the transport and matching are constrained by the space available in the South Hall, the optics functions and beam sizes at the LEIR extraction point, the matching of the dispersion and its derivative to zero at the target, and by the required beam sizes and uniformity at the target planes. The main beam parameters are given in table~\ref{Tab:BeamLines_Requ_Parameters}. The initial conditions at the first electrostatic extraction septum in the LEIR ring and at the target planes at the beamline extremities are given in table~\ref{Tab:BeamLines_Requ_InitialConditions}.

\begin{table}[ht]
\centering
\caption{Main beam parameters for extraction beamlines.}
\label{Tab:BeamLines_Requ_Parameters}
\begin{tabular}{l l r}
\hline
\textbf{Beam parameter} & \textbf{Unit} & \textbf{Value}\\
\hline
\rule{0pt}{3ex}Extraction energy ($^{12}$C$^{6+}$)	& MeV/u & 50-440 \\
Extraction rigidity	& T.m & 1.3-6.7 \\
Total intensity	& ions	& $1\times10^{10}$ \\
Spill duration & s & 1-10 \\
Broad beam size & mm$\times$mm &50$\times$50 \\
Broad beam uniformity & I/I$_{0}$ & 0.90 \\
Pencil beam size (FWHM)	& mm$\times$mm & 5$\times$5 \\
Max. flat-top length & s & 10.8 \\
Min. ramp-up/down time & s & 1.2 \\
Min. repetition period & s & 13.2 \\
\hline
\end{tabular}
\end{table}

\begin{table}[ht]
\centering
\caption{Initial and final conditions for beamline matching.}
\label{Tab:BeamLines_Requ_InitialConditions}
\begin{tabular}{l r r r r r}
\hline
  & \textbf{$\epsilon_{rms}$\,[$\pi$.mm.mrad]} & \textbf{$\beta$\,[m]} & \textbf{D\,[m]} & \textbf{D'} & \textbf{$\alpha$}\\
\hline
\multicolumn{6}{l}{\textit{Initial parameters at first electrostatic septum}} \\
Vertical & 0.6-4.2 & 15 & 0 & 0 & -2.8 \\
Horizontal & 2 & 15 & -4 & -1 & 0 \\
\hline
\multicolumn{6}{l}{\textit{Target parameters at end of beamlines}} \\
Pencil & 4.5 & 1 & 0 & 0 & 0 \\
Broad & 4.5 & 50 & 0 & 0 & - \\
\hline
\end{tabular}
\end{table}

\section{Transfer line design}
\label{Sec:BeamLines_3}

The proposed layout permits installation of the vertical beamline at the location of maximum available vertical space while respecting existing building walls.

\subsection{Dipole and quadrupole magnets}
\label{SubSec:BeamLines_Magnets}
The lines have been designed using 'sector' bends, where the beam enters the dipole perpendicular to the pole face, but in reality curved parallel face dipoles are preferred with a 'natural' pole-face rotation of half the bending angle in the bending plane. The gap field is 1.5\,T for both the vertical and horizontal dipoles. 

The resulting H-beamline dipoles are 3.415\,m (magnetic) along the curved beam trajectory and the V-beamline dipoles are 1.335\,m. The dipoles shall be laminated and curved to follow the beam trajectory, to reduce significantly the horizontal aperture required and the magnet size. The dipoles in each plane can be powered in series if advantageous for the power converters. Assumed parameters are given in table~\ref{Tab:BeamLines_Magnets_Dipoles}.

The quadrupoles are assumed to be 0.517 m long, with a maximum gradient of around 25\,Tm$^{-1}$. The actual strengths in operation vary considerably depending on the optics and on the beamline \cite{Abler:1742073}. A total of 20 installed quadrupoles are needed for the three beamlines. It is assumed in a first instance that all quadrupoles are individually powered, to give maximum flexibility and in view of the different optics needed. Some rationalisation may be possible for  similar strength quadrupoles by rematching and powering in series, to reduce this number slightly. Assumed parameters are given in table~\ref{Tab:BeamLines_Magnets_Quadrupoles}.

The possibility of reusing existing quadrupoles can be studied, e.g. from ISR beam transfer lines~\cite{NORMA:01,NORMA:02}. These would provide the required integrated gradient, but are physically longer than those used in the design, at 1.070\,m compared to 0.517\,m. A new optics layout would be needed using these quadrupoles, which has to be coupled to the integration study.

\begin{table}[ht]
\centering
\caption{Extraction beamlines main dipole parameters.}
\label{Tab:BeamLines_Magnets_Dipoles}
\begin{tabular}{l l r r}
\hline
\textbf{Extraction beamline dipoles} & \textbf{Unit} & \textbf{H-dipole} & \textbf{V-dipole} \\
\hline
Max beam rigidity & T.m & 6.7 & 2.6 \\
Dipole bend angle & deg & 44 & 45  \\
Gap dipole field & T & 1.5 & 1.5 \\
Magnetic length & m & 3.415 & 1.335 \\
Integrated field & Tm & 5.1 & 2.0\\
Horizontal full aperture & mm & 90 & 80\\
Vertical full aperture & mm & 80 & 90\\
Number installed (spare) &  & 5 (1) & 2 (1)\\
\hline
\end{tabular}
\end{table}

\begin{table}[ht]
\centering
\caption{Extraction beamlines main quadrupole parameters.}
\label{Tab:BeamLines_Magnets_Quadrupoles}
\begin{tabular}{l l r}
\hline
\textbf{Extraction beamline quadrupoles} & \textbf{Unit} & \textbf{Value} \\
\hline
Max beam rigidity & T.m & 6.7 \\
Magnetic length & m & 0.517 \\
Nominal gradient & Tm$^{-1}$ & 25 \\
Integrated gradient & T & 13 \\
Normalised gradient & m$^{-2}$ & 3.8 \\
Integrated normalised gradient & m$^{-1}$ & 2.0 \\
Inscribed pole tip radius & mm & 45 \\
Number installed (spare) &  & 20 (2) \\
\hline
\end{tabular}
\end{table}

\subsection{Trajectory correction}
\label{SubSec:BeamLines_Magnets_TrajCorr}

Trajectory correction has not been studied but some basic estimates for the numbers and strengths of the correctors have been extrapolated from other projects, see e.g.~\cite{Bauche:1753827}. From this, a tentative specification of 0.02\,Tm per corrector is proposed, giving a 3\,mrad deflection. A dual-plane corrector is proposed at each quadrupole group, with an additional corrector near the target to allow orthogonal steering, giving a total of about 11 correctors installed for the three beamlines (see table~\ref{Tab:BeamLines_Magnets_TrajCorr_Correctors}). A detailed error study is needed with an aperture model to validate these assumptions about the corrector numbers and maximum kick. The initial assumptions for the inputs for the error study are based on numbers for similar beamlines~\cite{Parfenova:1690026}, and tentative values are given in table~\ref{Tab:BeamLines_Magnets_TrajCorr_Tolerances}. These also provide a rough specification for the magnetic field errors and alignment requirements. It is likely, given the experience from other beamlines, that individual powering of the main dipoles represents an advantage for additional trajectory correction degrees of freedom.

\begin{table}[ht]
\centering
\caption{Extraction beamlines corrector parameters.}
\label{Tab:BeamLines_Magnets_TrajCorr_Correctors}
\begin{tabular}{l l r}
\hline
\textbf{Extraction beamline correctors} & \textbf{Unit} & \textbf{Value} \\
\hline
Max beam rigidity & T.m & 6.7 \\
Deflection angle & mrad & 3  \\
Gap dipole field & T & 0.1 \\
Magnetic length & m & 0.2 \\
Integrated field & Tm & 0.02\\
Horizontal full aperture & mm & 90\\
Vertical full aperture & mm & 90\\
Number installed (spare) &  & 11 (1) \\
\hline
\end{tabular}
\end{table}

\begin{table}[ht]
\centering
\caption{Assumed installed tolerances for beamline elements.}
\label{Tab:BeamLines_Magnets_TrajCorr_Tolerances}
\begin{tabular}{l l l r l}
\hline
\textbf{Error source} & & \textbf{Unit} & \textbf{Value} & \textbf{Distribution} \\
\hline
\rule{0pt}{3ex}Dipole field & $\Delta$Bdl/B$_{0}$dl & & 1$\cdot$10$^{-3}$ & uniform \\
Dipole roll angle & d$\psi$ & rad & 2$\cdot$10$^{-4}$ & uniform \\
Dipole long. position & dS & m & 1$\cdot$10$^{-3}$ & uniform \\
Quadrupole field & $\Delta$K/K$_{0}$ & & 1$\cdot$10$^{-3}$ & uniform \\
Quadrupole H,V shift & dX, DY & m & 2.5$\cdot$10$^{-4}$ & Gauss($\sigma$) \\
LEIR delivery precision & dX, DY & m & 5$\cdot$10$^{-4}$ & Gauss($\sigma$) \\
                  & dPX, DPY & mrad & 5$\cdot$10$^{-4}$ & Gauss($\sigma$)  \\
Monitor H,V shift & dX, DY & m & 2.5$\cdot$10$^{-4}$ & Gauss($\sigma$) \\
Monitor resolution & X, Y & m & 2$\cdot$10$^{-4}$ & uniform \\
\hline
\end{tabular}
\end{table}

\subsection{Scattering foil and beam size at target}
\label{SubSec:BeamLines_Magnets_Beamsize}

A thin scattering foil is foreseen after the triplet quadrupole in the first common beamline, to transform the extracted 'bar of charge' generated by the slow extraction process into an approximately Gaussian distribution. Approximately equal emittances in both planes can be obtained by adjusting the thickness of the scatter foil and  functions at its position. It is assumed that $\epsilon_{rms,x/y}=4.5\,\pi$.mm.mrad is achievable after scattering for all energies, using foils with several 10$^{-3}$ radiation lengths thickness. By strongly defocusing the beam in both planes, the rms beam half-width on target can be increased until the desired level of dose uniformity across a given sample area is reached, at the cost of collimating the portion of the beam exceeding the sample dimensions. A beam half-width of 77\,mm would be needed to provide I/I$_{0}$ of 0.9 over a 50$\times$50\,mm$^2$ sample, which would make use of 6.5\% of the beam particles and imply beta functions at the target of around 1500\,m \cite{Abler:1742073}. 

\subsection{Tail folding and octupoles}
\label{SubSec:BeamLines_Magnets_Tailfolding}

The possibility of tail folding to achieve a uniform transverse distribution, see e.g.~\cite{Batygin:2132685} has not been explored in detail but would require the addition of strong octupoles for the over-focusing; tentatively these would need a strength $\frac{\partial^3 B_y}{\partial x^3}$ of around $3\times{×}10^{4}$\,Tm$^{-3}$ and be 0.33\,m long. This approach would have the advantage of increasing the intensity on target, and hence reducing irradiation times, but needs further study; the integration of the non-linear lenses into the beamline(s) also may prove problematic.

As for the octupole magnets themselves, the parameters of table~\ref{Tab:BeamLines_Magnets_Tailfolding_Octupoles} are similar to the LODN octupoles installed in the SPS~\cite{NORMA:03}, although the latter are 0.7 m long. Although there are no spares of the LODN, many out of the 28 installed units currently installed in the ring are not powered, so it could be envisaged to utilise some of them here, or build new magnets with this design.

\begin{table}[ht]
\centering
\caption{Extraction beamlines octupole parameters.}
\label{Tab:BeamLines_Magnets_Tailfolding_Octupoles}
\begin{tabular}{l l r}
\hline
\textbf{Extraction beamline octupoles} & \textbf{Unit} & \textbf{Value} \\
\hline
Max beam rigidity & T.m & 6.7 \\
Magnetic length & m & 0.33 \\
Nominal strength & Tm$^{-3}$ & 30,000 \\
Integrated strength & Tm$^{-2}$ & 10,000 \\
Normalised strength & m$^{-4}$ & 2,000 \\
Integrated normalised strength &  m$^{-3}$ & 660 \\
Inscribed pole tip radius & mm & 45 \\
Number installed (spare) &  & 2 (1) \\
\hline
\end{tabular}
\end{table}

\subsection{Optics functions and beam envelopes}
\label{SubSec:BeamLines_Magnets_BeamEnvelopes}

\begin{figure}[ht!]
    \centering
    \begin{subfigure}{.4\linewidth}
 \hspace*{-45mm}\includegraphics[width=2.4\linewidth]{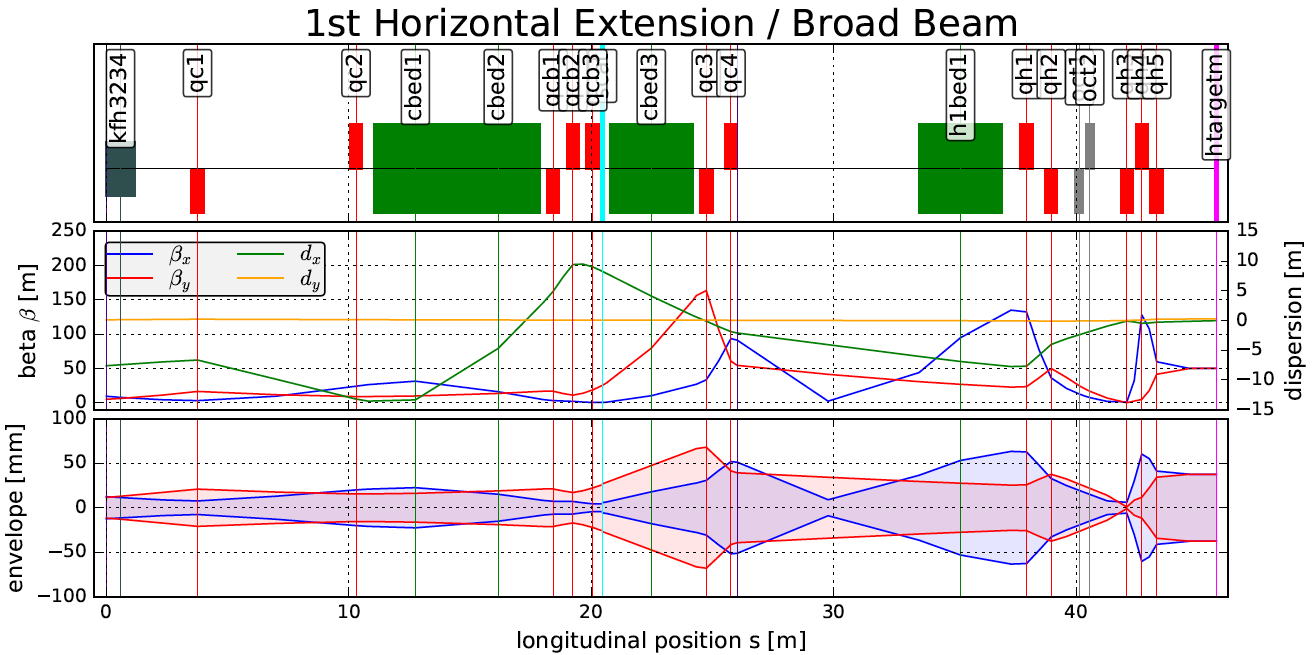}{}
        \caption{Horizontal beamline H1 broad beam}
    \end{subfigure}
    
    \begin{subfigure}{.4\linewidth}
   \hspace*{-45mm}\includegraphics[width=2.4\linewidth]{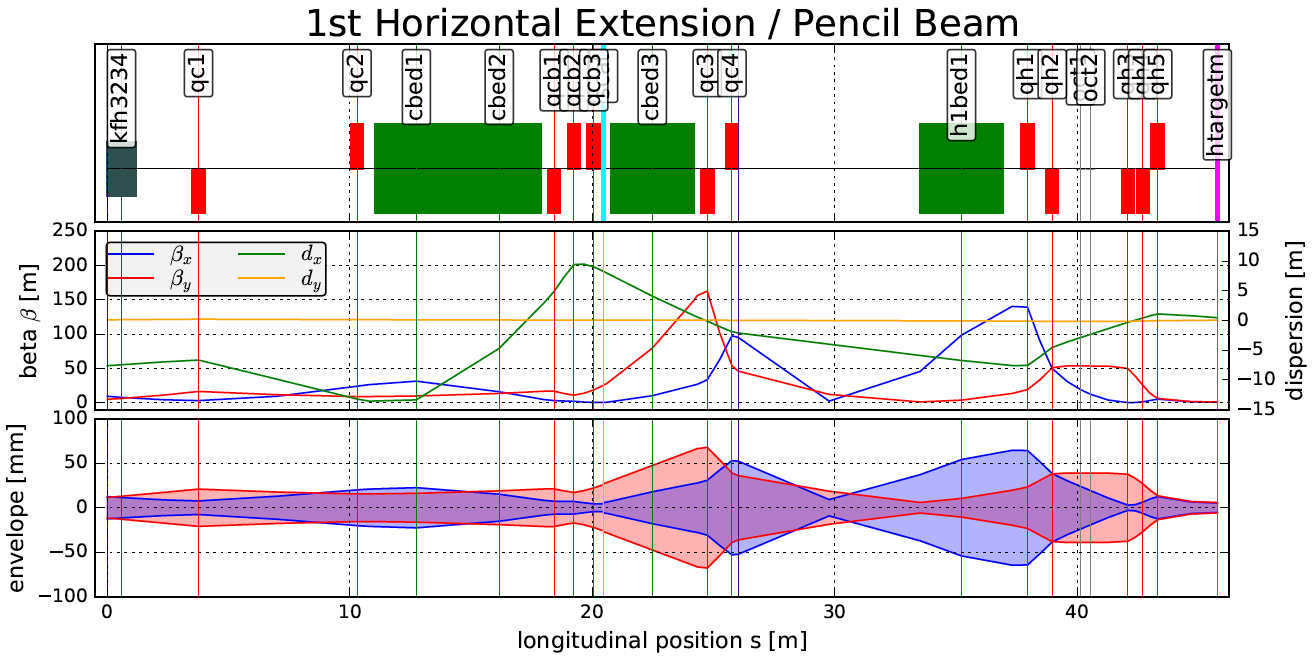}{}
    \caption{Horizontal beamline H1 pencil beam}
    \end{subfigure}
\caption{Optics and beam envelope plots for H1, for the broad beam and pencil beam optics with $\beta$\,=\,50\,m and $\beta$\,=\,1\,m in the target plane, respectively.}
\label{Fig:BeamLines_H1}
\end{figure}

The detailed optics calculations have been made for the vertical and two horizontal beamlines shown in figure~\ref{Fig:BeamLines_1.1}. The H1 beamline allows the optics functions in the target plane to be adjusted over a range of $\beta=1-50$\,m, as illustrated in figure~\ref{Fig:BeamLines_H1} and thus fulfill field size and homogeneity criteria outlined in table~\ref{Tab:BeamLines_Requ_Parameters}. Enforcement of achromatic beam transfer through the common horizontal bending segment defines the strengths of the initial quadrupoles, and the optical functions in this part of the line show little dependency on target parameters. The horizontal (vertical) $\beta$ functions at the scatter foil can be adjusted between 20-100\,m approximately, for tuning the blowup. 

\begin{figure}[ht!]
    \centering
    \begin{subfigure}{.4\linewidth}
 \hspace*{-45mm}\includegraphics[width=2.4\linewidth]{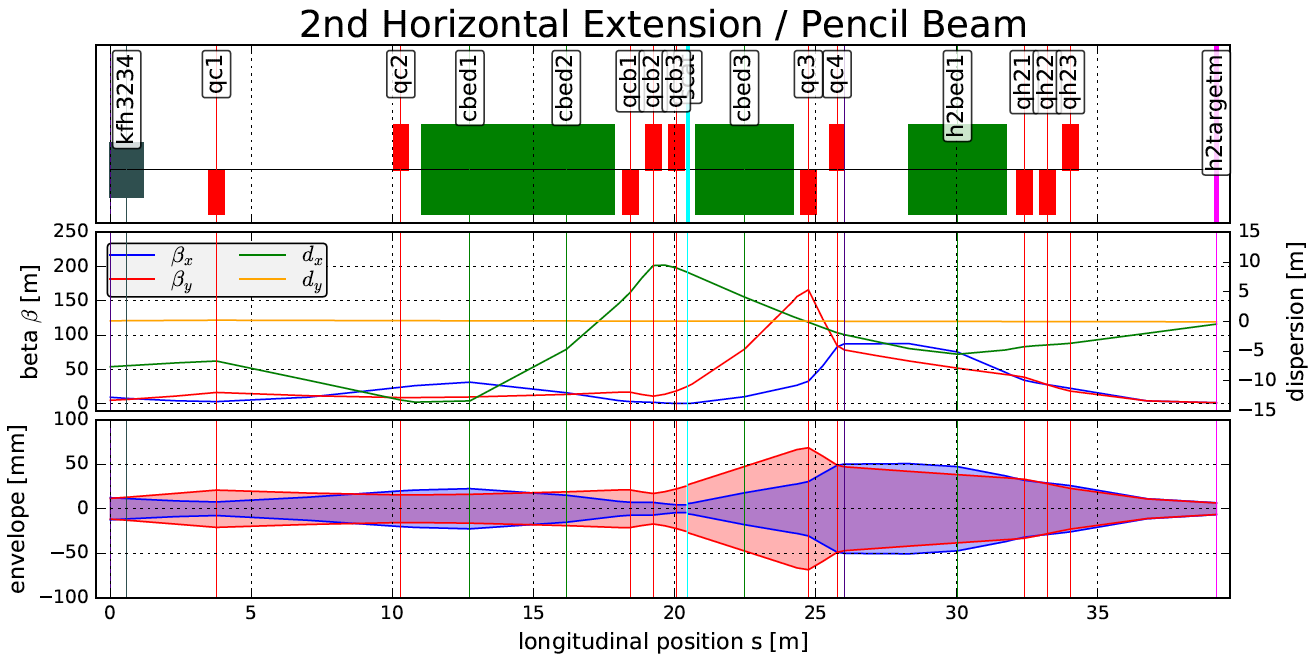}{}
        \caption{Horizontal beamline H2 pencil beam}
    \end{subfigure}
    
    \begin{subfigure}{.4\linewidth}
   \hspace*{-45mm}\includegraphics[width=2.4\linewidth]{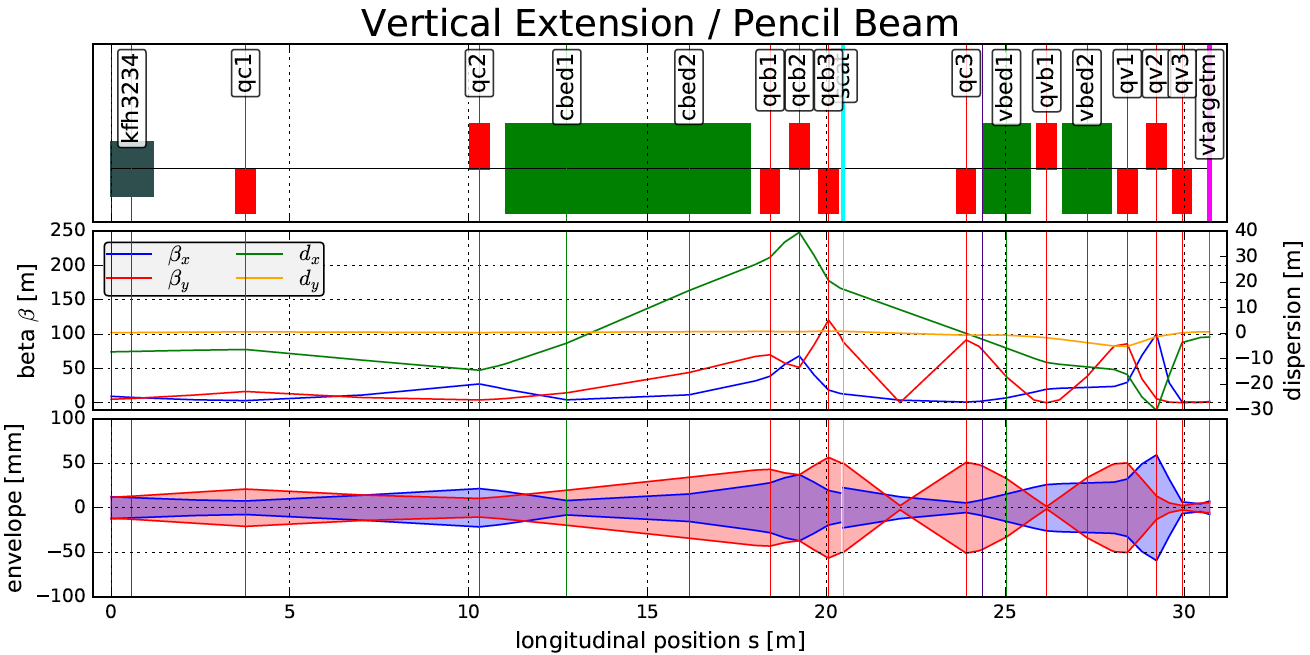}{}
    \caption{Vertical beamline V pencil beam}
    \end{subfigure}
\caption{Optics and beam envelope plots for H2 and V, for the pencil beam optics with $\beta$\,=\,1\,m in the target plane.}
\label{Fig:BeamLines_H2_V}
\end{figure}

The optics and envelope plots for the second horizontal beamline H2 and the vertical beamline V are shown in figure~\ref{Fig:BeamLines_H2_V}.




At this preliminary stage, the optics matching of all beamlines still needs to be improved, to reduce the envelope to $\pm$40\,mm, by controlling the beta functions to below about 100\,m. This is evident from the envelope plots which exceed $\pm$50\,mm for all beamlines and optics in the present design.  An improvement is expected to be possible with optimisation of the quadrupole strengths and locations.

\subsection{Beam purity}
\label{SubSec:BeamLines_Magnets_BeamPurity}
A beam purity of better than $10^{-4}$ is requested. This means that the scatterers and collimation elements must be upstream of the main bending dipoles, to limit the transmission for any unwanted secondaries or off-momentum primary ions or fragments. Detailed nuclear simulations need to be made for this aspect, coupled with the transfer line transport codes, to determine the expected contamination levels. This work requires the 'final' optics design, the error study, the design of the collimation system and the assumed operational scenarios (ion species, spot size, uniformity etc.) as input.

\section{Other beamline systems}
\label{Sec:BeamLines_OtherSystems}

\subsection{Powering cycle and power converters}
\label{SubSec:BeamLines_Othersystems_Power}

The magnets can be pulsed for energy economy, with a ramp-up/down time of the order of 1\,s and a maximum flat-top length of around 10\,s (see table~\ref{Tab:BeamLines_Requ_Parameters}). Fast changes of strength during an extraction are not considered in the power converter specifications - if a fast (10-15\,ms) change in the powering of the lines is required during a spill, more information is needed about the maximum energy change and also the allowed settling time or maximum allowable trajectory/optics errors. The minimum repetition period for this cycle is foreseen to be 13.2\,s.

All the main dipoles are envisaged to be powered individually as it is an advantage for trajectory correction. It is assumed that there are 5 H dipole converters and 2 V dipole converters.

A few of the quadrupole converters require polarity inversion to switch between pencil and broad beam optics or need to be bipolar. All of the corrector converters need to be bipolar with accurate regulation at zero current.
 The overall numbers of converters and preliminary requirements are given in table~\ref{Tab:BeamLines_Converters}.

\begin{table}[!ht]
\centering
\caption{Extraction beamlines power converter parameters.}
\label{Tab:BeamLines_Converters}
\begin{tabular}{l l r r r r r}
\hline
\textbf{Parameter} & \textbf{Unit} & \textbf{H-bend} & \textbf{V-bend} & \textbf{Quad.} & \textbf{8-pole} & \textbf{Corr.}\\
\hline
Magnets (spare) &  & 5 (1) & 2 (1) & 20 (2) & 2 (1) & 11 (1)\\
Converters (spare) &  & 5 (1) & 2 (0) & 20 (1) & 2 (0) & 22 (2)\\
Min. ramp time & s & 1 & 1 & 1 & 1 & 1\\
Max. flat-top time & s & 10 & 10 & 10 & 10 & 10\\
Max. repetition time & s & 13.2 & 13.2 & 13.2 & 13.2 & 13.2\\
Inductance & mH & 100 & 100 & 45 & 60 & 10\\
Resistance & m$\Omega$ & 100 & 100 & 62 & 30 & 100\\
Nom. current & A & 500 & 500 & $\pm$500 & 500 & $\pm$50 \\
Max. voltage & V & 120 & 120 & 120 & 120 & 30 \\
Current precision & ppm & $\pm$200 & $\pm$200 & $\pm$200 & $\pm$200 & $\pm$100 \\
AC input (3P-400Vac)& A & 125 & 125 & 125 & 125 & 4 \\
Dimensions & mm & \multicolumn{4}{c}{2300$\times$1600$\times$900} & 4Ux19'' \\
Cooling &  & \multicolumn{5}{c}{Forced air} \\
Peak power & kW & 100 & 100 & 100 & 100 & 1.5\\
\hline
\end{tabular}
\end{table}

The dipole, quadrupole and octupole magnets are envisaged to be powered using  COMET\_2P power converters. This option allows the reduction of the required spares and the use of standard power converters. COMET\_2P are 4-quadrants power converters composed of 2 units rated 30\,kW, connected in parallel. In normal operation, the converters are cycled to reduce power consumption. Moreover, during the ramping down of the current, the energy stored in the magnet is recovered in the converter. For the powering of the quadrupoles, further optimization of the matching of the converter and magnet designs could allow the use of a single COMET module (instead of two in parallel). If this option turns out to be feasible, it would lead to significant cost reduction. 

For the powering of the corrector magnets, CANCUN50 power converters are used. These converters are 4-quadrants converters that can be operated in cycled operation mode. The output current is bipolar with accurate regulation at zero current.

The converters are controlled using CERN standard FGC based control interface with Ehernet+. All the standard interfaces (magnet interlock, beam interlock etc.) are provided if required. All the requirements can be fulfilled by CERN standard power converters and no special designs are required allowing the costs and personnel requirements to be kept to a minimum.  

\subsection{Beam instrumentation}
\label{SubSec:BeamLines_Othersystems_BI}

The instrumentation for the beamlines has not been specified in detail, nor does a conceptual design exist. The proposed long spills and range of current pose technical difficulties, and mean that conventional solutions in the LEIR to PS transfer lines (see e.g.~\cite{Benedikt:823808, Roncarolo:971897}) like BPMs, BCTs, BTVs and SEM grids most likely do not work.
The technical solutions for the different monitors remain to be evaluated, the current ideas are described below.

For the present estimates the instrumentation is assumed to consist of four beam intensity instruments (one at extraction and one per line), three dual-plane profile monitors located in dispersion free regions to allow emittance measurement, plus profile monitors for beam size and position measurements just before each target plane, and finally around 12 beam position monitors for trajectory monitoring and correction (table~\ref{Tab:BeamLines_OtherSystems_Instruments}).

Beam position and profile monitoring in the experimental beamlines could be made with GEM monitors (gas electron multiplier) that are envisaged to be available from Antiproton Decelerator (AD) during LS2. GEMs have been deployed extensively in the present AD experimental areas and give excellent results for both position and profile measurements. Like multi-wire proportional chambers, these detectors are also gas-filled, and essentially the same physical phenomenon is exploited to multiply ionisation charge. A GEM is a 50-$\mu$m-thick foil of Kapton, copper clad on both sides, pierced with microscopic holes at a high density (our 10$\times$10\,cm$^2$ foils have about a million holes). A voltage of a few hundred volts applied to the top and bottom copper layers causes an electric field that focuses in the centre of these holes where it is just as strong as close to the wires of a wire chamber. Ionisation electrons enter the holes from one side, are multiplied inside the holes, and then exit on the other side where they are collected by a strip pattern that integrates the charge and reads out the profile.

The cathode window is a crucial element when absorption and multiple scattering of the beam are of concern. In the monitor design the cathode is also the gas enclosure, and it is stretched tight ($\sim$11\,MPa) to avoid any deformation by the slight overpressure in the chamber. It is made of the same base material GEMs are made of: copper clad polyimide. The copper is etched away in the active area of the detector, leaving just a thin ($\sim$100\,nm) layer of chromium which is there to act as a tie coat for a better adhesion of the copper layer to the polyimide substrate. The material traversed by the beam to enter the active volume of the detector thus amounts to 0.018\%\,X$_0$. On the other end of the chamber, the gas enclosure is made of a 25\,$\mu$m polyimide foil, adding 0.009\%\,X$_0$.

These recuperated monitors need an upgrade of their electronics to be compatible with the latest norms used by the Controls group. Their mechanical integration in the beamlines also needs to be studied by EN-MME group. The estimated cost including cabling and gas distribution is about 150\,kCHF.

The beam intensity in the beamlines could be measured with a scintillating fibre monitor (FISC) much like what is done in the SPS North Area. The cost is estimated to amount to 100\,kCHF.


\begin{table}[ht]
\centering
\caption{Required beam instrumentation.}
\label{Tab:BeamLines_OtherSystems_Instruments}
\begin{tabular}{l l l l l}
\hline
\textbf{Instrument} & \textbf{Observable} & \textbf{Unit} & \textbf{Resolution}&  \textbf{Installed (spare)} \\
\hline
Scintillator & Current & ions/s & $10^6$ &  3 (1) \\
MWPC & Position & mm & 0.25 & 12 (2) \\
 & Size ($\sigma$) & mm & 0.2 &  \\

\hline
\end{tabular}
\end{table}

\subsection{Vacuum systems}
\label{SubSec:BeamLines_OtherSystems_Vacuum}

The vacuum system for the beamlines needs to ensure the required pressure through the lines, and (probably more importantly), to allow the very low pressure of better than $4\times10^{-12}$\,mbar N$_{2}$ equivalent needed for the LEIR machine~\cite{Mahner:902810}. 
The detailed vacuum system design is found in chapter~\ref{Chap:Vacuum}.



\subsection{Interlocks}
\label{SubSec:BeamLines_OtherSystems_Interlocks}

Some type of interlock system is needed to cut beam extraction in case of e.g. power converter failure, vacuum fault or high beam loss signal. If very high reaction speed and reliability is required, this system can be based on the standard CERN Beam Interlock Controller system~\cite{Puccio:1277643}, linked to the Warm Magnet Interlock system which surveys the magnets. Otherwise a simple software solution may suffice. In the absence of clinical irradiation, the reliability and safety requirements can be met by these solutions.

\subsection{Supports and alignment}
\label{SubSec:BeamLines_OtherSystems_Supports}

Individual magnet supports are needed for the main dipoles, which can be assumed to be based on three separate jacks. The quadrupoles, beam diagnostics and correctors can be installed on common support girders, with one XYZ translation stages per machine element. Some special individual adaptations may be needed where the beamlines pass through the walls between the different halls, but no particular issues are expected for the support and alignment of the horizontal beamlines. 

The required alignment tolerances are standard for this type of beamlines~\cite{Parfenova:1690026}, with assumed tolerances listed in table~\ref{Tab:BeamLines_Magnets_TrajCorr_Tolerances}. However, with a target plane some 6.5\,m above the floor, the vertical beamline requires the installation of a main dipole and three quadrupoles, plus correctors and diagnostics, at a considerable height above the floor. This requires the development of special supports and alignment designs and techniques.

\subsection{Cabling}
\label{SubSec:BeamLines_OtherSystems_Cabling}

Electrical distribution, power and controls cabling are required for the magnets, power converters, vacuum systems, beam instrumentation and interlocks. In the absences of layouts for the rack, controls and power converter locations, and associated cable lengths, only very basic estimates of costs can be made.

\subsection{Cooling}
\label{SubSec:BeamLines_OtherSystems_Cooling}

All the power converters for the beamlines are air cooled but the magnets require water cooling. From table~\ref{Tab:BeamLines_Converters}, it can be computed that the magnets running at full current in DC require a maximum total of about 350\,kW of cooling power. 

\section{Resource estimate}
\label{Sec:BeamLines_Costing}

The estimated CERN personnel needs (Staff and Fellows) are given in table~\ref{Tab:BeamLines_Costing_Manpower}. The estimates include the remaining conceptual design phase, engineering design, prototyping, construction and installation.

\begin{table}[ht]
\centering
\caption{Preliminary personnel estimate for design, construction, installation and commissioning of extraction beamline elements. Fellow costs are included in table~\ref{Tab:BeamLines_Costing_Material}.}
\label{Tab:BeamLines_Costing_Manpower}
\begin{tabular}{l r r}
\hline
\textbf{System} & \textbf{Staff [MY]} & \textbf{Fellow [MY]} \\
\hline
Beam dynamics & 1.0 & 2.0 \\
Magnets & 2.0 & 2.0\\
Power converters & 0.7 & 1.4 \\
Diagnostics & 2.0 & 4.0 \\
Magnet supports & 0.2 & - \\
Power/control cabling & 0.3 & -\\
Water cooling & 0.2 & - \\
Alignment and survey & 0.2 & - \\
Transport and handling & 0.1 & - \\
Installation coordination  & 1.0 & - \\
Beam preparation and commissioning & 2.0 & 2.0 \\
\hline
\rule{0pt}{3ex}\textbf{Total} & \textbf{9.7} &\textbf{11.4}\\
\hline
\end{tabular}
\end{table}

The numbers required  and assumed costs for the different equipment subsystems described above are given in table~\ref{Tab:BeamLines_Costing_Material}. Note that the numbers given are extremely preliminary and that actual costs are likely to vary by as much as 50\% from the numbers given, due to the  precursory stages of the study, the lack of validation of some key aspects and the fact that engineering designs have not yet been started. Total cost includes prototyping, installation and one-off tooling costs not included in the unit cost. The costs include all non-CERN staff resources such as Field Support Units. Note that Fellows are not included in these costs, but are counted under CERN personnel.

No detailed exploitation cost or workforce estimates for the equipment are given, nor staff needed for maintenance and operation. This is typically of the order of 5\% of the investment cost of an installation, and so can be roughly estimated to about 300\,kCHF and 1\,FTE per year, spread across the technical groups concerned.

\begin{table}[!ht]
\centering
\caption{Preliminary cost estimate for design, construction, installation and commissioning of extraction beamline elements. Total costs include prototyping and overheads where relevant, e.g. for tooling or design.}
\label{Tab:BeamLines_Costing_Material}
\begin{tabular}{l r r r}
\hline
\textbf{System} & \textbf{Units} & \textbf{Unit cost} & \textbf{Total} \\
\textbf{} & \textbf{} & \textbf{[kCHF]} & \textbf{[kCHF]} \\
\hline
H dipole magnets  & 6 & 150 & 1100 \\
V dipole magnets  & 3 & 100 & 500 \\
Quadrupole magnets  & 22 & 20  & 700 \\
Corrector magnets  & 12 & 12 & 250 \\
Octupole magnets  & 3 & 40 & 200 \\
Dipole converters  & 8 & 60 & 480 \\
Quadrupole converters  & 21 & 60 & 1260 \\
Corrector converters  & 24 & 10 & 240 \\
Octupole converters  & 2 & 60 & 120 \\
Scintillators & 4 & 15 & 100 \\
MWPC detectors & 13 & 10 & 150 \\
Beam and magnet interlock system & 1 & 75 & 75\\
Dipole jacks & 21 & 5 & 105 \\
Support girders & 11 & 5 & 75 \\
XYZ stages & 50 & 2 & 100 \\
Magnet power cabling & 49 & 5 & 250 \\
Beam instrumentation cabling & 15 & 5 & 75\\
Magnet water cooling & 12 & 5 & 60 \\
Layout/integration (man-months) & 8  & 10 & 80 \\
Alignment and survey (man-months) & 3  & 10 & 30 \\
Transport and handling (man-months) & 12  & 10 & 120 \\
\hline
\rule{0pt}{3ex}\textbf{Total} & & &\textbf{6070}\\
\hline
\end{tabular}
\end{table}

\section{Potential risks and remaining conceptual issues}
\label{Sec:BeamLines_Risks}

Apart from the element design and the associated integration, especially for the vertical beamline, the main unresolved technical issue appears to be the required uniformity at the target, which is strongly linked to the collimation and beam purity which can be expected. This needs design and calculations, for both the achievable beam purity and for the beam loss and activation aspects, and also for the investigations of the tail-folding with non-linear lenses. Alternatives such as scanning and wobbling systems also need to be studied. 

The detailed studies still required are for the trajectory correction/aperture and the beam instrumentation systems. These are not expected to pose serious difficulties, although the very large beta functions for the broad beam spot introduce more sensitivity to errors, but the assumptions need to be validated and basic specifications and engineering concepts need to be established.

It shall be noted that the beam half-widths shown in figures~\ref{Fig:BeamLines_H1} and \ref{Fig:BeamLines_H2_V} are calculated using 2.5\,$\sigma$ envelopes using the rms emittances specified in~\cite{Abler:1742073} and assuming maximum momentum spread $\delta$p/p=0.2\% throughout the line. No allowance for mechanical tolerance nor trajectory excursions are included, which means that the realistic aperture for the beamline is likely to be closer to 2\,$\sigma$. This does not represent a particular issue for the intensity delivered to the targets, but needs to be investigated in terms of beam loss and activation, and (probably more crucially) in terms of collimation and beam purity at the targets. A thorough optics design optimisation of the lines is therefore mandatory to validate the assumed magnet gap openings and apertures.

%% file: Chapters/EA.tex
\chapter{Experimental Area}
\label{Chap:EA}
This chapter discusses all aspects of the biomedical experimental area for BioLEIR. The experimental area layout depends on: 
\begin{enumerate}
\item the 3D coordinate position of the two horizontal and the vertical beamline isocenters; 
\item the available infrastructure such as electricity (e.g. room lighting and electrical outlets), medical grade gases (e.g. N$_2$, O$_2$, CO$_2$ and Ar) and cooling and ventilation within the irradiation room around the isocenters; 
\item structural elements such as safe and easy access to the vertical irradiation point, two equipped counting rooms (with a set of racks) and fast Ethernet connections; 
\item the beamline requirements; 
\item requirement for radiation protection such as requirement for personnel access to the irradiation points, requirement for the cool-down area for sample stacks and requirement for radiation protection for the vertical beam irradiation point and its access; 
\item requirements on access control such as requirement for access to an irradiation room while the other two irradiation rooms are in operation, requirement for interlock/integration into the CERN accelerator system/beam delivery (veto while samples are not ready) and requirement for integration of the control irradiator operation into the overall access control system.
\end{enumerate}

The biomedical experimental area has an impact on radiation protection requirements around the irradiation points, infrastructure aspects and shielding footprint requirements for the 6\,MeV photon linac used as control irradiator.


\section{Outline of the experimental hall}
\label{Sec:EA_ExperimentalHall}

\begin{figure}[ht!]
    \centering
    \begin{subfigure}{.4\linewidth}
      \hspace*{-45mm} \includegraphics[width=2.2\linewidth]{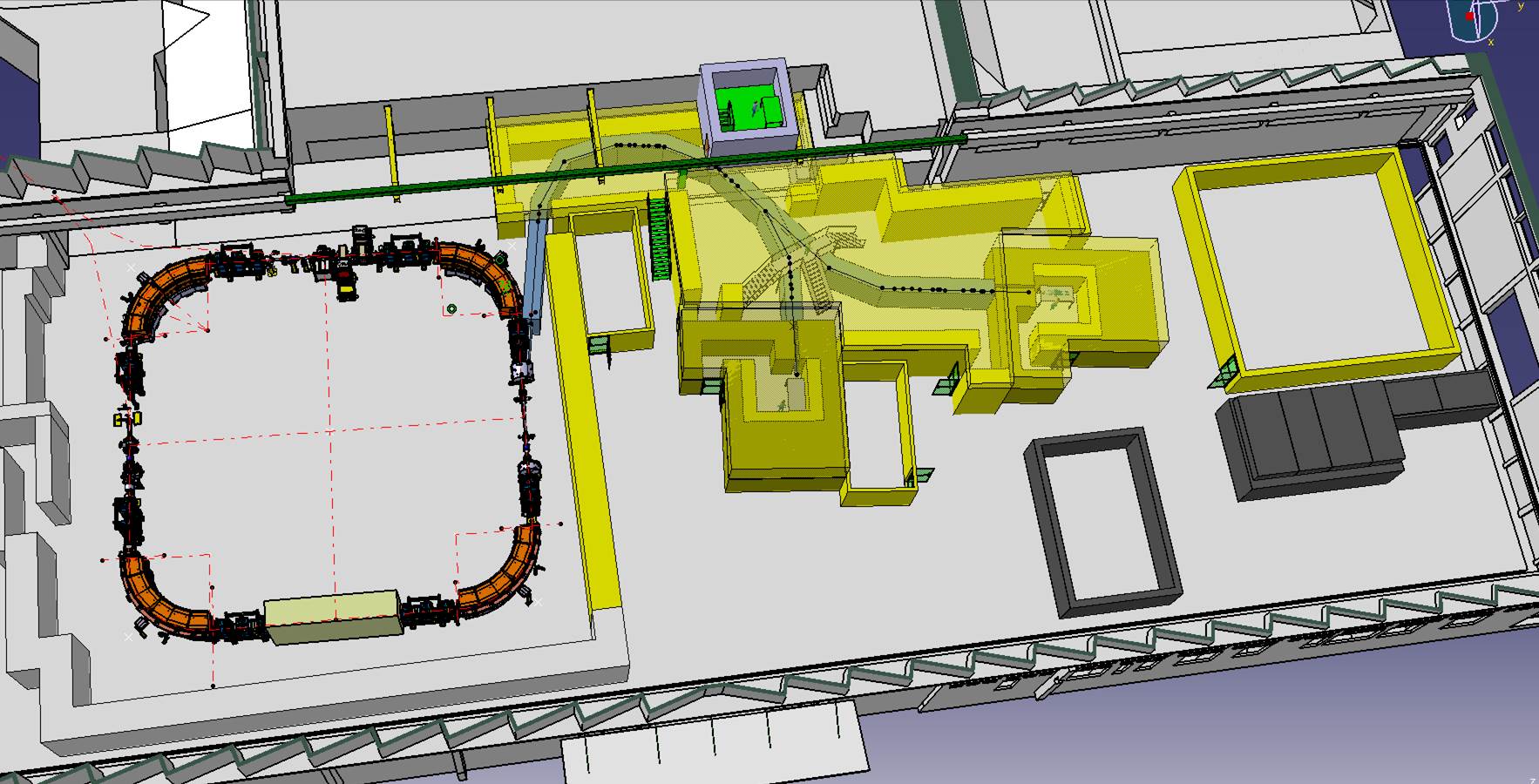}{}
        \caption{}
    \end{subfigure}
    
    \begin{subfigure}{.4\linewidth}
     \hspace*{-45mm} \includegraphics[width=2.2\linewidth]{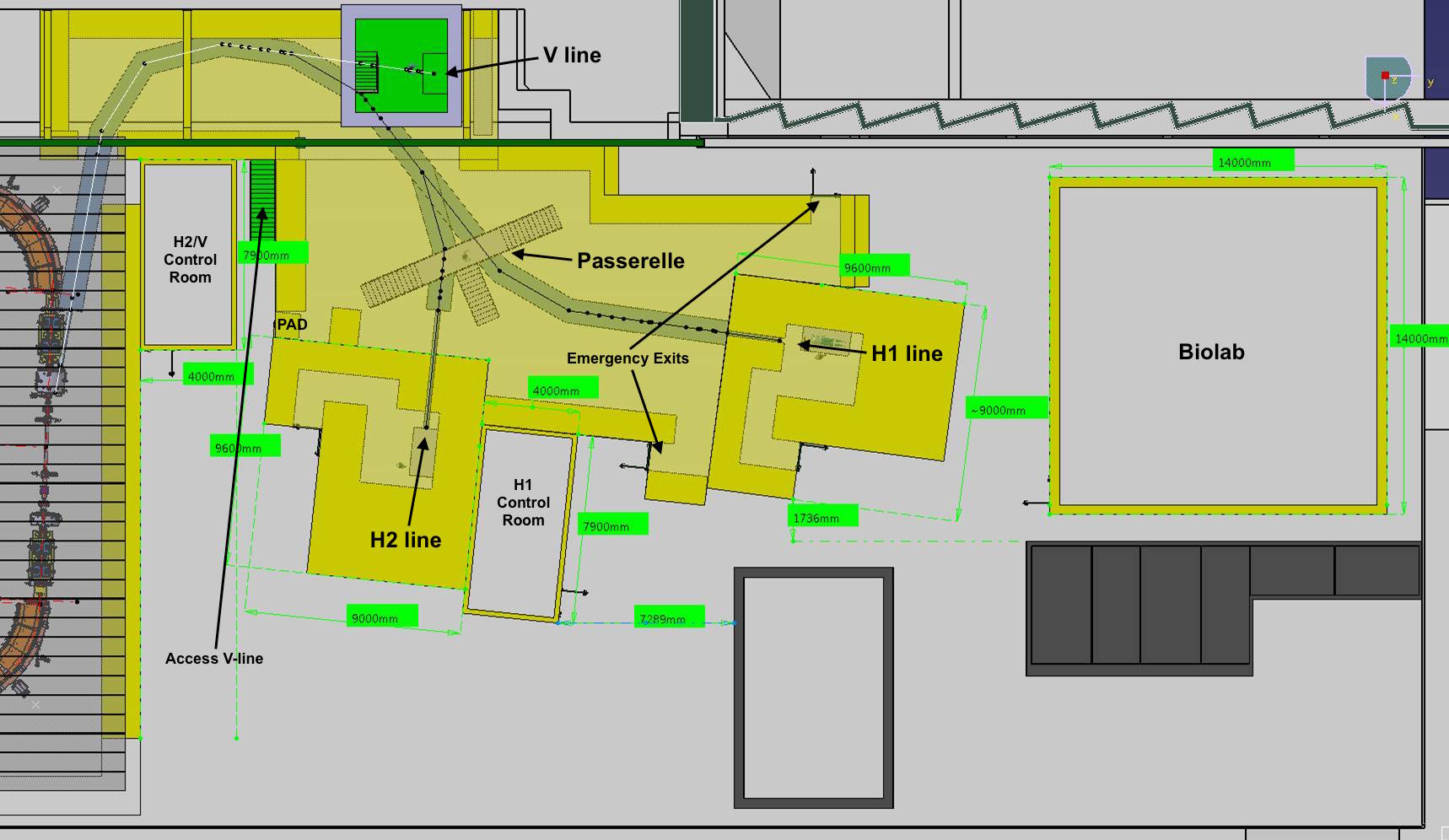}{}
    \caption{}
    \end{subfigure}
\caption[]{BioLEIR experimental area outline (a) and its nomenclature (b).}
\label{Fig:EA_Outline}
\end{figure}

The experimental irradiation area of BioLEIR can be located in the "South Hall" adjacent to the LEIR accelerator in building~150. A preliminary facility outline (see figure~\ref{Fig:EA_Outline}) shows the BioLEIR switchyard where the three experimental lines split, as well as the location of the irradiation rooms, their respective counting rooms, and the Biolab.

    


Three irradiation rooms are planned. They are operationally independent of each other, with independent access and respective dedicated purpose and applications.

Each of the three irradiation rooms have their specific constraints in terms of maximum particle energy and maximum beam intensity. It may become necessary to restrict the type of light ion that is available in a specific irradiation room. The constraints are dictated by an optimization of radiation protection requirements, the space availability for adequate shielding, as well as keeping a reasonable cost for shielding :
\begin{itemize}
   \item \emph{horizontal beamline, H1:} intended for biomedical experiments on cells, with ion energies up to 440\,MeV/u and intensities between $10^8$ and $10^{10}$ protons/s on target, as well as $10^6$ to $10^8$ ions/s on target for light ions. This irradiation room is closest to the Biolab, optimizing the distance for transport of the cell-samples. A pencil beam of 5-10\,mm\,FWHM as well as a homogeneous broad beam of 50$\times$50\,mm$^2$ are available at this irradiation point.
   \item \emph{horizontal beamline, H2:} reserved for phantom work, (micro-)dosimetry and detector development, with ion energies up to 440\,MeV/u and intensities up to $10^{10}$ protons/s on target, and up to $10^8$ ions/s on target for light ions.   
   \item \emph{vertical beamline, V:} ion energies up to 70\,MeV/u and ion intensities up to $10^8$ ions/s on target.
\end{itemize}

Table~\ref{Tab:EA_ExperimentalHall_Rooms} gives a summary of the characteristics of the three irradiation rooms for ease of reference.  Note that proton intensities up to $10^{11}$ protons/s on target and ion intensities up to $10^{9}$ ions/s are possible, if the shielding is adapted accordingly. The chosen shielding dimensions for the baseline BioLEIR facility represent a reasonable compromise between shielding cost and intensity and energy of the ion beams.\\


\begin{table}[ht]
\caption{Summary of ion species, energies and intensities available in each of the three irradiation rooms, together with the corresponding shielding dimensions. The numbers are given for Stage 1+2 (maximum 250\,Mev/u) and for Stage 3 (maximum 440\,MeV/u). The vertical line is always limited to a maximum of 70\, Mev/u.}
\label{Tab:EA_ExperimentalHall_Rooms}
\centering
\begin{tabular}{l c c c c c c c}
\hline
\rule{0pt}{3ex}\textbf{Room} & \textbf{Pencil} & \textbf{Broad} & \textbf{Max} & \textbf{Intensity} & \textbf{Intensity} & \textbf{Lateral} & \textbf{Longitudinal}\\
\textbf{} & \textbf{ beam} & \textbf{ beam} & \textbf{energy} & \textbf{[p]} & \textbf{[$<=$Ar]} & \textbf{shielding} & \textbf{shielding}\\
 &  &  & [MeV/u] & [ions/s] & [ions/s] & [m] & [m]\\
\hline
\rule{0pt}{3ex}H1 & x & x & 250 / 440 & $10^9$ / $10^{10}$ & $10^8$ / $10^8$ & 1.0 / 1.8 & 3.4 / 4.0 \\
H2 & x & - & 250 / 440 & $10^9$ / $10^{10}$ & $10^8$ / $10^8$ & 1.0 / 1.8 & 3.4 / 4.0 \\
V  & x & - &  70 & 10$^{8}$ & 10$^8$ & 0.6 & 1.6\\
\hline
\end{tabular}
\end{table}

A dedicated, fully equipped Biolab will be used to prepare and analyse sample cell cultures. It is located in close vicinity to the H1 irradiation point and has a footprint of about 200\,m$^2$ (see section~\ref{SubSec:EA_ExperimentalHall_BioLab} for more details).

\subsection{Irradiation rooms}
\label{SubSec:EA_ExperimentalHall_IrradiationRooms}
Irradiation at the maximum ion energies and intensities will produce air and material activation. Therefore, extensive shielding, access maze, a controlled under-pressure air ventilation system are needed (see chapter~\ref{Chap:RP} for details on radiation protection constraints). The irradiation rooms are planned to be built out of precast concrete shielding blocks with the following requirements: 
\begin{itemize}
   \item making the rooms reasonably airtight,
   \item designing the access control system such that the necessary air-exchange takes place before the room can be entered after irradiation.
\end{itemize}

Depending on the exact final position and access arrangements of the vertical beam irradiation point, and because of the immediate proximity of the PS accelerator enclosure, further radiation protection from stray radiation streaming during PS operation may be necessary.

\subsubsection{Biological irradiation rooms: H1 and V}
\label{SubSubSec:EA_ExperimentalHall_IrradiationRooms_Bio}
Two of the three irradiation rooms are dedicated to experiments on cell cultures. The horizontal irradiation room H1 can have both a homogeneous broad beam of 50$\times$50\,mm$^2$ and a pencil beam of 5-10\,mm FWHM, including a beam deflection system for scanning. This room is in immediate vicinity of the Biolab.

The vertical irradiation room is located further away from the Biolab at a height of about 4.5\,m above ground level of hall~150, and is accessible by a stairway. The vertical irradiation room will receive a pencil beam of 5-10\,mm FWHM without a scanning system\footnote{Note that this is to achieve an endpoint for the vertical beamline that remains below the necessary height restriction for the overhead crane from building 151 (7.95\,m).}. In the vertical irradiation room, the sample holding table is motorized to move the sample around the isocenter, rather than deflecting the beam for scanning purposes.

\begin{figure}[!htb]
\centering
\includegraphics[width=0.8\textwidth,height=0.5\columnwidth]{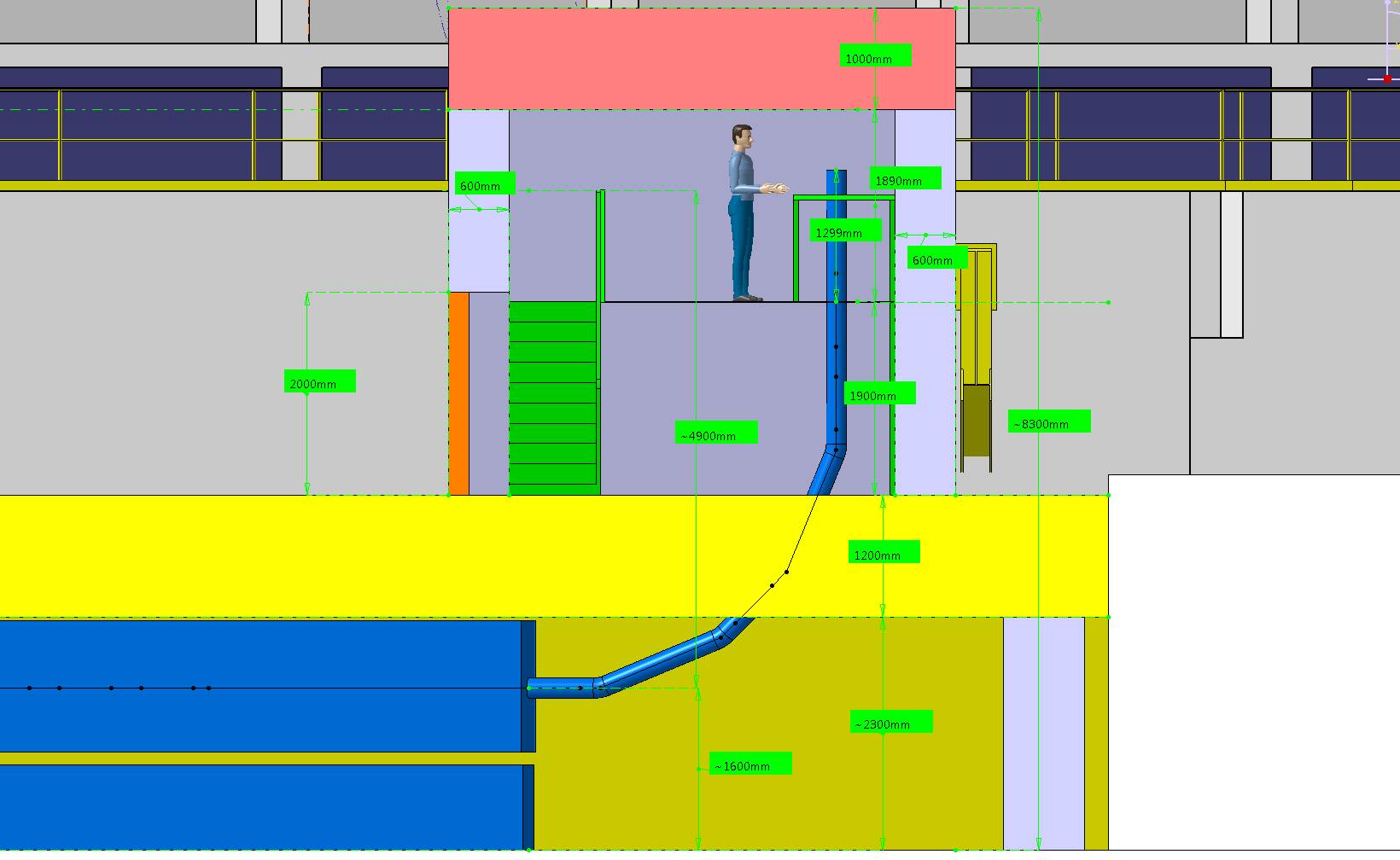}
\caption{Cross-section of the planned vertical irradiation room V.}
\label{Fig:EA_VLine}
\end{figure}

Space constraints for the vertical irradiation room are very stringent vertically, but also laterally as the mounting brackets for the rails of the overhead crane of building~150 delimit the available space. Figure~\ref{Fig:EA_VLine} shows a cross-section of the planned vertical irradiation room. In order to clear the overhead crane, the nominal shielding height of 1.6\,m with concrete needs to be reduced through the use of cast-iron.
The irradiation rooms H1 and V are temperature controlled to 37.5\,$\pm$\,0.5$^\circ$\,C and humidity-controlled to 40\,$\pm$\,10\%.

\paragraph{Automation of sample treatment:}
\label{Para:EA_ExperimentalHall_Robotization}
Irradiation times for the biological cell experiments are expected to be between 20~seconds and 8-10~minutes per sample. After irradiation, material and air at the irradiation point will be activated and access will be limited until air is flushed. In order to optimise throughput and avoid exposure of personnel, an automatic sample exchange and positioning system may be needed. 

An array of several cell samples is placed into a common sample holder which may be moved through the beam automatically. This sample tablet is expected to be $\sim$10\,$\times$\,10\,cm$^2$. A humidity and temperature controlled box (a large mobile incubator closet) is foreseen to hold up to ten such sample tablets, and to have a robotics-actionable door handle. Several times a day, the irradiation room is stocked with sample tablets to be irradiated  and the measurement programme is executed by a 6-axis custom-built robot. Ideally, the mobile incubator box containing the irradiated samples is transported towards the Biolab without the need for a person to access the irradiation room. Extensive experience for robotics applications is available within CERN's EN-STI group~\cite{DiCastro:2016}.

Inside irradiation rooms H1 and V, a robotic sample tablet placement system will be installed, with the following characteristics:
\begin{itemize}
	\item an X-Y table with mechanical adaptors for sample tablet placement with an accuracy of 1\,mm.
    \item a (re-)positioning system with an accuracy at the level of 0.1\,mm allows experiments at the cell level within the spread-out Bragg peak~\cite{Britten:2013}.
    \item a 6-axis custom-built robot that is programmed to automatically exchange the sample tablets stored in the mobile incubator box.
   \item the robot transports the mobile incubator without rail system towards the Biolab for further analysis, following a 1-hour cool-down period after irradiation. This movement needs to be done in a controlled way, avoiding any mechanical shocks and vibrations, using resilient materials.
   \item an automatic microscopic analysis of the sample tablets, still in the irradiation room would be an interesting augmentation of the facility, with subsequent more detailed microscopic studies performed in the Biolab area.
\end{itemize}

\subsubsection{Detector beamline: H2} \label{SubSubSec:_ExperimentalHall_IrradiationPoints_H2}
Irradiation room H2 is temperature-controlled to 22\,$\pm$\,1$^\circ$\,C and has a humidity below 40\%. This irradiation point has access to flushed gases typically used for propane-based detectors, as for example for GEM-detectors. A gas detection system is needed in this irradiation room.

\subsection{Biological laboratory}
\label{SubSec:EA_ExperimentalHall_BioLab}

The Biolab has a footprint of about 200\,m$^2$, shall have good lighting, air-conditioning and a floor that can be easily cleaned. It has all equipment specified in the chapter listing facility requirements (chapter~\ref{Chap:BeamParameters}), for which the procurement and acceptance is expected to be done by the biomedical community who have the corresponding specific competence.
The interior layout of the Biolab will be designed by the biomedical community.

We describe in this subsection the aspects that may have an impact on the design of the experimental area or the Biolab. It is necessary to consider:
\begin{itemize}
	\item safety closets for chemical storage (with independent air exhaust);
    \item safety closets for medical grade gas storage (with  independent air exhaust);
    \item access to purified and distilled water;
    \item access to an ice-machine;
    \item a cold-storage room with enough space for in-room work;
    \item power outlets, each with its own fuse;
    \item emergency eye shower;
    \item equipment for workbenches;
    \item sinks with three water taps (cold, warm and de-mineralized) with their corresponding drains;
    \item one or two dark rooms.
    \item fast Ethernet.
\end{itemize}

The cost of the Biolab is estimated from the recently built Biolabs at the MedAustron facility in Austria, scaled down for size~\cite{Schreiner:2016}. The estimated cost for BioLEIR Biolab is 530\,kCHF.

\subsubsection{Counting rooms}
The experimental area needs counting rooms with space for readout crates, with fast readout links and fast Ethernet access. The H2 and V irradiation rooms can share counting room C2, while irradiation room H1 uses counting room C1. The cost for two equipped counting barracks is estimated at 100\,kCHF, as detailed in table~\ref{Tab:Infrastructure_CountingRooms}.

\subsection{Control irradiator}
\label{SubSec:EA_ExperimentalHall_ControlIrradiator}
A control irradiator made of a horizontal 6\,MeV photon linac would ideally be available in close vicinity to the Biolab and the irradiation rooms. A 6\,MeV linac requires such a large shielding footprint that space constraints in hall~150 prohibit its installation at this stage of the study.

In order to still have fast access to a suitable photon or electron linac on the CERN site, albeit not in hall~150, one could investigate the feasibility of the parasitic use of the existing 9\,MeV electron linac of the GBAR experiment in building~393. 

Details on the importance of the control irradiator choice are given in appendix~\ref{App:EA_ControlIrradiator}. 

As an alternative to a linac irradiator, we plan to have a Cs-137 source in a vertical configuration inside hall~150. The Cs-137 source, for which commercially available solutions are available, is costed in this chapter, including shielding, monitoring and access controls.

\section{Resource estimate}
\label{SubSec:EA_Cost}

Tables~\ref{Tab:EA_Costing_Material} and \ref{Tab:EA_Costing_FTE} show a summary of the estimated cost (material and personnel) associated with the preparation of the BioLEIR experimental area.

\begin{table}[!htb]
\caption{Preliminary estimate of the material needed for the design, construction and furbishing of the BioLEIR experimental area. Infrastructure costs are included in infrastructure estimates (see chapter \ref{Chap:Infrastructure}).}
\label{Tab:EA_Costing_Material}
\centering
\begin{tabular}{l l c c r r}
\hline
\rule{0pt}{3ex}\textbf{System} & \textbf{Element} & \textbf{Units} & \textbf{Unit cost} & \textbf{Sub-total} & \textbf{Total} \\
\textbf{} & \textbf{} & \textbf{} & \textbf{[kCHF]} & \textbf{[kCHF]} & \textbf{[kCHF]} \\
\hline
\rule{0pt}{3ex}Robotics Stage 1  
 & \textit{Positioning table} & &  & & \\
  & H1 & 1 & 30 &  & \\
  & V & 1 & 30 &  & \\
  \textbf{Total} & & & & & \textbf{60} \\
\hline
\rule{0pt}{3ex}Robotics Stage 2 
 & \textit{Robotic sample change} & &  & & \\
  & H1 & 1 & 25 &  & \\
  & V & 1 & 25 &  & \\
  \textbf{Total}& & &  & & \textbf{50} \\
\hline
\rule{0pt}{3ex}Robotics Stage 3  
 & \textit{Telemanipulation} &  & & & \\
  & H1 & 1 & 20 &  & \\
  & V & 1 & 20 &  &\\
  \textbf{Total}& & &  & & \textbf{40} \\
\hline
\rule{0pt}{3ex}Irradiation rooms 
 & furbishings & 3 & 5 & 15 & \\
  \textbf{Total} & & & & & \textbf{15} \\
\hline
\rule{0pt}{3ex}Biolab 
 & furbishing & 200 & 2.6 & 530 & \\
 & Cs-137 source & 1 & 100 & 100 & \\
  \textbf{Total} & & & & & \textbf{630} \\
\hline
\rule{0pt}{3ex}\textbf{Total [kCHF]} & & & & & \textbf{795}\\
\hline
\end{tabular}
\end{table}

\begin{table}[ht]
\caption{Estimate for the personnel necessary for the design, construction and furbishing of the BioLEIR experimental area. Infrastructure estimates are included in the infrastructure cost (see chapter \ref{Chap:Infrastructure}).}
\label{Tab:EA_Costing_FTE}
\centering
\begin{tabular}{l r}
\hline
\rule{0pt}{3ex}\textbf{System} & \textbf{Personnel [PM]} \\
\hline
\rule{0pt}{3ex}Robotics Stage 1: Positioning table& 3 \\
Robotics Stage 2: Robotic sample change & 4 \\
Robotics Stage 3: telemanipulation  & 2 \\
Biolab & 6 \\
Coordination & 6 \\
\hline
\rule{0pt}{3ex}\textbf{Total experimental area [person-months]} & \textbf{21}\\
\hline
\end{tabular}
\end{table}

Note that the estimates are high-level engineering estimates with an uncertainty up to 30\%. A more reliable cost estimation for the experimental area requires more detailed integration and optimization work from several CERN groups (EN-CV, EN-ACE, EN-EA, EN-HE, EN-EL, etc.), as well as the biomedical community.


Table~\ref{Tab:EA_Maintenance_FTE} summarizes the estimates for required personnel for yearly operation and coordination of the experimental area. The material cost for maintenance of the area is estimated at 5\% of the initial investment cost.

\begin{table}[ht]
\caption{Estimate for personnel necessary to maintain and operate the BioLEIR experimental area. Note that infrastructure estimates are included in the infrastructure chapter (see chapter \ref{Chap:Infrastructure}).}
\label{Tab:EA_Maintenance_FTE}
\centering
\begin{tabular}{l r}
\hline
\rule{0pt}{3ex}\textbf{System} & \textbf{FTE} \\
\hline
\rule{0pt}{3ex}Robotics & 0.02 \\
Biolab & 0.5 \\
Coordination & 0.5 \\
\hline
\rule{0pt}{3ex}\textbf{Total [FTE]} & \textbf{1}\\
\hline
\end{tabular}
\end{table}

%% file: Chapters/Vacuum.tex
\chapter{Vacuum Aspects}
\label{Chap:Vacuum}

This chapter describes the vacuum system for the different machine sections of BioLEIR. In addition to the LEIR ring vacuum, four further main vacuum sections have been identified as follows:

\begin{itemize}
	\item LINAC5, including the source;
	\item Injection transfer line from LINAC5 to LEIR;
	\item Extraction transfer line;
	\item Experimental beamlines.
\end{itemize}

For all of the above lines, beam physics considerations call for an average pressure of around 10$^{-10}$\,mbar. In addition, the new extraction area shall match the 10$^{-12}$\,mbar vacuum of LEIR. Therefore, the components at the interface with LEIR have a Non-Evaporable Getter (NEG) thin film coating. 
The vacuum system includes vacuum chambers, as well as vacuum pumps, sector valves, vacuum instrumentation, vacuum controls and their interconnections. 
The design, procurement, commissioning, operation and maintenance of the vacuum system is under the responsibility of the TE-VSC group. 
In absence of a more detailed study, the cost estimate has been based on the experience from other projects, such as LINAC4, HIE-ISOLDE and ELENA. For the cost estimate, three contributions have been considered:
\begin{itemize}
	\item Material (including all the hardware, expressed in kCHF).
	\item Non-staff personnel (contractors, project associates etc., expressed in kCHF).
	\item Staff (including Fellows and expressed in FTE).
\end{itemize}

\noindent The vacuum system has interfaces with other equipment, such as:
\begin{itemize}
	\item Beam instrumentation (e.g. Beam Position Monitors BPMs);
	\item Alignment and supporting systems;
	\item Magnets, such as dipoles, quadrupoles and kickers;
	\item Safety system (e.g. high voltage for pumps and hot surfaces during bake-out).
\end{itemize}

The cost of the above systems as well as the necessary general infrastructure, is not included in the estimate for the vacuum system.

\section{Requirements}
\label{Sec:Vacuum_Requirements}
\textbf{Materials:} it has been decided to adopt austenitic stainless steel with low permeability as the material for fabrication of the TE-VSC chambers, and in particular for the flanges (AISI316 LN).  The bellows convolutions will be in stainless steel AISI 316 L.
\\
\textbf{Joints:} ConFlat-type joints are envisaged everywhere.
Vacuum firing of all stainless-steel parts is envisaged to reduce the thermal gas load contribution.
\\
\textbf{Valves:} For LINAC5, the cost estimate has been based on series 10 VAT valves as for the other CERN linacs. For the transfer lines, all metal gate valves are envisaged and the cost estimate is based on the Series 48 all-metal gate valves from VAT~\cite{VATValve}.
\\
\textbf{Gauges:} for the ultra-high vacuum in the LEIR ring, a SVT 305 gauge has been installed. 
In LINAC5 and in the different transfer lines, Pirani gauges are planned, at least one per vacuum sector. 
Movable Inverted-Magnetron Penning gauges, IKR 070, are used to monitor the pressure during bake-out, and avoid straining the filament of the SVTs, as done for LEIR. 
Due to the lack of space, only few Residual-Gas Analysers (RGA) are expected to be installed permanently. Movable RGAs are used during bake-out and NEG-activation.
\\
\textbf{Heating temperatures:} all vacuum components are expected to be bakeable at least up to 250$^\circ$C. 
\\
\textbf{Heaters:} the mechanical design of all magnets has to foresee a 10\,mm gap for bake-out equipment. 
Whenever possible, dedicated jackets with heaters are used. Ribbon heaters could also be used for complicated geometries.
\\
\textbf{NEG coating:} Non-Evaporable Getter (NEG) thin film coating is envisaged in the extraction line at the interface with the LEIR ring.
\\
\textbf{Bake-out control:} the bake-out cycle is done with movable standard bake-out units and is controlled by thermocouples that are placed at strategic locations to assure that temperatures cannot exceed safe limits.
\\
\textbf{Vacuum controls:} standard interface to existing LEIR vacuum equipment bays and machine/personnel interlocks are provided. 
\\
\textbf{Pumps:} the tight longitudinal space left between the elements of the different sections do not allow the installation of many pumps. Fixed turbo-molecular and ion pumps have been selected.
The rough-pumping stage and the pumping during bake-out is carried out by mobile turbo carts.

\section{LINAC5 vacuum section}
\label{Sec:Vacuum_Linac5}
\begin{figure}[!htb]
\centering
\includegraphics[width=0.8\columnwidth]{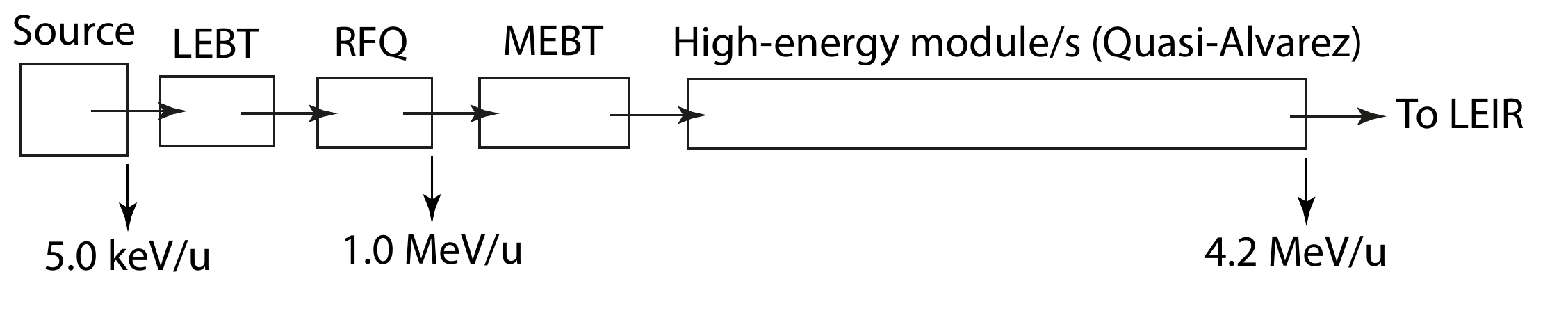}
\caption[]{LINAC5 schematic layout.}
\label{Fig:Vacuum_Linac5}
\end{figure}

LINAC5 is envisaged in the tunnel currently occupied by LINAC2 and as such uses all existing transfer lines to LEIR at minimal extra cost (see also chapter~\ref{Chap:Linac}). Once LINAC2 is decommissioned and removed, the space is available along with its existing power, cooling and access infrastructure. This has major cost saving advantages in the areas of civil engineering, transfer line component procurement and power and cooling facilities.
A preliminary basic schematic layout of LINAC5 can be seen in figure~\ref{Fig:Vacuum_Linac5}.


The main components of LINAC5 are:
\begin{itemize}
	\item The source and the Low Energy Beam Transport (LEBT);
	\item The Radio-Frequency Quadrupole (RFQ);
	\item The Medium Energy Beam Transport (MEBT);
	\item The Quasi-Alvarez linac section.
\end{itemize}

The Quasi-Alvarez linac section consists of one accelerating module with a length of about 5.2\,m (see table~\ref{Tab:Linac_Cellparams2} in chapter~\ref{Chap:Linac}).
The vacuum system is based on the layout adopted for LINAC4 (see \cite{LINAC4_Vacuum_system, LINAC4_Tech_description, LINAC4_RFQ}). From the source to the injection transfer line, LINAC5 is sectorised in vacuum sectors by all-metal gate valves. Each sector may be pumped down separately by turbo molecular pumping groups down to $10^{-6}$\,mbar and by sputter ion pump below this pressure. The pressure level of every sector is monitored with several vacuum gauges.

\subsection{Source and LEBT vacuum section}
\label{SubSec:Vacuum_Linac_Source}
Pumping of the source and LEBT is done by:
\begin{itemize}
	\item turbo-molecular pumps installed at the source and in the LEBT section;
	\item pairs of Pirani/Penning gauges to measure the total pressure, or other types of gauges depending on the functional specification of the source.
\end{itemize}
A gate valve separates the first vacuum sector from the RFQ one.

\subsection{RFQ vacuum section}
\label{SubSec:Vacuum_Linac_RFQ}
The vacuum elements for the RFQ are mainly located in the extremity segments:
\begin{itemize}
	\item sputter ion pumps (e.g. 240\,l/s);
	\item turbo-molecular pumps (e.g. 685\,l/s);
    \item pairs of Pirani/Penning gauges.
\end{itemize}
A gate valve separates the RFQ segment from the MEBT section.

\subsection{MEBT vacuum section}
\label{SubSec:Vacuum_Linac_MEBT}
The MEBT vacuum design is based on the same assumptions as the LEBT:
\begin{itemize}
	\item sputter ion pumps (e.g. 240\,l/s);
	\item pairs of Pirani/Penning gauges to measure the total pressure, or other types of gauges depending on the functional specification of the source.
\end{itemize}
A gate valve separates the MEBT vacuum part from the vacuum part of the Quasi-Alvarez linac.

\subsection{Quasi-Alvarez linac vacuum section}
\label{SubSec:Vacuum_Linac_QuasiAlvarez}
For this vacuum portion, the following vacuum components are considered:
\begin{itemize}
	\item sputter ion pumps (e.g. 240\,l/s);
	\item pairs of Pirani/Penning gauges.
\end{itemize}
A gate valve separates the Quasi-Alvarez linac vacuum section from the injection transfer line vacuum portion. \\

Drift vacuum chambers are considered between the different sections of LINAC5. 
The vacuum equipment for LINAC5 with the corresponding cost is given in table~\ref{Tab:Vacuum_Linac_Cost}. 
The number of bellows takes into account the need for the vacuum chambers themselves as well as the estimated bellows needed for beam instrumentation. 

\begin{table}[!h]
\caption{Cost estimate for LINAC5 vacuum equipment.}
\label{Tab:Vacuum_Linac_Cost}
\centering
\begin{tabular}{l r r}
\hline
\rule{0pt}{3ex}\textbf{} & \textbf{Total cost} \\
\textbf{} & \textbf{[kCHF]} \\
\hline
\rule{0pt}{3ex}Turbo-molecular pumps  & 240  \\
Sputter ion pumps & 40 \\
Pirani/Penning gauges & 20  \\
Valves & 148\\
Vacuum chambers & 16 \\
Bellows & 32\\
\hline
\rule{0pt}{3ex}\textbf{Total kCHF} &\textbf{496}\\
\hline
\end{tabular}
\end{table}

\section{Injection transfer line}
\label{Sec:Vacuum_InjectionTransferline}
The ions from LINAC5 need to be transferred into LEIR, using the existing infrastructure as much as possible. The possibility to inject protons into LEAR from LINAC2 existed in the past, using the E0 line, linking the LT and ITE line at the level of the 180$^\circ$ turnaround loop. The support and vacuum chamber elements of this short E0 section have been removed during the End-of-Year-Extended-Technical-Stop (EYETS) 2016-2017 \cite{Removal_E0}, and are stored in good condition for potential reuse by BioLEIR. One of the dipoles (LT.BHZ25) is found to be quite activated at 2.5\,mS/h at contact. Since a larger aperture dipole is needed for BioLEIR, and considering that no spare is available for that design, a new dipole and its vacuum chamber is needed. The other magnets' condition shall be assessed in the next stage of the project.  The power converters and cabling for the magnets in this section of the E0 line need to be supplied. These magnets comprise three quadrupoles of MQNAI type, and one geometric dipole of MDX type (see also chapter~\ref{Chap:Transferline}). 

A part of the section of the LT line, between the LINAC4 connection and the switch to
the LTE loop (between BHZ20 and BHZ25, inclusive), shall be upgraded: two geometric horizontal dipoles are concerned (LT.BHZ20 and LT.BHZ25), four quadrupoles, as well as their power converters and controls. 

\begin{figure}[ht!]
    \centering
    \begin{subfigure}{0.5\linewidth}
      \includegraphics[trim={4cm 0cm 0cm 0cm},clip, width=1.4\linewidth]{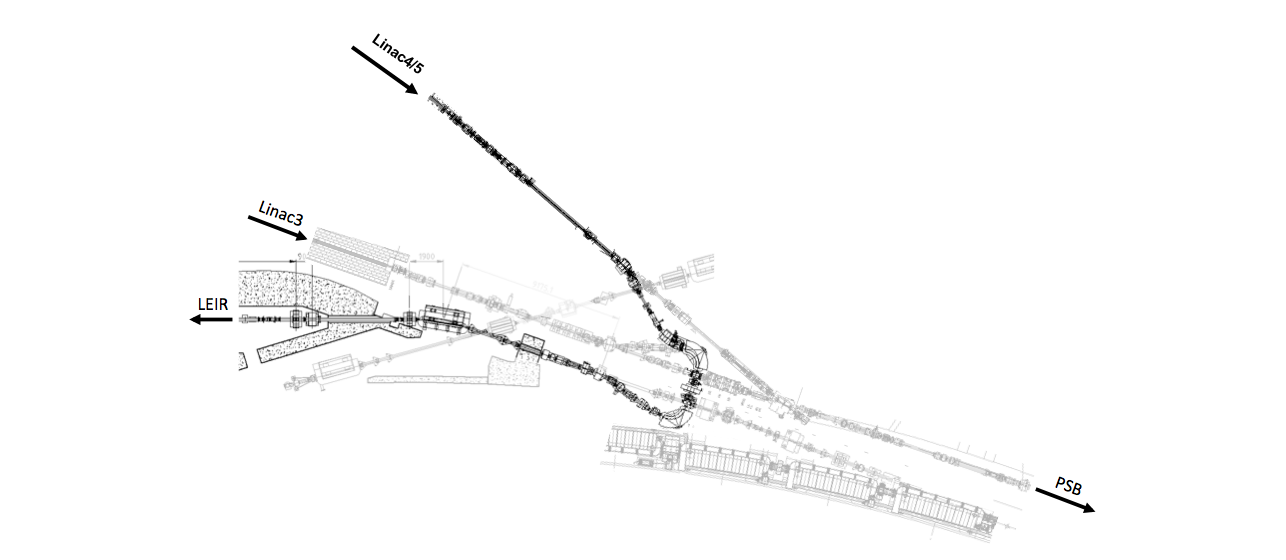}{}

        \caption{}
    \end{subfigure}
    \hskip3em
    \begin{subfigure}{0.4\linewidth}
    \includegraphics[trim={0cm 0cm 0cm 0.1cm},clip,width=1.1\linewidth]{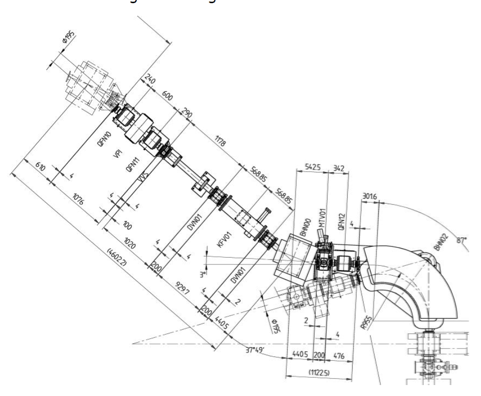}{}
    \caption{}
    \end{subfigure}
\caption{The injection transfer line from LINAC5 to LEIR (a) and a close-up of the E0 line (b).}
\label{Fig:Vacuum_InjectionTransferlines}
\end{figure}

The overall configuration of the line which would be used for injection of light ions from LINAC5 into LEIR for BioLEIR is shown in figure~\ref{Fig:Vacuum_InjectionTransferlines}(a). 
The layout of the E0 line and the junction with the LTE line is shown in figure~\ref{Fig:Vacuum_InjectionTransferlines}(b).

    


For the injection transfer line the following vacuum components are needed:
\begin{itemize}
	\item turbo-molecular pumps (e.g. 685~\,l/s);
	\item sputter ion pumps (e.g. 240\,l/s);
	\item pairs of Pirani/Penning gauges.
\end{itemize}

A gate valve separates the injection transfer line from the LEIR ring.
Vacuum chambers are needed for bending magnets, quadrupoles and for the injection into LEIR. In addition, the vacuum chamber at the interface to LEIR needs to be adapted or replaced. 
The vacuum equipment for the injection transfer line with the corresponding cost is given in table~\ref{Tab:Vacuum_Transferline_Cost}. 

\begin{table}[!ht]
\centering
\caption{Cost estimate for the injection transfer line vacuum equipment.}
\label{Tab:Vacuum_Transferline_Cost}
\begin{tabular}{l r r}
\hline
\rule{0pt}{3ex}\textbf{} & \textbf{Total cost} \\
\textbf{} & \textbf{[kCHF]} \\
\hline
\rule{0pt}{3ex}Turbo-molecular pumps & 60  \\
Sputter ion pumps  & 35  \\
Pirani/Penning gauges & 17.5 \\
Valves & 69\\
Vacuum chambers & 180  \\
Bellows & 40  \\
\hline
\rule{0pt}{3ex}\textbf{Total} &\textbf{401.5}\\
\hline
\end{tabular}
\end{table}

\section{LEIR}
\label{Sec:Vacuum_LEIR}
\begin{figure}[!hb]
\centering
\includegraphics[width=0.8\textwidth]{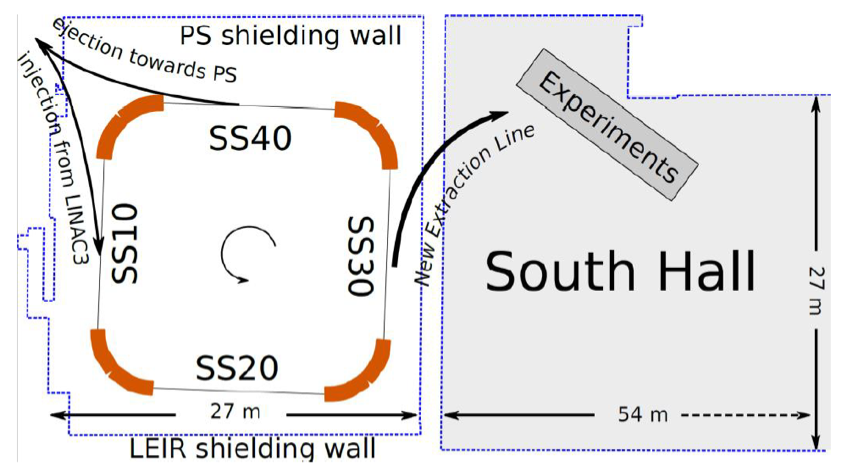}
\caption{Schematic view of LEIR.}
\label{Fig:Vacuum_LEIRSchematics}
\end{figure}

The LEIR synchrotron has a circumference of around 80\,m and is composed of four 13\,m-long straight sections (SS) numbered 10 to 40 and four 90$^\degree$ bending magnets, as shown in figure~\ref{Fig:Vacuum_LEIRSchematics}.
The first straight sections (SS10) is used for injection, the second one (SS20) houses the electron cooler, the third one (SS30) accommodates kicker magnets for fast extraction and the fourth one (SS40) houses the RF cavities and the extraction septum towards the Proton Synchrotron (see figure~\ref{Fig:Vacuum_LEIRSchematics}).
Some of the LEIR components have to be adapted to the injection and the extraction sections. Gate valves separate the LEIR ring from the injection and extraction sections. Two additional pairs of Pirani/Penning gauges are needed. 

The new vacuum equipment for LEIR with the corresponding cost is detailed in table~\ref{Tab:Vacuum_LEIR_Cost}. 

\begin{table}[!ht]
\centering
\caption{Cost estimate of additional LEIR vacuum equipment.}
\label{Tab:Vacuum_LEIR_Cost}
\begin{tabular}{l r r}
\hline
\rule{0pt}{3ex}\textbf{} & \textbf{Total cost} \\
\textbf{} & \textbf{[kCHF]} \\
\hline
\rule{0pt}{3ex}Pirani/Penning gauges & 5 \\
Valves & 74 \\
Vacuum chamber modifications & 25  \\
\hline
\rule{0pt}{3ex}\textbf{Total} &\textbf{104}\\
\hline
\end{tabular}
\end{table}

\section{Extraction transfer lines}
\label{Sec:Vacuum_Extraction}
\begin{figure}[!hb]
\centering
\includegraphics[width=0.8\textwidth]{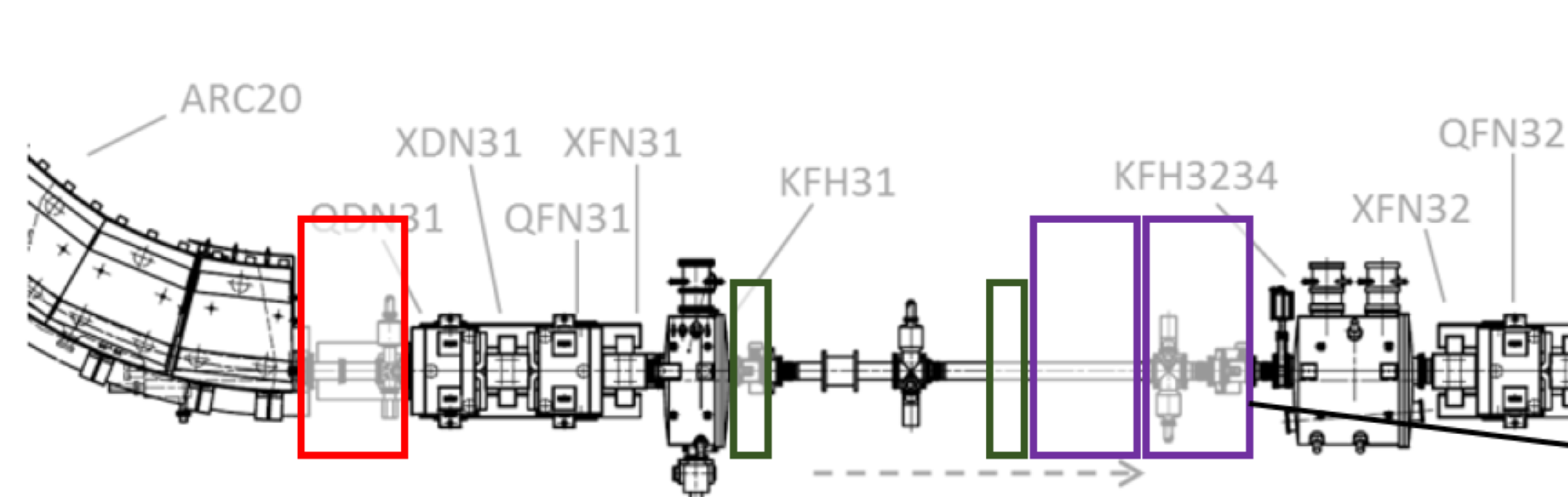}
\caption{Extraction area of LEIR into the BioLEIR experimental beamlines\cite{Garonna:1703468}. The rectangles identify the areas of LEIR where the vacuum equipment is modified.}
\label{Fig:Vacuum_LEIR_extraction_area}
\end{figure}

New electrostatic and magnetic septa are needed in SS30 to accomplish the extraction process (see figure~\ref{Fig:Vacuum_LEIR_extraction_area}). Some changes to the machine vacuum chambers are required to increase aperture in specific locations to accept the extracted beam.
A special extraction vacuum chamber (Y-chamber) needs to be designed and built together with a spare. This vacuum chamber shall incorporate the septum coil fixation. The vacuum chamber is expected to be bake-able up to 300$^\circ$C and supported via the septum magnet support structures. Vacuum chambers are also needed for the dipole and quadrupole magnets. 

The vacuum equipment for the extraction line with the corresponding cost is given in table~\ref{Tab:Vacuum_Extraction_Cost}. 
The number of bellows takes into account the need for the vacuum chambers and assumes a reasonable number for the beam instrumentation whose detailed design is not finalized at this stage. 

\begin{table}[!ht]
\centering
\caption{Cost estimate of the extraction line vacuum equipment.}
\label{Tab:Vacuum_Extraction_Cost}
\begin{tabular}{l r}
\hline
\rule{0pt}{3ex}\textbf{} & \textbf{Total cost} \\
\textbf{} & \textbf{[kCHF]} \\
\hline
\rule{0pt}{3ex}Turbo-molecular pumps & 60  \\
Sputter ion pumps & 25 \\
Pirani/Penning gauges & 10 \\
Valves & 63  \\
Vacuum chambers & 239  \\
Bellows & 40  \\
\hline
\rule{0pt}{3ex}\textbf{Total} &\textbf{437}\\
\hline
\end{tabular}
\end{table}

\section{Beamlines to the experimental area}
\label{Sec:Vacuum_Beamlines}
\begin{figure}[!hb]
\centering
\includegraphics[width=1.0\textwidth]{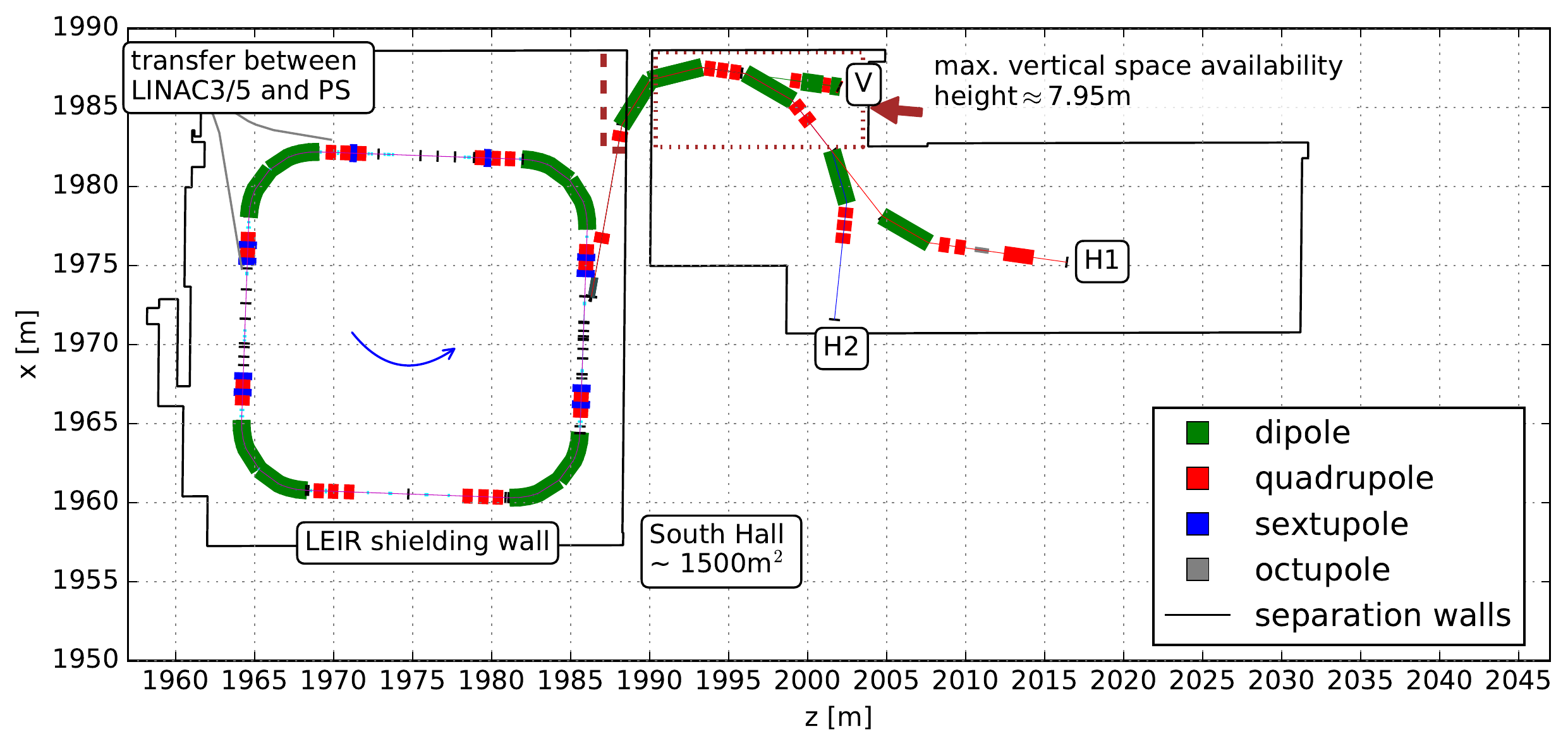}
\caption{Layout of LEIR and the three experimental lines of BioLEIR.}
\label{Fig:Vacuum_Beamlines_bioleir}
\end{figure}

In Stage 1 of BioLEIR, two horizontal and one vertical experimental beamlines are foreseen (see chapter~\ref{Chap:BeamLines}). The
beamline minimum full aperture is essentially 80\,mm at all magnetic elements, to accommodate the expected beam envelopes. The layout of the beamlines is shown schematically in figure~\ref{Fig:Vacuum_Beamlines_bioleir}. The common horizontal beamline has a 178$^\degree$ bending angle, accomplished with three identical dipoles, with a quadrupole triplet between the second and third dipoles, and an additional separated dipole for each beamline. The fourth dipole in the H1 beamline is inverted, as shown in figure~\ref{Fig:Vacuum_Beamlines_bioleir}. The vertical beamline has a 90$^\degree$ bend, using a split 45$^\degree$ dipole pair with an interposed quadrupole. For the vertical beamline, a single quadrupole is used to make the bend achromatic, while for the horizontal beamlines a quadrupole triplet is used between the second and third dipoles in the common part of the beamline to cancel the horizontal dispersion coming from the LEIR ring. The third and fourth dipoles in the H2 beamline are planned to be equipped with 'Y' chambers to enable switching between the different experimental stations. Overall there are about 65\,m of beamline to construct.

The resulting dipoles for the horizontal beamlines are 3.415\,m long (magnetic length), whereas the dipoles for the vertial beamline are 1.335\,m long. The quadrupoles are assumed to be 0.517\,m long. Seven dipoles, twenty quadrupoles and two octupoles are needed for the three beamlines (see chapter~\ref{Chap:BeamLines} for more details).\\

\begin{figure}[!hb]
\centering
\includegraphics[width=1.2\textwidth]{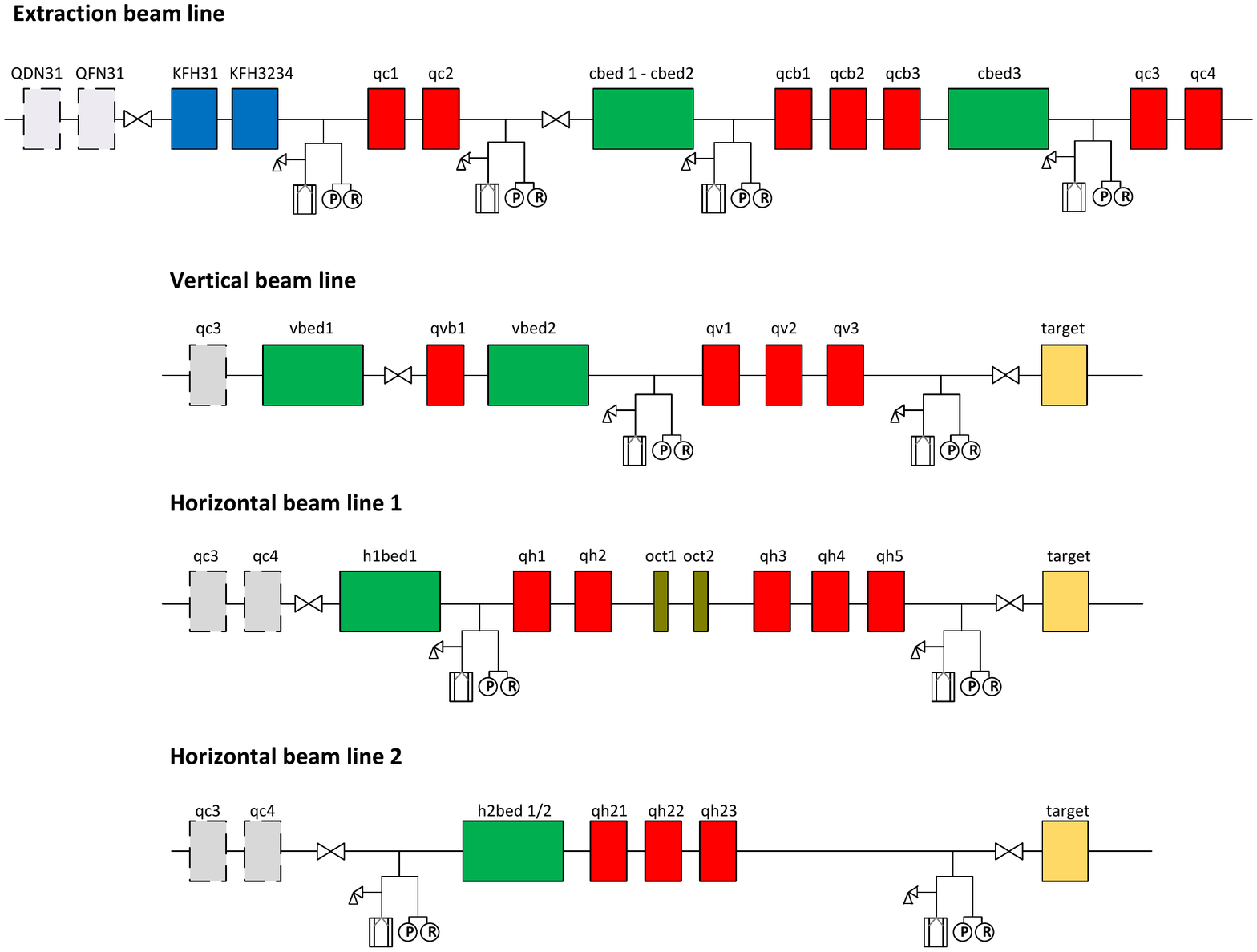}
\caption{Vacuum system for the BioLEIR extraction and beamlines.}
\label{Fig:Vacuum_Beamlines}
\end{figure}

The instrumentation for the beamlines has not been specified in detail. The vacuum system for the beamlines needs to ensure the required pressure of 10$^{-10}$\,mbar through the lines and to ensure the excellent vacuum of better than 4$\times$10$^{-12}$\,mbar N$_2$ equivalent achieved in the LEIR machine. The quadrupole, steerer and drift chambers are planned circular with all-metal seals. The dipole chambers are more complicated and can possibly be made of several straight segments welded in a pseudo-arc to facilitate construction.
\\
The vacuum chambers of the beamlines are expected to be made from austenitic stainless steel. To avoid magnetic field perturbations, AISI 316L without longitudinal welds can be used for the quadrupole and steerer circular chambers, while AISI 316LN can be used for the bending dipole chamber. \\
The vacuum system has been based on the preliminary optics design. The schematic layout of the vacuum system is given in figure~\ref{Fig:Vacuum_Beamlines}. The quantity of vacuum equipment for the beamlines with the corresponding cost is given in table~\ref{Tab:Vacuum_Beamlines_Cost}. 

\begin{table}[!ht]
\centering
\caption{Cost estimate for beamlines vacuum equipment.}
\label{Tab:Vacuum_Beamlines_Cost}
\begin{tabular}{l r}
\hline
\rule{0pt}{3ex}\textbf{} & \textbf{Total cost} \\
\textbf{} & \textbf{[kCHF]} \\
\hline
\rule{0pt}{3ex}Turbo-molecular pumps & 60\\
Sputter ion pumps & 30 \\
Pirani/Penning gauges & 15 \\
Valves & 67\\
Vacuum chambers & 447\\
Bellows & 60\\
\hline
\rule{0pt}{3ex}\textbf{Total} &\textbf{679}\\
\hline
\end{tabular}
\end{table}

\section{Controls and interlocks}
\label{Sec:Vacuum_Controls}
The cost for the vacuum control system has been estimated from the number of components in the different sections. The estimate includes contributions for the hardware (material), cabling as well as for installation and commissioning. The material contribution takes into account valves, gauges, interlocks, crates and pumps. 
Table~\ref{Tab:Vacuum_Controls_Cost} shows the overall vacuum control and interlock system cost for LINAC5 and for all transfer lines (injection, extraction and beamlines). In absence of a detailed integration, it is assumed that the racks are located in building 152. 

\begin{table}[!ht]
\centering
\caption{Cost estimate for vacuum system controls and interlocks.}
\label{Tab:Vacuum_Controls_Cost}
\begin{tabular}{l r r }
\hline
\rule{0pt}{3ex}\textbf{} & \textbf{Material \& Cabling} & \textbf{Installation \& Commissioning} \\
\textbf{} & \textbf{[kCHF]} & \textbf{[kCHF]} \\
\hline
\rule{0pt}{3ex}LINAC5 & 219  & 80    \\
Transfer lines & 158  & 70    \\
\hline
\rule{0pt}{3ex}\textbf{Total} & \textbf{377} & \textbf{150}\\
\hline
\end{tabular}
\end{table}

\section{Surface treatment}
\label{Sec:Vacuum_Surface Tests}
Surface treatment is foreseen for vacuum chambers (more than 60) as well as for non-vacuum components. It is estimated that 0.5\,person-years are needed for this work. The cost for NEG coating has been considered only for a few elements in the extraction line at the interface with the LEIR ring. The total estimated cost for surface treatment material therefore amounts to 75\,kCHF.

\section{Acceptance tests}
\label{Sec:Vacuum_Acceptance Tests}
Acceptance tests are foreseen for vacuum chambers (more than 60) as well as for non-vacuum components. 0.5\,person-years and 75\,kCHF are estimated for acceptance tests.

\section{Bake-out equipment}
\label{Sec:Vacuum_Bakeout equipment}
Bake-out equipment is mainly needed for vacuum chambers, ion pumps, valves and for other non-vacuum components such as magnets. In the absence of a detailed design, the cost estimate has been based on the recent experience of the ELENA project \cite{ELENA2015}. The overall estimated bake-out cost amounts to 300\,kCHF for the materiel and to 0.5\,person-years.  

\section{Installation and vacuum commissioning}
\label{Sec:Vacuum_Bakeout}
The resources needed for the installation are estimated based on the following assumptions:
\begin{itemize}
	\item 2019-20: installation of the extraction line and all three beamlines.
	\item 2021: running with LINAC3.
	\item 2021: dismantling of LINAC2.
	\item 2022: installation of LINAC5.
	\item 2022-23 (YETS): connection of LINAC5.
\end{itemize}
The summary of the estimated installation cost is shown in table~\ref{Tab:Installation_Cost}.

\begin{table}[!ht]
\centering
\caption{Cost estimate for vacuum installation and commissioning.}
\label{Tab:Installation_Cost}
\resizebox{\textwidth}{!}{%
\begin{tabular}{l r r r}
\hline
\rule{0pt}{3ex}\textbf{} & \textbf{Hardware} & \textbf{Non-staff workforce} & \textbf{Staff} \\
\textbf{} & \textbf{[kCHF]} & \textbf{[kCHF]} & \textbf{[Person-Year]} \\
\hline
\rule{0pt}{3ex}Installation of the extraction line and beamlines & 30  & 120  & 1  \\
Dismantling of LINAC2  & 20  & 40  & 0.2  \\
Installation of LINAC5 & 20 & 120 & 0.7 \\
Vacuum commissioning (Stage 1) & 25 & 25 & 0.35 \\
Vacuum commissioning (Stage 2) & 25 & 25 & 0.35 \\
\hline
\rule{0pt}{3ex}\textbf{Total} & \textbf{120} & \textbf{330} & \textbf{2.6}\\
\hline
\end{tabular}
}
\end{table}

\section{Resource estimate}
\label{Sec:Vacuum_Cost}
The cost estimate refers to the full project cost (from Stage 2 onwards). Therefore, it includes the extraction line and the three beamlines, as well as LINAC5.  
The cost estimates take into account all project phases from the design to the commissioning:
\begin{itemize}
	\item Engineering design.
	\item Procurement and fabrication.
    \item Controls.
	\item Surface treatment.
	\item Acceptance tests.
    \item Bake-out equipment.
	\item Installation and vacuum commissioning.
	\item Operation.
\end{itemize}

\vspace{-2mm}
The vacuum design for LINAC5 has been based on the LINAC4 design. As the detailed conceptual design and the related integration drawings are not available at this time, the estimates have been based on the quantities listed in tables~\ref{Tab:Vacuum_Linac_Cost}, \ref{Tab:Vacuum_Transferline_Cost}, \ref{Tab:Vacuum_LEIR_Cost}, \ref{Tab:Vacuum_Extraction_Cost} and \ref{Tab:Vacuum_Beamlines_Cost}. The cost includes hardware and personnel (CERN Staff and industrial support and collaborators) and is summarized in table~\ref{Tab:Vacuum_Cost_Total_1}. CERN Staff estimates include Fellows. The uncertainty of the estimate is at the level of 30-40\%.

\vspace*{-3mm}
\begin{table}[!ht]
\centering
\caption{Cost estimate of the BioLEIR vacuum system.}
\label{Tab:Vacuum_Cost_Total_1}
\resizebox{\textwidth}{!}{%
\begin{tabular}{l r r r}
\hline
\rule{0pt}{3ex}\textbf{} & \multicolumn{1}{c}{\textbf{Hardware [kCHF]}} & \multicolumn{1}{c}{\textbf{Non-staff workforce [kCHF]}} & \multicolumn{1}{c}{\textbf{Staff [PY]}} \\
\hline
\rule{0pt}{3ex}Engineering and vacuum design &90  &  &0.5  \\
Procurement and fabrication &2118 & &0.5 \\
Controls &377  &150  &3.5  \\
Surface treatment &75  &  &0.5  \\
Acceptance tests &75  &  &0.5  \\
Bake-out equipment &250  &50  &0.5  \\
Installation \& commissioning & 120 & 330 & 2.6 \\
\hline
\rule{0pt}{3ex}\textbf{Total} &\textbf{3104} &\textbf{530} &\textbf{8.6}\\
\hline
\rule{0pt}{3ex}Maintenance \& Operations (per year) &35  &  &0.5 \\
\hline
\end{tabular}
}
\end{table}

Table~\ref{Tab:Vacuum_Cost_Total} shows the resources mapped to Stages 1 and 2. Stage 3 does not require any additional vacuum system components.

\begin{table}[!ht]
\centering
\caption{Mapping of the estimated vacuum system cost to facility stages.}
\label{Tab:Vacuum_Cost_Total}
\resizebox{\textwidth}{!}{%
\begin{tabular}{l | r r r r r r}
\hline
\rule{0pt}{3ex}\textbf{} & \multicolumn{2}{c}{\textbf{Hardware [kCHF]}} & \multicolumn{2}{c}{\textbf{Non-staff Labor [kCHF]}} & \multicolumn{2}{c}{\textbf{Staff [PY]}} \\
\rule{0pt}{3ex}\textbf{} & \textbf{Stage 1} & \textbf{Stage 2} & \textbf{Stage 1} & \textbf{Stage 2} & \textbf{Stage 1} & \textbf{Stage 2}\\
\hline
\rule{0pt}{3ex}Engineering Design & 60 & 30 & - & - & 0.33 & 0.17 \\
Procurement and fabrication & 1622 & 496 & - & - & 0.33 & 0.17 \\
Controls & 158 & 219 & 70 & 80 & 1.75 & 1.75 \\
Surface Treatment & 50 & 25 & - & - & 0.33 & 0.17\\
Acceptance Tests & 50 & 25 & - & - & 0.33 & 0.17\\
Bake-out  & 167 & 83 & 33 & 17 & 0.33 & 0.17 \\
Installation \& commissioning & 75 & 45 & 185 & 145 & 1.55 & 1.05\\
\hline
\rule{0pt}{3ex}\textbf{Total} &\textbf{2181} &\textbf{923} &\textbf{288} &\textbf{242} &\textbf{4.97} &\textbf{3.63}\\
\hline
\end{tabular}
}
\end{table}

%% file: Chapters/Infrastructure.tex
\chapter{Infrastructure and Integration}
\label{Chap:Infrastructure}
This chapter discusses all aspects of the global infrastructure, integration, general services and civil engineering needed for BioLEIR, based on the biomedical requirements and the description of the experimental set-up from chapter~\ref{Chap:EA}. In particular, the infrastructure depends on: 
\begin{enumerate}
\item the biomedical experimental area; 
\item radiation protection: aspects that need to be studied and designed include shielding of the three irradiation rooms, of the access areas to the irradiation points, of the BioLEIR switchyard areas 1 and 2, of the extraction area, and finally a shielding roof above the LEIR accelerator; 
\item safety as input for structural elements; 
\item access control: access to the three irradiation points shall ideally be managed independently; 
\item CV and EL needs for the light ion injector, the transfer line from the frontend to LEIR, the slow extraction system, and the beamlines to the experimental area.
\end{enumerate}

The infrastructure has an impact on the biomedical experimental area, control systems, access systems and safety requirements.

\section{The experimental hall}
\label{Sec:Infrastructure_Hall150}
\begin{figure}[!hb]
\centering
\includegraphics[width=0.95\textwidth]{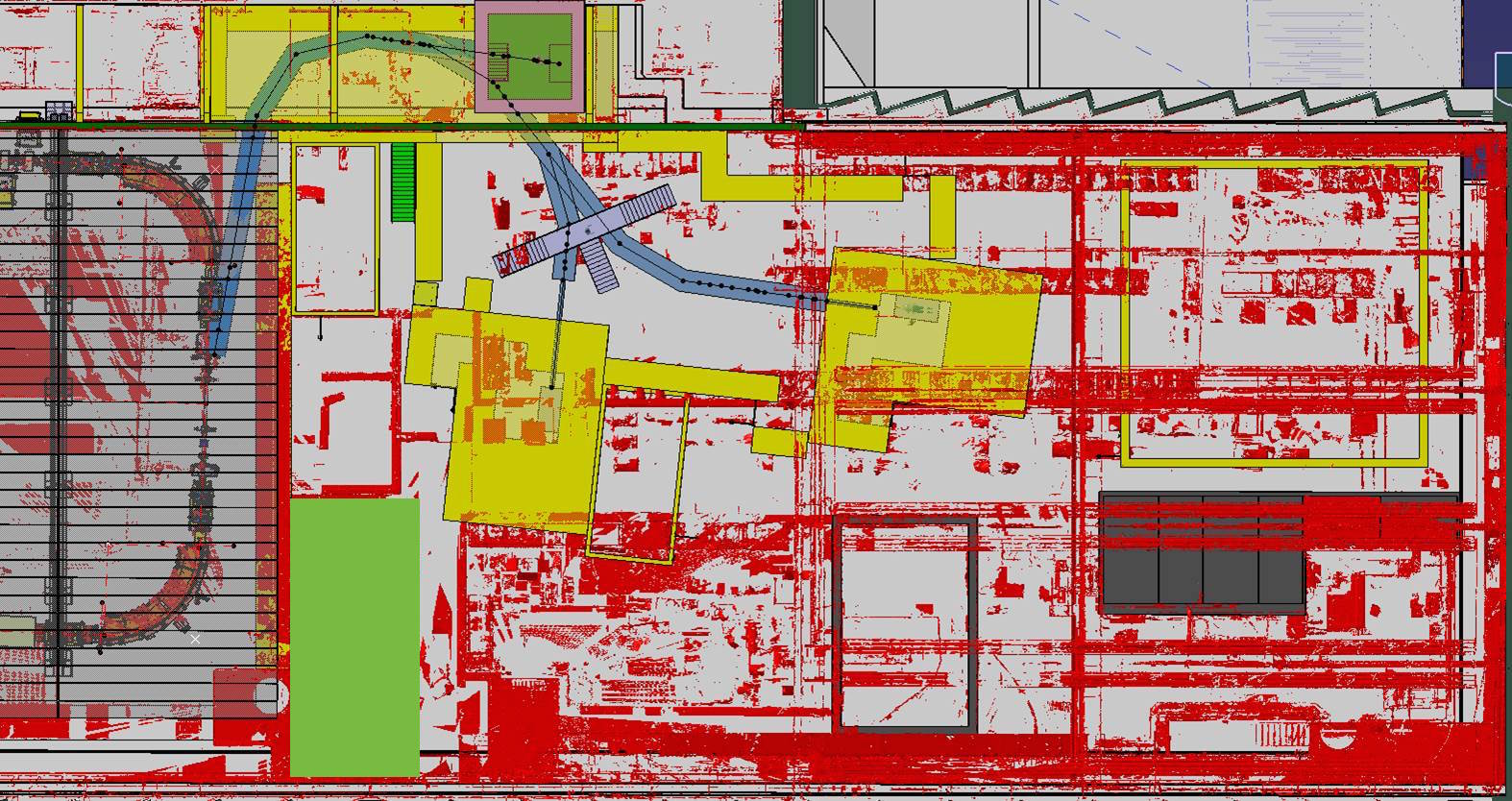}
\caption{A 3D scan of hall~150, in red, shows mostly stored material. The loading dock zone for the crane is indicated in green and the Reasearch and Development areas (XBOX and EBIS) are shown in grey. The layout of the new experimental area for the BioLEIR facility is superimposed in yellow.}
\label{Fig:EA_Scan}
\end{figure}

The hall of building~150 has been used as an experimental area in the past when antiproton experimental lines were fed by LEAR between 1983 and 1997. Currently, most of hall~150 is used as storage area for slightly radioactive material, as well as experimental equipment. A fraction of the hall is used for ongoing Research and Development on accelerators (XBOX, EBIS). Figure~\ref{Fig:EA_Scan} shows a 3D scan of hall~150 with the new beamlines and horizontal irradiation rooms superimposed. Various service lines are still routed in the hall, yet their condition has not been verified and it is likely that they no longer satisfy current safety requirements or have reached the end of their lifetime. The refurbishment of building 150 itself has been excluded from the estimates given in this document. 

\begin{figure}[!h]
\centering\includegraphics[width=0.9\linewidth]{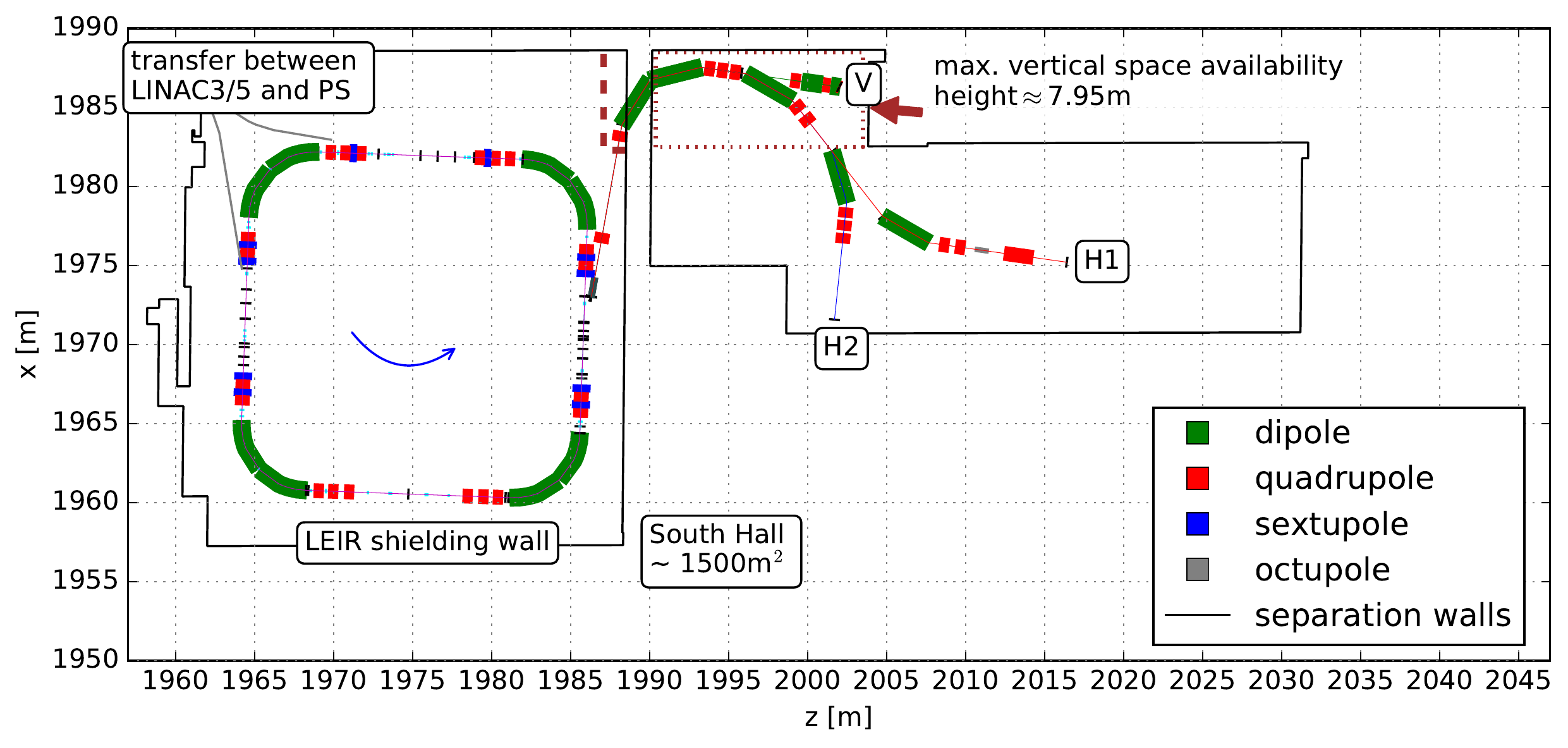}
\caption{Layout of LEIR and the experimental beamlines. The dotted, red rectangle indicates the area in hall~150 where the maximum vertical space under the two cranes is available.}
\label{Fig:Infrastructure_Hall_Beamlines}
\end{figure}

Figure~\ref{Fig:Infrastructure_Hall_Beamlines} shows the expected physical layout of the three experimental beamlines (i.e. Horizontal 1 (H1), Horizontal 2 (H2) and Vertical (V)) within hall~150.

\begin{figure}[!h]
\centering
\includegraphics[width=0.8\linewidth]{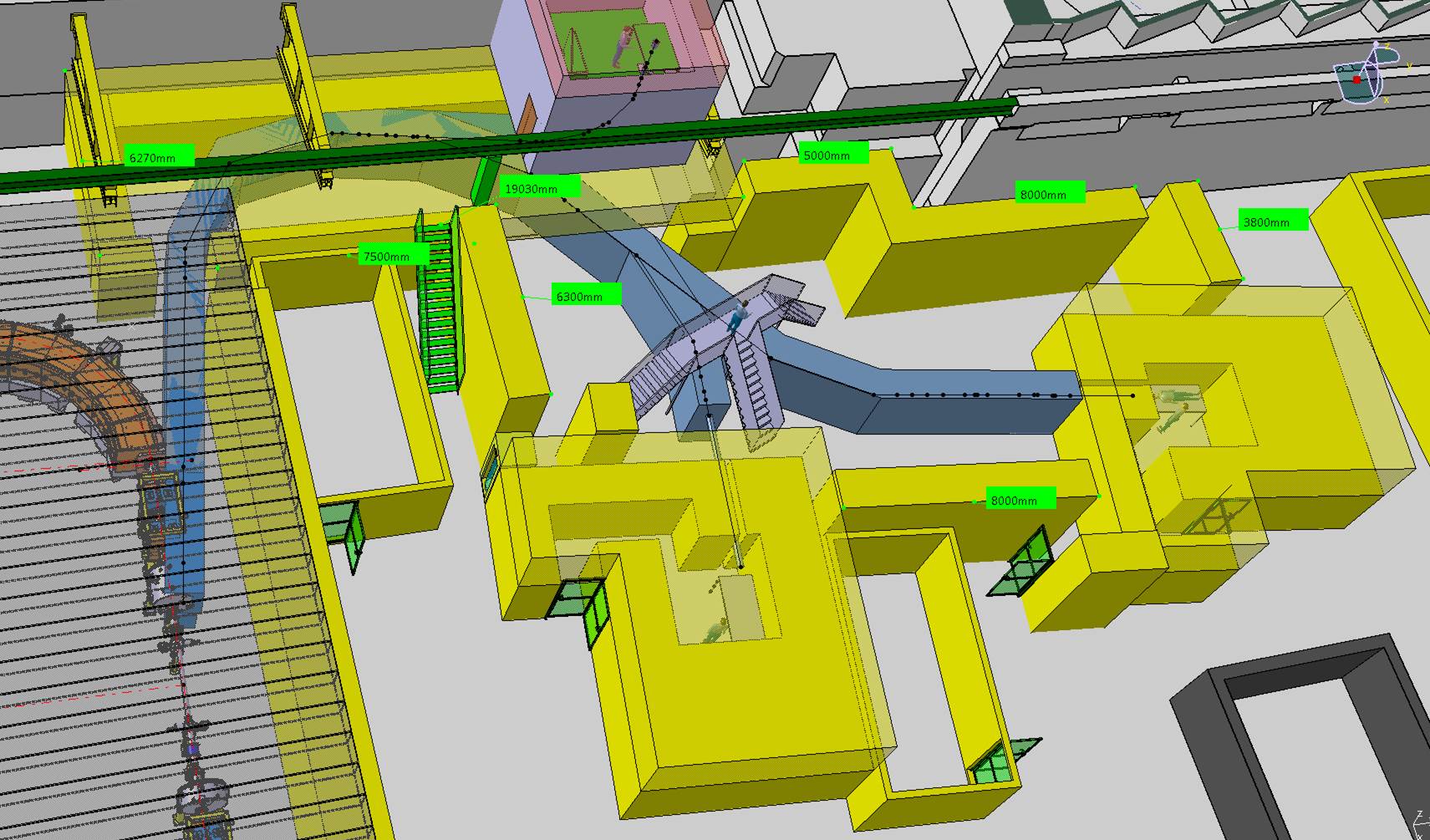}
\caption{Layout of the three beamlines and the corresponding irradiation rooms with their respective preliminary shielding designs.}
\label{Fig:Infrastructure_ExpArea}
\end{figure}

\begin{figure}[!h]
\centering
\includegraphics[width=0.8\linewidth]{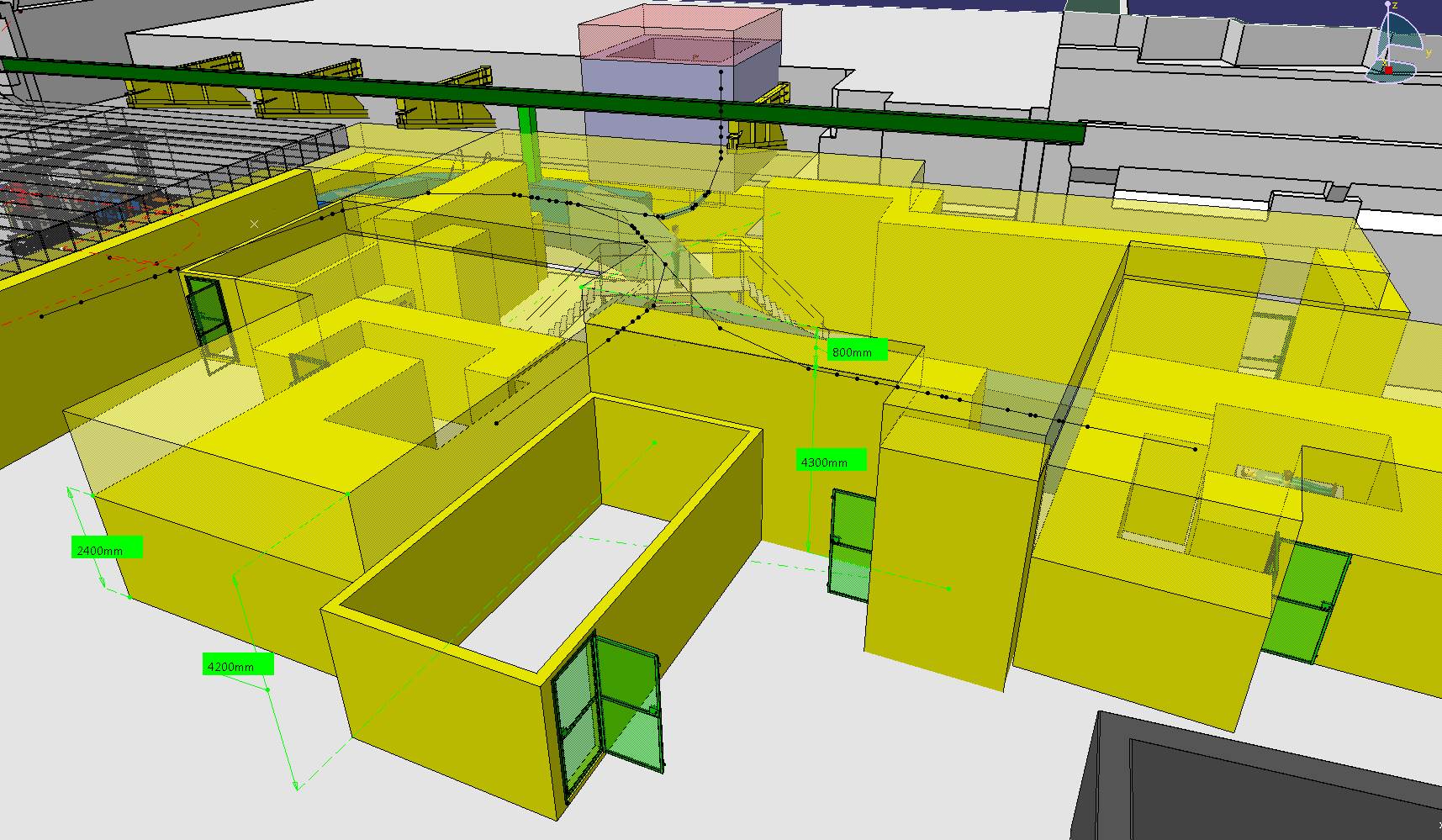}
\caption{Preliminary layout of the shielding in the experimental area.}
\label{Fig:Infrastructure_Shielding}
\end{figure}

\begin{figure}[!h]
\centering
\includegraphics[width=0.8\linewidth]{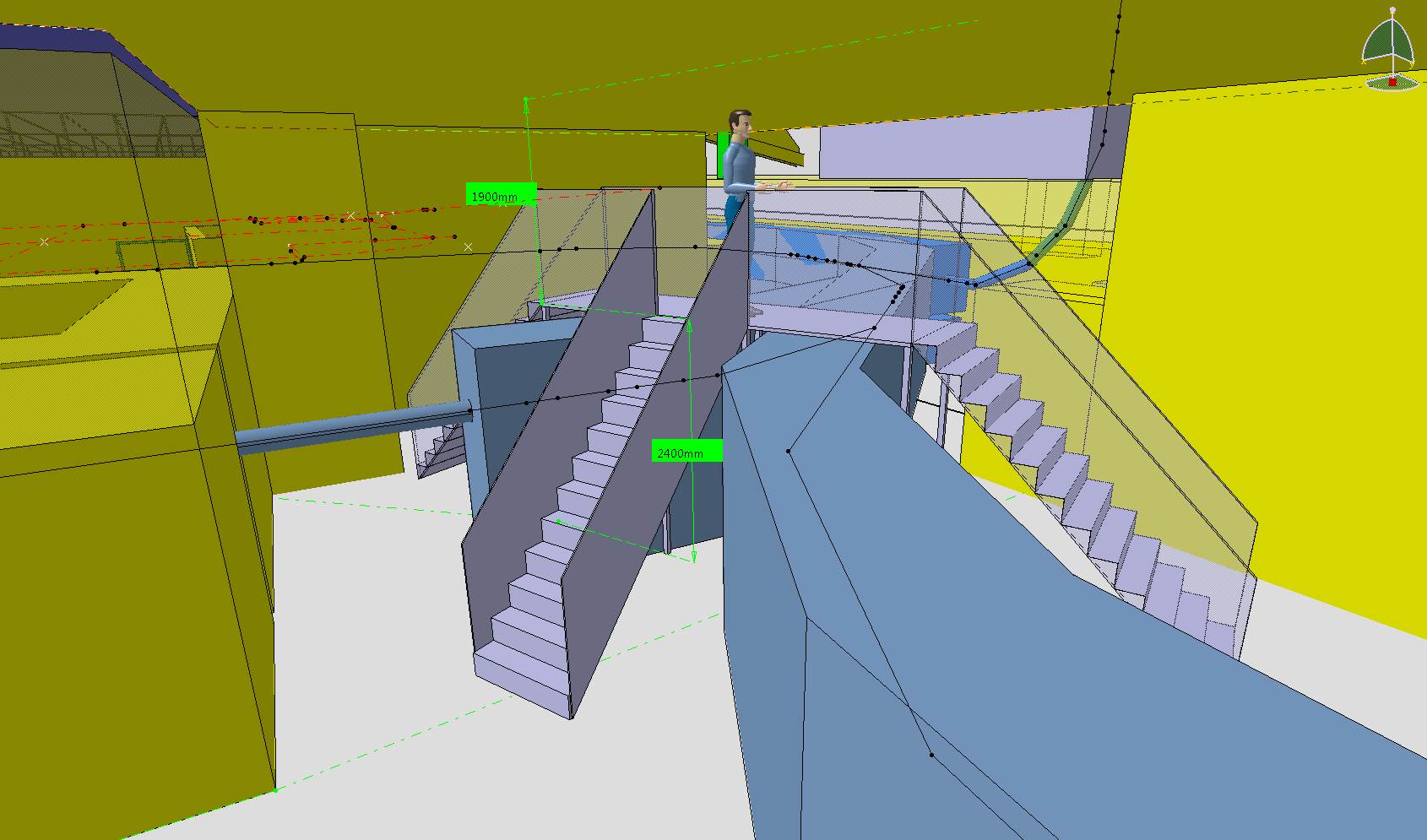}
\caption{Preliminary layout inside Switchyard 2 within the experimental area.}
\label{Fig:Infrastructure_Switchyard2}
\end{figure}

The proposed full layout of the experimental facility is described in more detail in section~\ref{Sec:EA_ExperimentalHall}. In this chapter, aspects pertaining to infrastructure are discussed. Figure~\ref{Fig:Infrastructure_ExpArea} shows the preliminary integration design relevant to infrastructure estimations. Figure~\ref{Fig:Infrastructure_Shielding} shows the substantial dimensions of the necessary shielding for the irradiation rooms, as well as the switchyard areas, whereas figure~\ref{Fig:Infrastructure_Switchyard2} shows the spatial situation within Switchyard 2.\\

The following aspects are considered in the integration process for the BioLEIR experimental area:
\vspace*{-0.5cm}
\begin{itemize}
   \item the proposed layout of the experimental area;
   \item the footprints for the different areas and their specific requirements;
   \item available space in hall 150;
   \item the clearance needed for the loading dock in hall~150;
   \item the clearance needed for the 35\,t crane of hall~150 (serving LEIR and hall~150);
   \item the clearance needed for the 60\,t crane of hall~151 (serving the PS complex);
   \item the requirements from the experimental area:
   \begin{itemize}
      \item electrical requirements (lights, low voltage distribution, mono- and three phase sockets distributed in hall 150);
      \item cooling and ventilation requirements;
	  \item fluids (water etc.);
      \item gas network requirements;
      \item safety;
      \item Radiation Protection;
      \item access control design;
      \item IT-requirements.
   \end{itemize}
\end{itemize}

Table~\ref{Tab:Infrastructure_Hall_Clearances} gives a summary of clearances that need to be respected for crane operation in halls~150 and 151.

\begin{table}[ht]
\centering
\caption{Summary of clearances to be respected for crane operations in halls~150 and 151.}
\label{Tab:Infrastructure_Hall_Clearances}
\begin{tabular}{l c c c}
\hline
\rule{0pt}{3ex}\textbf{Hall} & \textbf{Max structure height} & \textbf{Lateral rail clearance} & \textbf{Relevant for irradiation point(s)}\\
\textbf{} & \textbf{[mm]} & \textbf{[mm]} & \textbf{}\\
\hline
\rule{0pt}{3ex}150  & 5050 & 270 & H1, H2 \\
151  & 7950 & 270 & V\\
\hline
\end{tabular}
\end{table}

\section{Cooling and ventilation}
\label{Sec:Infrastructure_CV}
This section deals with aspects of the heating and ventilation system, as well as of the fluid supply for the whole of the BioLEIR facility, from the source to the experimental area.

\subsection{Demineralised water cooling}
\label{SubSec:Infrastructure_CV_DeminearlizedWater}

\subsubsection{Beamlines in the South Hall}
One demineralised water cooling circuit needs to be installed to serve all three beamlines from LEIR. The total heat load dissipated in water is estimated to be 130\,kW per beamline. The demineralised water cooling station includes one heat exchanger of 390\,kW, a duty and standby pump and a demineraliser. The water conductivity is maintained below 1~$\mu$S/cm. The water temperature is 27$^\circ$C for the supply and 37$^\circ$C for the return from equipment.

The equipment in the Biolab is expected to dissipate around 10\,kW. Its cooling is connected to the beamline cooling pipeline.

\subsubsection{Light ion source}
The baseline solution for the new light ion source requires about 50\,kW of demineralised water cooling (i.e. 4.3\,m$^3$/h @ 27\,-\,37$^\circ$C), see chapter~\ref{Chap:Source}. The existing ion source for LINAC2 is currently using a local closed cooling circuit which means that a new demineralized water cooling system is required. The overall cooling capacity from the existing station 234~ED~49 is deemed sufficient and no upgrade is foreseen. The existing distribution line is modified for the ion source equipment, with a cost estimated at 50\,kCHF.

\subsubsection{LINAC5 and transfer line to LEIR}
LINAC5 requires 100\,kW of demineralised water cooling (i.e. 8.6\,m$^3$/h @ 27\,-\,37$^\circ$C). The existing LINAC2 is currently using some 20\,m$^3$/h. The existing distribution line needs to be modified for the LINAC5 equipment. The overall cooling capacity from the existing station 234~ED~49 is also deemed sufficient for cooling of LINAC5 (in addition to cooling of the new light ion source) and no upgrade is foreseen.

\subsubsection{LEIR power converter upgrade}
As described in chapter~\ref{Chap:LEIR}, the LEIR accelerator needs an upgrade of its power converters in order for light ions to be accelerated up to 440\,MeV/u. The new power converters for the magnets ER.BHN and ER.QDN2040 require 26\,kW of demineralised water cooling (i.e. 2.2\,m$^3$/h @ 27\,-\,37$^\circ$C). At present, LEIR requires around 100\,m$^3$/h (measurements taken in 2007) of demineralised cooling water. The power upgrade may require up to 30\% additional cooling. This increase of cooling requirement can only be accommodated with an increase of the temperature difference between supply and return. It is not planned to upgrade the cooling station 234~ED~49 and its cooling towers. The pipework in building~250 requires some modifications, with an estimated cost of 30\,kCHF.

\subsection{Primary water}
\label{SubSec:Infrastructure_CV_PrimaryWater}
The primary water is supplied from the consolidated East Area cooling tower if it is technically feasible and approved. One duty and one standby distribution pumps are required, and where possible, the existing pipelines to South Hall are reused.

Alternatively, a new 2-cell cooling tower (n+1) of 400\,kW per cell can be installed in order to cool the demineralised water cooling station. A circuit distributes the primary water in a duty and standby (1+1) arrangement to the station. The new plant room houses the redundant pumps, the filtration, the water treatment station and the frost protection system. This alternative solution results in a higher cost for maintenance and water treatment, which is why the first proposal is considered the baseline solution.

\subsection{Ventilation systems}
\label{SubSec:Infrastructure_CV_Ventilation}

\subsubsection{LEIR}
\label{SubSubSec:Infrastructure_CV_Ventilation_LEIR}
For radiation protection reasons, LEIR shall need a concrete roof of 80\,cm thickness when operated with light ions, see chapter~\ref{Chap:RP}. A preliminary design for the roof above LEIR, dating from the original LEIR construction period, requires a 785\,m$^2$ surface area at a height of about 3\,m. Therefore, the volume to be ventilated is estimated around 2400\,m$^3$.
In addition, the BioLEIR switchyard areas (1 and 2) which also need to be covered by a roof, require a ventilation system. We assume the switchyard volume to represent 15\% of the LEIR volume. The required ventilation type is IIA according to the ISO17873 norm.

The temperature is maintained at 22$\pm$4$^\circ$C. The humidity is not controlled. The total dissipated heat load is yet to be defined. The HVAC system consists of a duty and a standby Air Handling Unit of 25000\,m$^3$/h, with dimensions 8\,m$\times$3\,m$\times$3\,m, recirculating the room air through filters. The air supply and return ducts have regular spaced duct mounted grids.
The necessary under-pressure is ensured using a duty and a standby extractor fan (3000\,m$^3$/h) connected to nuclear filters and a chimney. The air quality is monitored by sampling the air in the chimney stack.

\subsubsection{Irradiation room H1}
\label{SubSubSec:Infrastructure_CV_Ventilation_H1}
The H1 irradiation room (3\,m$\times$3\,m$\times$2.5\,m) is maintained at 37.5$\pm$0.5$^\circ$C and RH=40$\pm$10\%. The required ventilation type is IIA according to ISO 17873 norm. There is no sensitive or latent load dissipated in the air.

For this irradiation room, the HVAC system consists of an Air Handling Unit of 500\,m$^3$/h recirculating the room air through filters. The under-pressure is ensured using a duty and a standby extractor fans (150\,m$^3$/h) connected to nuclear filters and a common chimney.

\subsubsection{Irradiation room H2}
\label{SubSubSec:Infrastructure_CV_Ventilation_H2}
The H2 irradiation room (3\,m$\times$3\,m$\times$2.3\,m) is maintained at 22$\pm$1$^\circ$C and RH$<$40\%.
The required ventilation type is IIA according to the ISO 17873 norm. There is up to 1\,kW of sensitive load dissipated in the air.

For this irradiation room, the HVAC system consists of an Air Handling Unit of 2000\,m$^3$/h recirculating the room air through filters. The under-pressure is ensured using a duty and a standby extractor fans (150\,m$^3$/h) connected to nuclear filters and a common chimney.

\subsubsection{Irradiation room V}
\label{SubSubSec:Infrastructure_CV_Ventilation_V}
The vertical irradiation point V (3\,m$\times$3\,m$\times$4\,m) is maintained at 37.5$\pm$0.5$^\circ$C and RH=40$\pm$10\%. The required ventilation type is IIA according to the ISO 17873 norm. There is no sensitive or latent load dissipated in the air.
The HVAC system consists of one Air Handling Unit of 800\,m$^3$/h, recirculating the room air through filters. The under-pressure is ensured using a duty and a standby extractor fans (300\,m$^3$/h) connected to nuclear filters and a common chimney.

\subsubsection{Biolab}
\label{SubSubSec:Infrastructure_CV_Ventilation_Biolab}
The Biolab (14\,m$\times$14\,m) is maintained at 22$\pm$1$^\circ$C and RH=40$\pm$10\%. The acceptable quantity of airborne particles satisfies ISO 8 requirements in accordance with the ISO 14644-1 norm. The room pressure is not yet defined. The heat load dissipated by the equipment is estimated to be 5\,kW.

Several functional equipment (e.g. incubators, dye-process) require a separate air-exchange. The HVAC system includes an Air Handling Unit of 16000\,m$^3$/h with high efficiency filters.

\subsubsection{Counting rooms}
\label{SubSubSec:Infrastructure_CV_Ventilation_Countingrooms}
Both counting rooms are equipped with racks for electronics readout. The racks may need water cooling depending on the heat load which is not yet defined.

\subsubsection{Chilled water for air handling units}
\label{SubSubSec:Infrastructure_CV_Ventilation_ChilledWater}
The chilled water production plant for the building 150 area is located in building 355. The chilled water temperature regime is 6$^\circ$C for the supply and 12$^\circ$C for the return. The cooling water batteries of each Air Handling Unit is connected to this distribution pipework.

\subsubsection{Heating for air handling units}
\label{SubSubSec:Infrastructure_CV_Ventilation_Heating}
The heat losses of the new areas are negligible because the rooms to be built are located in the main hall (building 150) which is already heated and ventilated.
The heating required after the dehumidification process and the heating of the irradiation points is ensured using electrical heaters. The centralised heating system of Meyrin is shutting down during the summer, therefore it cannot be used for our purpose.
No active air-conditioning is currently provided to the general volume of the hall. This aspect remains the same for the BioLEIR project.

\subsection{Resource estimate}
Table~\ref{Tab:Infrastructure_CV_Material} shows the estimate of the cost for cooling and ventilation. The personnel needs for Cooling and Ventilation are estimated to be 1.5 person-years as detailed in table~\ref{Tab:Infrastructure_CV_FTE}. Experience from recent projects carried out at CERN has guided these resource estimates.

\begin{table}[!htb]
\caption{Material cost estimate for design, construction and installation of the cooling and ventilation systems for the BioLEIR experimental area.}
\label{Tab:Infrastructure_CV_Material}
\centering
\begin{tabular}{l l c r}
\hline
\rule{0pt}{3ex}\textbf{System} & \textbf{Element} & \textbf{Cost} & \textbf{Total} \\
\textbf{} & \textbf{} &  \textbf{[kCHF]} & \textbf{[kCHF]} \\
\hline
\rule{0pt}{3ex}Cooling  
  & Source & 50 &  \\
  & Frontend & 50 &  \\
  & Injection transferline & 50 &  \\
  & Beamlines &  400 &  \\
  & LEIR PC upgrade &  30 &  \\
  & Primary water & 100 &  \\
  & & \textbf{Total} & \textbf{680} \\
\hline
\rule{0pt}{3ex}Ventilation  
  & LEIR + Switchyards 1 and 2 & 550 &  \\
  & H1 & 75 &  \\
  & H2 & 50 &  \\
  & V & 75 &  \\
  & Biolab & 250 &  \\
  & Chilled water & 150 &  \\
  & Heating water & 150 &  \\
  & & \textbf{Total} &  \textbf{1300} \\
\hline
\rule{0pt}{3ex}\textbf{Total CV} & & &\textbf{1980}\\
\hline
\end{tabular}
\end{table}

\begin{table}[htb]
\centering
\caption{Estimated personnel needed for design, construction and installation of the cooling and ventilation systems for BioLEIR, given in integrated person-years.}
\label{Tab:Infrastructure_CV_FTE}
\begin{tabular}{l l r}
\hline
\rule{0pt}{3ex}\textbf{Aspect} & \textbf{} & \textbf{Staff [PY]}\\
\hline
\rule{0pt}{3ex}Design 		 & Engineer  	& 0.5 \\ 
Installation & Engineer 	& 0.5 \\ 
			 & Technician  & 0.5 \\
\hline
\rule{0pt}{3ex}\textbf{Total CV} & \textbf{Person-years}& \textbf{1.5}\\
\hline
\end{tabular}
\end{table}

\section{Fluids}
\label{SubSec:Infrastructure_Fluids}
This section consists of the extension of the fluid networks into the BioLEIR experimental area.
The fluids network extension includes:
\begin{itemize}
   \item the provision of a demineralised water network extension to the Biolab;
   \item an extension of the compressed air network with connection possibility in regular intervals;
   \item design and work supervision.
\end{itemize}
 
In the absence of more concise requirement specifications for fluids in the experimental area, the cost is roughly estimated at the level of 20\,kCHF for additional piping.
 
\section{Electrical infrastructure}
\label{Sec:Infrastructure_EL}

This section consists of all general power supply for the BioLEIR facility, including all machine aspects and the experimental area.

The electrical services within building 150 are at the end of their working life and require complete replacement in order to satisfy modern safety requirements. 
In addition to the need to replace immediate electrical equipment installed within building 150, the low voltage infrastructure supplying the South Hall and LEIR also requires replacement in order to remain serviceable for the lifetime of the BioLEIR project. Existing High Voltage infrastructure and transformers are in acceptable condition and can be reused.

Certain installations in the Biolab need high reliability power supplies (with an expressed preference for diesel generator based supply), at the level of a few kW. A review of the spare capacity on the existing diesel generator supplies to this area for critical and life safety applications shall be done in the next stage of the project. For the purpose of this report, it has been assumed that enough diesel generator capacity is available. However, the low voltage infrastructure into the experimental hall needs to be replaced in order to comply with modern safety standards. The cost for safe distribution of the diesel generator supply to the Biolab is estimated at about 80\,kCHF.

\subsection{Machine requirements for electrical infrastructure}
\label{SubSec:Infrastructure_EL_Machine}
The requirements for electrical power distribution for specific machine elements (for example cabling for magnets or controls) is discussed and estimated in the relevant machine chapters. This subsection is concerned with the electrical infrastructure available to the machine elements.

\subsubsection{Linac}
\label{SubSubSec:Infrastructure_EL_Linac}
The electrical installations serving LINAC2 are at the end of their working life and require complete replacement before a new installation can be made in order to satisfy modern safety requirements. 
A cost estimate for the electrical distribution has been made based on analysis of the recent LINAC4 installation, the cost of which was 240\,CHF/m$^2$ for underground installations, and 320\,CHF/m$^2$ for surface areas. Considering the footprint of LINAC2, LINAC5 is estimated to occupy 400\,m$^2$ of beamline areas and 500\,m$^2$ of technical installation area for which a cost of 250\,kCHF is assumed.


\subsubsection{Extraction and beamlines to the experimental area}
\label{SubSubSec:Infrastructure_EL_Beamlines}
Areas enclosed due to radiation protection and requirements for closed ventilation and air extraction require additional conventional and emergency lighting. In addition, the enclosed LEIR zone requires radiation resistant lighting throughout. This additional electrical infrastructure (lighting, emergency lighting, cable trays) is estimated at the level of 100\,kCHF. 

\subsubsection{LEIR power converter upgrade}
The LEIR power converter upgrade to allow for light ions to be accelerated up to 440\,MeV/u requires a refurbishment of the associated electrical distribution network. The cost for this is estimated, at 15\% of the total power converter material cost (670\,kCHF), to about 100\,kCHF.

\subsection{Requirements of the experimental area on electrical infrastructure}
\label{SubSec:Infrastructure_Electric_EA}
This subsection contains the requirements for general power supply for lighting, various electrical lab equipment, readout and computer equipment and low tension socket boxes distributed in experimental area. It also includes the power supplied to cooling- and ventilation plants including their need for secured power. The counting rooms are equipped with lights, single-phase 230~V sockets, power supplies and bare racks for the use by experiments. Dedicated circuits are provided for specific items of medical research equipment.

From the 1200\,m$^2$ surface area of BioLEIR, about 650\,m$^2$ are considered highly serviced spaces at 320\,CHF/m$^2$. The remaining 550\,m$^2$ are considered to be general areas with lower servicing requirements and are therefore estimated at a cost of 40\,CHF/m$^2$. The total electrical infrastructure cost for the experimental area is therefore estimated at the level of  230\,kCHF.

\subsection{Resource estimate}
Table~\ref{Tab:Infrastructure_EL_Material} shows the estimate of all electrical infrastructure. The corresponding staff needs are estimated to 2 person-years as detailed in table~\ref{Tab:Infrastructure_EL_FTE}. Experience from recent projects carried out at CERN has guided these resource estimates.

\begin{table}[!htb]
\caption{Material cost estimate for design, construction and installation of the electrical infrastructure for BioLEIR.}
\label{Tab:Infrastructure_EL_Material}
\centering
\begin{tabular}{l c }
\hline
\rule{0pt}{3ex}\textbf{Element} & \textbf{Cost} \\ 
\textbf{} &  \textbf{[kCHF]} \\
\hline
\rule{0pt}{3ex}Frontend & 250  \\
Extraction \& beamlines & 100   \\
Experimental area &  230  \\
Diesel generator distribution & 80 \\
LEIR PC upgrade &  100 \\
\hline
\textbf{Electrical infrastructure total} & \textbf{760} \\
\hline
\end{tabular}
\end{table}

\begin{table}[htb]
\centering
\caption{Estimated personnel needs for design, construction and installation of the electrical infrastructure for BioLEIR, given in integrated person-years.}
\label{Tab:Infrastructure_EL_FTE}
\begin{tabular}{l r}
\hline
\rule{0pt}{3ex}\textbf{Aspect} & \textbf{Staff [PY]}\\
\hline
\rule{0pt}{3ex}LV installation	& 1 \\ 
Cabling supervision 	& 1 \\ 
\hline
\rule{0pt}{3ex}\textbf{Total EL Person-years}& \textbf{2}\\
\hline
\end{tabular}
\end{table}

\section{Counting rooms}
\label{Sec:Infrastructure_CountingRoom}
The counting rooms are envisaged to be made of a light, sandwich structure that can be re-sized relatively easily if a reconfiguration is needed. Two counting rooms of about  9\,m$\times$4\,m are planned; one near H1 and the other near H2, intended for detector development or ballistics work.

Each counting room is equipped with a door with card reader. Part of the counting rooms is occupied with bare racks with space for readout electronics, specific to the relevant experiments. Each room should have a false floor such that readout or controls cables can be easily routed into the racks.

The counting rooms need electrical outfitting and air-conditioning. They shall also have fast Ethernet connections. The racks may need water cooling depending on the heat load, see section~(\ref{SubSec:Infrastructure_CV_Ventilation}) for more details.

The cost for two equipped counting barracks is estimated at 100\,kCHF as detailed in table~\ref{Tab:Infrastructure_CountingRooms}. Each equipped barrack has the following equipment:
\begin{itemize}
	\item electrical outfit including lights, four bare equipment racks, distribution boards, earthing, sockets, AUL and distribution boxes,
    \item air-conditioning.
\end{itemize}

\begin{table}[ht]
\centering
\caption{Cost estimate for two equipped counting rooms.}
\label{Tab:Infrastructure_CountingRooms}
\begin{tabular}{l r r r}
\hline
\rule{0pt}{3ex}\textbf{System} & \textbf{Units} & \textbf{Unit cost} & \textbf{Total} \\
\textbf{} & \textbf{} & \textbf{[kCHF]} & \textbf{[kCHF]} \\
\hline
\rule{0pt}{3ex}Standard bare barrack  & 2 & 15 & 30 \\
Electrical outfitting (lights, 4 bare racks) & 2 & 25 & 50 \\
Air-conditioning & 2 & 10 & 20 \\
\hline
\rule{0pt}{3ex}\textbf{Total} & & &\textbf{100}\\
\hline
\end{tabular}
\end{table}

The false floors under the counting rooms are part of the general structures detailed in section~\ref{Sec:Infrastructure_CE}.

\section{Gas networks}
\label{Sec:Infrastructure_Gas}
A gas supply panel for neutral gases is needed in the experimental area. The exact position is not defined, yet we assume that a standard panel is used with outlets for four gases (typically N$_2$, O$_2$, Ar and CO$_2$). This panel shall be supplied from a suitable location for gas bottles and connected via (approximately) 100\,m of thin stainless steel pipes. Medical grade gases for laboratory purposes in the Biolab are not included in this estimate but it is expected that the facility users shall requisition those according to their needs. 

For the EHN1 extension four neutral gas lines of small diameter are used, representing a cost of 65\,kCHF. We assume the same cost for the BioLEIR gas network.

\section{Civil engineering and structures within the experimental area}
\label{Sec:Infrastructure_CE}
\begin{figure}[!h]
\centering
\includegraphics[width=0.8\linewidth]{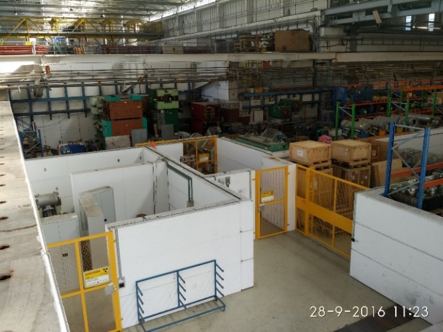}
\caption{Photograph of inside building 150 where the BioLEIR structures would be constructed.}
\label{Fig:Infrastructure_Hall_Photo}
\end{figure}

This section describes and estimates all structures that are needed for the experimental area of the BioLEIR facility. This includes some new civil engineering structures that need to be constructed in the existing building~150 (South Hall) that was built in the 1950's, as well as structures that are planned to be built out of precast concrete or cast iron shielding blocks. The photograph in figure~\ref{Fig:Infrastructure_Hall_Photo} shows the area of building~150 where the new structures are planned.

Table~\ref{Tab:Infrastructure_Structures} lists the new structures that are necessary (see section~\ref{Sec:EA_ExperimentalHall}) for shielding requirements, depending on the intensities and beam losses assumed within them (see chapter~\ref{Chap:RP} for details on the shielding requirements).

\begin{table}[!h]
\centering
\caption{Summary of shielding requirements for the different structures in the experimental area and the maximum energies and intensities available in the three irradiation rooms for Stage 1+2 and Stage 3 of the facility. The necessary shielding thickness is given in lateral and longitudinal directions with respect to the beam impinging on the target. The numbers are given for Stage 1+2, whereas the figures in parentheses indicate the shielding thickness necessary after the LEIR power converter upgrade to allow ion energies of 440\,MeV/u in Stage 3.}
\label{Tab:Infrastructure_Structures}
\resizebox{\textwidth}{!}{%
\begin{tabular}{l r r r c c c c }
\hline
\rule{0pt}{3ex}\textbf{Structure} & \textbf{Roof} & \textbf{Lateral wall} & \textbf{Long. wall} & \textbf{Protons} & \textbf{} & \textbf{Ions} & \textbf{}\\
\textbf{} & \textbf{[m]} & \textbf{[m]} & \textbf{[m]} & \textbf{[MeV]} & \textbf{[p/s]} & \textbf{[MeV/u]} & \textbf{[ions/s]}\\
\hline
\rule{0pt}{3ex}Shielding roof above LEIR & 0.8 & - & - \\
Shielding roof above extraction & 1.0 (1.8) & 1.0 (1.8) & 3.4 (4.0) \\
Switchyard 1 & 1.2 & 1 & - \\
Switchyard 2 & 0.8 & 1 & - \\
Horizontal irradiation rooms H1+H2 & 1.0 (1.8) & 1.0 (1.8) & 3.4 (4.0) & 250 (250) & $10^9$ ($10^{10}$) & 246/80 (440) & $10^8$ ($10^8$) \\
Vertical irradiation room V & 1.6 & 0.4 & - & 70 & $10^8$ & 70 & $10^8$ \\
\hline
\end{tabular}
}
\end{table}

\begin{figure}[!h]
\centering
\includegraphics[width=0.95\linewidth]{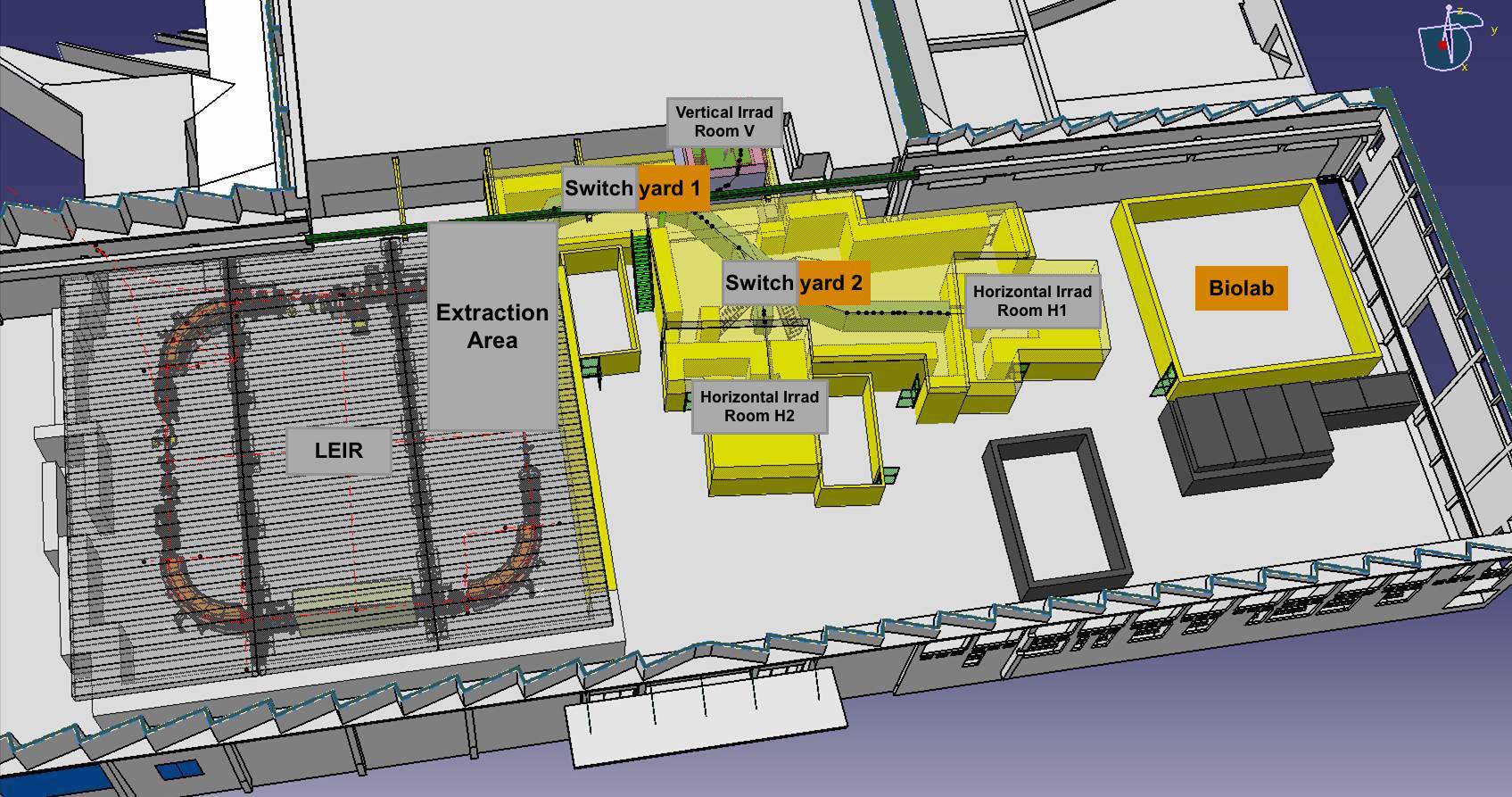}
\caption{Layout and nomenclature of the full BioLEIR facility. In the baseline Stages 1 + 2, the two horizontal irradiation areas H1 and H2 have the thinner shielding as indicated and for Stage 3 the shielding needs to be increased from the inside of the irradiation area to provide adequate radiation protection.}
\label{Fig:Infrastructure_Nomenclature}
\end{figure}

Figure~\ref{Fig:Infrastructure_Nomenclature} shows the conceptual layout for the new structures. Those colored in orange are within the scope of civil engineering (SMB department) and those colored in grey are under the responsibility of the experimental area group of EN (EN-EA).
Those new structures that need shielding are planned to be constructed of either in-situ reinforced concrete or of precast concrete or cast iron standard blocks with metallic structures where needed. Sandwich panels are planned for the Biolab (which does not need shielding).
The grey structures in figure~\ref{Fig:Infrastructure_Nomenclature} are costed with CERN standard-size precast concrete blocks, whereas the orange structures are costed assuming in-situ reinforced concrete. The two switchyard areas (colored in grey-orange) have a rather wide span which is why they have been costed with both methods and the precast concrete method has been chosen as baseline option. Should static evaluation by civil engineering experts show that precast concrete blocks cannot provide sufficient statics, in-situ poured concrete may have to be the method of choice including a number of undesired effects as discussed in the risk chapter of this document (see chapter~\ref{Chap:Risk}). The further optimization process in the next stage of this project may review this baseline.

\begin{figure}[!h]
\centering
\includegraphics[width=0.6\linewidth]{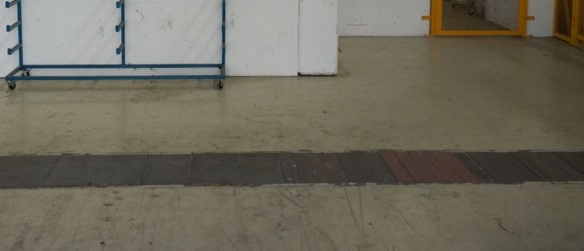}
\caption{Photograph of a typical service trench with steel covering.}
\label{Fig:Infrastructure_Hall_Trenches}
\end{figure}

The concrete floor of building~150 has many trenches crossing the hall with either steel or concrete removable covers. Often these trenches house existing services (electrical cables, water cooling pipes etc.). It may be necessary to deviate or modify these trenches for reasons of access or mechanical stability. This aspect has not been taken into account for this conceptual design stage. A typical floor trench is shown in figure~\ref{Fig:Infrastructure_Hall_Trenches}. Similarly, at this stage of the study, no detailed studies have been undertaken to determine the load capacity of the existing concrete floor slab. It is likely that structural strengthening of the floor may be required to withstand the loads of the substantial shielding, such as augured piling.  No cost estimates for such strengthening work nor modifications to the floor have been included in the cost estimates.

A seismic study to verify the stability of the various new structures needs to be performed in the next phase of this project.

\subsection{Switchyard 1 and Switchyard 2}
\label{Infrastructure_CE_Switchyards}
Both switchyard structures, which are non-standard shapes with relatively large spans, are to be built in areas with limited access and with rather thick roof slabs for radiation shielding. From recent experience in constructing shielding for experimental areas like GBAR or CHARM, we infer that, in principle, the switchyards can be produced more economically with precast concrete slabs and this option is therefore the baseline option. However, due to the large span widths, construction from in-situ reinforced concrete might become necessary.

The advantage of using in-situ reinforced concrete is that less supporting structures are needed for the thick roof slab, leaving more space for the various beamlines to enter and exit the rooms. Both switchyard areas are expected to have much less beam losses and material activation than the irradiation rooms which is why in-situ concrete can be a viable option. The obvious disadvantage of in-situ concrete is that it is difficult and expensive to modify and remove at a later stage, including any radiation protection issues at the dismantling stage.  Of course, the in-situ casting of concrete comes with a certain amount of nuisance such as dust and noise/vibration, which would have to be managed carefully during the work-site execution.

Comparing the pure material cost difference, precast concrete is more economical. However, due to necessary additional supporting structures for the wide spans, the full cost of in-situ concrete against off-site pre-fabrication is expected to be fairly similar. In the next stage of this study, cost and ease of installation/dismantling as well as stability issues need to be carefully optimized.

\subsection{Irradiation rooms H1, H2, V}
\label{Infrastructure_CE_IrradiationRooms}
High levels of radiation are expected in all three irradiation rooms, as well as in the extraction area (see chapter~\ref{Chap:RP}) and there is always a possibility that the room layout or design may need to be modified. Therefore, the shielding needs to be done with precast concrete and cast-iron blocks.\\

As the facility is planned to be built in stages (see section~\ref{Sec:Intro_ProjectOverview}) with different ion energies and intensities available in the horizontal irradiation rooms (see also table~\ref{Tab:Infrastructure_Structures}), the shielding layout of the irradiation rooms is designed to allow straightforward upgrade of the shielding thickness for Stage 3. For the baseline of Stage 1 and 2, the shielding thicknesses correspond to ion energies and intensities available in those stages within an outer footprint that is already adapted to the shielding thickness for Stage 3. This approach produces a slightly larger working area (4.6$\times$3.6\,m$^2$). Once the LEIR power converter upgrade is performed for Stage 3, the shielding is increased, taking space from the inside of the irradiation room to provide the corresponding radiation protection, and providing a final room footprint of 3$\times$3\,m$^2$ (see figure~\ref{Fig:Infrastructure_Nomenclature}).

For the current cost estimate, straight-forward rectangular shapes are assumed and the cost is derived from volume and weight estimates. In the next stage of the project, an optimization process for the detailed shielding thickness shape can be envisaged in close collaboration with the RP group. This effort should allow saving at the level of 10-20\% of the shielding cost.

\subsection{Biolab}
\label{Infrastructure_CE_Biolab}
The Biolab will be built from lightweight sandwich panels consisting of a mineral fiber insulation with external steel plates on the outer and inner faces. As the Biolab has a footprint of 14$\times$14\,m$^2$, a lightweight metallic structure is required to carry the weight of the roof panels. A typical thickness of the wall sandwich panels is 10\,cm, yielding a fire rating of 90~minutes.
If the Biolab is to be considered as a low-class clean room some remedial works is required on the existing floor of building 150. For costing purposes a low level false floor is included in the estimate, yet a simple vinyl flooring like the one around the GBAR experimental area could be considered fully adequate.
As the clean room class requirement is not clear at this conceptual stage, no special measures have been included in the cost estimate, such as airlock doors or resin floor treatment.

\subsection{Shielding roof above the LEIR accelerator}
\label{Infrastructure_CE_LEIRRoof}
Radiation protection studies found that running with light ions in the LEIR accelerator requires LEIR to be covered with an 80\,cm roof of concrete (see chapter~\ref{Chap:RP}). Original drawings for the LEIR accelerator were found, showing that operation was initially foreseen with a roof. The optimal pillar locations are identified in those drawings and could be used as baseline for the LEIR roof design, with metal I-beams connecting the pillars and precast concrete blocks making up the roof between the I-beams. A surface of about 820\,m$^2$ needs to be covered.

\subsection{LEIR extraction area}
\label{Infrastructure_CE_Extraction}
As most beam losses are expected to happen in the extraction region (40\% beam loss expected), the lateral shielding in this area needs to be the same as in the irradiation rooms. A roof surface area of 10$\times$14.5\,m$^2$ area shall cover the full extraction area.

\subsection{Small access structures}
\label{Sec:Infrastructure_Hall_Structures}
In addition to the shielding structures described above, complementary small structures are needed, predominantly to delineate the radiation controlled area and smaller access structures like stairways and access platforms. Such structures are listed below:

\begin{itemize}
   \item access to the vertical irradiation room, directly from the ground floor of hall~150 via a staircase with an estimated cost of about 6\,kCHF.
    \item additional walkways, staircases and possible ladders typically for access to technical installations, as e.g. the tripe stair with a small platform on top in Switchyard 2, estimated at the level of 12\,kCHF.
    \item gridded protection barriers (for material and detector storage of about 9\,m$^2$, which is estimated at $\approx$10\,kCHF.
	\item for ease of installation of services between the different parts of the experimental area (irradiation rooms, Biolab, counting rooms etc.) a false floor supported by a metal structure should be installed over an area of approximately 20\,m$\times$10\,m which includes also the area underneath the 2 counting rooms.
\end{itemize}
 
\subsection{Resource estimate}
\label{Infrastructure_CE_Cost}
At this early stage of the study, a civil engineering cost estimate has been established assuming an accuracy of about $\pm$30\%.

It is assumed that all services enter and exit the new structures from above e.g. via elevated cable trays, and as such, no new galleries in the concrete floor have been costed. 
The structural and civil engineering cost estimates have been based on similar works already performed on the CERN site (e.g. GBAR, CHARM). For the precast blocks, the costing includes delivery. About half of the overall transport and handling cost for the experimental area (estimated at 2 man-months) should cover the cost of installation of the precast blocks.

Table~\ref{Tab:Infrastructure_Costing_Structures} shows the expected baseline cost for all  structures and civil engineering aspects of the experimental area. The estimated cost for the switchyards built with in-situ concrete is shown in italics.

\begin{table}[!htb]
\centering
\caption{Material cost estimate for the infrastructure of the experimental area. Total costs include overheads where relevant, e.g. for tooling, delivery etc. The baseline solution calls for precast concrete slabs for Switchyard 1 and Switchyard 2 and their cost includes 20\% for installation. The estimated cost for an alternative in-situ solution is given in italics.}
\label{Tab:Infrastructure_Costing_Structures}
\resizebox{\textwidth}{!}{%
\begin{tabular}{l l r r r r}
\hline
\rule{0pt}{3ex}\textbf{System} & \textbf{Element} & \textbf{} & \textbf{} & \textbf{Cost} & \textbf{Total} \\
\textbf{} & \textbf{} & \textbf{} & \textbf{} & \textbf{[kCHF]} & \textbf{[kCHF]} \\
\hline
Small structures \& \\
False floors  &  &  & & & \\
  & floor (incl structure) & 200 & 150 & 30 & \\
  & triple staircase & & & 12 & \\
  & staircase Switchyard 1 & & & 6 & \\
  & storage barriers & & & 10 & \\
  & Vertical room structure & & & 10 & \\
  & & & & \textbf{Small Structures} & \textbf{68} \\
Shielding  &  &  & & & \\
  &  & \textit{concrete} & \textit{cast-iron} & &\\
  & H1 & 87 & - & 112 & \\
  & H2 & 87 & - & 112 &\\
  & V & 16 & 299 & 315 &\\
  & \underline{Switchyard 1} &  &  &  &\\
  & \quad Baseline: precast & 145 & - & 145 & \\
  & \quad \textit{in situ alt.} & \textit{289} &  &  &\\
  & \underline{Switchyard 2} &  &  &  &\\
  & \quad  Baseline: precast & 222 & - & 222 &\\
  & \quad \textit{in-situ} & \textit{401} &  &  &\\
  & Extraction & 17 & - & 17 &\\
  & \underline{LEIR roof} &  & & 433 &\\
  & \textit{concrete} & 392 &  &  &\\
  & \textit{12 pillars} &  & 12 &  &\\
  & \textit{24 I-beams} &  & 24 &  &\\
  & \textit{Installation: 2 person-weeks} &  & 5 &  &\\
  & & & & \textbf{Shielding Stage 1+2} & \textbf{1356} \\
  & \underline{Stage 3 energy upgrade} & & & & \\
  & \quad additional shielding Stage 3 & 161 & & & \textbf{161}\\
Biolab & & & & \\
  & Wall \& roof panels &  & & 168 &\\
  & False floor \& structure & 196 & 150 & 29 &\\
  & & & & \textbf{Biolab} & \textbf{197} \\
Design services CE & 10\% & & & & \textbf{89} \\
Detailed design of shielding walls \& roofs & & & & & \textbf{10}\\
Contingency & 10\% & & & & \textbf{89} \\
CE concept drawings & & & & & \textbf{50} \\
\hline
\rule{0pt}{3ex}\textbf{Structures} & & & & &\textbf{2020}\\
\hline
\end{tabular}
}
\end{table}

\begin{figure}[!htb]
\centering
\includegraphics[width=1.0\linewidth]{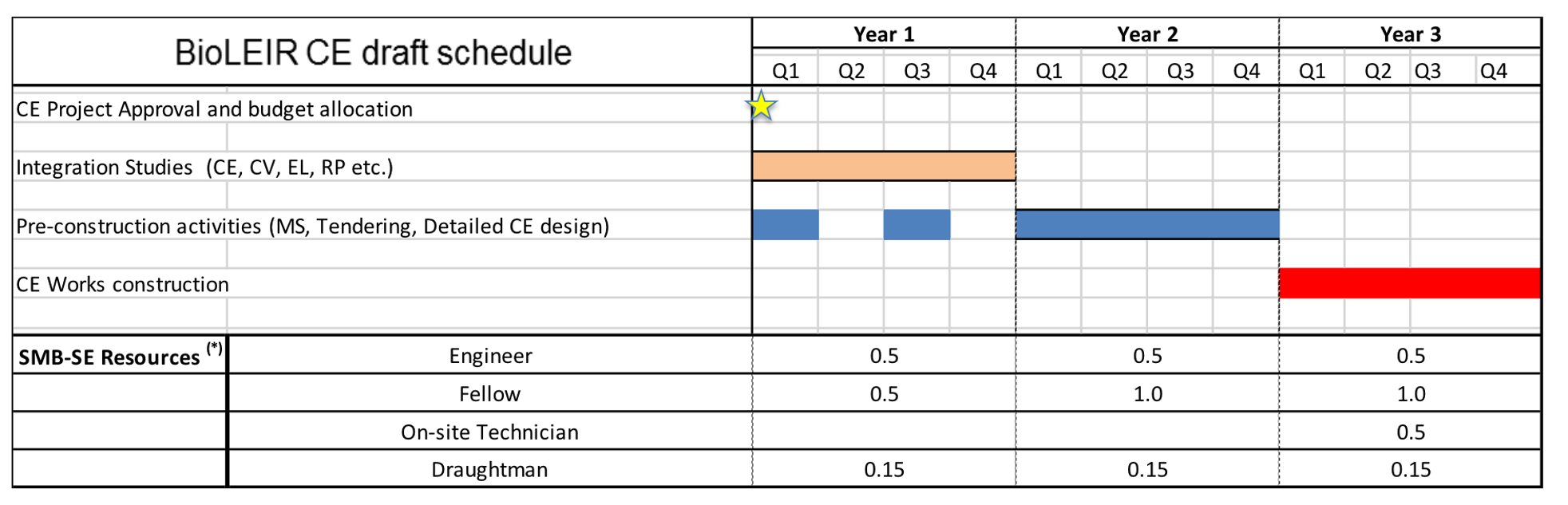}
\caption{Simplified civil engineering schedule for design/procurement/construction and required resources.}
\label{Fig:Infrastructure_CE_Resources}
\end{figure}

The civil engineering construction work is expected to take about one year to complete on site with a lead time for integration studies and procurement procedures of 1.5 to 2 years. A simplified schedule is shown in figure~\ref{Fig:Infrastructure_CE_Resources} and table~\ref{Tab:Infrastructure_CE_FTE} shows the personnel estimate for this part of the work.\\
Obviously, to allow structural and civil works to start, the existing building needs to be completely emptied. Any costs associated with the clearance of this equipment is not included in the estimate.

\begin{table}[!htb]
\centering
\caption{Staff estimate for the design, construction and installation of the structural elements for the experimental area, given in person-years.}
\label{Tab:Infrastructure_CE_FTE}
\begin{tabular}{l r r}
\hline
\rule{0pt}{3ex}\textbf{Structural Aspect} & \textbf{Staff} & \textbf{Fellow} \\
\textbf{} & \textbf{[PY]} & \textbf{[PY]} \\
\hline
\rule{0pt}{3ex}On-site technician & 0.5 & \\
Engineering & 2.5 & 1.5 \\
Precast block work supervision & 0.2 & 0.2\\
\hline
\rule{0pt}{3ex}\textbf{Total structures/civil engineering [person-years]} & \textbf{3.2} & \textbf{1.7}\\
\hline
\end{tabular}
\end{table}

\section{Doors}
\label{Sec:Infrastructure_Doors}
This section contains the supply and installation of personnel access doors. These doors must accommodate the standard badge-reading system which is required for access to a supervised area (locks operated by programmable badge reader). One such door is needed for each of the 2 horizontal irradiation rooms. In addition, a normal door (without card reader) is needed closer to the irradiation point at the inside maze endpoint in order to delimit the volume for which closed air ventilation is needed. The cost for those doors is estimated at 3\,kCHF for one outer and 1\,kCHF for one inner door. The two necessary escape doors from the Switchyard 2 area are estimated at 3\,kCHF each. For the vertical irradiation room V, a shielded sliding door is needed - equivalent to the shielding of a 0.6\,m concrete wall. This door should have a width of at least 1\,m. The cost is estimated to be about 30\,kCHF, based on a recent estimate for the AWAKE experiment. A total of 44\,kCHF is estimated for doors in the experimental area of BioLEIR.

\section{IT-networks}
\label{Sec:Infrastructure_Hall_IT}

 There is a need to have an Ethernet network and sockets for the counting barracks as well as for all services with the possibility to connect to either the technical- or the general purpose network at CERN. It is envisaged to install about 30 UTP sockets. In addition, it is important to have a local Wi-Fi system for the counting barracks and the Biolab (3 bases). The IT-services are estimated at the level of 25\,kCHF.

\section{Transport and handling} \label{Sec:Infrastructure_Transport}
Transport and handling for the BioLEIR experimental area is expected to cost at least at the level of 20\,kCHF and require 0.1\,person-years of Staff. This includes installation and handling of all equipment and shielding blocks.

\section{Resource Estimate}
\label{Sec:Infrastructure_Cost}

The volume of the irradiation rooms should be kept as small as reasonable to minimize cost. They are temperature- and humidity-controlled, and need a closed, monitored air circulation.

For those aspects of infrastructure where no specific cost estimate could be provided by the relevant equipment groups, a rough cost estimate was derived from the estimates submitted for the EHN1 extension~\cite{ENH1:EMDS}, scaling with a factor of 5.5 for the surface ratio between EHN1 (3600\,m$^2$) and BioLEIR (650\,m$^2$).

Table~\ref{Tab:Infrastructure_Costing_Material} shows the estimate of the cost for design, construction and installation of the infrastructure for the BioLEIR experimental area, while the personnel needs for the infrastructure are detailed in table~\ref{Tab:Infrastructure_Costing_FTE}.

\begin{table}[ht]
\centering
\caption{Material cost estimate for design, construction and installation of the infrastructure for the BioLEIR experimental area. Total costs include prototyping and overheads where relevant, e.g. for tooling or design.}
\label{Tab:Infrastructure_Costing_Material}
\begin{tabular}{l r}
\hline
\rule{0pt}{3ex}\textbf{System} & \textbf{Total} \\
\textbf{} & \textbf{[kCHF]} \\
\hline
Cooling  &  680 \\
Ventilation  &  1300 \\
Fluids  &  20 \\
EL  & 760 \\
Counting rooms  & 100 \\
Structures \& civil engineering  & 2020 \\
Transport \& handling & 20 \\
Gas network  &  65 \\
Doors  &  44 \\
Cabling & 20\\
IT-services & 25 \\
\hline
\rule{0pt}{3ex}\textbf{Total} &\textbf{5054}\\
\hline
\end{tabular}
\end{table}

\begin{table}[ht]
\centering
\caption{Personnel estimate for design, construction and installation of the infrastructure for the BioLEIR experimental area, given in person-years.}
\label{Tab:Infrastructure_Costing_FTE}
\begin{tabular}{l r}
\hline
\rule{0pt}{3ex}\textbf{System} & \textbf{[PY]} \\
\hline
\rule{0pt}{3ex}CV & 1.5 \\
EL (incl. cabling) & 2 \\
CE & 4.9 \\
Shielding design (RP) & 0.2 \\
Gas Network  & 0.1 \\
Counting room + false floor + doors  & 0.1 \\
Layout/integration & 0.5 \\
Alignment and survey  & 0.1 \\
Transport and handling & 0.1 \\
IT-services & 0.1 \\
\hline
\rule{0pt}{3ex}\textbf{Total infrastructure [person-years]} & \textbf{9.7}\\
\hline
\end{tabular}
\end{table}

%% file: Chapters/MO.tex
\chapter{Operations}
\label{Chap:MO}

\section{Operational scenarios}
\label{Sec:MO_Scenarios}

In order to start delivering beams to BioLEIR as early as possible, the facility is proposed to be built in stages. A specific operational scenario corresponds to each of the stages that are increasing in flexibility: initially, BioLEIR runs with ions from LINAC3 outside the period when LHC and/or the North Area (NA) take ions, and the energy range is limited to 246\,MeV/u for light ions. Then, as soon as LINAC5 is commissioned, it is possible to switch between BioLEIR and North-Area/LHC operation in a matter of minutes. Finally, upgrading the main power converters of LEIR will allow delivering light ions at energies up to 440\,MeV/u.
In an optional upgrade, several transfer line and injection elements could be upgraded to support pulse-to-pulse modulation (PPM) allowing service to both BioLEIR and the LHC or the North Area within the same LEIR super-cycle.

It is noted explicitly that operational priority will always be given to the LHC. However, some form of beam sharing cycle-to-cycle  can be envisaged to increase beamtime to BioLEIR beyond the 4 months of dedicated operation, with a gradual increase in complexity.

\subsection{Stage 1}
\label{SubSec:MO_Scenarios_Stage1}
In the initial stage, only the new slow extraction system and the 3 extraction lines towards the South Hall are added to the existing LEIR machine. The ions for BioLEIR are provided by LINAC3. Since there is only one ion source connected to LINAC3, it normally takes several weeks to switch from one ion species for BioLEIR to another one for the LHC. This imposes BioLEIR operation to take place outside the periods of heavy ion collisions in the LHC, including outside the full preparation time.
An exception is the case where the ions desired by BioLEIR are used as a support gas for the ions requested by the LHC, in which case the switching time between species can be reduced to less than one day.  Typically, Oxygen is used as a support gas for Lead and Xenon. We suggest to start BioLEIR operation with Oxygen ions during Stage 1 to have maximum flexibility. Machine developments with Oxygen ions can also be carried out before the BioLEIR extraction line is installed, in order to assess beam performance.
The other ions that could be supplied by LINAC3 are typically Carbon, Argon, Indium and Xenon.

In Stage 1, the maximum energy of the ions accelerated in LEIR is limited by the power converters of LEIR. Therefore, the energy at BioLEIR in Stage 1 is limited to 80\,MeV/u for Argon ions and to 246\,MeV/u for Oxygen ions.

\begin{figure}[!htb]
\centering
\includegraphics[width=0.8\columnwidth]{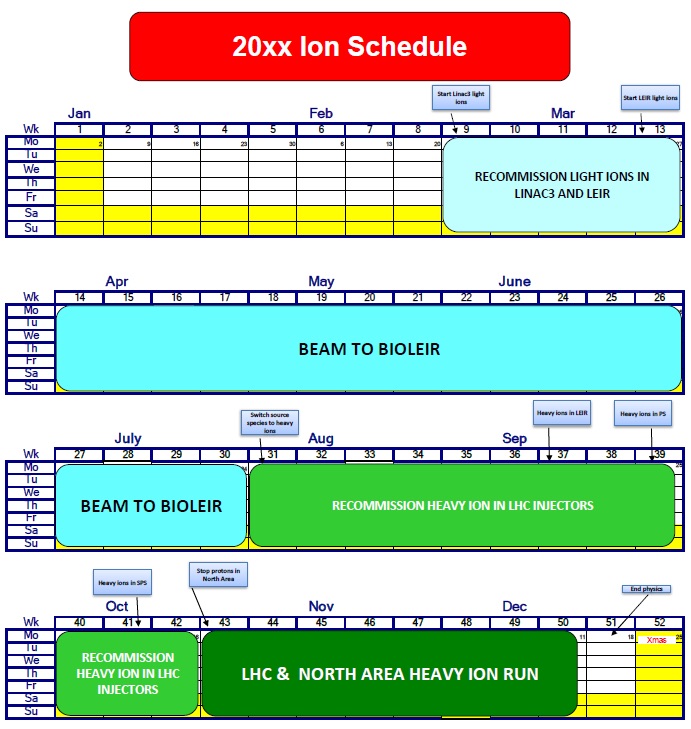}
\caption[]{Typical expected yearly schedule of LINAC3 for BioLEIR Stage 1 operation when LINAC3 is not operating with Oxygen. Whenever LINAC3 is operated with Oxygen, switching between BioLEIR and North-Area/LHC operation can be done within a few hours, not months.}
\label{Fig:MO_Stage1_Schedule}
\end{figure}

Figure~\ref{Fig:MO_Stage1_Schedule} shows how a typical operation schedule of the facility could look like when LINAC3 delivers non-Oxygen beam to BioLEIR. This schedule represents the worst-case scenario and yields four full months of beam in dedicated BioLEIR operation mode. The switching time between operation for BioLEIR and the North-Area/LHC, including the source maintenance, is at the level of weeks in this scenario. For the periods that BioLEIR requests Oxygen ions, beam delivery can be switched over to BioLEIR within a few hours, and integrated beam-time availability at BioLEIR can be increased in this way.

\subsection{Stage 2}
\label{SubSec:MO_Scenarios_Stage2}
Once LINAC5 is made operational in Stage 2, LINAC3 is freed for exclusive heavy ion operation. In Stage 2, light ions can be accelerated to a maximum energy of 246\,MeV/u and up to 80\,MeV/u for Argon ions.
Some of the LEIR injection elements, as well as part of the magnets in the injection line, can work only with a single ion type at a time. It is possible to switch between heavy and light ions within minutes. Outside of the periods of heavy ion beam preparation, beam can be sent to BioLEIR on request, including during LHC fillings. 
BioLEIR operation are foreseen to take place from Easter to early November, when the North Area takes ions at every LEIR super-cycle. 

\subsection{Stage 3}
\label{SubSec:MO_Scenarios_Stage3}
For Stage 3, the power converters of LEIR are upgraded to allow light ion acceleration up to 440\,MeV/u. This energy upgrade of the LEIR machine does not have any impact on the BioLEIR machine operation.

\subsection{Optional upgrade to interleaved operation}
\label{SubSec:MO_Scenarios_Upgrade}
In order to be able to interleave BioLEIR and LHC/NA cycles, LINAC5, the LEIR machine and their connecting beamlines undergo a number of modifications:
\begin{itemize}
\item the switching magnet ITE.BHN20 needs to be replaced and its power supply modified to support pulse-to-pulse modulation.
\item LEIR injection elements (magnetic and electrostatic septa) and injection line magnets need to be modified to support pulse-to-pulse modulation, unless LINAC5 is able to provide particles at exactly the same magnetic rigidity as the ones sent by LINAC3. 
\item a safety system needs to be put in place to avoid particles from being sent to the wrong destination (heavy ions destined to LHC or the North Area sent to BioLEIR or vice-versa).
\end{itemize}

\noindent \textbf{Non-interleaved operation} 

\noindent The advantages of non-interleaved operation are:
\begin{itemize}
	\item  For non-interleaved operation, the 2 linacs are uncoupled from a timing point of view.
	\item It is the fastest and easiest implementation.
	\item It is the cheapest implementation.
\end{itemize}

\noindent The disadvantages of non-interleaved operation are:
\begin{itemize}
	\item Running mode only per destination/user: 
	\begin{itemize}
		\item Mode for physics: only destinations to the North Area, PS East Hall, LHC \&  LEIR machine developments cycles can be concurrently in a LEIR super-cycle.
		\item Mode for biology: only destinations to BioLEIR experiments \& LEIR machine developments cycles can be concurrently in a LEIR super-cycle.
	\end{itemize}
	Switching back and forth between these 2 modes takes around 15 minutes at best.
	\item LINAC3 or LINAC5 have idle time.
	\item A more interdependent organization and scheduling between all users is needed which complicates operation.
	\item The non-interleaved operation can still be used to fill the LHC with 2 different ions species but one beam after the other, as in the proton-ion scheme.
\end{itemize}

\noindent \textbf{Interleaved operation}

\noindent Interleaved operation entails the following complexities and costs:
\begin{itemize}
	\item Full timing synchronisation between LEIR, LINAC3 \& LINAC5 is needed.
	\item All elements of the EI and ITE injection lines need to be fully pulse-to-pulse modulated (PPM) which entails additional cost. 
	\item The LEIR injection septa which are presently fully DC static, non-PPM devices need to be upgraded at additional cost.
\end{itemize}

\noindent The interleaved option assures the following benefits:
\begin{itemize}
\item A maximized and optimized use of the 3 machines with less linac downtimes (''on the fly'' LEIR super-cycle update to meet as closely as possible the user requests with an infinite number of cycle combinations \& duty cycles in the super-cycle), increasing operational flexibility.
	\item An optimized and more flexible way to fill the LHC with 2 different ion species.
\end{itemize}

        


\begin{figure}[ht!]
    \centering
    \begin{subfigure}{.4\linewidth}
      \hspace*{-45mm}  \includegraphics[width=2.4\linewidth]{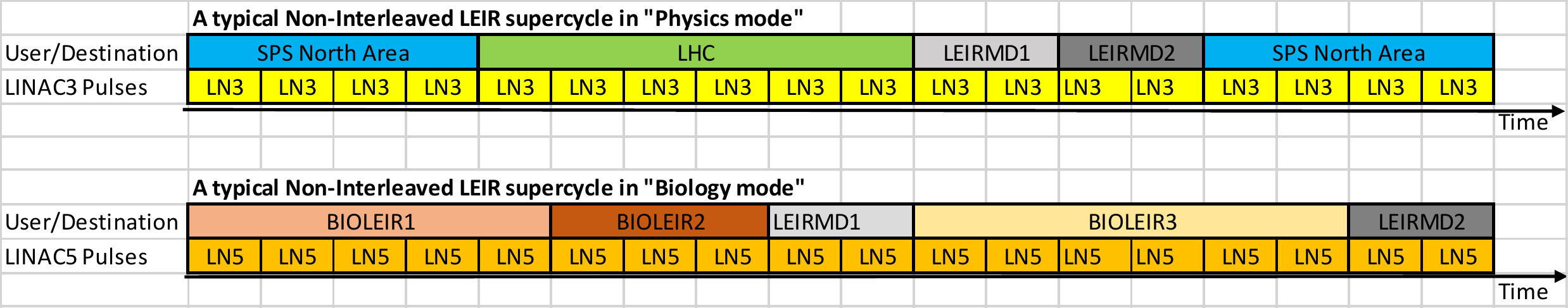}{}
        \caption{}
    \end{subfigure}
    
    \begin{subfigure}{.4\linewidth}
    \hspace*{-45mm} \includegraphics[width=2.4\linewidth]{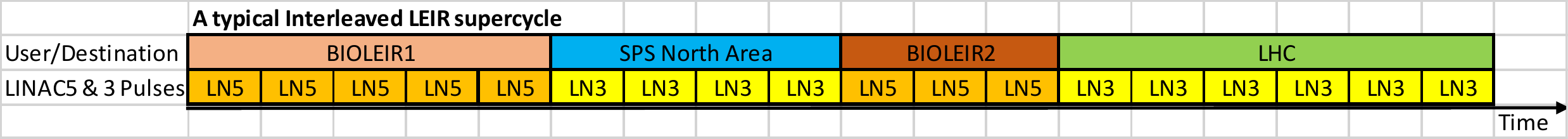}{}
    \caption{}
    \end{subfigure}
\caption[]{Typical expected super-cycles for (a) non-interleaved and (b) interleaved operation of LEIR.}
\label{Fig:MO_LEIRSupercycles}
\end{figure}

Figure~\ref{Fig:MO_LEIRSupercycles} shows how the physics- and biology-mode could look like in non-interleaved operation mode and for interleaved operation.

Due to the stringent constraints outlined above, interleaved operation may prove too costly compared to the benefits gained, although it would make BioLEIR operation possible all year round without limitations.

We conclude that non-interleaved operation is the easiest and cheapest way to operate LINAC3, LINAC5 and the LEIR machine. Interleaved operation can be considered a potential upgrade of the facility, should available beamtime in dedicated mode ever become an obstacle.

\section{LEIR's magnetic cycle}
\label{Sec:MO_MagneticCycle}
\begin{figure}[!ht]
\centering\includegraphics[trim={6cm 3cm 6cm 3.4cm},clip, width=1.0\linewidth]{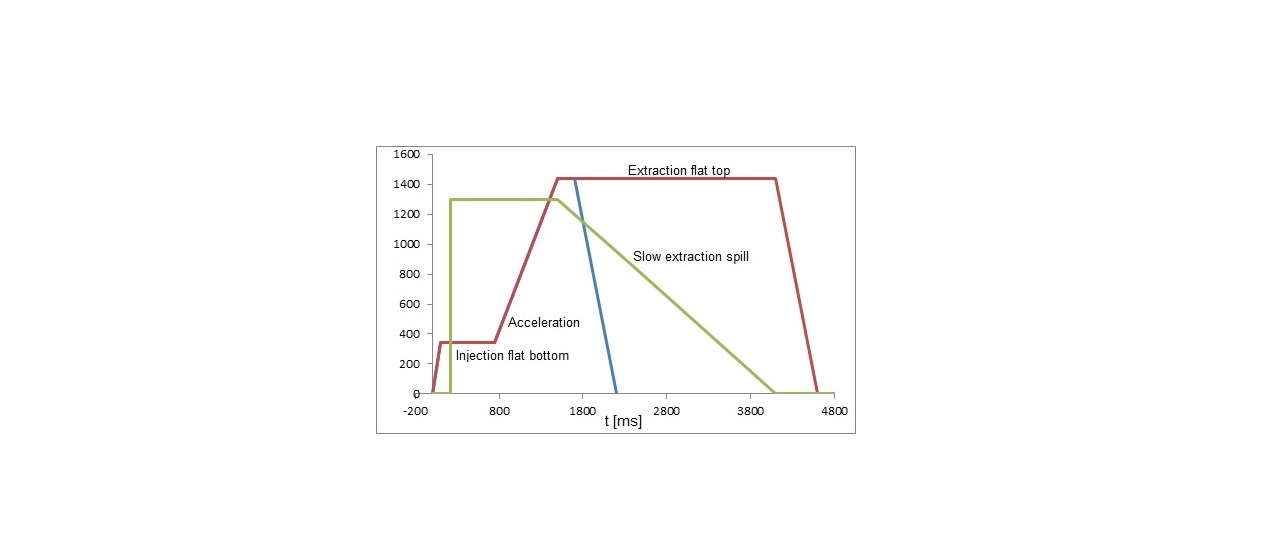}
\caption{Operational cycle of LEIR. The magnetic rigidity in red is expressed in MeV/c per charge, the intensity in green is shown in arbitrary units. The short cycle used for the LHC is printed in blue for comparison.}
\label{Fig:MO_MagneticCycle}
\end{figure}

The proposed magnetic cycle is similar to the one currently used for the LHC pilot or the North Area, completed with an extended flat top for a 2.4 second long slow extraction, see figure~\ref{Fig:MO_MagneticCycle}. The field is first ramped to the injection energy. After a single injection and a short ($\sim$300\,ms) cooling time at the flat bottom, the beam is bunched and accelerated during 500\,ms to the extraction energy. After the slow extraction, the magnetic field is ramped down to the minimum value.\\
Compared to the short cycle used for the LHC, the heat load increases due to the long flat top. This limits the number of available cycles per unit time to what the available cooling power permits.

\section{Preliminary machine development periods}
\label{Sec:MO_MDS}
The slow extraction section will not be installed in LEIR before LS2 at the earliest. 
However, as Oxygen is used as support gas for Xenon and Lead operation, it is already possible to carry out some machine studies with Oxygen in 2017. 

In order to benchmark the expected radiation levels during the running of LINAC5 for BioLEIR, a radiation measurement campaign should be scheduled in 2017 during the running of LINAC3 with Oxygen ions. 

In order to better understand the beam dynamics with light ions in LEIR, we recommend to initiate in 2017 a series of machine development studies on LEIR. 
A non-exhaustive list of studies would include: 
\begin{itemize}
	\item detailed characterisation of the LINAC3 Oxygen beam;
    \item injection trajectories and efficiencies, and correction;
    \item optimisation of the acceleration cycle with minimal losses;
    \item transverse beam dynamics;
    \item cooling rates with and without electron cooler.
\end{itemize}
    
In addition to their interest for BioLEIR, studies with Oxygen could benefit the rest of the CERN programme in case LHCf (or another experiment) is approved for O-O or p-O collisions at low luminosity in the LHC, and for the delivery of fixed target Oxygen beams in the North Area.

\section{Resource estimate}
\label{Sec:MO_Cost}
During routine operation, the SPS operation team, working 24/7 in shift-mode, is in charge of monitoring LEIR, keeping its performance up-to-date, and acting as first line support in case of problems. One dedicated LEIR machine coordinator is on-call for one week at a time for setting-ups, further optimisation, as well as diagnosis and resolution of complex issues.  The same approach can be kept during Stage 1 of BioLEIR, provided an additional 0.5 FTE completes the team as LEIR would have to run all year round. From Stage 2 onwards, an additional team of 4 technical engineers is needed to ensure smooth operation of the LEIR machine with 2 injectors, LINAC3 and LINAC5.\\

In addition, during the commissioning phase, a dedicated team of accelerator physicists has to be assigned to the running-in. The necessary resources for the running-in are estimated at 1.5-2 person-years (i.e. 6 physics engineers for a duration of 3-4 months).



%% file: Chapters/Controls.tex
\chapter{Control System}
\label{Chap:Controls}

The BioLEIR control system will use standard tools and frameworks already implemented in the existing accelerator complex. In particular, no development of new hardware or software components is foreseen, but rather adaptation or extension of existing controls solutions is favoured.
As part of the control system, an industrial solution may be used to operate some of the equipment. Details of industrial solution installations are discussed with the BE-ICS group and other units at CERN.

\section{Requirements}
\label{SubSec:Controls_Requirements}
The BioLEIR facility is installed and operated in 3 distinct stages (see chapter~\ref{Chap:MO} for more details). This approach allows delivery of useful beam to experiments as early as possible. Each succeeding facility stage requires a different level of investment from the controls group.

\subsection{Stage 1}
\label{SubSec:Controls_Requirements_Stage1}
In Stage 1, the control system has to manage the new LEIR extraction elements and the three beamlines towards the experiments in the South Hall. Standard BE-CO hardware modules and software libraries suffice to provide the needed functionality.

Timing-wise, the LINAC3 and LEIR machines are operated in coupled mode. This means that they are controlled simultaneously and it is impossible to cycle them independently. For Stage~1 of BioLEIR it is sufficient to keep the two machines in coupled mode and therefore, the current timing system is considered adequate.

\subsection{Stage 2}
\label{SubSec:Controls_Requirements_Stage2}
In Stage 2, the control system has to steer the new LINAC5 machine. Provision of the full control stack for a brand new accelerator is a considerable task. It is therefore expected that standard BE-CO hardware modules and software libraries are used wherever possible to provide most of the needed functionalities. The largest upgrade needed concerns the timing system.

From an operational point of view, the control system is able to provide heavy ion beams from LINAC3 together with LEIR, as well as light ion beams from LINAC5 in conjunction with LEIR. Switching between the two types of beam can be done within minutes. When one of the two linacs produces beam with LEIR, the other linac should be able to cycle separately for testing purposes.

\subsection{Stage 3}
The power converter upgrade for Stage 3 has no impact on the controls system.

\subsection{Optional upgrade to interleaved operations}
\label{SubSec:Controls_Requirements_Upgrade}
It is possible to change between heavy-ions and light-ions cycles in a matter of seconds, once the LEIR transfer line is upgraded to support multiplexing and therefore interleaved operations (pulse-to-pulse modulation, PPM - see section~\ref{SubSec:MO_Scenarios_Upgrade}).
The standard BE-CO stack supports multiplexing, and such multiplexing is expected to be a straight-forward implementation of the control system. However, the timing system may require further changes to be able to control the 3 machines (LINAC3, LINAC5 and LEIR) independently, the details of which depend on the implementation chosen for Stage 2.

\section{Implementation}
\label{Sec:Controls_Implementation}
Standard BE-CO hardware and software tools are used to control the new machine equipment as well as magnets needed for the new slow extraction from LEIR and for the beamlines towards the experimental areas (see chapters~\ref{Chap:Extraction} and \ref{Chap:BeamLines} for details on equipment to be controlled). The same holds true for each of the three beam stoppers, the beam dump, the scanning magnets for the pencil beams, and any specific equipment of the biological laboratory requiring controls.

\subsection{Dumping the beam}
\label{SubSec:Controls_Implementation_DumpingBeam}
The decision to dump a beam may be taken at almost any time. In such a case, no beam shall reach any of the BioLEIR irradiation areas. Lead ions can be safely dumped inside the LEIR machine, or an extraction kicker between LEIR and the PS machine can be used to dump the beam on a concrete block between the two machines. For light ions operation, the beam shall be dumped in a dedicated beam dump. The beam dump implementation details need to be clarified in the next stage of the project. It is assumed that the beam dump control system is based on standard BE-CO components.

\subsection{Dedicated control room}
\label{SubSec:Controls_Implementation_CR}
It is expected that a dedicated, local BioLEIR control room be established in the vicinity of the experimental area. We presume that two new workstations and four screens are needed for local controls applications.

\subsection{Robotic table for samples}
\label{SubSec:Controls_Implementation_Robotics}
As described in section~\ref{Para:EA_ExperimentalHall_Robotization}, it is planned to have a robotic table that positions and exchanges cell samples in irradiation room H1 and V. The automation system is provided by EN-STI-ECE group and is expected to be an industrial system, comprising equipment and its specific controls that need to be integrated in the controls system. Exact requirements and installation details shall be defined in the next project stage.

\subsection{Timing system and cycling of the machine}
\label{SubSec:Controls_Implementation_Timing}
One of the most challenging control system modifications concerns the timing system ensuring synchronisation of equipment elements and beam scheduling (sequencing).\\

In Stage 1 the timing system is expanded to control the three new beamlines. 
Stage 2 requires a bigger adaptation. First, the control system has to be extended to control the new LINAC5. Second, LINAC3 and LEIR are currently controlled by the same central timing. Their cycles are defined and executed together. Introduction of LINAC5 that also needs to work in conjunction with LEIR necessarily alters that concept. Two potential solutions exist: (1) extending the coupled/decoupled mode, or (2) implementing separate central timings for each of the accelerators. Stage 3 requires no further change of the timing system. \\

The optional upgrade requires the second solution for Stage 2, as the control system at that point has to provide enough flexibility to operate the three accelerators independently.

\subsection{Open analogue signal information system -- OASIS}
\label{SubSec:Controls_Implementation_OASIS}
It is estimated that the following analogue system set-up is needed to control BioLEIR:
\begin{itemize}
\item up to 32 analog inputs (sources) to monitor.
\item possibility to monitor up to 16 of them in parallel (i.e. 4 oscilloscopes each with 4 digitizers).
\item the fastest signals to monitor (fastest sampling) are needed for beam position monitors and beam current transformers.
\end{itemize}

To provide the above functionality, the OASIS installation consists of 48 multiplexed analogue signals and four ~1\,GS/s digitisers. Power converters use FGC3 modules, and therefore do not need integration into OASIS. Requirements on the control system from LINAC5 are not specified at this time.

\subsection{Injector controls architecture -- InCA}
\label{SubSubSec:Controls_Implementation_INCA}
It is expected that the high level controls infrastructure (e.g. settings management) is integrated into the LEIR operation configuration. The settings are configured in a standard way using knobs and working sets.

\subsection{Status display}
\label{SubSubSec:Controls_Implementation_Display}
For the moment no new status displays (fixed display) are requested. Should such a fixed status display become useful, it can be developed relatively rapidly based on existing software frameworks.

\subsection{Other controls services}
\label{SubSubSec:Controls_Implementation_Other}
It is assumed that other standard controls services are used as needed. These may include: the controls configuration database, the machine data logging service, the alarm service and monitoring of equipment.

\section{Resource estimate}
\label{Sec:Controls_Cost}
In this early study stage, the cost estimate for the controls system (detailed in table~\ref{Tab:Controls_Cost}) is provided with an uncertainty at the level of 10\%.\\

The estimation for Stage 1 is based on a similar installation done recently for the new AD extraction line towards ELENA. LINAC5 estimations are based on LINAC4 cost for controls, scaled to the relative machine size. The indicated cost concerns only what is needed for implementation of the BE-CO systems (dedicated to BioLEIR, LEIR or LINAC5). In particular, industrial controls, or Ethernet cabling (for example for LINAC5) are not taken into account.

\begin{table}[!ht]
\centering
\caption{Cost estimate for the BioLEIR controls system.}
\label{Tab:Controls_Cost}
\begin{tabular}{c|ccr}
\hline
\rule{0pt}{3ex}\textbf{Stage} & \textbf{HW cost} & \textbf{OASIS cost} & \textbf{Total cost} \\
\textbf{} & \textbf{[kCHF]} & \textbf{[kCHF]} & \textbf{[kCHF]} \\ \hline
\rule{0pt}{3ex}1 & 55 & 75 & 130 \\
2 & 80 & 100 & 180 \\
3 & - & - & - \\
Optional upgrade & 20 & & 20 \\ \hline
\rule{0pt}{3ex}\textbf{Total} & & & \textbf{330} \\ \hline
\end{tabular}
\end{table}

No technical show-stopper for the controls system has been identified. The necessary BioLEIR controls implementation tasks are executed within corresponding BE-CO units as part of the usual structure. Table~\ref{Tab:Controls_Cost_FTE} summarizes the resource estimations, mapped to the different facility stages.

\begin{table}[!ht]
\centering
\caption{Personnel estimate for controls implementation of BioLEIR.}
\label{Tab:Controls_Cost_FTE}
\begin{tabular}{c|cccr}
\hline
\rule{0pt}{3ex}\textbf{Task} & \textbf{Stage 1} & \textbf{Stage 2-3} & \textbf{Optional upgrade} & \textbf{Total cost} \\ \hline
\rule{0pt}{3ex}Hardware modules and cabling & 0.5 & 0.5 & 0.0 & 1.0 \\
OASIS modules and cabling & 0.5 & 0.5 & 0.0 & 1.0 \\
Timing & 0.0 & 1.0 & 0.5 & 1.5 \\
Software & 0.5 & 0.5 & 0.5 & 1.5 \\ \hline
\rule{0pt}{3ex}\textbf{Total} & 1.5 & 2.5 & 1.0 & \textbf{5.0} \\ \hline
\end{tabular}
\end{table}

\section{Outstanding questions and issues}
\label{Sec:Controls_Issues}
Part of the transfer line from LINAC5 to LEIR is shared with LINAC4, which accelerates negative polarity particles (H$^-$). This aspect may present stringent operational limitations due to necessary polarity switching of the transfer line magnets when switching from one beam to the other. The magnet polarity swtiching may render operations and controls rather complex for this part of the transfer line. It therefore needs to be investigated if doubling those $\sim$20\,m of transfer line might not be a more reasonable, robust and efficient operation solution, even if that may present additional investment cost.

Detailing of controls for the experiments will need to be properly integrated into the controls design for BioLEIR. 
Requirements on the control system from LINAC5 are not specified at this stage of the project.

%% file: Chapters/RP.tex
\chapter{Radiation Safety and Radiation Protection}
\label{Chap:RP}

This chapter is dedicated to radiation safety and radiation protection aspects of the BioLEIR facility.
After defining the parameters of the scenarios that have been used for the radiation protection assessment in section~\ref{Sec:RP_Assumptions-Scenarios}, the estimate for the prompt radiation and the shielding requirements derived from it are presented in section~\ref{Sec:RP_Prompt-Shielding}.
Section~\ref{Sec:RP_Activation} is devoted to the analysis of activation of the facility and notably the induced air-activation in the closed irradiation stations. 
The radiation protection requirements for the access system are detailed in section~\ref{Sec:RP_Access} and an outline of the radiation protection monitoring system is presented in section~\ref{Sec:RP_Monitoring}.
Finally, a first estimate for the costs related to radiation protection is given in section~\ref{Sec:RP_Cost}.


\section{Scenarios considered for the radiation protection assessment}
\label{Sec:RP_Assumptions-Scenarios}

The particle species and beam intensities used for the radiation protection assessment are presented in table~\ref{Tab:RP_Beam-parameters}. Once the LEIR power converters are upgraded, the BioLEIR facility can provide beams fulfilling the beam specification requirements as described in table~\ref{Tab:BeamParameters_WhatCanCERNProvide_IonIntensities}. These intensities correspond to the maximum intensities to be delivered to the irradiation rooms. 

\begin{table}[ht]
\caption{Particle species and maximum beam intensities used for the radiation protection assessment.}
\label{Tab:RP_Beam-parameters}
\centering
\begin{tabular}{l c c c}
\hline
\rule{0pt}{3ex}\textbf{Particle species} & \textbf{Kinetic energy} & \textbf{Intensity during spill} & \textbf{Average intensity}\\
& [MeV/u] & [particles/s] & [particle/s] \\
\hline
\rule{0pt}{3ex}Argon & 440 & $10^{9}$  & 5$\times 10^{8}$\\
 & 80 & $10^{9}$  & 5$\times 10^{8}$\\
 & 70 & $10^{9}$  & 5$\times 10^{8}$\\
Oxygen & 440 & $10^{9}$  & 5$\times 10^{8}$ \\
 & 246 & $10^{9}$  & 5$\times 10^{8}$ \\
Carbon & 440 & $10^{9}$  & 5$\times 10^{8}$ \\
 & 246 & $10^{9}$  & 5$\times 10^{8}$ \\
Protons & 250 & $10^{11}$ & 5$\times 10^{10}$ \\
\hline
\end{tabular}
\end{table}

Two generic target cases have been investigated to cover both extreme cases of beam interaction inside an irradiation room:
\begin{description}[leftmargin=0cm]
	\item[Cell culture:] the beam traverses a 4-cm-thick water equivalent target before hitting the beam dump. This case corresponds to the irradiation of cell cultures, and the main radiation source is the interaction of the beam with the beam dump. 
	\item[Phantom for dosimetric studies:] a water cylinder with a radius of 20\,cm and a length of 40\,cm is placed in the beamline such that the beam impinges in the center of the cylinder face. In this case, the main radiation source is the phantom target, and the Bragg peak is fully contained inside the phantom.
\end{description}

For both cases, a water beam dump with the same dimensions as the phantom is placed inside the downstream shielding wall. The beam area is enclosed by a concrete shielding.

It is expected that 40\% of the beam is lost at the extraction from LEIR, which means that the loss rate at extraction is comparable to the intensities delivered to the irradiation rooms. Therefore, the shielding requirements for the irradiation rooms also apply to the LEIR extraction region.

\section{Prompt radiation and derived shielding requirements}
\label{Sec:RP_Prompt-Shielding}

FLUKA Monte Carlo simulations~\cite{FLUKA, FLUKA2} have been performed to assess the prompt and residual ambient dose equivalent rates for both target cases and all particle species defined in table~\ref{Tab:RP_Beam-parameters}.

\begin{figure}[ht!]
    \centering
    \begin{subfigure}{.4\linewidth}
       \hspace*{-27mm} \includegraphics[trim={0 0.5cm 0 0},clip, width=2.0\linewidth]{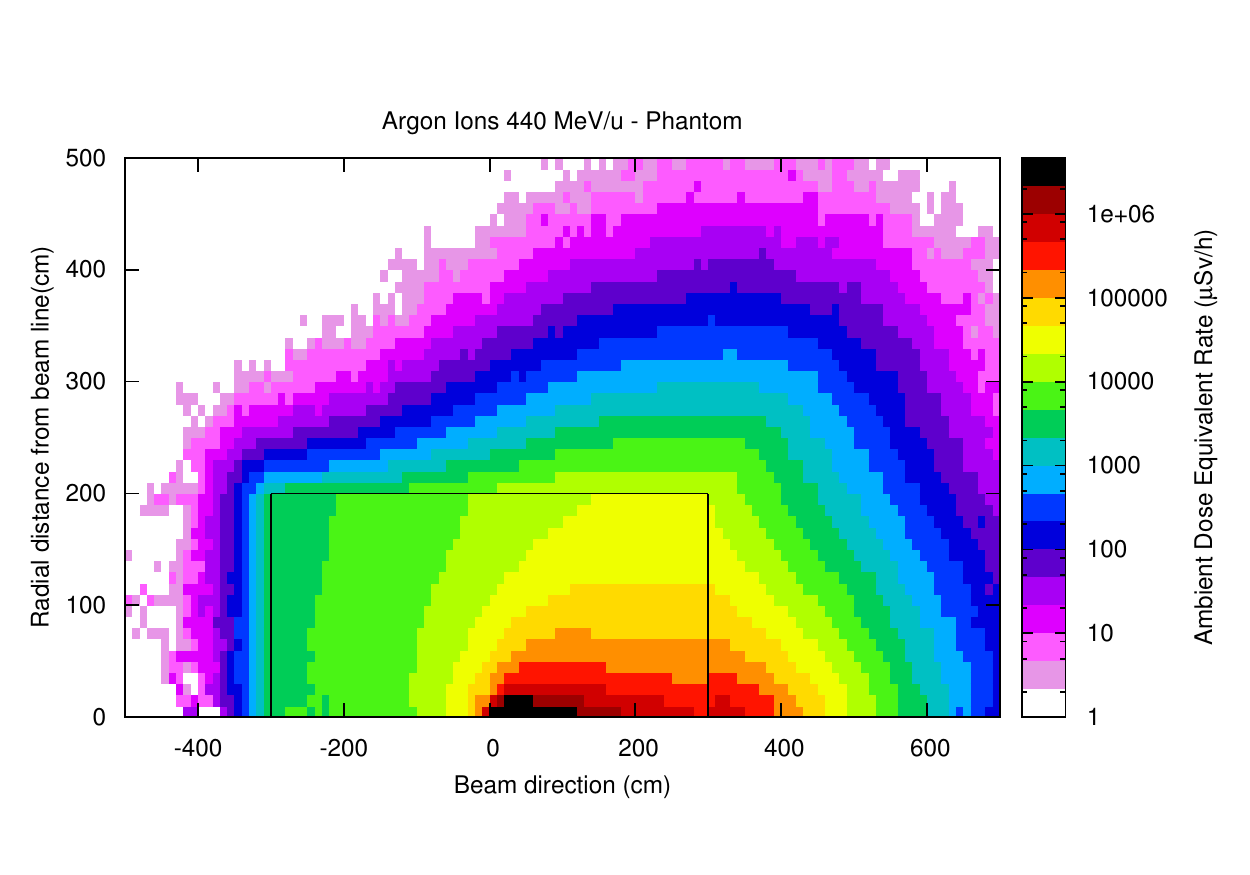}{}
        \caption{}
    \end{subfigure}
    
    \vspace*{-0.8cm}
    \begin{subfigure}{.4\linewidth}
     \hspace*{-27mm}\includegraphics[trim={0 0.5cm 0 0},clip,width=2.0\linewidth]{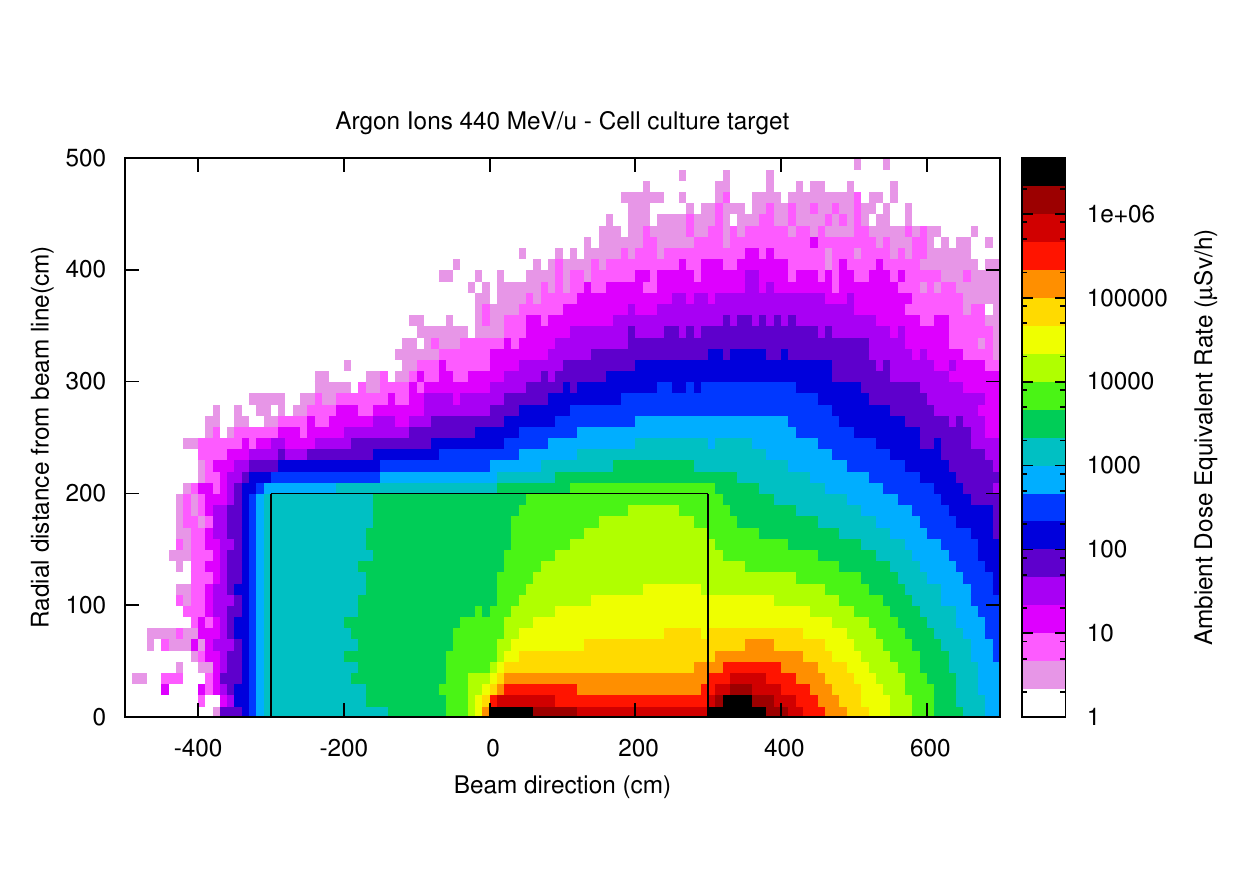}{}
    \caption{}
    \end{subfigure}
\caption[]{Prompt ambient dose equivalent rates from $5\times10^{8}$ Argon ions per second impinging with 440 MeV/u onto a water phantom (a) and a cell culture and subsequent beam dump (b).}
\label{Fig:RP_PromptDR_Ar}
\end{figure}

The prompt ambient dose equivalent rate distribution for Argon ions is shown in figure~\ref{Fig:RP_PromptDR_Ar} for phantom and cell culture targets. The prompt ambient dose equivalent rate at 1\,m laterally from the target can be up to $\sim$50\,mSv/h. The prompt ambient dose equivalent rate for Oxygen ions operation as well as for protons operation at 250\,MeV is lower than that for Argon ions operation.




\vspace*{-0.8cm}

To respect the radiological area classification limit of 15\,$\mu$Sv/h for supervised radiation areas~\cite{supervisedAreas}, 2.8\,m of lateral concrete shielding and 5\,m of downstream concrete shielding are required for Argon ions operation with 440\,MeV/u and an average beam intensity of $5\times10^{8}$ ions/s. For protons at energies of 250\,MeV and intensities of $5\times10^{10}$\,protons/s, the corresponding thicknesses are 2.6\,m  and 4.6\,m, respectively. It should be noted that the downstream shielding thickness could be significantly reduced by shielding the beam dump locally with cast iron. The shielding thickness requirements are shown in table~\ref{Tab:RP_shielding} for the various particle species, kinetic energies and spill intensities.

In order to allow a progressive enhancement of the facility capabilities, it has been decided to select an initial shielding configuration of 1.6\,m in lateral and 3.2\,m in longitudinal thickness for Stage 1 and 2 of BioLEIR. This shielding allows the intensities and energies of protons and light ions as summarised in table~\ref{Tab:Infrastructure_Structures}. The shielding thicknesses of the horizontal irradiation rooms is increased to 1.8\,m laterally and 4.0\,m longitudinally (from the inside of the irradiation rooms) for Stage 3 of BioLEIR, permitting protons of energies up to 250\,MeV and intensities up to $10^{10}$\,protons/s, as well as light ions up to Oxygen with energies up to 440\,MeV/u and intensities up to $10^8$\,ions/s in the irradiation rooms. For the case that higher intensities are desired, the shielding thicknesses need to be adapted accordingly. The chosen baseline shielding characteristics are similar to those implemented in existing non-clinical biomedical research facilities.

The final shielding configuration needs to be optimized in terms of minimal dose for permanent workplaces and offices in the vicinity, such as the CMS control room.

The lateral shielding thickness for the switchyard areas and for the transfer lines to the irradiation rooms could be reduced to 0.8\,m, provided that adequate radiation monitoring is provided as described in section~\ref{Sec:RP_Monitoring}.

The detailed design for the mazes to access the switchyard areas and the irradiation rooms has to be foreseen for the next stage of this study.

A shielding roof of 80\,cm concrete shall be foreseen for the LEIR machine. The final thickness of the shielding roof for LEIR has to be decided when the expected beam loss pattern in the LEIR machine for the future operation modes can be estimated more precisely.

\begin{table}[ht]
\caption{Shielding thicknesses for the various particle species and spill intensities. A cycle repetition time of 4.8\,s with a spill length of 2.4\,s, i.e. a duty factor of 50\%, has been assumed.}
\label{Tab:RP_shielding}
\centering
\begin{tabular}{lcccc}
\hline
\rule{0pt}{3ex}\textbf{Particle species} & \textbf{Kinetic energy} & \textbf{Intensity during spill} & \multicolumn{2}{c}{\textbf{Shielding thickness}}\\
& & &  \textbf{Lateral} & \textbf{Downstream}\\
& [MeV/u] &  [particle/s] & [m] & [m]\\
\hline
\rule{0pt}{3ex}Protons & 

          250 & $10^{11}$ &  2.6 & 4.6 \\
        & 250  & $10^{10}$ &  1.8 & 4.0 \\
        & 250  & $10^{9}$ &  1.0 & 3.4 \\
\hline
\rule{0pt}{3ex}Oxygen & 
		 440  & $10^{9}$  & 2.6  & 5.0 \\
       & 440  & $10^{8}$  & 1.6  & 4.0 \\
       & 246  & $10^{9}$  & 1.6  & 3.2 \\
       & 246  & $10^{8}$  & 1.0  & 2.2 \\
\hline
\rule{0pt}{3ex}Argon & 
		440  & $10^{9}$  & 2.8 & 5.0 \\
      & 440  & $10^{8}$  & 1.8 & 4.0 \\
      & 80  & $10^{9}$  & 0.8 & 2.2 \\
      & 80  & $10^{8}$  & 0.4 & 1.6 \\
      & 70  & $10^{9}$  & 0.8 & 2.0 \\
\hline
\rule{0pt}{3ex}Carbon & 
		  440  & $10^{9}$  & 2.6 & 5.0 \\
		& 440  & $10^{8}$  & 1.6 & 4.0 \\
		& 246  & $10^{9}$  & 1.6 & 3.2 \\
		& 246  & $10^{8}$  & 1.0 & 2.2 \\
\hline
\end{tabular}
\end{table}

\vspace*{-0.4cm}
\section{Activation}
\label{Sec:RP_Activation}

\subsection{Air activation}
\label{Sec:RP_Activation-Air}

The desired high beam intensities and energies (see table~\ref{Tab:RP_Beam-parameters}) produce air activation in the irradiation rooms at levels requiring
a ventilation system to provide dynamic confinement and a flush before access. The radioactivity that is released by the ventilation system via a dedicated exhaust has to be monitored as described in section~\ref{Sec:RP_Monitoring}. This is in accordance with best practices applied at clinical facilities of similar design. 

\subsection{Residual radiation}
\label{Sec:RP_Activation-Residual}

Based on the results of the FLUKA Monte Carlo simulations, the residual ambient dose equivalent rates at 40\,cm from the phantom and the beam dump are estimated to be below 15\,$\mu$Sv/h after one hour of cool-down time after Argon ions operation. This means that for access during Argon ions operation, the irradiation rooms could be classified as supervised radiation areas, according to reference~\cite{supervisedAreas}. The residual ambient dose equivalent rates after one hour of cool-down time after protons operation could exceed 50\,$\mu$Sv/h. For this case, the irradiation room has to be classified as limited stay radiation area~\cite{supervisedAreas}. 

\section{Radiation safety requirements for the personnel protection system}
\label{Sec:RP_Access}

The switchyard areas and the irradiation rooms have to be considered as primary beam areas, notably for the Important Safety Elements (EIS-b) requirements, due to the high prompt ambient dose equivalent rates (see section~\ref{Sec:RP_Prompt-Shielding}). This implies that these areas are not accessible during beam operations. 

Before access to an irradiation room, the ventilation system has to flush the irradiation room. Only after the flushing has been verified by the access system, access to the irradiation room can be granted. The possibility for applying a radiation protection veto at the access points shall be foreseen.

\section{Radiation protection monitoring system}
\label{Sec:RP_Monitoring}

Based on the preliminary BioLEIR beamlines and shielding layout, the following monitors have to be installed to provide adequate radiation protection monitoring:
\begin{itemize}
\item Ionization chambers of type IG5--H20 to monitor neutron stray radiation:
a total of 10 IG5--H20 monitors are needed: 3 around the LEIR shielding, 1 above the roof of the switchyard, 2 around the side walls of the switchyard and 2 per irradiation room - 1 laterally and 1 downstream. The ionization chambers are equipped with one alarm unit per chamber. It is necessary to foresee the possibility to interlock these ionization chambers with the access system in case of a radiation alarm.
\item Ionization chambers of type PMI to monitor residual radiation:
a total of 3 PMI monitors are needed: 1 for each horizontal irradiation room and 1 for the LEIR extraction area.
\item Ventilation monitoring station to monitor the airborne radioactivity release:
for this study, it is assumed that it is possible to have one common air release point for the whole BioLEIR facility such that only 1 ventilation monitoring station is required.
\end{itemize}

All these radiation monitors are integrated into the REMUS (Radiation and Environment Monitoring Unified Supervision) system.

\section{Control irradiator}
\label{Sec:RP_control_irradiator}

A control irradiator in the form of a Cs-137 source in a vertical configuration shall be available within the BioLEIR experimental area. At a later stage, a 6\,MeV control irradiation facility is required as a horizontal beam (see appendix~\ref{App:EA_ControlIrradiator} for more details).
The control irradiators shall be properly integrated into the CERN safety, control and radiation protection framework.

\section{Resource estimate}
\label{Sec:RP_Cost}

The estimated material cost directly related to radiation protection comes to a total of 410\,kCHF and is detailed in table~\ref{Tab:RP_costs}. 

\begin{table}[ht]
\caption{Cost estimate for radiation protection for the full BioLEIR facility.}
\label{Tab:RP_costs}
\centering
\begin{tabular}{lr}
\hline
\rule{0pt}{3ex}\textbf{Item} & \textbf{Cost} \\
 & \textbf{[kCHF]} \\
\hline
\rule{0pt}{3ex}Monitoring system & 330 \\
Preparation \& installation work & 30\\
Bufferzone & 50\\
\hline
\rule{0pt}{3ex}\textbf{Total} & \textbf{410}\\
\hline
\end{tabular}
\end{table}

The radiation protection monitoring system is estimated to cost 330\,kCHF. In addition, 30\,kCHF shall be foreseen for contractor work during the area preparation and installation phase. A further 50\,kCHF are needed for refurbishing of the buffer zone. This cost estimate does not include the shielding structures and the ventilation system.

The estimated CERN personnel needs (Staff and Fellows) for the remaining conceptual design phase, engineering design, construction, installation and commissioning are given in table~\ref{Tab:RP_Manpower}.

During the exploitation phase of BioLEIR, 0.5 CERN Staff FTE have to be foreseen to ensure adequate radiation protection.

\begin{table}[ht]
\centering
\caption{Personnel estimate for design, construction, installation and commissioning of the radiation protection system for BioLEIR, given in person-years.}
\label{Tab:RP_Manpower}
\begin{tabular}{l r r}
\hline
\rule{0pt}{3ex}\textbf{} & \textbf{Staff [PY]} & \textbf{Fellow [PY]} \\
\hline
\rule{0pt}{3ex}Radiation Protection & 1.0 & 3.0 \\
\hline
\end{tabular}
\end{table}

%% file: Chapters/Safety.tex
\chapter{Safety Aspects}
\label{Chap:Safety}

BioLEIR is similar in scope and use to other experimental facilities at CERN, for example ELENA at the Antiproton Decelerator. The beam and radiation hazard is the main concern. Other risks are identified primarily as industrial risks that are already covered by CERN rules and practices, and taken into account in the design and specification processes. This chapter considers global safety aspects, focussing on hazard monitoring and alarms.

\section{Operational safety of the ion sources and the linacs}
In Stage 1 of the facility, LINAC3 with its source is used to provide ions for BioLEIR, and its safety aspects are already covered. The modification of this frontend to produce lighter ions introduces new chemical substances. The associated risks have to be assessed and mitigated. It is expected that these risks are easily alleviated by complying with existing CERN rules.

In Stage 2, the design, construction, commissioning and operation of a new linac introduces risks that CERN is used to deal with on a regular basis: mainly electricity, radio-frequencies and magnetic fields. This part of the Stage 2 facility is similar in scope and use to other beam facilities at CERN. The chemical risk has to be assessed and mitigated for all newly introduced light ion species.


\section{Operational safety of the transfer lines (from LEIR to experimental areas)}
\label{Sec:Safety_Operational}
As expressed above, BioLEIR is similar to other CERN beam facilities, and to be designed, built, commissioned, and operated according to CERN rules and practices. Most of the operational procedures are very similar to the ones already applied around beam facilities at CERN.

The main issues to be considered for personal safety around the beamlines are electrical and radiation hazards. The electrical circuits used for powering the beamline elements as well as for the infrastructure shall be designed according to CERN rules and best professional practices. HSE-SEE shall inspect and validate the beamlines and equipment with respect to electrical risk.

The beam and radiation safety is ensured by the Personnel Protection System (PPS, see section~\ref{Sec:Safety_PPS}), complemented by the beam permit process put in place by the BE DSO (Departmental Safety Officer) and followed by the BE-OP team. The PPS design is based on the results of a risk analysis and takes into account that BioLEIR is classified as a primary beam area (see section~\ref{Sec:RP_Access}) for which CERN practices impose two independent beam interruption mechanisms (see section~\ref{SubSec:Infrastructure_PPS_SIF}).

In order to safely dispose of the beam during setting up or as an emergency measure during operation, an external dump is foreseen to be installed at the beginning of the extraction channel, still at a location inside the shielding of the LEIR machine. The second EIS protecting the BioLEIR switchyard area is an upstream bending magnet.

As the three irradiation rooms will be independently accessible, each beamline is protected by two EISs: one beam stopper and one bending magnet. 
The dump and stoppers have to be dimensioned for the respective beam intensities at the horizontal and vertical irradiation rooms, H1, H2 and V (see table~\ref{Tab:EA_ExperimentalHall_Rooms} for the planned ion intensities and energies).

The beam dump and the 3 beam stoppers are estimated to cost at the level of 400\,kCHF.

To cover appropriate mitigation of the radiation risk, HSE-RP participated to the risk analysis and validated the design. They also monitor, check and validate the operation (see section~\ref{Sec:RP_Monitoring}). To cover the radiation risk associated with potentially activated materials, an ''RP bufferzone'' currently exists in hall~150. This bufferzone has to be displaced and possibly refurbished.

Prior to beam transfer towards the switchyards, the zone is patrolled (by BE-OP), and the PPS generates a beam-imminent warning signal (pseudo-BIW, with flashing lights). Because of the geometry and size of the area, there is no need to implement a more robust system as it is done for other primary beam areas at CERN. In particular, sirens shall be avoided as they could be heard from one area to another and generate confusion.

A dedicated emergency system is implemented in the form of an emergency push-button which sends a veto on the EIS-beam via the PPS. This automatically cuts all beams in the switchyard area and generates a signal in the control room. An autonomous escape lighting device is installed above the access door.

The switchyard is such a small area, with low occupancy of only specialists expected, that no dedicated automatic evacuation system is required. On the other hand, the general evacuation signal from hall~150 must be audible in the shielded switchyard areas. Eventually, a siren coupled to this system has to be added. Evacuation signs and lights need to be put at appropriate places.

The fire risk is considered as negligible for the personnel and for the environment. The highest probability for a fire to start is during operation, when magnets and equipment are powered and no one is allowed into Switchyard 1 or 2 (see nomenclature in section~\ref{Fig:Infrastructure_Nomenclature}). Magnets and sensitive equipment are usually equipped with temperature sensors and monitoring. Detection of abnormally high temperatures provokes powering off the equipment, and possibly generates a technical alarm.

A starting fire during presence of personnel would immediately be detected by the smell and tackled by means of a fire extinguisher (to be made available). No fire detection system is currently foreseen, and a fortiori no automatic fire-fighting system. However, this aspect shall be reviewed at a more detailed risk assessment in the next project stage.

It is noted that the present fire detection system in LEIR has to be renewed once a roof covers the machine (as foreseen for operation with light ions in Stage 2, for radiation safety reasons). When this new fire detection system is implemented for LEIR, it will be useful to reassess at the same time the requirements for a fire detection system in the transfer line area, as well as in the experimental areas.

The walls, radiation shielding, mechanical structures, staircases, walkways and shielded doors have to be designed and built according to the CERN mechanical safety regulation (SR-M). In particular, design calculations have to be provided to HSE-SEE for validation.

\section{Operational safety of the experimental area}
\label{Sec:Safety_ExpArea}
The experimental area consists of three independently accessible irradiation points: H1, H2 and V, as well as a Biolab, and the shielded switchyard areas 1 and 2 (see detailed descriptions in chapter~\ref{Chap:EA}).

H2 houses propane-based gases for dosimetry development. This area needs a flammable gas detection system with alarm. A detailed risk analysis based on actual gas or gases to be used and detected, as well as on the final configuration of the area determines whether the system should generate a level 3 alarm (for personnel protection) or only a technical alarm (for equipment protection). In addition, the ventilation system might have to be ATEX (Atmosph\'eres Explosibles) rated.\\

The storage of gases must be adapted outside of the building. Control panels and gas lines have to be built according to rules. In particular, flammable-gas pipes have to be metallic, inspected and validated by HSE before being put in operation.\\

In the three irradiation areas, the main issues are radiation, electrical and fire hazards. For electricity and fire, the same mitigation measures as for the beamlines apply. In particular, during Stage 1 of BioLEIR, no fire detection system is imposed, as it would not be for personnel protection. For further facility stages, when activation of materials in the areas could become significant, a fire could have an impact on the environment. The need to implement a fire detection system must reviewed at that time.\\

The same emergency system as in the switchyards is envisaged to be implemented in the irradiation areas. The activation of one of the emergency push-buttons sends a veto on the EIS-beam via the PPS. This automatically cuts the beam in the area and generate a signal in the control room. Due to the small size of the irradiation rooms and the Biolab, no dedicated emergency system is necessary, but sirens coupled to the evacuation system of hall~150 may have to be added. Autonomous escape lighting devices shall be installed above the door of each irradiation room, as well as of the Biolab.

\subsection{Operational safety specifics of the Biolab}
A new risk potentially imported by BioLEIR users is the chemical hazard. This is obviously the main risk in the Biolab. The CERN regulation concerning chemical hazards is SR-C. In particular, dedicated chemical cabinets are provided for users in the Biolab (according to Safety Guideline C-1-0-1). Three different cabinets may be needed: one for acids, one for bases, and one for other chemicals, possibly flammable.

As it is expected that the quantities of chemicals brought to CERN are relatively low, no whole-body safety shower is foreseen. An emergency fixed eye wash is foreseen in the Biolab.

To help mitigate specific chemical hazards, the proposed experiments have to be presented (with a safety file) to a Technical Board who assesses the risks and their mitigation measures, and finally give their approval or request additional safety measures. Users have to write specific procedures concerning their brought-in hazards such as chemicals.

The biomedical research shall be done exclusively on cell work related to cancer research. Absolutely no biohazard (viruses and other) will be on site.

\section{Personnel access - the Personnel Protection System PPS}
\label{Sec:Safety_PPS}

This section introduces the principles of the Personnel Protection System (PPS) that secures the facility by means of access control and safety interlock functions.
The PPS regulates the accesses of personnel to areas where a potential hazard may be present, by means of controlled physical barriers such as access points, doors, signalization warnings, personnel identification, as well as verification of their work authorization and adequate training level for the area concerned. User identification is done via the RF-ID chip located in their personal dosimeter.

In addition to the above, the PPS is responsible for:
\begin{itemize}
\item preventing access when a hazard (radiation, electricital power etc.) is present, 
\item interlocking any potentially dangerous action (e.g. extraction of beam) when personnel is detected inside a controlled area.
\end{itemize}

The safety functions of the PPS cover the full BioLEIR accelerator chain from the ion production and transport, through LINAC3 and LINAC5, through the PS switchyard, LEIR, the BioLEIR switchyard area, and then finally the three experimental irradiation rooms. A specific zone is always secured by diversely redundant EISs located in the upstream zone. 

To ensure this objective, an update of the PPS for LINAC3, LINAC5, the PS-switchyard is required, and a new PPS is devised for the BioLEIR experimental zones.

In this document, only the characteristics of the BioLEIR experimental zone PPS are described. The outstanding PPS aspects shall be detailed in the next stage of the project.

The design of the PPS is driven by the normative context of the IEC-61508/615013 Functional Safety standards, covering all project stages from the preliminary risk analysis, the identification of all protection and mitigation barriers, allocation of their Safety Integrity Level (SIL), the testing, up to the organisation of the operation and maintenance. The system also integrates the specific nuclear safety requirements such as compliance with diversity, redundancy, common mode of failure, single failure criteria, and robustness to external aggression.

\subsection{Sectorization}
\label{SubSec:EA_PPS_Sectorization}
The accessible areas of the BioLEIR facility are divided into 3 zones, according to their radiological classification from chapter~\ref{Chap:RP}:

\begin{itemize}
\item \underline{\textbf{ZONE 1}}: this area includes Switchyards 1 and 2, where the ion beam is extracted from the LEIR accelerator and dispatched  towards the three irradiation rooms. The RP analysis(chapter~\ref{Chap:RP}) showed that this zone is the most exposed to ionizing radiation from beam loss during the extraction process.

The presence of personnel and beam in this zone are mutually exclusive. Zone 1 is fully shielded and access of specialists is regulated via a specific Access Point (AP) when the zone is safe, i.e. when both LEIR extraction EISs are in the ACCESS position. 

\item \underline{\textbf{ZONE 2}}: this area comprises the three irradiation rooms. Even though the simulations of the radiological impact inside this zone is estimated lower than Zone 1 (chapter~\ref{Chap:RP}), the presence of personnel in any of the three irradiation rooms is not allowed during the irradiation. Each room has shielding walls, a shielding roof and an access point (AP). Access to any irradiation room will is possible only when the EISs are in the ACCESS position on the corresponding beamline.

\item \underline{\textbf{ZONE 3}}: this area represents the remainder of building 150, comprising the BioLEIR counting rooms and the Biolab. 

\end{itemize}

\subsection{Access modes and operation}
\label{SubSec:Infrastructure_AccessModes}
An important strategy for the risks mitigation consists on the alternation of specific exploitation modes: 
\begin{itemize}
\item when the absence of radiological hazards is ensured, the ACCESS mode is allowed; 
\item  when the absence of personnel inside the radiological areas is ensured, by means of a PATROL mode, then BEAM mode is allowed.
\end{itemize}

The different exploitation modes are controlled by the operators via a hard-wired command console (CC), installed in the local BioLEIR counting room. The operators are responsible for the safety.

In addition to the PPS, the person responsible for the overall safety of the facility defines the population of users allowed to access and coordinates the various parallel activities, minimizing potential dangerous circumstances. 

\subsection{Main principle of protection}
\label{SubSec:Infrastructure_PPS_SIF}
This section summarizes the main principles of the BioLEIR PPS:

\begin{enumerate}
\item The BioLEIR facility is completely independent from the LEIR accelerator. The PPS ensures autonomously the safety of the facility and interlocks the LEIR accelerator EISs when needed.

\item The protection of BioLEIR is based on the following 4 safety elements:
\begin{itemize}
\item Chain 1: prevent extraction of beam from LEIR into Zone~1 by operating on a FAILSAFE beam stopper and a magnetic deflecting EIS located in the BioLEIR extraction.
\item Chain 2: prevent extraction of beam from  Zone~1 into irradiation room H1 by operating on a FAILSAFE beam stopper and a magnetic deflecting EIS located in the H1 beamline.
\item Chain 3: prevent extraction of beam from Zone~1 into irradiation room H2 by operating on a FAILSAFE beam stopper and a magnetic deflecting EIS located in the H2 beamline.
\item Chain 4: prevent extraction of beam from  Zone~1 into the vertical irradiation room V by operating on a FAILSAFE beam stopper and a magnetic deflecting EIS located in the vertical beamline.
\end{itemize}

\item The three irradiation rooms can be operated independently - one room can be in ACCESS mode while the others are in BEAM mode. This works under the condition that the beam stopper of the room in ACCESS mode remains in safe (closed) position and its magnetic EIS powered down. It is therefore possible to individually control the other two magnetic deflecting EIS. Whenever one of the safety conditions is lost, an automatic interlock acts on Chain 1, stopping beam in every irradiation room (safety retro-action).

\item Considering the proximity and the small user community, access to Zone~1 does not require a personnel safety token delivery (as it is usually done for primary beam zones), neither the presence of a sophisticated Personnel Access Device (PAD) preventing tailgating or piggybacking intrusions. However, in order to heighten security, the patrol state of Zone 1 is systematically lost during every access. The same principle of access is applied to the three irradiation rooms.

\item When the ACCESS mode is set for Zone~1 or Zone~2, the corresponding doors can be unlocked after RF-ID identification.    

\item The PPS shall interface with the ventilation system to ensure both: the ventilation flush prior to every access and the maintaining of the atmospheric depression inside all areas where air can be activated. 

\item Warning flashing lights are activated in every BioLEIR area prior to switching into BEAM mode. This is equivalent to the activation of the BIW for the primary zones. 

\item The PPS monitors local emergency buttons and interlocks all extraction chains in case of activation of one button. 

\end{enumerate}

\subsection{Resource estimate of the personal protection system}
\label{SubSec:Infrastructure_PPS_Cost}
Table~\ref{Tab:Infrastructure_PPS_Cost} shows the estimated material cost for equipment and external services provided by contractors, while table~\ref{Tab:Infrastructure_PPS_FTE} shows the estimated CERN personnel needed. 
It shall be noted that the provision and installation of the two escape doors of Zone 1 and the three access doors to the irradiation rooms are not included in the PPS cost estimation. Their cost estimate is included in the infrastructure chapter (see section~\ref{Sec:Infrastructure_Doors}).

\begin{table}[!h]
\centering
\caption{Cost estimate for the PPS material.}
\label{Tab:Infrastructure_PPS_Cost}
\resizebox{\textwidth}{!}{%
\begin{tabular}{l r r r}
\hline
\textbf{List of Material} & \textbf{Units} & \textbf{Unit cost} & \textbf{Total} \\
\textbf{} & \textbf{} & \textbf{[kCHF]} & \textbf{[kCHF]} \\
\hline
Access point frame structure  (recuperation from  PS complex) & 1 & BE/ICS & 0 \\
Access point refurbishing (painting, instrumentation) & 1 & 2 & 2 \\
Patrol boxes & 9 & 0.3 & 2.7 \\
Signaling panels & 4 & 0.4 & 1.6 \\ 
Access modes console & 1 & 2 & 2\\
Local HMI Display & 1 & 1 & 1 \\
Cables/connectors + installation & 200m & 15 CHF/m & 3 \\
Logic controller + I/O cards & 1  & 5 & 5 \\
RD-ID badge reader & 4  & 0.15 & 0.6 \\
Rack 19'' CERN standard & 1 & 1.5 & 1.5 \\
Fixation/electrical material + junction boxes & 1 & 2.4 & 2.4 \\
Installation of PPS components/connections  & 1 & 10.5 & 10.5 \\
Electrical drawings/documentation & 1 & 10 & 10 \\
PPS main rack construction/cabling & 1 & 5.7 & 5.7 \\
\hline
\rule{0pt}{3ex}\textbf{Total} & & &\textbf{48}\\
\hline
\end{tabular}
}
\end{table}

\begin{table}[!h]
\centering
\caption{Cost estimate of contracted labour.}
\label{Tab:Infrastructure_PPS_FTE}
\begin{tabular}{l r r r}
\hline
\textbf{Services provided by BE/ICS} & \textbf{FTE} \\
\textbf{} & \textbf{[PY]} \\
\hline
Safety documentation preparation (SIF, SIL allocation, architecture)  & 0.1 \\ 
PPS software prototyping/development & 0.1 \\ 
PPS software verification and validation & 0.04\\
Final site tests/commissioning  & 0.04 \\
\hline
\rule{0pt}{3ex}\textbf{Total} & \textbf{0.28}\\
\hline
\end{tabular}
\end{table}

\section{Resource estimate}
\label{Sec:Safety_Cost}
Safety considerations require installation of a personnel protection system, an RP monitoring system, a beam dump for LEIR, one beam stopper per irradiation line, and the reduced cost of five fire extinguishers (one per switchyard and one per irradiation room). No further safety expenditures are to be foreseen.

Should the users require a smoke detection system, the related cost would be born from their budget. The cost for possible additional measures to cover the chemical and biological risks is also to be covered by team budget codes.

The fire detection system to be implemented in LEIR for Stage 2, and possibly in other BioLEIR areas, would cost more than 100\,kCHF. The cost could be partially covered by the consolidation budget, an estimated 100\,kCHF is considered to be contributed by the BioLEIR project.

The cost of the flammable gas detection and alarm system depends on the result of the risk analysis. The minimum cost would be 6\,kCHF for a local system generating a technical alarm, and up to more than 60\,kCHF for an integrated system generating a level-3 alarm. For the purpose of providing a cost estimate, we use the median value of 33\,kCHF as a cost estimate for this safety aspect.

The evacuation paths and signalization shall be refurbished and is estimated to cost at the level of 10\,kCHF.\\

\begin{table}[!h]
\centering
\caption{Material cost estimate for safety aspects of BioLEIR.}
\label{Tab:Safety_Cost_Material}
\begin{tabular}{l l r r}
\hline
\textbf{Equipment/System} & Comments & \textbf{Cost} & \textbf{Total}\\
\textbf{} & \textbf{} & \textbf{[kCHF]} & \textbf{[kCHF]} \\
\hline
\rule{0pt}{3ex}\textbf{Stage 1:} & & \\
Fire extinguishers & & 2 \\
Smoke detection & risk analysis needed & 33 \\
Emergency push-button &  & 4 \\
5 emergency lighting devices &  & 6 \\
Additional sirens &  & 6 \\
Personnel Protection System  &  & 48 \\
Evacuation paths & & 10 \\
Chemical cabinets & & 10\\
Fixed eye-wash & & 10 \\
Beam dump \& stoppers & & 400 \\
   & \textbf{Total Safety Stage 1} & & \textbf{519} \\
\textbf{Stage 2:}\\
Fire detection & & 100 \\
   & \textbf{Total Safety Stage 2} & & \textbf{100} \\

\hline
\rule{0pt}{3ex}\textbf{Total Safety [kCHF]} & & &\textbf{619}\\
\hline
\end{tabular}
\end{table}

\begin{table}[!h]
\centering
\caption{Personnel estimate for design, supervision, installation and commissioning of safety related aspects for BioLEIR, given in person-years.}
\label{Tab:Safety_Cost_FTE}
\begin{tabular}{l r}
\hline
\rule{0pt}{3ex}\textbf{System} & \textbf{[PY]} \\
\hline
\rule{0pt}{3ex} PPS & 0.3 \\
overall safety coordination & 0.5 \\
\hline
\rule{0pt}{3ex}\textbf{Total Safety [FTE*year]} & \textbf{0.8}\\
\hline
\end{tabular}
\end{table}

Table~\ref{Tab:Safety_Cost_Material} summarizes the expected cost related to conventional risk mitigation that amounts to a total of 619\,kCHF of which 100\,kCHF is needed only for Stage 2 of the facility. The estimated CERN workforce (Staff and Fellows) for design, supervision and installation of safety related aspects for Bio-LEIR is given in table~\ref{Tab:Safety_Cost_FTE}.

%% file: Chapters/Planning.tex
\chapter{Installation Planning}
\label{Chap:Planning}

This chapter summarizes the foreseen installation schedule for the BioLEIR facility, taking into account the following aspects:
\begin{itemize}
	\item CERN machine schedules with respective scheduled stops that can be used to install BioLEIR elements;
    \item design and production lead times for the different elements;
    \item minimum times necessary for installing and commissioning facility elements;
    \item identification of staging scenarios, aiming to have beams of interest for biomedical research available as soon as possible;
    \item overall CERN schedules regarding the consolidation programmes and resources availability.
\end{itemize}

\section{CERN schedules until 2035}
\label{Sec:Planning_CERN}
\begin{figure}[!htb]
\centering\includegraphics[width=0.85\linewidth, height=0.37\linewidth]{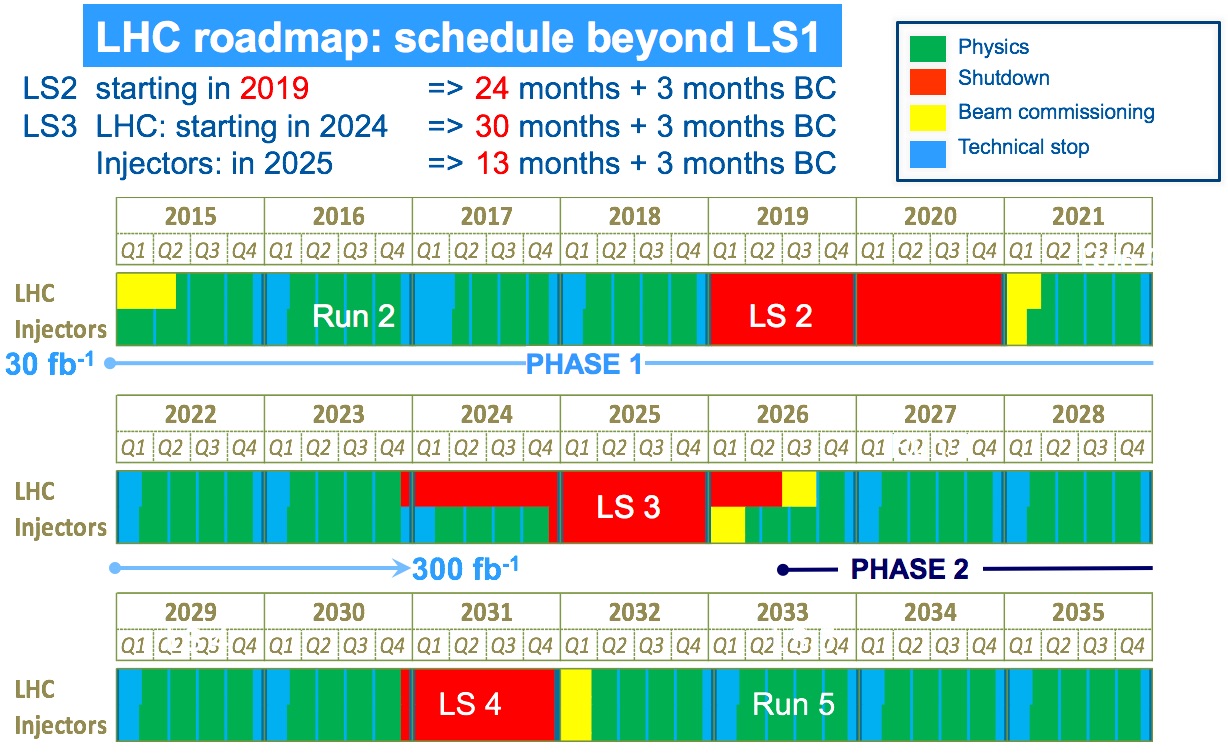}
\caption{The overall CERN machine schedule.}
\label{Fig:Planning_CERN}
\end{figure}

Figure~\ref{Fig:Planning_CERN} shows the planned schedule for the overall CERN accelerator complex. Beyond LS2 in 2019/2020, the schedule is only indicative.

\section{BioLEIR installation scenario}
\label{Sec:Planning_Sequence}
This section lists the different possible installation periods, the installation works that could be foreseen in any specific period and the resulting possible operation scenarios.\\

\noindent The following assumptions and inputs have been used to establish the proposed schedule:
\begin{enumerate}
	\item Identification of alternative space for storage of material currently stored in hall~150 and its subsequent removal.
	\item At least the slow extraction and one beamline need to be installed for the BioLEIR facility to be useful for initial biomedical studies, using light ions produced with  LINAC3.
    \item A shutdown of at least 3 months would be necessary to install all elements up to the LEIR shielding wall. The duration of a normal Year-End-Technical-Stop (YETS) would not be sufficient while an Extended-YETS (EYETS) would provide adequate time.
    \item Work on the beamlines in the experimental area (behind the LEIR shielding wall) as well as infrastructure work for the experimental area is possible while LEIR is running for the CERN Heavy Ion programmes.
    \item Recent experience in beamline design, procurement, installation and commissioning of HIE-ISOLDE showed that the minimum time from project start to fully commissioned state is 3.5\,years. Some time could be gained by using in-kind material contributions from collaborating institutes.
	\item Construction of the new frontend (light ion source, RFQ and LINAC5) can proceed in parallel with the normal CERN accelerator operation.
    \item Connection of LINAC5 to the LEIR injection transfer line, as well as necessary modifications to the transfer line, can take place during a normal YETS.
	\item The upgrade of the LEIR power converters can be achieved during a normal YETS, provided it is well prepared.
    \item Installation of a roof over LEIR can take place during a normal YETS, before LINAC5 ions are injected.
\end{enumerate}

Combining the above assumptions and inputs together with the overall CERN schedule results in the following BioLEIR installation and operation scenario. It is mapped to 3 facility stages:\\

\noindent \textbf{Preparation phase (2017-2020)}:\\

\vspace*{-2mm}
\noindent \textit{2017 Run period}
\vspace*{-2mm}
\begin{itemize}
	\item Profit from the planned Xenon run to gain experience with Oxygen in LEIR, as Oxygen is used as support gas for Xenon (and Lead) operation. A few weeks of a dedicated Oxygen run shall be foreseen.
    \item The new light ion source could be procured as soon as funds are available. Source characterization on a test stand would provide useful input for the optimization of the LINAC5 design.
\end{itemize} 

\noindent \textit{YETS 2017/2018 (2 Months)}
\vspace*{-2mm}
\begin{itemize}
	\item No BioLEIR-related work is foreseen as there is not enough time to complete design and procurement of equipment to be installed.
\end{itemize}

\noindent \textit{2018 Run period}
\vspace*{-2mm}
\begin{itemize}
	\item Study beam dynamics in the LEIR machine, no modifications to the LEIR machine are needed.
	\item Empty and prepare the hall~150 for the BioLEIR experimental area.
\end{itemize} 

\noindent \textit{LS2 (2019/2020) (18 Months)}
\vspace*{-2mm}
\begin{itemize}
	\item Install the slow extraction system.
	\item Modify the shielding wall between LEIR and the new experimental area.
	\item Install the 3 beamlines (2 horizontal and 1 vertical)
	\item Install equipment in the new experimental area.
	\item Separate LINAC2 from the PS complex to allow for dismantling of LINAC2 and installation of LINAC5, while the accelerator complex is running.
\end{itemize}


\noindent \textbf{Stage 1 (2021-2022): Running with ions from LINAC3 up to 250\,MeV/u}\\

\vspace*{-2mm}
\noindent \textit{2021 Run period}
\vspace*{-2mm}
\begin{itemize}
	\item Operate BioLEIR using LINAC3 as injector, producing the lightest ion species accessible (Argon, Oxygen).
	\item Dismantle LINAC2.
    \item Install the new light ion source.
\end{itemize}

\noindent \textit{YETS 2021/2022 (2 Months)}
\vspace*{-2mm}
\begin{itemize}
	\item Continue dismantling LINAC2.
\end{itemize}

\noindent \textit{2022 Run period}
\vspace*{-2mm}
\begin{itemize}
	\item Operate BioLEIR using LINAC3 as injector, producing the lightest ion species accessible (Argon, Oxygen).
	\item Install LINAC5 in the LINAC2 tunnel.
\end{itemize} 

\noindent \textit{YETS 2022/2023 (2 Months)}
\vspace*{-2mm}
\begin{itemize}
	\item Connect LINAC5 to LEIR, requiring work in the PS Switchyard.
	\item Include LINAC5 into the access control and safety system of the PS complex.
\end{itemize} 


\noindent \textbf{Stage 2 (2023- ): Running with ions from LINAC5 up to 250\,MeV/u}\\

\vspace*{-2mm}
\noindent \textit{2023 Run period}
\begin{itemize}
	\item Operate BioLEIR using LINAC5 with all light ion species available.
\end{itemize} 

\noindent \textit{YETS 2023/2024 (2 Months)}
\vspace*{-2mm}
\begin{itemize}
	\item Deploy the LEIR power converter upgrade.
\end{itemize}


\noindent \textbf{Stage 3 (2024- ): Running with ions from LINAC5 up to 440\,MeV/u}\\

\vspace*{-2mm}
\noindent \textit{2024 Run period}
\begin{itemize}
	\item Operate BioLEIR using LINAC5 with all light ion species available up to an ion energy of 440\,MeV/u.
\end{itemize} 
\vspace*{-5mm}
\noindent\rule{6cm}{0.4pt}
\vspace{5mm}

Staging of the 3 experimental beamlines was considered with the goal to provide beam as early as possible to BioLEIR. However, the construction of the experimental beamlines is not on the critical path.

\begin{figure}[ht]
\begin{adjustbox}{addcode={\begin{minipage}{\width}}{
\caption{The overall preliminary BioLEIR schedule.
\label{Fig:Planning_BioLEIR_new}
}\end{minipage}},rotate=90,center}
\centering\includegraphics[trim={2mm 2mm 2mm 2mm}, width=1.5\linewidth]{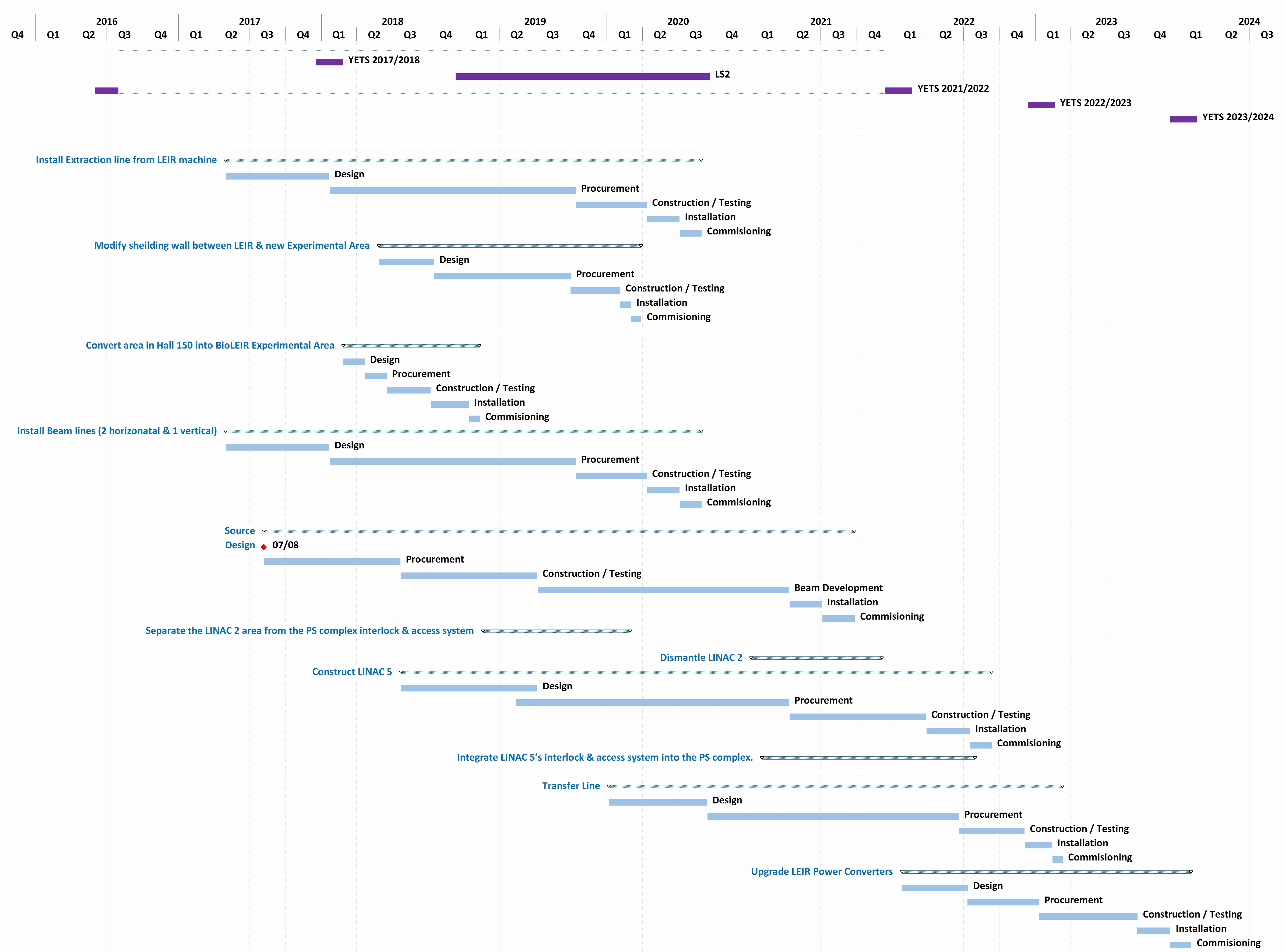}
\end{adjustbox}

\end{figure}

Figure~\ref{Fig:Planning_BioLEIR_new} shows a preliminary schedule for the preferred installation scenario for BioLEIR. Assuming a project start without delay, with funds and experienced resources committed by mid-2017 at the latest, Stage 1 of BioLEIR could come on-line in 2021, after LS2. Stage 2 of BioLEIR can be operational 2 years after LS2, in 2023. Finally, Stage 3 of the facility can be made available any time when LEIR is not running for a short period of time (e.g. during a Year End Technical Stop).

Should the start date be postponed to later than mid-2017, the project will miss the window for installation of the slow extraction in LEIR during LS2. Based on the current CERN schdules, the next possible installation window would be LS3 (2025) with first beams to BioLEIR in 2026 - at which time the facility could come on-line with Stage 1, Stage 2 and Stage 3 directly.\\


%% file: Chapters/Cost.tex
\chapter{Resource Estimate}
\label{Chap:Cost}

This chapter brings all subsystem cost estimates together to arrive at a global cost and personnel estimate at the engineering level. It must be noted that at this stage of the study the cost estimate cannot be better than 20-30\%. In some areas, the uncertainty is at the level of 50\%.

Table~\ref{Tab:Cost_Material} summarizes the estimated material cost for BioLEIR together with their uncertainties, whereas table~\ref{Tab:Cost_FTE} summarizes the projected integrated workforce needs in person-years to realize the facility.
Please refer to the individual technical chapters for details on cost estimates and/or technical decisions.
We add 1.5~FTE per year for project management, including technical coordination, schedule and costing, for the duration of the project implementation of about 7 years. These are included in table~\ref{Tab:Cost_FTE}.

\begin{table}[!htb]
\centering
\caption{Preliminary estimates of the material cost needed for the design, construction and furbishing of the BioLEIR facility. The numbers are rounded to 100\,kCHF.}
\label{Tab:Cost_Material}
\begin{tabular}{l r c}
\hline
\rule{0pt}{3ex}\textbf{System} & \textbf{[kCHF]} & \textbf{Uncertainty}\\
\hline
\rule{0pt}{3ex}Ion source baseline & 1100 & \\
LINAC5 & 7000 & \\
Transferline to LEIR & 1800 & \\
LEIR power converter upgrade & 700 & \\
Slow extraction & 1200 & \\
BioLEIR beamlines & 6100 & \\
Experimental Area & 800 & \\
Infrastructure & 5100 & \\
Vacuum & 3600 &\\
Radiation protection & 400 & \\
Controls system & 300 & \\
Safety system & 600 & \\
\hline
\rule{0pt}{3ex}\textbf{Total [kCHF]} &\textbf{28700} & 30\%\\
\hline
\end{tabular}
\end{table}

\begin{table}[!htb]
\centering
\caption{Preliminary estimates of integrated workforce needs (in person-years) for the design, procurement, construction, testing, installation and commissioning of the BioLEIR facility. The numbers are rounded to the closest integer.}
\label{Tab:Cost_FTE}
\begin{tabular}{l r c}
\hline
\rule{0pt}{3ex}\textbf{System} & \textbf{[PY]} & \textbf{Uncertainty}\\
\hline
\rule{0pt}{3ex}Ion source & 7 &\\
LINAC5 & 23 &\\
Transferline to LEIR & 8 &\\
LEIR power converter upgrade & 2 &\\
LEIR beam dynamics studies & 5 &\\
Slow extraction & 9 &\\
BioLEIR beamlines & 21 &\\
Experimental area & 2 &\\
Infrastructure & 10 &\\
Vacuum & 9 &\\
Radiation protection & 4 &\\
Operations & 2 &\\
Controls system & 5 &\\
Safety system & 1 &\\
Project management & 11&\\
\hline
\rule{0pt}{3ex}\textbf{Total [Person-Years]} &\textbf{119}&10-15\%\\
\hline
\end{tabular}
\end{table}

The BioLEIR facility is planned to come online in stages (see section~\ref{Sec:Intro_ProjectOverview}), thereby increasing the capabilities and flexibility of the facility, as well as its research possibilities gradually with time. The staging aspect also eases the investment for the facility which is expected to be spread over 5-7 years. Tables~\ref{Tab:Cost_Material_Staging} and~\ref{Tab:Cost_FTE_Staging} show the material and personnel investment needed for each of the planned stages respectively.

\begin{table}[!htb]
\centering
\caption{Mapping of the material investment to each of the 3 facility stages. The numbers are rounded to 100\,kCHF.}
\label{Tab:Cost_Material_Staging}
\begin{tabular}{l l r r}
\hline
\rule{0pt}{3ex} \textbf{Stage} & \textbf{Element} & & \textbf{Stage Cost} \\
\textbf{} & & \textbf{[kCHF]} & \textbf{[kCHF]}\\
\hline
\rule{0pt}{3ex} \textbf{Stage 1:} & & & \\
 & Slow extraction & 1200 & \\
 & BioLEIR beamlines & 6100 & \\
 & Experimental area & 800 & \\
 & 70\% of Infrastructure (- PC upgrade) & 3300 & \\
 & Vacuum & 2500 & \\
 & Radiation protection & 400 & \\
 & Controls system & 100 & \\
 & Safety system & 500 & \\
 & & \underline{\textbf{Total Stage 1}} & \textbf{14900}\\
\textbf{Stage 2:} & & & \\
 & Ion source baseline & 1100 & \\
 & LINAC5 & 7000 & \\
 & Transferline to LEIR & 1800 & \\
 & Controls system & 200 & \\
 & 30\% of infrastructure (- PC upgrade) & 1400 & \\
 & Vacuum & 1200 & \\
 & Safety system & 100 & \\
 & & \underline{\textbf{Total Stage 2}} & \textbf{12800}\\
\textbf{Stage 3:} & & & \\
 & LEIR power converter upgrade & 700 & \\
 & Associated EL \& CV & 100 & \\
 & Associated shielding & 200 & \\
 & & \underline{\textbf{Total Stage 3}} & \textbf{1000}\\
\\
\hline
\end{tabular}
\end{table}

Stage 1 is estimated to cost about 15\,MCHF and 61\,person-years and includes a new slow extraction from LEIR, as well as all three new beamlines and the full corresponding experimental area. In this stage, the lightest ions accessible with LINAC3 are available to BioLEIR during about 4 months per year.

Stage 2 is projected to cost around 13\,MCHF and 55\,person-years and will have a new light ion injector chain that can produce the full range of light ions requested by the biomedical community.

During Stage 1 and 2, the ion energies will be limited to 250\,MeV/u.

Stage 3 will allow ion energies up to maximum values of 440\,MeV/u and cost an additional 1\,MCHF and about 3 person-years.

The cost for radiation shielding, magnets, power converters, as well as electrical power and cooling requirements scale with the maximum ion energies requested. We assume maximum ion energies of 440\,MeV/u, which is consistent with the realisation of the facility up to and including Stage 3. If the decision can be made upfront to limit the maximum ion energies to 250\,MeV/u, and to forego Stage 3 entirely, additional savings beyond the cost of Stage 3 itself, could be realised for Stage 1 and Stage 2 by reducing the scope for radiation shielding, magnets, power converters, electrical power and cooling systems.
The current rough estimate of this cost reduction for reduced project scope is of the order of 20\% of the total project cost.

\begin{table}[!htb]
\centering
\caption{Mapping of the number of person-years for realisation of the 3 stages. The numbers are rounded to the closest integer.}
\label{Tab:Cost_FTE_Staging}
\begin{tabular}{l l r r}
\hline
\rule{0pt}{3ex} \textbf{Stage} & \textbf{Element} & & \textbf{Personnel/Stage} \\
\textbf{} & & \textbf{[PY]} & \textbf{[PY]}\\
\hline
\rule{0pt}{3ex} \textbf{Stage 1:} & & & \\
 & LEIR beam dynamics studies & 2 \\
 & Slow extraction & 9 & \\
 & BioLEIR beamlines & 23 & \\
 & Experimental area & 2 & \\
 & 70\% of infrastructure & 7 & \\
 & Vacuum & 5 & \\
 & Radiation protection & 4 & \\
 & Operations & 1 & \\
 & Controls system & 2 & \\
 & Safety system & 1 & \\
 & 60\% of project management & 7 & \\
 & & \underline{\textbf{Total Stage 1}} & \textbf{61}\\
\textbf{Stage 2:} & & & \\
 & LEIR beam dynamics studies & 3 \\
 & Ion source baseline & 7 & \\
 & LINAC5 & 23 & \\
 & Transferline to LEIR & 8 & \\
 & Controls system & 3 & \\
 & 30\% of infrastructure & 3 & \\
 & Vacuum & 4 & \\
 & Operations & 1 & \\
 & 30\% of project management & 3 & \\
 & & \underline{\textbf{Total Stage 2}} & \textbf{55}\\
\textbf{Stage 3:} & & & \\
 & LEIR power converter upgrade & 2 & \\
 & Associated EL \& CV & $<$0.5 \\
 & Associated shielding & $<$0.5 & \\
 & 10\% of project management & 1 & \\
 & & \underline{\textbf{Total Stage 3}} & \textbf{$<$4}\\
\\
\hline
\end{tabular}
\end{table}

Table~\ref{Tab:Cost_Maintenance} summarizes the estimates of the maintenance material cost and personnel requirements (in FTE) per year where it cannot be absorbed in normal machine operations duty.\\

\begin{table}[!h]
\centering
\caption{Preliminary estimate of the yearly material and personnel cost for maintenance and operation of the facility.}
\label{Tab:Cost_Maintenance}
\begin{tabular}{l r r r r}
\hline
\rule{0pt}{3ex} \textbf{System} & \multicolumn{2}{c}{\textbf{Material [kCHF]}} & \multicolumn{2}{c}{\textbf{Personnel [FTE]}}\\
\textbf{} & \textbf{Stage 1} & \textbf{Stage 2/3} & \textbf{Stage 1} & \textbf{Stage 2/3}\\
\hline
\rule{0pt}{3ex}Ion source & - & 250 & - & 0.7\footnote{averaged over 10 years of operation} \\
Frontend \& linac & - & 370 & - & 1 \\
LEIR injection line & - & - & - & - \\
LEIR \& beam dynamics & - & - & 1.2 & 1.2 \\
Extraction & 70 & 70 & 0.3 & 0.3 \\
Beamlines & 300 & 300 & 1 & 1 \\
Experimental area & 40 & 40 & 1 & 1 \\
Infrastructure & 30 & 50 & 0.1 & 0.1 \\
Vacuum & 35 & 35 & 0.5 & 0.5 \\
Radiation protection & - & - & 0.5 & 0.5 \\
Operations & - & - & 0.5 & 4 \\
Controls system & 15 & 30 & 0.1 & 0.2 \\
Safety & 20 & 30 & 0.1 & 0.2 \\
\hline
\rule{0pt}{3ex}\textbf{Total [kCHF]} & \textbf{510} & \textbf{1175} & \textbf{5.3} & \textbf{10.7}\\
\hline
\rule{0pt}{3ex}$^1$averaged over 10 years of operation\\
\end{tabular}
\end{table}

It is instructive to compare the cost of the BioLEIR facility with the cost of an equivalent biomedical facility if it were built from green-field elsewhere. A survey of clinical facilities in Europe that are capable of providing different ions and protons found a construction cost at a conservative level of 140\,MCHF (excluding personnel cost).
In order to account for the higher cost of a clinical facility, we scale this cost down by about 30\% to a level of 100\,MCHF needed to establish a centre dedicated to R\&D with a scientific outcome comparable to that which can be reached with BioLEIR.\\

Constructing BioLEIR at CERN for 28.7\,MCHF and 119\,person-years therefore represents a significant cost saving through the re-use of existing CERN infrastructure, as well as the LEIR synchrotron and buildings that are already available at CERN.\\

%% file: Chapters/Risk.tex
\chapter{Risk Assessment}
\label{Chap:Risk}

This chapter lists and describes all identifiable risks for the BioLEIR project, as it is known and defined in this document. Details of proposals and ideas to mitigate a particular risk are elaborated where this information is already available. For some areas, we can only establish a list of issues at this point. 

This chapter focuses on major unlikely events that entail high consequences at low probability of occurrence. 

We identified the following areas and types of potential risk:
\begin{enumerate}
	\item Safety of personnel and of the installation 
  	\item Schedule
  	\item Cost
    \item Undesired beam characteristics
    \item Application of new technologies
  	\item Infrastructure and integration
\end{enumerate}

\section{Safety}
\label{Sec:Risk_Safety}
The risks pertaining to the safety of personnel and of the installation are  essentially: exposure to ionizing radiation, fire or explosion from flammable gases, chemical hazards, electrical hazards, environmental risks, heavy handling.  

They are generally well understood at CERN and appropriate mitigation measures can be implemented as an intrinsic part of the project. The corresponding mitigation actions are discussed in detail in the chapters on safety (chapter~\ref{Chap:Safety}),  radiation protection (chapter~\ref{Chap:RP}) and infrastructure (chapter~\ref{Chap:Infrastructure}).

A detailed risk analysis, particularly regarding beam safety aspects for the full accelerator chain, is required and must be undertaken in the next stage of the project. 

We have identified a risk of biological contamination which is unknown at CERN at this time.  This risk is however well known in the biomedical field and in hadron-therapy clinical facilities in particular. A full risk analysis will be done also for biological contamination and mitigation measures elaborated and implemented. 


\section{Schedule}
\label{Sec:Risk_Schedule}
Several aspects in this study assume best-case scenarios. A few additional detailed studies have been suggested in chapter~\ref{Chap:PointsFlagged}. They might reveal technical issues that may require more work and potentially a schedule delay.
In particular, the risk that the ageing infrastructure that BioLEIR plans to use, would require an extensive overhaul or refurbishment, entails not only a budgetary risk, but also a considerable schedule risk.

The biggest risk to the proposed schedule is the availability of the committed funds and experienced resources by mid-2017, in order to have first beam to BioLEIR after LS2 in 2021. CERN experienced staff are already committed to CERN's core missions and cannot take on a project of the magnitude of BioLEIR. Therefore, CERN relies on the in-sourcing of experienced staff for this project. 
Also, even the training phase of newly hired or in-sourced inexperienced personnel would certainly delay the project delivery date.

The CERN schedule that we have taken into account to establish the BioLEIR installation and commissioning schedule is based on hypotheses that are regularly reconsidered and changed. Hence the BioLEIR installation schedule could also benefit from a change in the CERN schedule that would allow some installation to be carried out between LS2 and LS3. 

\section{Cost}
\label{Sec:Risk_Cost}
The cost of the facility has been evaluated in a very short time with minimal resources. In most areas, high-level engineering estimates have been given and no detailed studies have been performed. This aspect is reflected in the large uncertainty of the cost of the order of $\pm$30\%.

Since BioLEIR relies largely on existing infrastructure, there is a possibility that certain infrastructure and integration aspects have not been included in the cost estimates. In particular, we assumed that buildings and infrastructure require only small scale modifications, and no large infrastructure investment has been considered. A more detailed study of infrastructure and integration will allow better identification and possible mitigation of those risks.

\section{Undesired beam characteristics}
\label{Sec:Risk_BeamCharacteristics}
It is essential to have accurate knowledge of the beam characteristics all along the irradiation process such that irradiation with unwanted beam (i.e. incorrect energy, incorrect intensity, incorrect ion species) can be avoided.

Since LEIR will be used for both, heavy ions for the North Area and LHC physics programmes as well as light ions for BioLEIR, an unwanted event would be the accidental extraction of heavy ions towards the BioLEIR experimental beam-lines. At worst the heavy ion beams could actually hit the biomedical samples. This type of risk is present at other places in the CERN accelerator chain and mitigation measures can be identified and implemented.

\section{Application of new technologies}
\label{Sec:Risk_NewTechnologies}

In general there are  very few risks linked to the application of new technologies  for BioLEIR. One of the main design criteria was to reuse proven and tested technologies in order to ensure  minimum cost, maximum operability and availability for beams to the experimental areas. 

However, in the case of LINAC5 we have chosen to use the technology of permanent magnet quadrupoles (PMQ) for beam focusing in the linac structures. Although PMQs have already been used in linacs \textemdash very notably in LINAC4 which is being commissioned at CERN \textemdash we have no evidence that they have been used for a linac that accelerates different types of particles and therefore should nominally require tuning of the optics.  Preliminary studies show that the stability of the optics is ensured with PMQs for all ions requested, but additional studies are required. The resistance of PMQ material to radiation is also identified for detailed study in chapter~\ref{Chap:PointsFlagged}.

\section{Infrastructure and integration aspects}
\label{Sec:Risk_Integration}

\subsection{Ageing infrastructure}
This feasibility study assumes that building 150 and the LINAC2 area can be re-used with only small investments into the electrical, cooling and piping infrastructures, as well as the structure of the buildings.

\pagebreak
However, due to the age of the infrastructure, the risk exists that more investment has to be considered. In particular, we are aware that the following aspects shall be examined in a more detailed study:
\begin{itemize}
\item adequacy of electrical distribution and compliance 
with modern safety standards;
\item adequacy of current LINAC2 tunnel ventilation (AHU);
\item necessity of refurbishment of cabling and water piping for the LINAC2 tunnel and building 150;
\item rain water leakage from roofs;
\item asbestos in areas that need to be repaired or modified.
\end{itemize}
The above aspects are not part of the current CERN consolidation project.


\subsection{Availability of hall~150}
Before any new facility can be installed in building~150, the material stored in the hall needs to be evacuated. The material is slightly radioactive. A new and appropriate storage area for this material needs to be identified or constructed. This represents potentially a schedule risk for BioLEIR.

\subsection{Concrete radiation shielding in switchyard}
The radiation protection shielding for the BioLEIR switchyard structures should be built using precast concrete blocks. However, due to space constraints in the hall, the structures are of non-standard shapes with relatively large spans, and are to be built in areas with limited access and with very thick roof slabs for radiation shielding. Building with precast concrete blocks requires support structures to uphold a heavy concrete or cast-iron roof. A full static calculation should confirm whether this solution is adequate. 

Alternatively,  the structures could be built from reinforced concrete that would be cast in-situ, similar to the construction of a building.
Several risks are related to pouring concrete in-situ: 
\begin{itemize}
\item a cost increase of about 330\,kCHF for the construction of the switchyard structures.;
\item an increase in construction time and potential delay of the project.;
\item modifications to fixed concrete shielding structures are difficult and costly;
\item deconstruction and proper disposal of the irradiated concrete shielding at the end of life of the facility is also difficult and costly.
\end{itemize}

%% file: Chapters/PointsFlagged.tex
\chapter{Points Flagged for Further Study and Investigation}
\label{Chap:PointsFlagged}

This chapter lists and describes some topics that have been raised during discussions in the preparation of this document. These specific topics require significant ad-hoc studies, potential Research and Development activities and specific measures that shall be investigated and considered when the study becomes a project.

\section{Beam Requirements}
\label{Sec:PointsFlagged_BeamRequirements}
The requirements for the BioLEIR beams shall be reviewed and refined with the biomedical community, taking due account of the various upstream instruments and techniques being used to produce the beams. Indeed, some of these techniques might have an adverse effect on some beam parameters and we need guidance and validation of the balance between different requirements, including ease of operation. For example, the requested beam purity of 10$^{-4}$ might be adversely affected by any range shifter placed into the extraction line. 

\section{Machine developments in 2017 and 2018}
\label{Sec:PointsFlagged_MD}
In order to benchmark the expected radiation levels during the running of LINAC5 for BioLEIR, a radiation measurement campaign should be scheduled in 2017 or 2018 during the running of LINAC3 with Oxygen and Lead ions. 

In order to better understand the beam dynamics with light ions in LEIR, we recommend to initiate in 2017 a series of machine development studies on LEIR. 
A non-exhaustive list of studies would include: 
\begin{itemize}
	\item detailed characterisation of the LINAC3 Oxygen beam;
    \item injection trajectories and efficiencies, and correction;
    \item optimisation of the acceleration cycle with minimal losses;
    \item transverse beam dynamics;
    \item cooling rates with and without electron cooler.
\end{itemize}
    
\section{Frontend and linac}
\label{Sec:PointsFlagged_Linac}
The choice of PMQs within the Quasi-Alvarez high-energy accelerating modules reduces the power consumption and complexity of the linac since there is no feed-through for electrical power or cooling channels. However, the PMQs will have to withstand significant radiation levels from particle losses and the materials used in the  production of the PMQs should be validated for radiation hardness.

\section{Transfer line from LINAC5 to LEIR injection}
\label{Sec:PointsFlagged_TL}
A very preliminary analysis of the injection transfer beamline has been carried out. However, no actual detailed design could be produced in the time frame of this study. The baseline design (see chapter~\ref{Chap:Transferline}) of the injection transfer line from LINAC5 to LEIR includes a short section ($<$ 20\,m and about 6 magnets) which should be shared between beams coming from LINAC4 and from LINAC5, before they branch respectively towards the PS Booster and LEIR. This situation may present operational limitations due to the necessity to switch magnet polarity between LINAC4 H$^-$ ions and LINAC5 positive ions. 

The design should be refined from a technical and operational point of view. The option to have completely separated beamlines for LINAC4 on the one hand, and LINAC5 on the other hand, should be studied, although the integration in a small tunnel presents some challenges. 

Further downstream, the recommissioning of the beamline connecting to the loop used for injection into LEIR requires prior field characterisation of the dipoles and detailed analysis of the beam transport through the fringe field. This factor will be a driver in the optics performance.

\section{Beamlines to the BioLEIR experimental area}
\label{Sec:PointsFlagged_Beamlines}
Apart from the element design and the associated integration, the main unresolved technical issue appears to be the required uniformity at the target, which is strongly linked to the collimation and beam purity which can be expected. This needs design and calculations, for both the achievable beam purity and for the beam loss and activation aspects. Further investigations of the tail-folding with non-linear lenses and alternatives such as beam scanning and wobbling systems also need to be studied. 
f

%% file: Appendices/appendix_Introduction.tex
\chapter{Biomedical motivation for a biomedical research facility hosted at CERN}
\label{App:Intro_A}
{\centering \textit{Bleddyn Jones}

}

CERN is internationally recognised for probing the nature of matter and its interactions, by studying the behaviour of sub-atomic particles, and for playing a major role in establishing the most precise physics model available to us today - the Standard Model of particle physics. Fundamental research in particle physics needs powerful colliders excelling at the energy frontier with ever higher collision rates. CERN is providing accelerators that supply the global particle physics research community with reliable, performance-record-breaking data intensities. It is thanks to particle physics' unique way of collaborating globally, and with the generous support of many governments, that the impressive advances of particle physics in the past 50 years were possible. For the next fifty years, experiments on  fundamental aspects of physical systems at the smallest scales are likely to continue, with opportunities to make more discoveries that provide a deeper understanding of our universe.

An added perspective for future basic research would be to concretely investigate how the technological advances needed for particle physics - most notably accelerator technology, particle detector equipment and IT expertise - might create a positive impact in other fields. In particular, there is interest to use CERN's core technological competencies to improve medical diagnosis and therapy, as well as contribute to better radiation protection in its various applications.

To understand why this is useful, a brief description of the evolution of radiation physics based applications in medicine is given, with insights into radiation biology and how better knowledge of fundamental interactions could lead to improved clinical outcomes. In so doing, the evolution of detector technology may lead also to better diagnostic tools for clinicians.

\section{Brief history of medical applications that followed advances in radiation physics}
\label{App:Intro_Motivation_History}
The discovery of X-rays and natural radioactivity were soon followed by their application in medicine. Diagnostic imaging became widespread within a few decades, and therapeutic advances in cancer irradiation, as an alternative to surgery, became increasingly important and remain so. Technical developments, notably the achievement of higher energies and tissue ranges, improvements in dosimetry and computational interactions yielded further advances in medical diagnosis, including nuclear medicine (radioisotope emissions), computerised tomography (CT scans), positron emission applications, as well as imaging advances from other branches of physics (Magnetic Resonance Imaging (MRI) and ultrasonics).

Naturally occurring radioisotopes were used in an increasingly controlled manner with excellent results when the radioactivity could be concentrated within a cancer. The use of rare and manufactured isotopes and various methods of enhancing their localisation continue to be developed. The best example is thyroid cancer, which although rare, has distinct subtypes each with different propensities to trap iodine. Cure rates are very high for the avid iodine trapping cancers, even when these have spread to distant parts of the body. Sealed radioisotope sources also became important for applications where the source could be placed within or close to many forms of primary cancers and remain in use. The present status and potential role for radiobiological modelling in "nuclear medicine" is considered in another section below. 

External radiation beams became increasingly dependent on electron linear accelerators to produce megavoltage X-rays, after the potential of the cavity magnetron had been fully realised, and to an extent that cyclotron accelerators were only rarely used, for example in fast neutron clinical studies. More recently, there has been an expansion in cyclotron and synchrotron acceleration to deliver protons and light ions for cancer therapy. These modalities capitalise on the Bragg peak effect so that with good energy selection coupled with detailed imaging, energy deposition can be better restricted to the volume of a cancer and its immediate surroundings compared to X-ray/photon beams.

Both radioisotopes and charged particle beams rely on preferential radiation dose distribution, respectively due to inverse square dose fall-off with distance, and the Bragg peak effect. 

Following advances in radiation dosimetry\footnote{Therapeutic radiation dosimetry is specified by Fluence (tracks per unit area), KERMA (energy released per unit mass), Absorbed dose (energy retained per unit mass) and LET (energy released per unit distance in the forward track direction).}, research studies showed that the biological effects of radiation varied not only with dose but also the dose rate, the degree in which the dose can be split in time (fractionation), the chemical environment of the cells (some chemicals protecting and others sensitising radiation by influencing the yield of free radicals), and also the "quality" of the radiation. The latter refers to the linear energy transfer (LET) characteristics of a radiation, which depends not only on the nature of the radiation (e.g. photon or hadron), but also its energy (lower energies confer higher LET) and nuclear charge. 

\begin{figure}[!hbt]
\centering
\includegraphics[width=0.9\columnwidth]{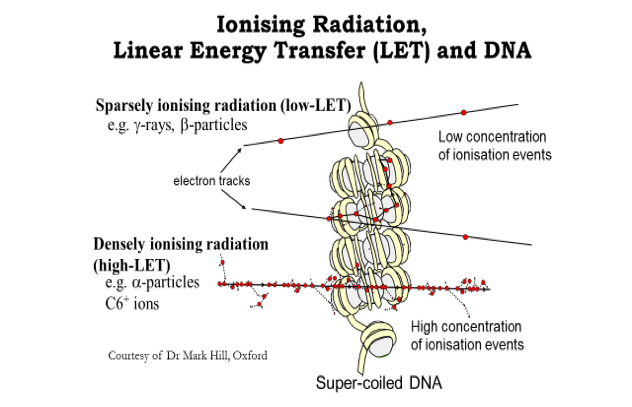} 
\caption{Diagram of radiation tracks of different LET values superimposed on supercoiled-DNA. The ionisation events are shown as red dots.}
\label{Fig:Intro_LET_DNA}
\end{figure}

\begin{figure}[!hbt]
\centering
\includegraphics[width=0.9\columnwidth]{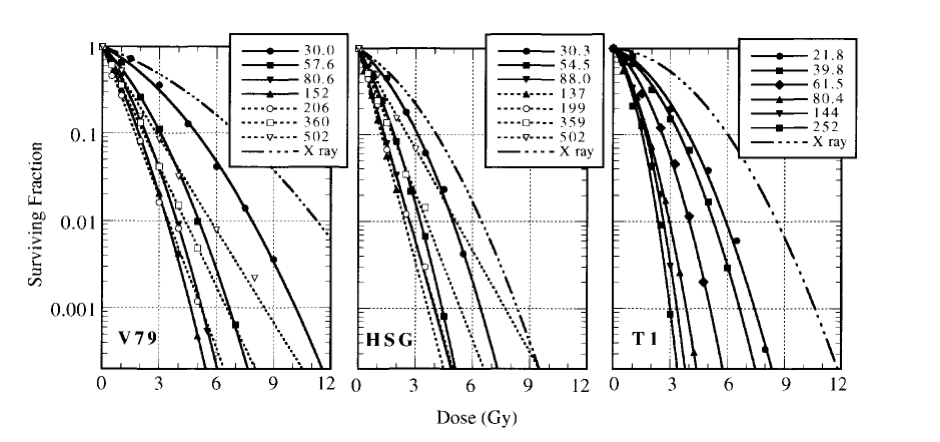} 
\caption{Examples of cell survival data expressed as the Surviving Fraction (SF) from the data of Furusawa \textit{et. al.} 2000~\cite{Furusawa:2000}, for three cell types (V79, HSG and T-1) exposed to increasing doses of Carbon ions at different LETs in [keV/$\mu$m]. The curves and lines are fitted with a Poisson-distribution, for no lethal events per cell being $SF=exp[-\alpha$d -- $\beta$d$^2$], where $d$ is the dose. Also included are the X-ray (control) curves which shows the most inefficient cell kill. The doses for a horizontal iso-effect for all LET values can be found. The control dose provides the numerator, and any higher LET dose the denominator to yield the relative biological effectiveness (RBE).}
\label{Fig:Intro_LET_CellSurvival}
\end{figure}

Increasing the LET results in more clustered DNA damage (see figure~\ref{Fig:Intro_LET_DNA}) which is less amenable to cellular DNA repair mechanisms, so that biological killing becomes more efficient, until eventually inefficiency dominates due to over-localisation of ionisation within a cell nucleus (an overkill effect). These effects can be seen in terms of cell survival in figure~\ref{Fig:Intro_LET_CellSurvival}.

\begin{figure}[!hbt]
\centering
\includegraphics[trim={0 0 0 0},clip, width=0.45\columnwidth]{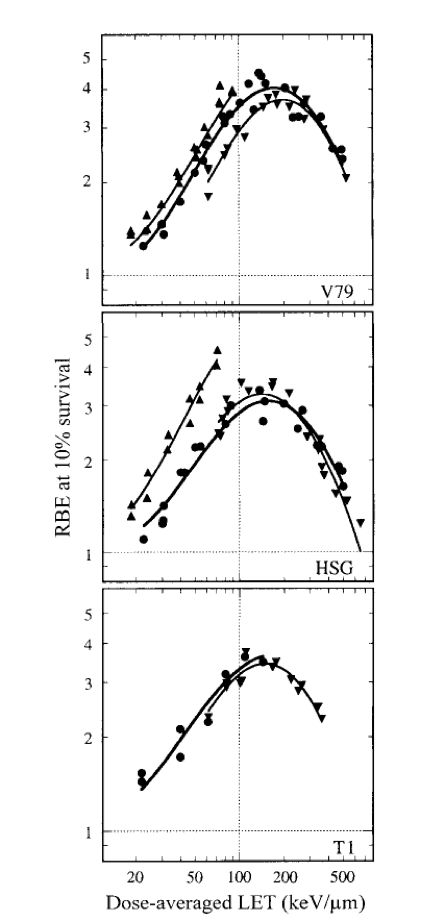} 
\caption{Distribution of RBE at the 10\% survival level under aerobic conditions for V79, HSG and T1 cells as a function of the dose-averaged LET. The cells were exposed to Helium ($^3$He: $\blacktriangle$), Carbon ($^{12}$C:$\bullet$) and Neon ($^{20}$Ne: $\blacktriangledown$) ions. RBE values were obtained from D$_{10}$ values calculated from $\alpha$ and $\beta$ values by a curve-fitting method. All data are the ratio to survival after irradiation with 200~kVp X-rays under aerobic conditions.}
\label{Fig:Intro_RBE_Changes_With_LET}
\end{figure}

The resultant increase in cell killing can be expressed as the relative biological effect (RBE\footnote{RBE is the ratio of dose required for the same bioeffect using two different qualities of radiation, expressed as dose of the lower LET radiation divided by the dose of the higher LET radiation. RBE, in most practical circumstances will be greater than one.}) which is non-linearly related to LET and dose, and is further modified by the type of cells or biological system being irradiated. The RBE changes with increasing LET (relative to the control low LET radiation) are shown in figure~\ref{Fig:Intro_RBE_Changes_With_LET}.

At the present time there are at least five variants of LET definitions in use; it remains to be decided which is best applied to biological work. 
So far, the RBE concept has not been used within the medical prescription process for radioisotopes, but is used to determine the clinical radiation dose in hadrontherapy, although there is controversy as to what value of RBE should be used: should it be a constant value as used for proton therapy worldwide, or an LET-dependent variable? Imprecise knowledge of RBE can result in an inappropriate dose being delivered to a patient. Whereas physical dose from an accelerator is subject to legal monitoring and should be within 2\% accuracy, the RBE variation exceeds this limit often by more than an order of magnitude in percentage terms.

The above is the reason why improved LET detectors and Monte Carlo simulations at lower energies would be worthy research goals at the BioLEIR facility. Improving our knowledge of RBE and its prediction from LET, dose, and perhaps other beam characteristics such as average inter-track distances, and its relation to radiobiological characteristics connected to the baseline radiosensitivity characteristics for low LET radiation (reflected in the intrinsic ability to repair the DNA radiation damage and the $\alpha/\beta$ ratio term in the extensively used linear-quadratic model\footnote{The linear quadratic model is based on the overall effect being proportional to $\alpha d + \beta d^2$, where $d$ is the dose. The biological effect is regulated by the $\alpha/\beta$ ratio.} of radiation effect).

\section{Radiobiological influences in hadrontherapy}
\label{App:Intro_Motivation_Hadrontherapy}
Physicists have made major contributions to radiobiological modelling. Initial power law models proved too inflexible in biological systems and were replaced by simple target theory models linked to Poisson statistics. Presently, the linear-quadratic (LQ) model is used to guide many clinical decisions. It is based on the original finding of Douglas Lea (a pupil of Rutherford) that the yield of significant (lethal-type) chromosomal perturbations varied with dose and dose-squared. The LQ can also be derived from spatial track considerations, repair kinetics and chromosomal perturbation effects, all of which seem to reflect a biological tendency. Correlations of fractionation experiments with transformed LQ equations showed that the $\alpha/\beta$ ratio and dose per fraction dominated the outcomes. Application of limit theory was used to derive the concept of Biological Effective Dose - BED (Barendsen, Fowler, Dale), which is the notional dose required for a specified bioeffect is given in ultra-small fractions. It provides a maximum ceiling, like the speed of light, below which all other dose fractionation, dose rate and even high LET treatments can be compared, or equated in terms of their bioeffectiveness. 

 
Also the BED-equivalent of other cancer treatments, (surgery chemotherapy, molecular approaches) can be assessed. For combinations of different forms of radiation treatment the respective BED's can be simply added, since BED units of dose are essentially logarithmic.
The general equation for bioeffectiveness is given by:

\begin{equation}
BED=nd(RBE_{max} + RBE_{min}^2  d/(\alpha/\beta)),
\label{Equ_App:Intro_BED}
\end{equation}
where $n$ is the number of treatments, $d$ the dose of each treatment, $RBE_{max}$ is the ratio by which the alpha parameter increases with LET and $RBE_{min}$ the ratio by which the square root of beta increases, and the alpha/beta  ratio is obtained from the alpha and beta obtained using control LET values. Essentially, the two RBE limits scale the effective RBE at any particular dose since alpha-related cell kill predominates at low dose, with the proportion of beta-related cell kill dominating at high doses. To estimate these RBE limits at any particular LET value,  further models are used. There are many existing models, of varying degrees of complexity, but with different degrees of success  when fitted to data examples. All remain to be verified using comprehensive data sets, as could be generated at BioLEIR. 
%
%

At the present time there are several models that attempt to link LET with RBE, each with varying degrees of success. Some are based on microdosimetry (and involve track structure considerations), while others concentrate more on phenomenological aspects. They also differ in their mathematical complexity and in the various assumptions made. In all cases they are hindered by a relative lack of extensive data gathered for a specific purpose: the classical experiments in this area have been done as long as 50 or more years ago, using a small variety of cell/tissue types, in different laboratories, with different beam characteristics and unique local uncertainties. The experiments were laborious, and were successfully designed to show overall phenomena, but not to produce accurate parameters for modelling purposes. Consequently, there are gaps in our knowledge and attempts at parametrisation should be considered to be tentative.  The beautiful symmetry of the LET-RBE relationships are amenable to further study in order to elucidate mechanism and provide robust predictions. 

There are around 5-6 ways in which LET can be specified (see appendix~\ref{App:EA_ControlIrradiator} for more information). It will be important to determine which are the best definition for which precise situation. Such data sets where not only the physics is studied closely, but also the biological effects are studied with high throughput to improve the data confidence limits, could only be achievable in a well-organised and resourced laboratory with ample beam time and resources to do the necessary experimental work. The BioLEIR facility is capable of doing this, and more - as will be discussed below.

To make further and definitive progress, the ideal situation would be for there to be a central laboratory which had sufficient beam time to test a range of ions at varying energies in appropriate panels of cells with known bio-characteristics, and which can work in harmony with the existing more clinically orientated treatment centres. Experiments would need to be designed to confer eventual greater predictive accuracy and inform the models, and so require larger experiments than previously performed or envisaged, with goodness of fit tests to determine which models are best suited for eventual clinical applications. The potential applications would not only be in the field of hadrontherapy, but also in nuclear medicine, radiation protection etc.

As has been very well demonstrated in particle physics following the discovery of the electron, a blend of theoretical and experimental work is required to make further and faster progress: such an approach is now required to solve remaining hadrontherapy issues such as in RBE. The CERN model is particularly important in terms of cooperative working, strategic thinking, and in a setting where impartial and collective decisions may be taken. The example of the fact that over the first 6 weeks of energy ramp up in the search for the Higgs Boson, the entire laws of particle physics were "re-discovered" and to a greater level of accuracy than in the past, needs to be applied to biology and encourage further mechanistic experiments with time.

\section{Radiobiological influences in nuclear medicine}
\label{App:Intro_Motivation_NuclearMedicine}
A significant number of cancer treatments involve the administration of unsealed radionuclides which are introduced into the body either orally or intravenously.  The first such therapies (developed in the 1940's) usually involved iodine-131 or phosphorus-32 (both of which are $\beta$ emitters), typically for the respective treatment of thyroid cancer or metastatic tumour deposits in bone. 

In recent years a wider range of radionuclides have emerged as potential therapeutic candidates, usually attached to pharmaceuticals that can target the specific type of cancer requiring eradication. Of particular interest in such "targeted radiotherapy" (or radiopharmaceutical therapy) are relatively short-lived $\beta$-emitters such as (amongst others) astatine-211, bismuth-213 and actinium-235. If such radionuclides can be properly targeted to the tumour then the associated high LET ensures that local radiation damage to malignant cells is likely to be lethal, whilst the limited particle range ($10-100\mu\,m$) helps limit collateral damage to adjacent normal tissues. In principle such therapy has the capacity to act as a "magic-bullet" treatment, but 100\% targeting of all tumour cells is rarely achievable and neither can the uptake of activity be entirely uniform, both these possibilities resulting in dose "cold-spots" which can allow some tumour cells to escape sterilisation. Additionally, the radionuclide activities required for successful radionuclide therapy are very much greater than those associated with nuclear medicine imaging, meaning that a fuller understanding of the complex dosimetry issues is of fundamental importance in such therapy. 

Of equal relevance, tumour cure and organ toxicity end-points require that calculated physical doses be used to predict biological response, for which an understanding of the associated radiobiology is also essential. Dose-response relationships derived from external beam radiotherapy cannot be directly applied to radio-pharmaceutical therapy because dose delivery in the latter occurs over a greater time-period and with a different temporal pattern (exponentially decaying versus multiple short burst). However, radiobiological models are already available to help make inter-comparisons between differing dose-delivery patterns, for example by converting absorbed dose to a biologically effective dose (BED), a parameter which provides more reliable prediction of tumour outcomes and toxicity in all types of radiotherapy.
   
All of the radionuclides used in targeted radiotherapy deliver the majority of their radiation dose via particulate interactions and, especially in the case of $\beta$-emitters, microscopic local effects (expressed as LET) or more macroscopic effects (expressed as RBE). Both LET and RBE play a very significant role in determining overall biological impact. Consideration of RBE and its variation with radionuclide type, LET along tracks, absorbed dose, dose-rate and targeted tissue is thus of fundamental importance in this type of therapy, but remains an under-researched area. The theoretical models which allow examination of such inter-dependencies are yet to be properly validated. Accurate radiobiological dosimetry of the $\beta$ emitters used in radionuclide therapy thus remains a major challenge at present and, without better quantitative understanding, wider use of this potentially valuable mode of treatment is still some way off.
  
Carefully-designed laboratory experiments, primarily designed to explore the nature of RBE effects when using accelerated charged particle radiation, have the potential to provide additional information of direct relevance to radionuclide therapy. This is a consideration which opens additional avenues for other useful and clinically-relevant research and is achievable on the proposed modifications to the CERN LEIR synchrotron.

\section{Diagnostics and detectors}
\label{App:Intro_Motivation_Diagnosticcs}
It is self-evident that a test bed facility such as BioLEIR at CERN would need to incorporate state-of-the-art and innovative diagnostic instruments, coupled with extensive Monte Carlo FLUKA and GEANT based simulations. To study beam ballistics extensively and correlate beam track separations with LET and the way RBE may change with depth in biological material depending on incident energy will be challenging and can only be attempted when the physics issues have been settled. Remote detection of Bragg peak positions and so LET distributions in matter will be essential as well as accurate prediction and measurement of LET and inter-track distances. Nuclear fragmentation studies, and remote gamma ray and positron detection will be essential. Advanced and novel detector systems for these purposes could be designed and developed in such a facility. Again such research would generate improvements not only in hadrontherapy, but also in nuclear medicine. 

The use of positively charged hadrons for imaging purposes is attractive in terms of reduction of dose to the matter/tissue being studied, improved imaging resolution due to the reduction in scatter compared with photons. Further research is necessary with different ionic beams, especially to determine material composition by remote analysis. Potential applications are in diagnostic and therapeutic medicine, security screening etc. and will depend on fundamental interactions with matter and their accurate prediction, with three dimensional reconstructions. There are opportunities for improving the diagnosis and serial assessment of a multitude of medical conditions if tissue resolution is improved and provided the radiation exposure itself is sufficiently low. In this last respect charged particle beam imaging would confer lower tissue doses than conventional low keV X-rays since the Bragg peaks would occur outside the tissue itself (in the detector) and the LET of the intra-patient portion of the beam will be low. In this respect it is important to remember that the lower the energy, the greater is the LET, RBE and carcinogenic potential of any diagnostic medical beam.

\section{IT support challenges}
\label{App:Intro_Motivation_IT}
To plan, record and analyse the extensive simulation and experimental data that would be associated with the BioLEIR project, the data handling and storage facilities of CERN would furnish open access results to participating Universities, that will also hold data on the detailed biological characteristics of the cell systems and can perform further molecular based investigations remotely. The huge expansion in knowledge of the molecular basis of life and its signalling processes inevitably require large data storage capabilities for interactions between experiments using particle physics and biological interactions, if important mechanistic correlations are to be found. The use of Geant4 and Fluka and their inter-comparisons will be essential for observed versus expected studies for beam ballistics and cell survival studies, as well as in the simulation of 3D LET, dose and eventually RBE maps using a variety of models in order to determine which fits the data best and in which circumstances (low or high dose, ionic Z number, energy etc.). Geant4 is presently incorporating radiation transport interactions with DNA, rather than in water. In order for the simulations to be as precise as possible, further extensive studies are required to determine the most appropriate parameters, especially at lower energies, for photons, protons and the relevant ions. The BioLEIR laboratory would be capable of contributing to these challenging objectives in collaboration with other laboratories worldwide.

\section{Conclusions}
\label{App:Intro_Conclusions}
In the various brainstorming meetings held under the auspices of CERN, and in the medical applications International Scientific Committee, extensive support has been given to prioritise the BioLEIR project, with an emphasis on the fundamentals of RBE as being of high priority. The knowledge gathered would have a positive future impact on clinical decision making and so for quality of life in many clinical situations including cancer, and also along a wide diagnostic front.

The entire project requires advanced accelerator and detector development, particle physics expertise and extensive computing power, all of which exist in the domain of CERN.

%% file: Appendices/appendix_AlternativeUseCases.tex
\chapter{Potential interest of availability of light ions to the CERN research programme and the existing CERN users' community}
\label{App:Intro_AlternativeUseCases}

{\centering \textit{Marco Silari}

}

In addition to the necessity of light ions for biomedical research, light ions could also be of interest to the CERN research programme and the CERN user community at large. A first, non-comprehensive feedback and interest has been gauged.

\section{Accelerator ion programmes}
\label{Sec:AppAlternative_NonBioResearch_Accelerators}
Although nothing is presently foreseen in the future LHC programme, there has been some interest in the LHC community in, for example, Argon-Argon, proton-Argon and proton-Nitrogen collisions. The availability of deuterons is of potential interest for electron-deuteron collisions at the LHeC. Different types of light ions might also be of interest to physics programmes currently underway in the North Area, for example at NA61. There is interest to study atmospheric cosmic rays for which ions from Iron to Nitrogen could be of interest. Similarly, the community studying the effect of radiation on electronics used in space exploration could be interested in Iron ions. Should BioLEIR become a project, availability of light ions could spark research project requests from the CERN physics community and such interest could be investigated in more detail in a later project phase.

\section{Detector development and testing}
\label{Sec:AppAlternative_NonBioResearch_Detectors}
Advances in radiation therapy technologies call for an increasing reliability and accuracy in dosimetric techniques. In hadrontherapy, it also implies the verification of the position of the Bragg peak in a humanoid phantom. GEMPix is a novel detector developed at CERN by coupling two CERN technologies, the Gas Electron Multiplier (GEM) and the Timepix ASIC. It offers great potential for medical dosimetry to provide a 3D image of a particle beam and for improved Quality Assurance (QA) procedures. First measurements performed with the GEMPix at CNAO (Pavia, Italy) showed promising results, allowing off-line reconstruction of the 3D energy deposition by a Carbon ion beam in water. Following this test, CERN just started the development of an integrated system consisting of a water phantom (a professional tool used for daily QA in radiation therapy, donated by a hospital), the GEMPix, a PTW ion chamber as reference monitor for normalisation of the GEMPix data, and associated hardware and software for phantom control, data acquisition and on-line data analysis. BioLEIR would be the ideal source of a range of ions for tests in clinical conditions and possible industrialisation and commercialisation of this system.

GEMPix also has the potential to act as a "tracking microdosimeter" as it provides the possibility of measuring the track structure of ionizing radiation down to the scale of tens of nanometres. Studies of this potential application are about to start and BioLEIR would be the ideal supplier of beams for testing GEMPix, as well as for testing classical instruments like Tissue Equivalent Proportional Counters (TEPCs) as microdosimeters with clinical ion beams.

In general, BioLEIR would offer the possibility of extensive tests with various ion beams of detector technologies that are currently not yet in routine use at particle therapy facilities; such as GEMs, scintillating crystals, diamond detectors, the Timepix etc., allowing thorough investigation of their characteristics and response to ion beams of present and future clinical relevance. A preliminary test of a triple-GEM as real-time beam monitor in hadrontherapy developed at CERN has been performed at CNAO, for measuring the beam spot dimensions and the homogeneity of the scanned irradiation field, which are daily QA tasks commonly performed using radiochromic films. 

A very low material budget GEM-based detector that is read out optically is also under development at CERN. It is ideally suited for beam monitoring and dosimetry for hadrontherapy, providing real time beam position and shape, possibly even during treatment. Tests are being carried out around Europe and in collaboration with clinics and industries.

Light ion beams from BioLEIR would expedite the developments of active target TPCs for the study of low energy nuclear reactions, easing the benchmark and qualification of the detector prototypes. TPCs are suitable devices also for the study of low energy nuclear interaction with fixed targets (see section~\ref{Sec:AppAlternative_NonBioResearch_MC}). Accessible ion beam would be beneficial in tailoring the TPCs for specific measurements.

It should also be mentioned that with the slow but steady development of proton therapy, the issue of the unwanted dose received by the patient outside the treatment volume becomes increasingly relevant. Quantification of the global advantage of net benefit offered by protons over conventional radiation therapy with photon and electron beams is of particular interest. Measurements of secondary neutron dose generated in the water phantom by different light ions and energies are important data that will be needed in the future to decide on the best treatment approach (e.g. for paediatric patients). BioLEIR will also be beneficial for testing detectors for this type of application.

\section{Nuclear physics studies and benchmarking of Monte Carlo codes}
\label{Sec:AppAlternative_NonBioResearch_MC}
The availability of light ion beams is a fundamental condition for the investigation of the fragmentation process at intermediate/low energies, which plays a critical role in hadrontherapy. The possibility of carrying out experiments aimed at the measurement of double differential cross sections, such as fragment spectra produced at different angles in ion-nucleus interactions at tens/hundreds MeV/u, where data are still scarce, would be highly beneficial for benchmarking predictive Monte Carlo codes such as FLUKA and GEANT in use for clinical purposes.
Furthermore, double differential cross section measurements of neutron production from Oxygen beams on Hydrogen-rich targets would also be valuable to integrate and possibly correct the information presently available in evaluated cross section libraries.

\section{Radiation damage to electronics and microelectronics studies}
\label{Sec:AppAlternative_NonBioResearch_R2E}
The user community interested in radiation effects on electronics would clearly profit from a several hundred MeV/u light ion beam for both practical and radiation hardness assurance reasons. Standard heavy ion ground test facilities in Europe (e.g. UCL in Louvain-la-Neuve, Belgium, and RADEF in Jyv\"askyl\"a, Finland) provide heavy ion beams of 10 MeV/u, resulting in ranges from 90 to 200\,$\mu$m(Si) depending on the ion species. For a variety of components and effects, such penetration values may not be large enough to reach the component's sensitive volume, or at least to cover it with a sufficiently constant LET value. This is particularly true for flip-chip layouts, in which the active part of the component can in general not be made accessible to the ion beams through un-packaging or thinning. Therefore, the radiation effects community could benefit from high energy heavy ion beams such as those proposed at BioLEIR that would enable testing with ranges large enough to penetrate into the sensitive area even of complex components or boards. 

In addition, weaknesses related to the LET metric for Single Event Effect (SEE) probability representation have manifested with the scaling of the components down to sensitive volume sizes comparable to the radial structure of the ionization track. Likewise, hardened devices for which nuclear interactions are expected to dominate the error rate in the space environment will also be affected differently by ions of similar LET values but different energies. In both cases, using relatively low energy ions to mimic the effects of the much more energetic Galactic Cosmic Ray (GCR) spectrum may lead to invalid prediction results. Therefore, testing in the more realistic several hundred MeV/u range in which the GCR flux peaks would clearly be an asset.

The CERN microelectronics section uses a facility in Louvaine-la-Neuve (Belgium) for checking integrated circuits for sensitivity to Single Event Upset (SEU). This is of growing interest for many chips used in CERN's experiments. If ions with well characterised LET and similar fluxes were available at BioLEIR there would be significant interest from the extended community of microelectronics groups active in HEP.

\section{Shielding data and induced radioactivity in accelerator and beamline components}
\label{Sec:AppAlternative_NonBioResearch_RP}
There is a growing number of proton and light ion therapy facilities in operation or at the planning stage worldwide. The light ion presently used in clinical practice is $^{12}$C. However, different ions (anything in the range from protons to Oxygen) may be considered in the future. A synchrotron generating $^{12}$C ion beams can, in principle, accelerate any ion with charge-to-mass ratio Q/A=1/2. For example, the Heidelberg Ion Therapy Center (HIT) can already produce $\alpha$-particles and Oxygen ions. There are appropriate computational tools for the design of the radiation shielding, ranging from simple analytical line-of-sight models to Monte Carlo codes, but there is a need for shielding data (source terms, attenuation lengths, thin and think targets neutron yields) for light ion facilities, in particular for ions other than $^{12}$C. The properties of various shielding materials to the secondary neutron spectra produced by light ions can be measured at BioLEIR.

Although there is a growing trend in using active beam scanning for dose delivery, cyclotrons using an Energy Selection System (ESS) are widespread. These accelerators and passive scattering techniques imply higher activation of accelerator and beamline components and radiation exposure of personnel in contrast to synchrotron-based facilities. Thus there is a need for more data on activation induced by light ions, which can be measured at BioLEIR.

\section{Real-time tumour tracking and dose delivery in hadrontherapy}
\label{Sec:AppAlternative_NonBioResearch_MovingTarget}
Dose delivery in hadrontherapy follows more or less the same protocol as conventional (electron and photon) radiation therapy that is as follows: identification of the target volume with proper imaging techniques, referencing the tumour location with markers in the patient's body, patient set-up with Computer Tomography (CT) images in the treatment position, radiation delivery with an isocentric gantry to optimise the distribution of dose to the tumour while maintaining the dose to normal tissue within tolerable limits. An integrated dose delivery system having the capability of real-time tumour imaging/tracking, spot scanning with real-time feedback on the beam delivery system in order to continuously adjust the beam spot (in depth and X/Y position) on the target voxel to be irradiated (tumour) - properly integrated within the treatment planning to spare sensitive healthy tissues - would allow treating patients with much more accurate dose delivery and without the necessity of isocentric irradiation. The beam could come from a fixed direction (either horizontal or vertical), and it would no longer be important that the patient is in the same position (supine) used to acquire the CT images for the treatment planning, as the position of the target to be irradiated is continuously known in real-time and the beam continuously adjusted to hit it. Such a system would have two benefits: 1) improved dose delivery in hadrontherapy, and thus expected improvement in the ratio dose-to-tumour versus dose-to-healthy tissue; 2) a substantial decrease in cost of particle therapy, as an isocentric gantry would no longer be needed. This would be even more beneficial for Carbon ion therapy (and any light ion in general), as commercial isocentric gantries do not exist.

The above requires developments and innovation in: real-time tumour imagining techniques, tumour tracking, feedback on beam control for tracking the beam to the tumour on-line, using spot scanning beam delivery and dose monitoring. BioLEIR could be used for developing and testing part or all of these technologies using CERN expertise in accelerator technologies, beam control, imaging detectors and dosimetry.

%% file: Appendices/appendix_EA.tex
\chapter{Control irradiator choice for the BioLEIR project}
\label{App:EA_ControlIrradiator}

{\centering \textit{Bleddyn Jones}

}

The study of the relative biological effectiveness (RBE) of different radiation qualities is important, since the RBE concept is used in the dose prescription process of hadrontherapy. RBE is defined as:

\begin{equation}
RBE = \frac{Dose~of~reference~radiation~(low~LET)}{Dose~of~hadronic~test-radiation~(higher~LET)},
\label{Equ_App:EA_A_RBE}
\end{equation}

\noindent such that the bioeffect of a specific class of radiation is scaled to a given reference radiation. The numerator dose becomes the number by which RBE is scaled from values of 1 upwards, the denominator being expected to be always less than one. 

RBE is dependent on the linear energy transfer (LET) of the radiation, with an initial rise of RBE, followed by a reduction at higher LET values due to overkill or cell killing inefficiency. The turnover point depends on the nuclear charge $Z$ of the ion species, but apparently not on the cell type. RBE is also inversely related to dose due to differences in cellular radio-sensitivities to the low LET control radiation.

The RBE concept was included in the dose prescription system for fast neutron therapy and continues to be used in proton and ion beam therapy. It is also potentially relevant in nuclear medicine and some forms of brachytherapy, but has not yet been used explicitly in these clinical applications.

The control radiation has varied considerably in experiments performed to determine RBE: examples include use of low-voltage X-rays (100-200\,keV peak) and orthovoltage (200-350\,keV peak), along with variable degrees of filtration to remove the lowest keV~X-rays of the full X-ray spectrum. Other examples include 137-Caesium or 60-Co $\gamma$-rays, megavoltage electrons and more rarely megavoltage X-rays in the clinical range of 4-10\,MeV.  The lower the energy of any beam, the higher the LET: lower energies result in a more localised energy release and absorption - a higher LET - and result in more clustered DNA damage. This, in turn, will increase RBE up to a saturation level beyond which RBE falls at even higher LET values. Conversely, the local LET and RBE are reduced for higher energies of a particle or photon beam  because the ionisations are not as spatially clustered. The above process applies to X-rays and $\gamma$-rays, electrons, protons, and all ions.

Since the control irradiation supplies the numerator dose for the RBE estimation, it must exceed the control hadron irradiation dose for the same bioeffect, for the RBE to be greater than one. For cases where the control  irradiation has an LET that is greater than the LET for the hadronic test irradiation, the RBE will be less than one.

Choosing a control irradiator that has a higher LET than that of the photons/X-rays used in hospital clinics, will give a lower estimate of RBE than would be the case in the hospital setting for the same cellular system.

\section{Choice of the ideal control irradiator}
\label{App:EA_A_Ideal}
As demonstrated in the section above, the choice of the control irradiation has an important effect on the measured RBE ratio. For example, Hall and Gaccia~\cite{Hall_Giaccia:2012} quote typical LET values  of 2\,keV/$\mu$m for 250\,keV irradiation (but the filtration is unspecified), 0.2\,keV/$\mu$m for 60-Co, 4.7\,keV/$\mu$m for 10\,MeV protons and 0.5\,keV/$\mu$m for 150\,MeV protons, while megavoltage photons are not mentioned.

More comprehensive data can be found in the ICRU report 16~\cite{ICRU:1970}, yet it must be considered that even these data are incomplete. The ICRU data suggest by extrapolation that megavoltage photons in the clinical range of 4-8\,MeV would have an LET value of around 0.2\,keV/$\mu$m. A useful collation of data in~\cite{Andrews:1978} shows an average LET of 0.5\,keV/$\mu$m for 3\,MeV photons from 60-Co irradiation, but stopping powers (or LET) of 0.18\,keV/$\mu$m for 2\,MeV electrons, which also corresponds to the minimal energy loss for any charged particle traversing matter, representing the lower energy transfer limit for relative biological effect studies.

The average Compton electrons produced from 60-Cobalt $\gamma$-rays have an LET of 0.26\,keV/$\mu$m which in turn means that a slightly lower value around 0.2\,keV/$\mu$m should be expected as energy transfer from clinical range megavoltage photons.

Further information is contained in some publications \cite{SpadingerPalcic:1992} where an LET range of 5.5\,-\,6\,keV/$\mu$m is quoted for 25\,keV X-rays  and 1.7\,keV/$\mu$m for 122\,keV  X-rays. The average value is obtained with 250\,keV X-rays using 0.35\,mm Cu and 0.4\,mm Sn filtration. For 11\,MeV electrons, an LET value of 0.2\,keV/$\mu$m has been measured, which is likely to closely resemble megavoltage photons in the clinical range. 

From the above data and depending on the experiment that shall be performed, it seems reasonable to use a megavoltage control irradiation of either electrons of photons with an LET value of around 0.2\,keV/$\mu$m as would be achieved with a clinical photon or electron beam. The added value of having a megavoltage photon beam would be for comparative combined ballistic and radiobiological studies in humanoid phantoms between charged particles and clinically relevant photon energies.

Below we quote some examples of the unsuitability of orthovoltage or low-voltage X-rays beams as control irradiation:
\begin{enumerate}
\item For the highly influential collated proton RBE data of Paganetti \textit{et al.}~\cite{Paganetti:2002}, 14 low-voltage electron or orthovoltage X-ray irradiations had an RBE less than 1 and only one such experiment showed an RBE just above 1. The probability that these results are due to chance or experimental error is low: use of the sign test provides the probability of there being no difference as being p=0.00061; a highly significant result.

\item The literature contains a report by Amols \textit{et. al.} 1986~\cite{Amols:1986} who measured cell survival data in DLD-1 human tumour cells. Their results demonstrate a statistically significant RBE difference between orthovoltage and megavoltage irradiation (p=0.001). A small difference is also measured in RBE between megavoltage photons and megavoltage electrons, but the difference is not statistically significant (p=0.25). All biological, dosimetric, and microdosimetric data were obtained under nearly identical geometric conditions. The measurements were performed on only a single cell line.
\end{enumerate}

\section{Discussion}
\label{App:EA_A_Discussion}

The literature contains many contributions which concentrate on the influence of photon energies in the low-keV range on RBE. However, these papers contribute mainly to the field of radiation protection, because this class of X-rays are used extensively in routine diagnostic medicine. Such work has confirmed that RBE varies inversely with photon and particle energy (see for example the review by Hill~\cite{Hill:2004}). 

RBE should be taken into account in clinical radiotherapy using positively charged particle beams, where RBE effects can be significant even in the case of protons: papers by~\cite{Britten:2013, Belli:2000, Marshall:2016} and others are examples of such work, containing well documented experimental detail. There is presently a potential shift of opinion to use variable RBE values rather than the constant values previously assumed in proton therapy - in order not to (a) under-dose certain radiosensitive tumour classes and (b) over-dose late-reacting normal tissues of clinical importance.

The urgent need to clarify RBEs in critical experimental normal tissues exposed to proton and other ionic beams requires careful choice of control irradiation if the purpose of such studies is to provide information that will be useful for ultimate clinical purposes.

Some potential forms of control irradiation can easily be eliminated: orthovoltage X-rays, even though inexpensive and convenient, should be avoided, since they lead to under-estimating RBE.

60-Cobalt units also have some significant disadvantages. These include the changing dose rate with time, since it is advisable to use the same exposure duration (with differences of only 1-2 minutes at the most) for the control and test irradiations, due to ongoing repair of sub-lethal radiation damage. Longer treatments have slightly lower bioeffectiveness and this influences the estimated RBE ratio. The need to compensate for radio-active decay on a weekly or monthly basis may sometimes be forgotten or wrongly applied. The LET can be lower than for megavoltage photons and, drift of dose rate with time, and modification of source treatment distances to compensate for radiation decay with accompanying changes in depth dose distributions, must be accounted for. There are also issues associated with radiation protection of staff, shielding requirements, as well as source replacements that are required with time, are expensive, and have their own difficulties.

Some laboratories do have a 137-Caesium irradiator, which is more compact, and delivers monoenergetic photons close to 0.5\,MeV, which should have an LET close to that of Cobalt beams. Again, radioactive decay has to be compensated for and dose rate effects may accrue with time, although this can be overcome by using shorter source to target distances. Such a unit can be fitted into the size occupied by a large double refrigerator without additional shielding. 
 
Megavoltage electrons are a reasonable choice for the control irradiation in RBE studies. However, electron dosimetry, although much improved in recent years, has limitations due to the electron's higher scattering especially for small field sizes, as well as the energy-dependent relationship of electron build-up with depth, which must be taken into account in any laboratory experiment. Electron fields are essentially spread-out Bragg peaks, often with lower RBEs than one in megavoltage electron plateau regions, but there remains uncertainty as to their higher RBE values with fall-off of dose and energy towards the end of their range. It must not be forgotten that Auger electrons (from low energy radioactive emissions) can have very high RBEs. However, a broad electron beam operating at say 8 -- 10\,MeV would give a satisfactory, uniform dose for radiobiological experiments, yet could not be used for beam ballistics combined with radiobiology comparisons.

There is interest in some clinical particle therapy centres across Europe, including international laboratories like CERN, to address the clinically important RBE issue. For there to be comparable results, it is very important that the control irradiation be standardised. The best standard to choose, despite the additional financial and spatial expense, would undoubtedly be megavoltage X-rays from a linear accelerator operating at typical energies used in clinics. It is already the case that national radiation standards laboratories such as NPL, Teddington, UK, contain a standard hospital linear accelerator for dosimetry comparisons and maintenance of standards. Likewise, laboratories that aim to contribute useful RBE data should use a control megavoltage linear accelerator with optional use of 4 -- 10\,MeV photons or electrons for their studies. A single 137-Caesium source could also be useful.

Control irradiations that are performed at another geographical site, would cause difficult logistical problems for visiting scientists and the results could be viewed with scepticism owing to inevitable time displacements, travel disruption and lack of overall control.

Interestingly, since hospital linear accelerators are normally replaced every ten years, it is relatively easy to obtain one which can be refurbished for experimental use. These would have a gantry for variable beam angles and collimation down to small field sizes.

\section{Conclusions}
\label{App:EA_A_Conclusion}
We conclude that a 137-Caesium source in a vertical configuration should be located in hall~150. A megavoltage electron linac is also necessary as a horizontal beam.

It would be of great interest to have access to a photon or electron linac onsite - as for example the GBAR 9\,MeV electron linac in building~393. It may be useful to investigate the practical feasibility of parasitic use of this existing facility at CERN.

%% file: FrontBackMaterial/10_backmatter.tex
\chapter*{Acknowledgements\markboth{Acknowledgements}{Acknowledgements}}
\addcontentsline{toc}{chapter}{Acknowledgements}
We appreciate the kind welcome we have received and invaluable input we have gained from our contacts at various clinical and non-clinical biomedical centres, most notably: Eleanor Blakely (LBL) on the Brookhaven Radiobiology facility and requirements; Boris Vojnovic (Oxford University); Peter Ursch\"utz and Thomas Schreiner from MedAustron; Marco Pullia from CNAO; Thomas Haberer, Andreas Peters and colleagues from HIT; Marko Schippers and Martin Grossmann from PSI; Denis Dauvergne and colleagues from FranceHadron.

We acknowledge the fruitful collaboration with Roger Barlow (University of Huddersfield) and his PhD students.

We wish to thank Christian Joram, Etiennette Auffray, Leszek Ropelewski, Florian Maximilian,  Eraldo Oliveri, Filippo Resnati, Francesco Cerutti, Michael Campbell, Markus Brugger and Ruben Garcia for providing input from the CERN user community. We further acknowledge valuable input from our colleagues Silvia Grau and Suitbert Ramberger on several technical aspects.

Daniel Abler, Manjit Dosanjh and Adriano Garonna wish to acknowledge the PARTNER Network (Project ID: 215840) and ULICE (Project ID: 228436) that have provided support for earlier work on which this study has been building. 

We wish to thank the quick and untiring support of the Overleaf online collaboration platform team (www.overleaf.com) - most notably LianTze Lim, John Lees-Miller and John Hammersley.

Finally, we thank the CERN publishing team and CREB to support us in realizing this report in an unusually short time frame.